%% file: reactor13.tex
\documentclass{article}
\usepackage{graphicx}
\usepackage{amsmath}
\usepackage{amssymb}
\usepackage{epsfig}
\include{header}                   
\begin{document}
\pagenumbering{roman}
\include{titlepage}
\include{executive}                
\tableofcontents
\pagenumbering{arabic}
\setcounter{page}{0}
\include{intro} 
\include{phyopp} 
\include{superbeams} 
\include{theory} 
\include{baseline_lum} 
\include{chooz} 
\include{paloverde} 
\include{kamland} 
\include{detector} 
\include{calibration} 
\include{background} 
\include{systematics} 
\include{systematics2} 
\include{sites} 
\include{other} 
\include{tunnel} 
\include{safety} 
\include{outreach} 
\newpage
\appendix
\include{brazil} %
\include{chooz2} 
\include{daya} %
\include{diablo} 
\include{illinois}
\include{kash} 
\include{krproj} 

\include{bib}
\end{document}

%% file: header.tex
\newcommand{\quq}{\theta_{13}}

\newcommand{\nme}{\nu_\mu \rightarrow \nu_e}

\newcommand{\neeb}{\bar{\nu}_e \rightarrow \bar{\nu}_e}

\newcommand{\dmsq}{\Delta m^2}
\newcommand{\dmsqat}{\Delta m^2_{atm}}
\newcommand{\mxang}{\sin^2(2\theta)}
\newcommand{\mxangg}{\sin^2(2\theta_{13})}

\newcommand{\nue}{$\nu_{e} $}
\newcommand{\numu}{$\nu_{\mu} \:$}
\newcommand{\nutau}{$\nu_{\tau} \:$}

\def\gsim{\:\raisebox{-0.5ex}{$\stackrel{\textstyle>}{\sim}$}\:}

%% file: titlepage.tex
\title{
\huge {WHITE PAPER REPORT on \\ Using Nuclear Reactors to 
Search for a value of $\theta_{13}$ \\ January 2004}
\nopagebreak
}

\author{
K.~Anderson$^{12}$ \and
J.C.~Anjos$^{7}$ \and
D.~Ayres$^3$ \and
J.~Beacom$^{14}$ \and
I. Bediaga$^{7}$ \and
A.~de~Bellefon$^{15}$ \and
B.E.~Berger$^{4}$ \and
S.~Bilenky$^{30}$ \and
E.~Blucher$^{12}$ \and
T.~Bolton$^{18}$
% ...
}
\maketitle

\begin{enumerate}
\item {\bf Academia}  Sinica, Taiwan
\item University of { \bf Alabama}
\item { \bf Argonne} National Laboratory
\item Lawrence { \bf Berkeley} National Lab (Nuclear Science)
\item Lawrence { \bf Berkeley} National Lab (Physics)
\item University of { \bf Bologna} and INFN-Bologna, Italy
\item Centro { \bf Brasileiro} de Pesquisas Físicas
\item University of { \bf California}, Berkeley
\item { \bf California} Institute of Technology
\item Universidade Estadual de { \bf Campinas}
\item { \bf Catholic} University of Rio de Janero
\item University of { \bf Chicago}
\item { \bf Columbia} University
\item { \bf Fermi} National Accelerator Laboratory
\item College de { \bf France}
\item { \bf Illinois} Institute of Technology
\item {\bf IHEP} Beijing
\item { \bf Kansas} State University
\item { \bf Kurchatov} Institute
\item { \bf Louisiana} State University
\item { \bf Max-Plank-Institut} f\"ur Kernphysik (Heidelberg)
\item { \bf Max-Plank-Institut} f\"ur Physik (Munich)
\item University of { \bf Michigan}
\item University of { \bf Minnesota} at Crookston
\item { \bf Niigata} University
\item { \bf Northwestern} University
\item University of { \bf Pittsburgh}
\item { \bf Rikkyo} University
\item { \bf Saclay}
\item {\bf SISSA} Trieste
\item State University of New York, { \bf Stony Brook}
\item University of { \bf Tennessee}
\item { \bf Technical} University Munich
\item University of { \bf Texas} at Austin
\item { \bf Tohoku} University
\item { \bf Tokyo} Institute of Technology
\item { \bf Tokyo} Metropolitan University
\item INFN { \bf Trieste}
\item {\bf Virginia } Tech
\item University of { \bf Washington}
\end{enumerate}

\vspace*{0.2in}
\hrule

\par  This document is available at\\
http://www.hep.anl.gov/minos/reactor13/white.html \\
or by writing: \\
Maury Goodman\\
HEP 362\\
Argonne Illinois 60439\\

%% file: executive.tex

{\noindent\bf\Large Executive Summary}
\\

\vspace*{12pt}

{\noindent\bf\large Purpose of this White Paper}\\
There has been superb progress in understanding the neutrino sector
of elementary
particle physics in the past few years.  It is now widely recognized that
the possibility exists for a rich program of measuring CP violation and
matter effects in future accelerator $\nu$ experiments, which has led
to intense efforts to consider new programs at neutrino superbeams,
off-axis detectors, neutrino factories and beta beams.  However,
the possibility of measuring CP violation can be fulfilled only
if the value of the neutrino mixing parameter
$\quq$ is such that $\mxangg$ greater than or equal to on the order
of 0.01.  The authors
of this
white paper are an International Working Group of physicists who
believe that a timely new experiment at a 
nuclear reactor sensitive to 
the neutrino mixing parameter $\quq$ 
in this range has a great opportunity for an exciting discovery,
a non-zero value to $\quq$.
This would be a compelling next step of this program.
We are studying possible new reactor
 experiments at a variety of sites around the world,
and we have collaborated  to prepare this document to advocate
this idea and describe some of the issues that
are involved.

\vspace*{12pt}

{{\noindent\bf\large Purpose of the Experiment}\\
In the presently accepted paradigm to describe the neutrino sector, there are
three mixing angles.  One is measured by solar neutrinos and the KamLAND
experiment,  one by atmospheric neutrinos and the long-baseline
accelerator projects.  Both angles are large, unlike mixing angles
among quarks.  The third angle, $\quq$, has not yet been measured to
be nonzero but
has been constrained to be small in comparison by 
the CHOOZ reactor neutrino experiment.

The basic feature of a new reactor
 experiment is to search for energy dependent $\bar{\nu}_e$
disappearance using two (or more) detectors, to see $\neeb$ disappearance.
The detectors need to be
located underground in order to reduce backgrounds from cosmic rays and
cosmic ray induced spallation products.  The detectors
need to be designed identically in order to reduce systematic errors to 1\% or less.
Control of the relative detector efficiency, fiducial volume, and good energy
calibration are needed.

\par
A measurement of or stringent
limit on $\quq$ would be
crucial as part of a long term 
program to measure CP parameters at accelerators, even though a
 reactor $\neeb$ disappearance experiment does not measure any CP 
violating parameter.  
A
sufficient value of $\quq$ measured in a reactor experiment would strongly
motivate the investment required for a new round of accelerator $\nu$
experiments.   A reactor experiment's unambiguous measurement of $\quq$
would also strongly
support accelerator
measurements by helping to resolve degeneracies and ambiguities.
The combination of measurements from reactors and neutrino results from
accelerators will allow early probes for CP violation without the necessity
of long running at accelerators with anti-neutrino beams.

\newpage

{{\noindent\bf\large Anticipated Sensitivity}\\
The best current limit on $\quq$ comes from the CHOOZ experiment and is a function
of $\dmsqat$, which has been measured using atmospheric neutrinos
by Super-Kamiokande and others.  The latest reported value of $\dmsqat$ from 
Super-Kamiokande is 
$1.2~<~\dmsqat~<~3.0~\times~10^{-3}$eV$^2$ 
with a best fit reported at 2.0.  The
CHOOZ limits for $\dmsqat$ of 2.6 and 2.0 $\times~10^{-3}$eV$^2$
are $\mxangg~<~$
0.14 and 0.20.  Global fits using the solar data limit the value for small
$\dmsqat$ to less than 0.12.  
In order to improve on the CHOOZ experiment, a new
reactor experiment needs  more statistics and better control of 
systematic errors.  The relative sensitivity at low $\dmsqat$ can be
improved by locating the far detector further than 1 km.  Increased statistics can
be achieved by running longer, using a larger detector, and judicious choice
of a nuclear reactor.  The dominant systematic errors in an absolute measurement
of the reactor neutrino flux, such as cross-sections, flux uncertainties,
and the absolute target volume, will be largely eliminated in a relative 
measurement with two or multiple detectors.  
Good understanding of the relative detector response and the backgrounds
is required for a precise relative measurement of the reactor neutrino 
flux and spectrum.  
Experiments are being considered which
increase the luminosity from the CHOOZ value of 12~t~GW~y (ton-Gigawatt-years)
to 
400~t~GW~y or more.  This will allow a mixing angle sensitivity of
$\mxangg~>~0.01$.  
For example, 400 t~GW~y would be obtained with a 10 (40) ton far
detector, and a 14 (3.5) GW reactor in 3 years.
One design consideration of the new experiment is the possibility for
upgrades to achieve even greater luminosity and sensitivity.
The ability to phase upgrades to achieve a luminosity
of 8000~t~GW~y is being considered.  

\vspace*{12pt}

{{\noindent\bf\large Major Challenges}\\
A new reactor experiment will build on the experience of several
previous reactor experiments, such as CHOOZ, Palo Verde and KamLAND
(described in Section~4 of this white paper).  
These experiments had different goals, mostly being designed for signals due
to large mixing.  
Important experience on calibration, control of systematic errors and
the reduction of background has also been obtained by the 
Super-Kamiokande, SNO and Borexino collaborations.
\par A next-generation reactor experiment will be designed to make a precision 
measurement of the reactor electron anti-neutrino survival
probability at different distances from the reactor and search for subdominant
oscillation effects associated with the mass splitting of the m$_1$ and
m$_3$ mass eigenstates.  A measurement at the
$\mathcal O$(1\%) level will require careful control of possible systematic
errors.  Most of the technical requirements of this experiment are well
understood but the details of the detector design still need to be
optimized.  Some of the open questions under consideration are the following:
liquid scintillator loaded with 0.1\% of gadolinium has been 
used in the past, but there are concerns regarding its stability in
solution and possible attenuation length degradation which need to be fully
understood.  
If movable detectors are chosen, there must be 
confidence that moving the detector does not introduce additional
time-dependent effects.  The use of a second detector will certainly
help to control many systematic errors, but also will present a
challenge in maintaining a known 
relative calibration over time.  Another challenge is reduction of 
cosmic ray associated backgrounds such as
neutrons and $^9$Li spallation and their accurate estimation.  The reduction of gamma ray background
is also important because it will affect the ability to
reduce the threshold to below 1~MeV.
These and other design issues are discussed in Sections~5-8 of this white paper.

\vspace*{12pt}

{{\noindent\bf\large Experimental Prospects}\\
The International Working Group
has held two workshops (April 30-May 1, 2003 at the University of
Alabama and October 9-11, 2003 at Technical University of Munich)
and we are planning a third one (March 20-22, 2004 at Niigata
University.)  
During the past year, the International Working Group has identified 
a large number of reactors 
as possible sites for a new experiment.  
Many of these sites are discussed in Section~9, and a few
are described in more detail in seven Appendices.
These include the Angra reactor in Brazil;
the possibility of a new experiment
at CHOOZ, called Double-CHOOZ (or CH$\quq \quq$Z); Daya Bay near Hong
Kong in China, Diablo Canyon in 
California; a reactor in Illinois; the reactor complex at
Kashiwazaki in Japan, and the Krasnoyarsk reactor underground
at Zhelezhnogorsk
in Russia.  

\par It is not the role of this document to provide a cost estimate or
schedule for
any of the experiments which will be proposed.  But it is appropriate to
try to set the scale of the endeavor
in order to compare to other kinds of 
initiatives in neutrino physics.
A two-detector system as described in this document seems
to cost in the range \$5M to \$15M.  The civil construction costs to place
these detectors underground will be very site dependent and require a
detailed engineering cost estimate as described in Section~11.  
Estimates are in the range of several tens of millions of dollars, depending
on site condition and tunnel length.
Since 
reactors with an underground site already exist, such as those
at CHOOZ and
Krasnoyarsk, there is  a strong  incentive to
consider those sites for the earliest experiment, though there may be
physics trade-offs which must be considered.
Some of the envisioned reactor experiments might start taking data
in 2007-2008.  First results could be achieved as early as 2009.

\par None of these efforts has yet resulted in a proposal to
a funding agency, but site specific proposals and R\&D proposals
will be submitted during
2004.  This white paper is a step in that direction.
Given the importance of the measurement of $\quq$ and the 
enthusiasm of the proponents, we are hopeful that two or more of 
these experiments will move forward on a favorable time scale.


%% file: intro.tex

\section{Introduction}

The discovery of neutrino oscillations is a direct indication of
physics beyond the Standard Model and it provides a unique new 
window to explore physics at high mass scale including unification, 
flavor dynamics, and extra dimensions. The smallness of neutrino 
masses and the large lepton flavor violation associated with neutrino
mixing are both fundamental properties that give insights
into modifications of current theories. Other possibilities that
may reveal themselves in the neutrino sector include extra
``sterile'' neutrinos, CP violation in the neutrino mixing matrix,
and CPT violation associated with the neutrino mass hierarchy. 
Since neutrino oscillations have now been established, the next step 
is to map out the parameters associated with neutrino masses and 
mixings. The experimental program to accomplish this goal will 
require a wide range of experiments using neutrinos from solar, 
atmospheric, reactor, and accelerator sources.  Due to the relations 
between these various measurements, it will be important for the 
world-wide community to set up a structured program to work through 
the experimental measurements in a coherent and logical manner.

\begin{figure}[t]
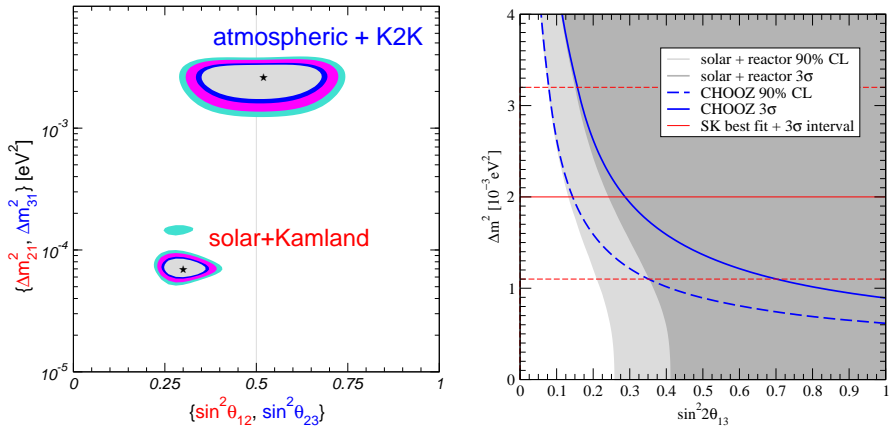

\begin{center}
\includegraphics[width=0.48\textwidth] {sol+atm.eps} \quad
\includegraphics[width=0.45\textwidth] {th13current.eps} 
\caption{Status of neutrino oscillation parameters from a combined analysis of
current global data~\cite{Maltoni:2003da}. Left panel: allowed regions of
solar ($\Delta m^2_{21}, \sin^2\theta_{12}$) and atmospheric ($\Delta
m^2_{31}, \sin^2\theta_{23}$) parameters at 90\%, 95\%, 99\% and $3\sigma$ CL.
Right panel: upper bound on $\sin^2 2\theta_{13}$ from the CHOOZ experiment at
90\% (dashed) and $3\sigma$ (solid) CL for 1 DOF as a function of $\Delta
m^2_{31}$. The light (dark) shaded region is excluded from CHOOZ + solar +
KamLAND data at 90\% ($3\sigma$) CL for 1 DOF. The horizontal lines indicate
the current best fit value and the $3\sigma$ allowed regions for $\Delta
m^2_{31}$.}\label{fig:osc_status}
\end{center}
\end{figure}

The existing experimental results fit rather nicely into a picture with three
massive neutrinos, which corresponds to the simplest scenario for three
generations (for recent global analyses see, {\it e.g.},
References~\cite{Maltoni:2003da,fogli}). Neutrino oscillations 
then involve two mass-squared differences 
    ($\Delta m^2_{21}$ and $\Delta m^2_{32}$, where $\Delta m^2_{ij} = 
    m(\nu_i)^2 - m(\nu_j)^2$), three mixing angles
($\theta_{12}$, $\theta_{23}$, and $\theta_{13}$), and a
CP-violating phase ($\delta$). The present status of these parameters is
summarized in Figure~\ref{fig:osc_status}. 
Atmospheric neutrino data~\cite{skatm} and the first results from the K2K
    long-baseline accelerator experiment~\cite{k2k}
 determine $|\Delta m^2_{32}| 
= (2^{+1.2}_{-0.9})
\times10^{-3}$~eV$^{2}$ (errors at $3\sigma$) and
$\theta_{23}\approx45^{\circ}$ \cite{skatm,fogli}, whereas 
most solar
data~\cite{solar,bib:SNOsalt}, combined with the results from the
KamLAND reactor experiment~\cite{Eguchi:2003prl}, 
lead to $\Delta m_{21}^{2} =
(6.9^{+2.6}_{-1.5}) \times 10^{-5}$~eV$^2$ and $\sin^2\theta_{12} = 0.3
^{+0.09}_{-0.07}$ at $3\sigma$ \cite{Maltoni:2003da}.

%
%
The neutrino sector may contain more than three neutrinos by including mixing
to sterile neutrinos (for example to account for the LSND
\cite{Aguilar:2001ty} anomaly), but in these cases the mixing matrix most
likely factors to a good approximation into a $(3\times3)$ submatrix with the
parameters given above. The investigation of oscillations involving sterile
neutrinos will demand measurements such as MiniBooNE as well as improved
disappearance measurements at high $\Delta m^{2}$. 

The current experimental situation can thus be summarized by two more or less
decoupled oscillations governed by the ``atmospheric'' and ``solar/reactor''
quadratic mass splittings $\Delta m^2_{atm}=\Delta m^2_{13}$ and $\Delta
m^2_{sol}=\Delta m^2_{21}$, respectively, and the corresponding mixing angles
$\theta_{12} = \theta_{sol}$ and $\theta_{23} = \theta_{atm}$, which turned
out to be surprisingly large. This leads in the future to two equally
important experimental directions: The first task is to improve the knowledge
of the above (leading) oscillation parameters and to make precision
measurements.  Conceptually at least equally important is the fact that three
flavors imply also three flavor oscillations and thus one further mixing
angle, $\theta_{13}$ as well as a CP violating phase $\delta$ \footnote{Note
that Majorana neutrinos imply also two further CP violating phases, but these
do not enter into neutrino oscillations.}.  The CP phase $\delta$ is a very
interesting, but so far a completely unknown, parameter. The fact the LMA
solution has been confirmed means that $\delta$ is in principle accessible in
future experiments if $\theta_{13}$ is not too small.  In many models of
neutrino masses the see-saw mechanism leads to connections of the leptonic CP
phase $\delta$ to the CP phases in the heavy Majorana sector and thus to
leptogenesis, one of the best known mechanism to explain the baryon asymmetry
of the universe (see e.g. \cite{Branco:2002xf}). Neutrino masses may therefore
explain a second indication for physics beyond the Standard Model, since the
observed baryon asymmetry cannot be generated from CP violation in the
Standard Model with massless neutrinos. Future neutrino experiments aim
therefore indirectly at another key question in physics, namely what causes
the baryon asymmetry in the Universe. 

The mixing angle $\theta_{13}$, the parameter relevant for three flavor
effects in neutrino oscillations, is known to be small from the CHOOZ
\cite{bib:chooz,Apollonio:1999ae} and also from the Palo Verde experiment
\cite{Boehm:1999gk}. The current bound from global data is summarized in the
right hand panel of Figure~\ref{fig:osc_status}. It depends somewhat on the true
value of the atmospheric mass squared difference, since the bound from the
CHOOZ experiment gets rather weak for $\Delta m^2_{31} \lesssim 2\times
10^{-3}$~eV$^2$. However, in that region an additional constraint on
$\theta_{13}$ from global solar neutrino data becomes
important~\cite{Maltoni:2003da}. At the current best fit value of $\Delta
m^2_{31} = 2\times 10^{-3}$~eV$^2$ we have the bounds at 90\% ($3\sigma$) CL
for 1 DOF
\begin{equation}
  \sin^2 2 \theta_{13} \le 0.16~(0.25) \,,\quad
  \sin^2\theta_{13} \le 0.053~(0.066) \,,\quad
  \theta_{13} \le 10.8^\circ~(14.9^\circ) \,.
\end{equation}
Genuine three flavor oscillation effects occur only for a 
finite value of $\theta_{13}$ and establishing a finite value 
of $\theta_{13}$ is therefore one of the next milestones 
in neutrino physics. Leptonic CP violation is also a three 
flavor effect, but it can only be tested if $\theta_{13}$ 
is finite. There is thus a very strong motivation to establish 
a finite value of $\theta_{13}$ in order to aim in the long run 
at a measurement of leptonic CP violation (see e.g. 
\cite{Lindner:2002pj,Freund:2001ui}). 

Future measurements of $\bar\nu_{e}$ disappearance using a two 
detector reactor experiment and long-baseline 
$\nu_{\mu}\rightarrow\nu_{e}$ and
$\bar{\nu}_{\mu}\rightarrow \bar{\nu}_{e}$ experiments will be
crucial in determining $\theta_{13}$, the sign of $\Delta m_{32}^{2}$, 
and the CP phase $\delta$. 
If $\theta _{13}\gtrsim 0.01$, the design of experiments to
measure the sign of $\Delta m_{32}^{2}$ and the CP phase $\delta$
become straight forward extensions of current experiments. For 
this reason, there is general agreement that a $\theta_{13}$ 
measurement should be the prime goal of the next round of experiments.
On the theoretical side these experiments could test if the small
value of $\theta_{13}$ could be a numerical coincidence or if e.g.
some symmetry argument is required to explain a tiny value.



Future measurements of $\theta_{13}$ are possible using reactor
neutrinos and accelerator neutrino beams. As will be shown in
subsequent sections, reactor measurements have the
property of determining $\theta_{13}$ without the ambiguities
associated with matter effects and CP violation. In addition, the
needed detector for an initial reactor measurement is small
($\lesssim50$ tons) and the construction of a neutrino beam is not
necessary. For this reason, a precision reactor experiment could
lead the way in  establishing the future
oscillation program by setting the scale of the $\theta_{13}$
mixing angle. The previous most accurate measurements were  by
 the CHOOZ and Palo Verde experiments where a single
detector was placed about 1 km from the reactor. Future reactor
experiments using two detectors ($\sim 50$ tons) at near (100 -
200 m) and far (1 - 2 km) locations will have significantly
improved sensitivity for $\theta_{13}$ down to the 0.01 level.
With $\theta_{13}$ determined, measurements of
$\nu_{\mu}\rightarrow\nu_{e}$ and $\bar{\nu}_{\mu
}\rightarrow\bar{\nu}_{e}$ oscillations using accelerator neutrino
beams impinging on large detectors at long baselines will improve
the knowledge of $\theta_{13}$ and also allow access to 
matter or CP violation effects. For the field to exploit the
physics opportunities available for neutrino oscillation
measurements, it is clear that a suite of experiments including
both reactor and long-baseline accelerator measurements will be
necessary. 

In addition to the general physics arguments, there are two factors that
lend
urgency to this initiative.  Our studies indicate that a reactor experiment
to
measure $\sin^2 2\theta_{13}$ to the level of 0.01 could be done at
significantly less cost and on a more rapid time scale than an accelerator
long-baseline neutrino experiment with comparable sensitivity.  This
conclusion
is influenced by several recent developments, including the High Energy
Physics Roadmap for the future, February 2002 \cite{bib:roadmap}, the prioritizations
made in ``High Energy Physics Facilities on the DOE Office of Science
Twenty-Year Roadmap" issued by the U.S. Department of Energy in March, 2003
\cite{bib:hepfac}
and
the ``Facilities for the Future of Science: a Twenty-year Outlook" issued by
the
Office of Science of the DOE in November, 2003~\cite{bib:facilities}.  In particular, the latter
document envisions that a high-intensity neutrino beam is more than 15 years
in
the future.  For comparison with an off-axis long baseline experiment,
we
use cost and time estimates based on current work for the
Fermilab
proposal P929, the NuMI Off-Axis Experiment.  We emphasize that a
new reactor experiment does
not
reduce the motivation for the latter experiment; it obtains information
complementary to that obtained by the reactor $\theta_{13}$ experiment, such
as
the mass hierarchy between $m(\nu_3)$ and $m(\nu_1)$.  Instead, we would
envision that, since the reactor experiment can be performed more quickly,
its
findings concerning $\theta_{13}$ will provide very important guidance for
the
long baseline program.


In this White Paper, we outline the capabilities of next generation 
reactor experiments and summarize the design considerations that 
groups are considering in developing this program. 
The International Working Group on $\quq$ is sharing ideas on 
how to best design a new reactor experiment, and one goal of this 
White Paper is to document the present status of our understanding 
of these issues.  

In the next section, we discuss in more detail the physics
opportunities and the motivation for a new reactor experiment. 
The following Section~\ref{sec:location} deals with the optimal
baseline, luminosity scaling and the impact of systematic errors.
Previous reactor experiments are described in Section~\ref{sec:previous},
and in Section~\ref{sec:design} we present some 
thoughts about the general layout of the detector, a multi-layered 
volume of scintillator designed to define the fiducial volume well, 
and also carefully control other potential systematic errors.  
In Section~\ref{sec:calibration}, the calibration requirements for 
the detector are reviewed.  Section~\ref{sec:bac} considers the 
issues of backgrounds and how they affect the required overburden.  
Depths that provide an overburden of 400 mwe to 1100 mwe are desirable.  
The goal of carefully minimizing systematic errors is qualitatively different
than has been required of neutrino experiments at reactors in the past.
We are confident that the two detector concept will provide lower systematic
errors than have been previously achieved, but the ultimate limit
on achievable systematic error has yet to be identified.  A discussion
of a variety of systematic errors is presented in Section~\ref{sec:sys}.  
%
%
Characteristics of a large number of sites are reviewed in 
Section~\ref{sec:site} and some more detailed experimental site plans for 
seven
of the possible locations are included in the Appendices to
this document.
Next we discuss other physics that can be done in Section~\ref{sec:other}.
Depending on the site, the costs of a new reactor experiment will potentially
be dominated by the civil construction of a shaft or tunnel.   Those civil
engineering issues are reviewed in Section~\ref{sec:tun}.  
Safety issues are discussed in Section~\ref{sec:safetysection}. 
Section~\ref{sec:outreach} is finally devoted to outreach and educational issues. 
The appendices contain further details of potential sites. 

%% file: phyopp.tex
\section{Physics Opportunities and Motivation}
\label{sec:physopp}

\subsection{Road Map for Future Neutrino Oscillation Measurements}

There is now a world-wide experimental program underway to measure
the parameters associated with neutrino oscillations. 
The
current experiments include K2K that measures $\nu_\mu$ disappearance
over a 250 km baseline from KEK to SK. Another experiment is
MiniBooNE that is searching for $\nme$
appearance signal  in
the LSND $\dmsq$ region from 0.2 to 1 eV$^2$. 
Upcoming longer-baseline ($\sim 700$ km)
experiments are NuMI/MINOS at Fermilab and CNGS at CERN that will
study $\nu_{\mu}$ oscillations in the atmospheric $\Delta m^{2}$
region. Groups in all the world-wide regions are also pursuing
sites and experiments for a precision reactor experiment using
detectors with fiducial volumes of 5 to 50 ton. 
Several near-term new long-baseline experiments
are planned which will use off-axis beams including the approved
J-PARC (previously called JHF) to Super-K (22.5 kton) 
experiment and the developing NuMI
off-axis experiment (50 kton detector). Following these
experiments, the next stage might be neutrino superbeam
experiments with even longer baselines that could possibly be
combined with large proton decay detectors. Four such projects
under consideration are: (i)~BNL with an AGS upgrade,
(ii)~Fermilab with a proton driver upgrade, (iii)~J-PARC (phase II),
and (iv)~a CERN Superconducting Proton LINAC experiment. Future
neutrino factories, using a muon storage ring, will provide the
ultimate in sensitivity and precision in oscillation measurements.

It is clear that developments in the field will dictate how the
community should proceed through these studies. As stated
previously, the size of $\theta_{13}$ is the small parameter that
sets the scale for further studies in a three neutrino scenario.
It is also clear that the final resolution of the LSND anomaly by
MiniBooNE could significantly affect the direction for new
investigations. To bring this information together in a coherent
way, we present a roadmap for neutrino oscillations which tries to
point out the  relations 
between the various measurements:

\begin{itemize}
\item  Stage 0: The Current Program

\begin{itemize}
\item  There are improved measurements of $\Delta m_{12}^{2}$ (5-10\%) by
solar neutrino and the KamLAND experiments.

\item  NuMI, CNGS, and K2K experiments check the atmospheric
oscillation phenomenology and measure $\Delta m_{23}^{2}$to $\sim
10$\%.

\item  MiniBooNE makes a definitive check of the LSND effect and measures the
associated $\Delta m^{2}$ if the effect is confirmed.
\end{itemize}

\item  Stage 1: Measurement or tight constraint's on the $\theta_{13}$ 
angle\footnote{The combination of all these experiments may give the first
indications of matter and CP violation effects.}\newline

\begin{itemize}
\item  The NuMI/MINOS on-axis experiment probes $\sin^{2}2\theta_{13}>0.06$ at
90\% CL.

\item  Two-detector, long-baseline reactor experiments probe $\sin^{2}%
2\theta_{13}>0.01$ at 90\% CL.

\item  The NuMI and J-PARC off-axis experiments with 20-50 kton
detectors investigate $\nu_{\mu}\rightarrow\nu_{e}$ transitions
for oscillation probabilities greater than 1\%.
\end{itemize}
\end{itemize}

\begin{itemize}
\item  Stage 2: Measurements of the sign of $\Delta m_{23}^{2}$ and CP violation using
superbeams and very large detectors (500 to 1000 kton)\newline (This is
feasible if $\sin^{2}2\theta_{13}>0.01$ and if $\delta$ is large enough.)

\begin{itemize}
\item  Measurements of $\nu_{\mu}\rightarrow\nu_{e}$ at several baselines need
to be combined with either precision reactor measurements of $\nu
_{e}\rightarrow\nu_{e}$ or with $\bar{\nu}_{\mu}\rightarrow\bar{\nu}_{e}$

\item  Increased neutrino beam rates are needed, especially for
the $\bar{\nu}_{\mu}$ running, which make high intensity proton
sources necessary.
\end{itemize}

\item  Stage 3: Measurements with a Neutrino Factory

\begin{itemize}
\item  New facilities probe a mix of $\overset{(-)}{\nu}_{\mu/e}\rightarrow\overset{(-)}{\nu
}_{e/\mu}$ transitions with sensitivities below the 0.001 level

\item  They also map out CP violation with precision for $\sin^{2}2\theta_{13}>0.001$.
\end{itemize}
\end{itemize}

\begin{figure}[ht]
\vspace*{0.5in}
\epsfxsize=5.0in\epsffile[0 97 529 673 ]{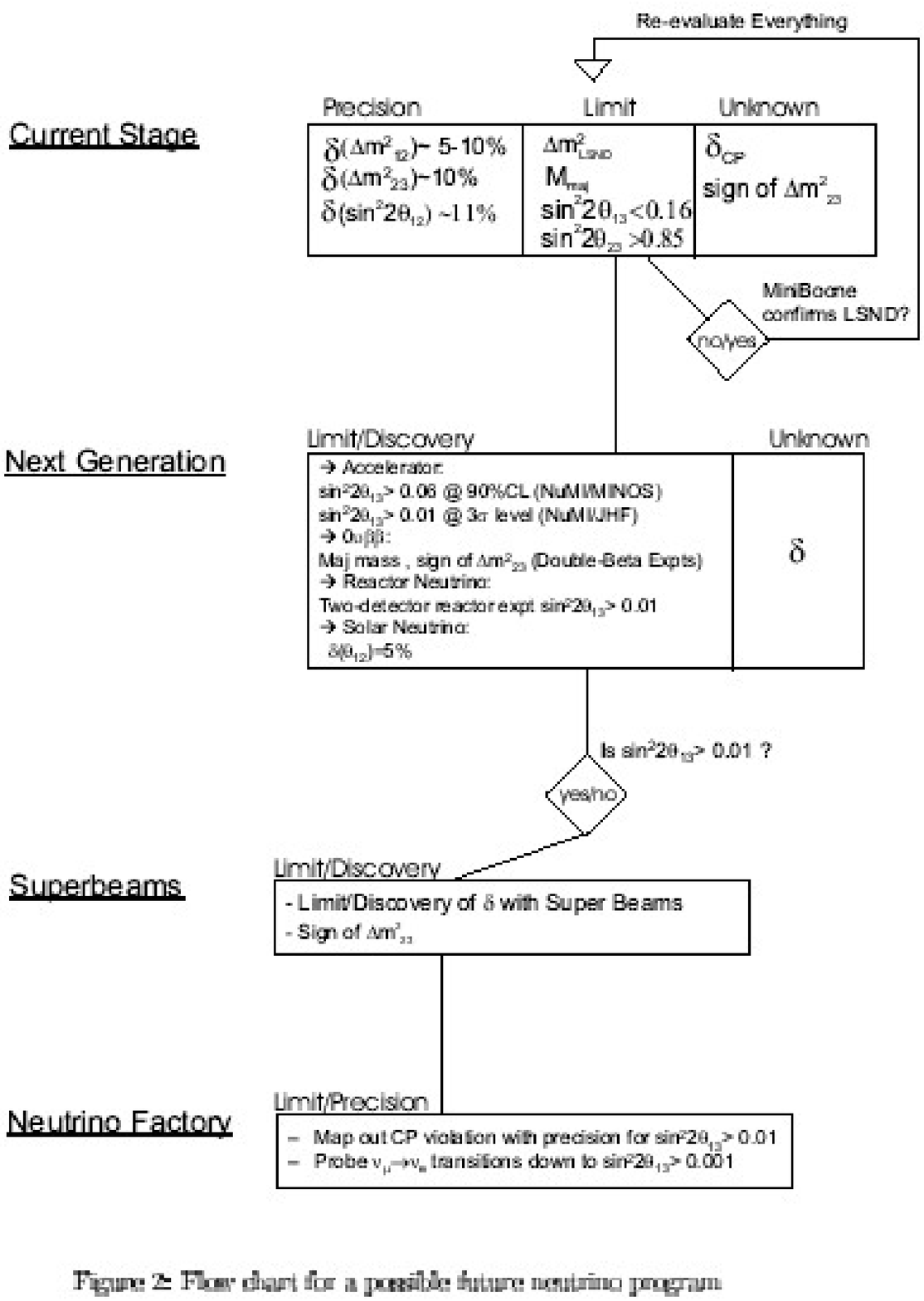}
\caption{Flow chart for a possible future neutrino program}
\label{fig:flow}
\end{figure}

A flow chart with these ideas is shown in Figure~\ref{fig:flow}.

\subsection{Where do reactor oscillation experiments fit in?}

\begin{figure}
\begin{center}
\includegraphics[width=12cm]{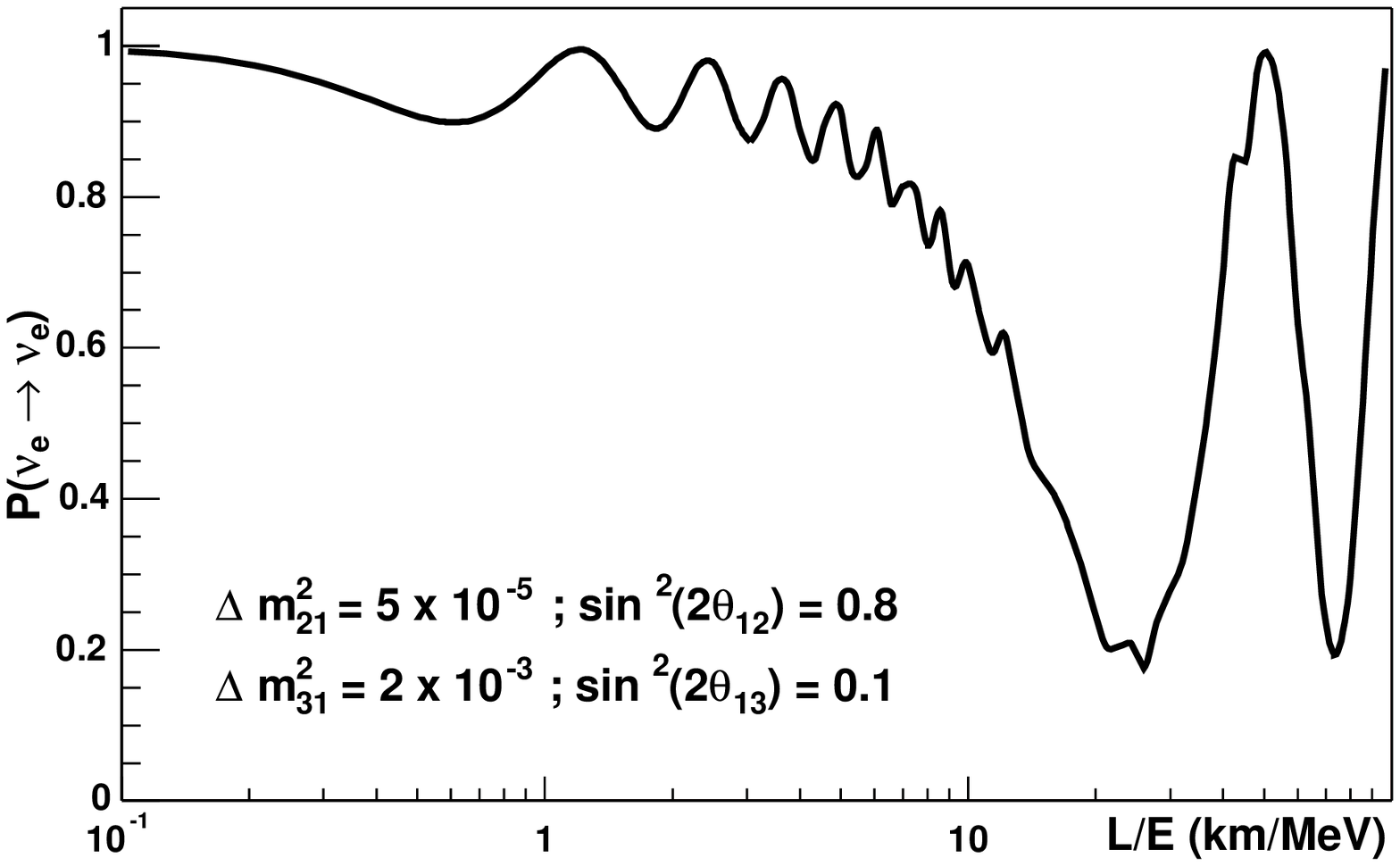}
\caption{Probability of $\nu_e$ disappearance versus $\mathrm{L/E}$ for 
$\quq$ at its current upper limit}
\label{fig:P}
\end{center}
\end{figure}

Any oscillation effect in $\bar{\nu}_e$ survival is governed, assuming
three flavor mixing, by the equation
\begin{equation}
\label{eq:p}
P(\bar{\nu}_e \rightarrow \bar{\nu}_e) \cong 1
- \sin^2 2 \theta_{13} \sin^2(\frac{\Delta m^2_{atm} L}{4E})  
- \cos^4 \quq \sin^2 2 \theta_{12} 
\sin^2(\frac{\Delta m^2_{12} L}{4E}) .
\end{equation}
This equation is plotted in Figure~\ref{fig:P} as a function of $\mathrm{L/E}$ with the 
current best values for 
the $\Delta m^2$s and mixing angles ($\sin^2(2\theta_{13})$ is set to
the maximum value allowed by current limits).  
One can clearly see the 
two oscillations governed by the two $\Delta m^2$s.  Experimentally, 
a judicious choice of $\mathrm{L/E}$ should be able to distinguish one effect 
from the other.
The KamLAND experiment is the first reactor experiment to see oscillation
effects, by measuring a 40\% disappearance of
$\bar{\nu}_e$.  Given that the average baseline for KamLAND is 180 km, 
the detected deficit is 
presumably associated with the third ($\Delta m^2_{12}$) term in 
Equation~(\ref{eq:p}).

The current best limit on $\quq$ comes from a lack of observed oscillations
at CHOOZ and Palo Verde ($\sin^2 2 \theta_{13} < 0.20$ for
$\dmsqat~=~2.0 \times 10^{-3}$eV$^2$).  These experiments 
were at a baseline distance of about 1 km
and thus more sensitive to the second ($\Delta m^2_{atm}$) term of 
Equation~(\ref{eq:p}).  Those experiments could not have had greatly improved 
sensitivity
to $\quq$ because of uncertainties related to knowledge of the flux of 
neutrinos from the reactors.  They were designed to test
whether the atmospheric neutrino anomaly might have been due to
$\nme$ oscillations, and hence were searching for large oscillation effects.

Since the effective disappearance will be very small (see Figure~\ref{fig:P}),
any new experiment which is designed to look for non-zero values of 
$\quq$ would need to move beyond the previous systematic limitations.
This could be achieved by utilizing the following properties:
\begin{itemize}
\item two or more detectors to reduce uncertainties to the reactor flux
\item identical detectors to reduce systematic errors related to detector
acceptance
\item carefully controlled energy calibration
\item low backgrounds and/or reactor-off data
\end{itemize}
Note that
CP violation does not affect a disappearance experiment, and that the
short baseline distances involved in a reactor measurement of oscillations
at the atmospheric $\Delta m^2$ allow us to safely ignore matter effects.

\begin{figure}
[ptb]
\begin{center}
\includegraphics[
height=1.0in, width=5.0in ]{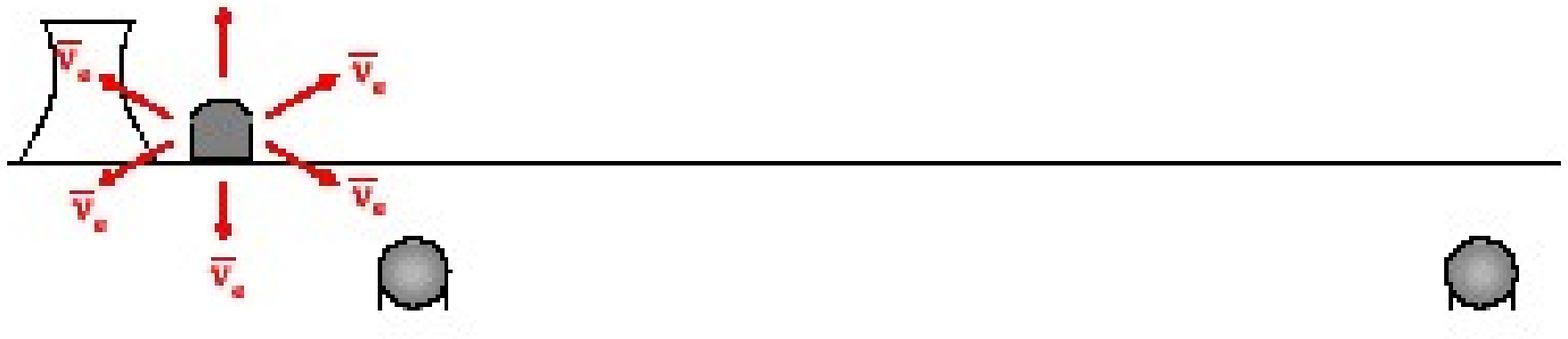} \caption{Schematic layout
of a two detector reactor neutrino oscillation experiment.
}%
\label{layout}
\end{center}
\end{figure}

\begin{figure}
[ptb]
\begin{center}
\includegraphics[
height=4.0in, width=4.0in ] {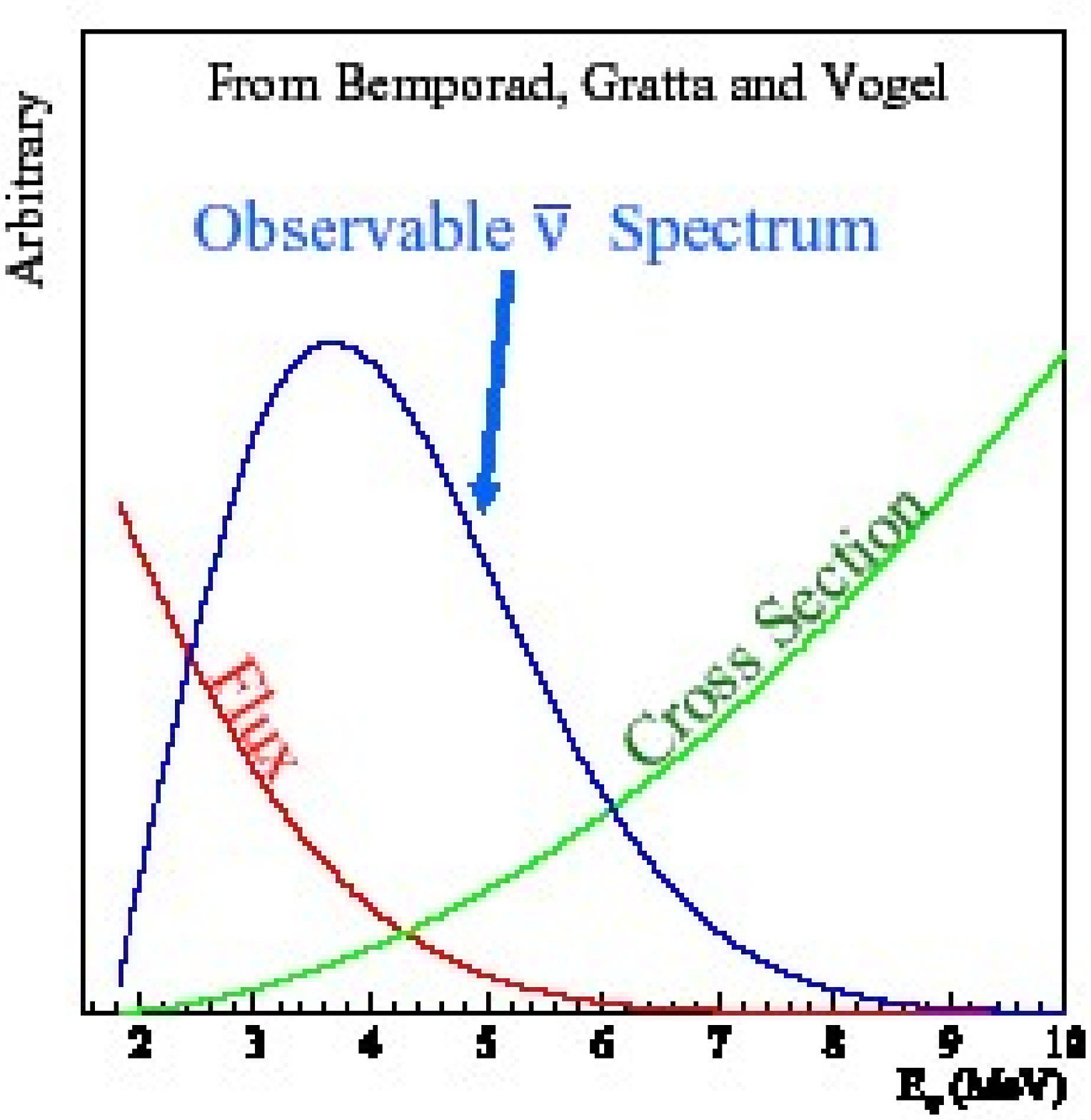}
\caption{Antineutrino flux, cross section, and relative event 
rate using a reactor source~\cite{bib:gratta}.}
\label{nuspec}
\end{center}
\end{figure}

A next generation reactor oscillation experiment would use at
least two detectors placed at various distances from a high power
reactor (Figure~\ref{layout}).  The reactor provides a high
intensity, isotropic source of neutrinos with a well-known spectrum
as shown in Figure~\ref{nuspec}.  The neutrino cross section for
this process is well known as
described in Reference~\cite{bib:beacom}.
Antineutrinos are detected through
the inverse-$\beta$ decay process followed by neutron capture.
\begin{align*}
\bar{\nu}_e + p \rightarrow e^{+} + n.
\end{align*}
The detector would most likely be composed of a vat of
scintillator oil viewed from its surface by an array of
photomultipliers. 
In order to reduce the background from
cosmic-ray spallation, the detectors will need to be underground
with at least 300 mwe of shielding. 
A detected event would correspond to a coincidence
signal of an electron and capture neutron. The incident neutrino
energy is directly related to the measured energy of the outgoing
electron. The search for oscillations would then involve comparing
the neutrino rate in the two detectors and looking for a
non-$1/r^2$ dependence.

As stated above, $\theta_{13}$ is a key parameter in developing
the future neutrino oscillation program. Reactor experiments offer
a straightforward and cost effective method to measure or
constrain the value of this parameter. The sensitivity of a two
detector experiment is comparable to that of the proposed initial
off-axis long-baseline experiment. Since a reactor experiment would
be much smaller and use an existing reactor neutrino source with a
well understood neutrino rate, the experiment should be able to be
done fairly quickly and at reduced costs. It is likely that an
early measurement of $\theta_{13}$ will be necessary before
the community invests a large amount of resources for a
full off-axis measurement. For the longer term, a reactor
experiment would be complementary to the off-axis experiments in
separating the measurement of $\theta_{13}$ from other physics
parameters associated with matter effects and CP violation. A
follow-up reactor experiment with much larger detectors at various
baselines will continue to be an important component of the
neutrino oscillation program.  

%% file: superbeams.tex
 
\def\GWth{\text{GW}_{\text{th}}} 
\def\eV{{\text{eV}}} 
\def\MeV{{\text{MeV}}} 
\def\GeV{{\text{GeV}}} 
\def\best{{\text{best}}} 
\def\sys{{\text{sys}}} 
\def\JPARC{{\text{JPARC}}} 
\def\reac{{\text{reac}}} 
\def\rCP{{\text{rCP}}} 
\def\tot{{\text{tot}}} 
\def\rate{{\text{rate}}} 
\def\sigDB{\sigma_{\text{DB}}} 
\def\sigDb{\sigma_{\text{Db}}} 
\def\sigdB{\sigma_{\text{dB}}} 
\def\sigdb{\sigma_{\text{db}}} 
\def\sigD{\sigma_{\text{D}}} 
\def\sigd{\sigma_{\text{d}}} 
\def\sigB{\sigma_{\text{B}}} 
\def\sigb{\sigma_{\text{b}}} 
\def\sigBG{\sigma_{\text{BG}}} 
\def\BG{{\text{BG}}}

\subsection{Reactor experiment as a clean laboratory for 
the $\theta_{13}$-measurement}

In this section, we demonstrate that a reactor measurement of
$\theta_{13}$ is a clean measurement which is free from 
any contamination, such as from effects 
of the other mixing parameters or from the Earth matter effect~\cite{bib:kash}. 
This key feature is one of the 
most important advantages of the reactor experiments. 
We use the standard notation of 
the lepton flavor mixing matrix:
\begin{eqnarray}
U=\left[
\begin{array}{ccc}
c_{12}c_{13} & s_{12}c_{13} &   s_{13}e^{-i\delta}\\
-s_{12}c_{23}-c_{12}s_{23}s_{13}e^{i\delta} &
c_{12}c_{23}-s_{12}s_{23}s_{13}e^{i\delta} & s_{23}c_{13}\\
s_{12}s_{23}-c_{12}c_{23}s_{13}e^{i\delta} &
-c_{12}s_{23}-s_{12}c_{23}s_{13}e^{i\delta} & c_{23}c_{13}\\
\end{array}
\right].
\label{MNSmatrix}
\end{eqnarray}

Due to the low neutrino energy of a few MeV, the reactor 
experiments are inherently disappearance experiments, i.e.,  
they can only measure the survival probability 
$P(\bar{\nu}_{e} \rightarrow \bar{\nu}_{e})$.
Unlike the case of the $\nu_e$ appearance 
probability, it is well-known that the survival probability does not
depend on the CP phase $\delta$ in arbitrary matter 
densities (for more details, see~\cite{bib:kash}). 
For reactor experiments, the matter effect is very small because 
the energy is quite low and the effect can be ignored to a 
good approximation. This can be seen by the comparison of the  matter 
and the vacuum effects
\begin{eqnarray}
\frac{a}{|\frac{\Delta m^2_{31}}{2E}|} = 3.4 \cdot 10^{-4}
\left(\frac{|\Delta m^2_{31}|}{2.5 \cdot 10^{-3}\,\mbox{eV}^2} \right)^{-1}
\left(\frac{E}{4\,\mbox{MeV}} \right)
\left(\frac{\rho}{2.8\,\mbox{g}\cdot\mbox{cm}^{-3}} \right)
\left(\frac{Y_e}{0.5} \right).
\end{eqnarray}
Here $E$ is the neutrino energy.
In addition, $a = \sqrt{2} G_F N_e$ denotes the index of refraction
in matter with the Fermi constant $G_F$ and the
electron number density $N_e$ in the Earth (which is related
to the Earth matter density $\rho$ by
$N_e = Y_e\rho/m_p$ with the proton fraction $Y_e$).

Since we know that the matter effect is negligible, we
immediately understand that the survival probability
is independent of the sign of $\Delta m^2_{31}$.
Therefore, one can use the vacuum probability formula 
for the analysis of a reactor measurement of $\theta_{13}$.
The expression for 
$P(\bar{\nu}_{e} \rightarrow \bar{\nu}_{e})$
in vacuum is given by~\cite{Minakata:2002jv}
\begin{eqnarray}
1-P(\bar{\nu}_{e} \rightarrow \bar{\nu}_{e}) 
& = & \sin^2{2 \theta_{13}}
\sin^2 \Delta_{31} \nonumber \\
& + & 1/2 \, c^2_{12}
\sin^2{2 \theta_{13}} \sin{2 \Delta_{31}}
\sin{ 2 \Delta_{21}}\nonumber \\
& + & c^4_{13} \sin^2{2 \theta_{12}} \sin^2{\Delta_{21}}  \nonumber \\
& +& c^2_{12} \sin^2{2 \theta_{13}} \cos{2 \Delta_{31}} \sin^2{\Delta_{21}},
\label{Pvac}
\end{eqnarray}
where $\Delta_{ij} \equiv \Delta m_{ij}^2 L/(4 E)$ and $c_{ij} = \cos \theta_{ij}$.
Defining the mass hierarchy parameter $\alpha$ 
as $\alpha \equiv \Delta m^2_{21}/\Delta m^2_{31}$, where 
$|\alpha| \simeq 0.03$, the second term in Equation~(\ref{Pvac}) is suppressed 
relative to the main depletion term (first term) by a factor of 
$| \alpha |\,  \sin^2{2 \theta_{13}} \leq 6 \times 10^{-3}$, the
fourth term even by a factor of $\alpha^2\,  \sin^2{2 \theta_{13}}$. Thus, we
can re-write Equation~(\ref{Pvac})  for $\Delta_{21} \ll 1$  (for 
the baselines considered) as
\begin{equation} 
1-P(\bar{\nu}_{e} \rightarrow \bar{\nu}_{e})  \quad \simeq \quad \sin^2 2 \theta_{13} \, \sin^2 \Delta_{31} +  \alpha^2 \, \Delta_{31}^2 \, c^4_{13} \, \sin^2 2 \theta_{12}.
\label{equ:PROBREACTOR} 
\end{equation} 
Though the second term on the right-hand side of this equation could be of the order of the first term for very large $| \alpha |$, it can be neglected for the first atmospheric oscillation maximum (where the first term is large) and $\sin^2 2 \theta_{13}$ larger than about $10^{-3}$.
Therefore, the disappearance probability can be well approximated by 
the two-flavor depletion term in vacuum, which is the first term in 
Equation~(\ref{Pvac}). 
Assuming that $|\Delta m^2_{31}|$ is accurately determined by 
a long-baseline $\nu_{\mu}$ disappearance measurement, 
the reactor experiments thus serve for a clean measurement of 
$\theta_{13}$ independent of other mixing parameters.

\subsection{Comparison to superbeams}

We have demonstrated in the last section that reactor measurements allow a 
degeneracy-free measurement of $\sin^2 2 \theta_{13}$. In order to qualitatively discuss the 
difference between reactor experiments and superbeams, we can compare 
the oscillation probabilities of the dominant oscillation channels. For 
the superbeams, one can expand the appearance probability $P_{\mu e}$ 
(or $P_{\bar{\mu} \bar{e}}$) in terms of the small mass hierarchy 
parameter $\alpha 
\equiv \Delta m_{21}^2/\Delta m_{31}^2$ and the small mixing angle $\sin 2\theta_{13}$ using 
the standard parameterization of the leptonic mixing matrix 
$U$ in Equation~(\ref{MNSmatrix}). As a first approximation for a qualitative discussion, 
one can use the vacuum formula  
from References~\cite{Freund:2001ui,Cervera:2000kp,Freund:2001pn} with the terms 
up to the second order (i.e., proportional 
to $\sin^2 2\theta_{13}$, $\sin 2\theta_{13} \cdot \alpha$, and 
$\alpha^2$): 
\begin{eqnarray}
P_{\mu e} & \simeq & | \sin 2\theta_{13} \, \sin \theta_{23} 
\sin {\Delta_{31}} e^{i(\Delta_{32}\pm\delta_{\mathrm{CP}})}
+ \cos \theta_{13} \, \cos \theta_{23} \sin 2\theta_{12} \sin {\Delta_{21}}|^2
\nonumber \\[0.25cm] 
& \approx & \sin^2 2\theta_{13} \, \sin^2 \theta_{23} 
\sin^2 {\Delta_{31}} \nonumber \\ 
& &\mp  \alpha\; \sin 2\theta_{13} \, \sin\delta_{\mathrm{CP}}  \, 
\cos\theta_{13} \sin 2\theta_{12} \sin 2\theta_{23} \, 
\Delta_{31} \sin^2{\Delta_{31}} \nonumber \\ 
& &+  \alpha\; \sin 2\theta_{13}  \, \cos\delta_{\mathrm{CP}} \, \cos\theta_{13} \sin 
2\theta_{12} \sin 2\theta_{23} \,
\Delta_{31} \cos {\Delta_{31}} \sin {\Delta_{31}} \nonumber  \\ 
& &+ \alpha^2 \, \cos^2 \theta_{23} \sin^2 2\theta_{12} \,\Delta^2_{31}. 
\label{equ:PROBVACUUM} 
\end{eqnarray} 
Here $\Delta_{ij} \equiv \Delta m_{ij}^2 L/(4 E) \equiv (m_i^2-m_j^2) 
L/(4E)$ and the sign of the second term refers to neutrinos (minus) or antineutrinos (plus). 
We have used the approximations that 
$\sin \Delta_{21} \simeq \alpha \Delta_{31} \ll 1 $
and that  $\Delta_{32} \simeq \Delta_{31}$.
 
For the reactor experiments, we have, up to the same order 
in $\sin 2\theta_{13}$ and $\alpha$, Equation~(\ref{equ:PROBREACTOR}). Comparing 
Equation~(\ref{equ:PROBREACTOR}) to Equation~(\ref{equ:PROBVACUUM}) 
clearly demonstrates 
that the superbeams are quite rich in physics
and much more complex to analyze.  
Depending on the true values of $\alpha$ and $\sin 2 
\theta_{13}$, each of the individual terms in 
Equation~(\ref{equ:PROBVACUUM}) obtains 
a relative weight. The result is 
then determined by the mutual interaction of the four terms in 
Equation~(\ref{equ:PROBVACUUM}) leading to multi-parameter correlations and degeneracies.  
Correlations and degeneracies are degenerate solutions in parameter space, where the correlations are  
connected solutions and the degeneracies are disconnected solutions in parameter space (at the chosen confidence level). For example, many of the degeneracy problems originate in the summation of 
the four terms in Equation~(\ref{equ:PROBVACUUM}) especially for 
large $\alpha$ and $\sin 2 \theta_{13}$, since changes of one 
parameter value can be often compensated by adjusting another one in 
a different term. This leads to the well-known $(\delta, 
\theta_{13})$~\cite{Burguet-Castell:2001ez}, $\mathrm{sign}(\Delta 
m_{31}^2)$~\cite{Minakata:2001qm}, and 
$(\theta_{23},\pi/2-\theta_{23})$~\cite{Fogli:1996pv} 
degeneracies, i.e., an overall ``eight-fold'' 
degeneracy~\cite{Barger:2001yr}, which can 
severely affect the potential of many experiment types~\cite{Huber:2002mx}. 
On the other hand, the reactor 
Equation~(\ref{equ:PROBREACTOR}) contains 
the product $\sin^2 2 \theta_{13} \cdot \sin^2 \Delta_{31}$ as the 
main contribution, which leads to a simple two-parameter correlation 
between  $\sin^2 2 \theta_{13}$ and $\sin^2 \Delta_{31}$. In this 
correlation, $\sin^2 2 \theta_{13}$ acts as the (energy independent) 
amplitude of the modulation and $\sin^2 \Delta_{31}$ contains the 
spectral information. Thus, with sufficiently good spectral information and 
the current knowledge about $\Delta m_{31}^2$, it is easy 
to disentangle these two parameters. In addition, the reactor 
measurement hardly depends on the true value of $\Delta m_{21}^2$.
 
\begin{figure}[t] 
\begin{center} 
\includegraphics[width=12cm]{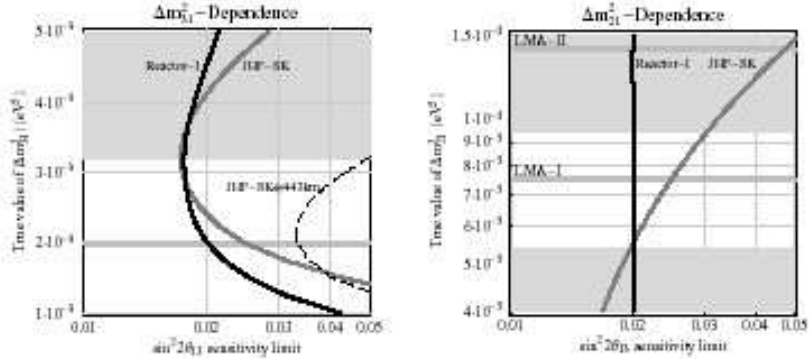} 
\end{center} 
\caption{\label{fig:dmdepnew} The $\sin^2 2 \theta_{13}$ sensitivity 
limit as function of the atmospheric (left) and solar 
(right) mass squared differences for the JPARC-SK (black 
curves) and Reactor-I (gray curves) experiments from 
References~\cite{Itow:2001ee,Huber:2003pm} (five years of neutrino 
running for JPARC-SK and an integrated luminosity for Reactor-I of $400 \, 
\mathrm{t \, GW \, y}$). In addition, JPARC-SK$@443 \, 
\mathrm{km}$ is shown for a modified baseline of $1.5 
\times 295 \, \mathrm{km} \simeq 443 \, \mathrm{km}$ in the 
left-hand plot (dashed curve). For the oscillation parameters, $\Delta 
m_{31}^2 = 2.0 \cdot 10^{-3} \, \mathrm{eV}^2$, $\sin^2 2 
\theta_{23}=1.0$, $\Delta m_{21}^2=7.5 \cdot 10^{-5} \, \mathrm{eV}^2$, and  
$\sin^2 2 \theta_{12} = 0.8$ are used 
(if not varied)~\cite{Fogli:2003am,Maltoni:2003da}. Furthermore, the 
$3 \sigma$ excluded regions are gray shaded and the current best-fit values 
are marked. The analysis includes systematics, multi-parameter correlations, and 
degeneracies as in Refs.~\cite{Huber:2002rs,Huber:2003pm}.} 
\end{figure} 
 
The dependence on the true values of the atmospheric and solar 
mass squared differences is, for the current best-fit values and 
ranges, illustrated in Figure~\ref{fig:dmdepnew}. The figure compares a 
reactor experiment with an integrated luminosity of 
$400 \, \mathrm{t \, GW \, y}$ to the JPARC to 
Super-Kamiokande first-generation superbeam experiment with a 
running time of five years (neutrino running only). 
The two plots illustrate that the considered reactor experiment would 
be better than the JPARC-SK superbeam at the current 
best-fit values of $\Delta m_{31}^2$ and $\Delta m_{21}^2$, as well 
as in most of the still allowed parameter ranges. Since the energy 
spectrum of a reactor experiment is broader than the one of 
an off-axis (narrow band) superbeam, the reactor experiment is less 
affected by the smaller value of $\Delta m_{31}^2$ after the 
Super-Kamiokande re-analysis \cite{bib:reanal}. In addition, it is hardly affected by the 
true value of $\Delta m_{21}^2$ as we discussed above. In the left plot 
of Figure~\ref{fig:dmdepnew}, we also show the JPARC-SK experiment for the 
(artificial) baseline of $1.5 \times 295 \, \mathrm{km} \simeq 443 \, 
\mathrm{km}$, which means that for this baseline the oscillation peak is 
shifted to $\Delta m_{31}^2 = 2.0 \cdot 10^{-3} \, \mathrm{eV}^2$. The figure 
clearly demonstrates that this shifting would not solve the problem due 
to the $1/L^2$ luminosity scaling. Because of the over-proportional loss of 
events for a lower neutrino energy due to the production mechanism and the 
cross section energy dependence, a lower energy instead of the 
longer baseline would also not help.

We have now demonstrated that the reactor measurement 
could provide a more robust limit on $\sin^2  2 \theta_{13}$ with 
respect to the (within certain ranges) true parameter values of 
$\Delta m_{31}^2$ and $\Delta m_{21}^2$. However, it is obvious 
from Equation~(\ref{equ:PROBREACTOR}) that reactor experiments at a baseline
of a few kilometers 
are not sensitive to the mass hierarchy or $\delta_{\mathrm{CP}}$, which 
means that superbeams will still be needed to test these parameters. On 
the other hand, a large reactor experiment could help to resolve the 
degeneracies in Equation~(\ref{equ:PROBVACUUM}) by measuring 
$\sin^2 2 \theta_{13}$. In this case, one could talk about synergies 
between a reactor experiment and a superbeam. For example, it has 
been demonstrated in Reference~\cite{Huber:2003pm} that there are several 
advantages from a large reactor experiment (e.g., with an integrated 
luminosity of $8000 \, \mathrm{t \, GW \, y}$). First of all, a 
reactor experiment could help to determine the mass hierarchy very well 
independently of the true value of $\Delta m_{21}^2$. Secondly
it could improve the 
CP sensitivity by allowing to operate the superbeam with neutrinos only 
(instead of using a fraction of antineutrino running). 
A reactor experiment performed on 
a shorter timescale than a superbeam would 
change the main goal of a superbeam from finding a non-zero value for
$\sin^2 2 \theta_{13}$ to measuring $\delta$ and the sign of $\Delta m_{21}^2$.

%% file: theory.tex
\subsection{Theoretical Motivation for non-zero $\theta_{13}$}

One may ask if there exist theoretical reasons why $\theta_{13}$ 
should be within the reach of a new experiment, with a sensitivity 
down to $\sin^2 2\theta_{13} \simeq 0.01$. 
This question is of course connected to the origin of neutrino masses.
For example, there exist apparent regularities in the 
fermionic field content which make it very tempting to introduce 
right-handed neutrino fields leading to both Dirac and Majorana 
mass terms for neutrinos. The diagonalization of the resulting 
mass matrices yields Majorana mass eigenstates and generically 
very small neutrino masses. This is the well known see-saw 
mechanism \cite{seesaw}. It can be nicely accommodated in embeddings 
of the SM into a larger gauge symmetry, such as SO(10).  

A reason for expecting a particular value of $\theta_{13}$ does clearly 
not exist as long as one extends the SM only minimally to accommodate 
neutrino masses. $\theta_{13}$ is then simply some unknown parameter 
which could take an arbitrarily small value, including zero. 
The situation changes in models of neutrino masses. Even then one 
should acknowledge that in principle any value of $\theta_{13}$ can 
be accommodated.
Indeed, before the discovery of large leptonic mixing, many theorists 
who did consider lepton mixing expected it to be similar to quark 
mixing, characterized by small mixing angles. Experiment led theory 
in showing the striking results that $\sin^2 2\theta_{23} \simeq 1$ 
and $\tan^2 \theta_{12} \simeq 0.44$, while $\theta_{13}$ is small. 
Indeed, the most remarkable property of leptonic mixing is that two 
angles are large. Therefore, today there is no particular reason to 
expect the third angle, $\theta_{13}$, to be extremely small or even 
zero. This can be seen in neutrino mass models which are able to 
predict a large $\theta_{12}$ and $\theta_{23}$. They often have a
tendency to predict also a sizable value of $\theta_{13}$.
This is both the case for models in the framework of Grand Unified 
Theories and for models using flavor symmetries. There exist also 
many different texture models of neutrino masses and mixings, which 
accommodate existing data and try to predict the missing information 
by assuming certain elements of the mass matrix to be either zero or 
equal. Again one finds typically a value for $\theta_{13}$ which is 
not too far from current experimental bounds. A similar behavior is 
found in so-called ``anarchic mass matrices''. Starting essentially 
with random neutrino mass matrix elements one finds that large mixings 
are actually quite natural. 

An overview of various predictions is given in Table~\ref{tab:ThPredictions}.  
For more extensive reviews, see for example 
\cite{Barr:2000ka,Altarelli:2002hx,Barbieri:2003qd,Chen:2003zv}.
The conclusion from all these considerations about neutrino mass 
models is that a value of $\theta_{13}$ close to the CHOOZ bound 
would be quite natural, while smaller values become harder and harder 
to understand as the limit on $\theta_{13}$ is improved.

\begin{table}[htb]
\begin{tabular}{lcc}
Reference & $\sin\theta_{13}$ & $\sin^2 2\theta_{13}$ \\
\hline
\emph{SO(10)} & & \\
Goh, Mohapatra, Ng \cite{Goh:2003hf} & 0.18 & 0.13 \\
\hline
\emph{Orbifold SO(10)} & & \\
Asaka, Buchm\"uller, Covi \cite{Asaka:2003iy} & 0.1 & 0.04 \\
\hline
\emph{SO(10) + flavor symmetry} & & \\
Babu, Pati, Wilczek \cite{Babu:1998wi} & $5.5 \cdot 10^{-4}$ &
 $1.2 \cdot 10^{-6}$ \\
Blazek, Raby, Tobe \cite{Blazek:1999hz} & 0.05 & 0.01\\
Kitano, Mimura \cite{Kitano:2000xk} & 0.22 & 0.18 \\
Albright, Barr \cite{Albright:2001uh} & 0.014 & $7.8 \cdot 10^{-4}$ \\
Maekawa \cite{Maekawa:2001uk} & 0.22 & 0.18 \\
Ross, Velasco-Sevilla \cite{Ross:2002fb} & 0.07 & 0.02\\
Chen, Mahanthappa \cite{Chen:2002pa} & 0.15 & 0.09 \\
Raby \cite{Raby:2003ay} & 0.1 & 0.04 \\
\hline
\emph{SO(10) + texture} & & \\
Buchm\"uller, Wyler \cite{Buchmuller:2001dc} & 0.1 & 0.04 \\
Bando, Obara \cite{Bando:2003ei} & 0.01 .. 0.06 &
 $4 \cdot 10^{-4}$ .. 0.01 \\
\hline
\emph{Flavor symmetries} & & \\
Grimus, Lavoura \cite{Grimus:2001ex,Grimus:2003kq} & 0 & 0 \\
Grimus, Lavoura \cite{Grimus:2001ex} & 0.3 & 0.3 \\
Babu, Ma, Valle \cite{Babu:2002dz} & 0.14 & 0.08 \\
Kuchimanchi, Mohapatra \cite{Kuchimanchi:2002fi} & 0.08 .. 0.4 & 
 0.03 .. 0.5 \\
Ohlsson, Seidl \cite{Ohlsson:2002rb} & 0.07 .. 0.14 & 0.02 .. 0.08 \\
King, Ross \cite{King:2003rf} & 0.2 & 0.15 \\
\hline
\emph{Textures} & & \\
Honda, Kaneko, Tanimoto \cite{Honda:2003pg} & 0.08 .. 0.20 &
 0.03 .. 0.15 \\
Lebed, Martin \cite{Lebed:2003sj} & 0.1 & 0.04 \\
Bando, Kaneko, Obara, Tanimoto \cite{Bando:2003wb} & 0.01 .. 0.05 &
 $4 \cdot 10^{-4}$ .. 0.01 \\
Ibarra, Ross \cite{Ibarra:2003xp} & 0.2 & 0.15 \\
\hline
\emph{$3 \times 2$ see-saw} & & \\
Appelquist, Piai, Shrock \cite{bib:s1, bib:s2} & 0.05 & 0.01 \\
Frampton, Glashow, Yanagida \cite{Frampton:2002qc} & 0.1 & 0.04 \\
Mei, Xing \cite{Mei:2003gn} (normal hierarchy) & 0.07 & 0.02 \\
\hphantom{Mei, Xing \cite{Mei:2003gn}} (inverted hierarchy) & $>0.006$ &
 $> 1.6 \cdot 10^{-4}$ \\
\hline
\emph{Anarchy} & & \\
de Gouv\^{e}a, Murayama \cite{deGouvea:2003xe} & $>0.1$ & $>0.04$ \\
\hline
\emph{Renormalization group enhancement} & & \\
Mohapatra, Parida, Rajasekaran \cite{Mohapatra:2003tw} & 0.08 .. 0.1 &
 0.03 .. 0.04 \\
\end{tabular}
\caption{
 Incomplete selection of predictions for $\theta_{13}$.
 The numbers should be considered as order of magnitude statements.
}
\label{tab:ThPredictions}
\end{table}

Besides, neutrino masses and mixing parameters are subject to quantum 
corrections between low scales, where measurements are performed, and 
high scales where some theory predicts $\theta_{13}$. Even in the 
``worst case'' scenario, where $\theta_{13}$ is predicted to be exactly 
zero, they cause $\theta_{13}$ to run to a finite value at low energy.  
Strictly speaking, $\theta_{13}=0$ cannot be excluded completely by this
argument, as the high-energy value could be just as large as the change
due to running and of opposite sign. However, a severe cancellation of 
this kind would be unnatural, since the physics generating the value 
at high energy are not related to those responsible for the quantum 
corrections. The strength of the running of $\theta_{13}$ depends on the 
neutrino mass spectrum and whether or not supersymmetry is realized.
For the Minimal Supersymmetric Standard Model one finds a shift 
$\Delta \sin^2 2\theta_{13} > 0.01$ for a considerable parameter range,
i.e.\ one would expect to measure a finite value of $\theta_{13}$
\cite{Antusch:2003kp}.  Conversely, limits on model parameters would 
be obtained if an experiment were to set an upper bound on 
$\sin^2 2\theta_{13}$ in the range of 0.01. In any case, it should be 
clear that a precision of the order of quantum corrections to neutrino 
masses and mixings is very interesting in a number of ways.

Altogether there exist very good reasons to push the sensitivity
limit from the current CHOOZ value by an order of magnitude and to hope
that a finite value of $\theta_{13}$ will be found. But as already 
mentioned, at this precision even a negative result would be very
interesting, since it would test or rule out many neutrino mass models 
and restrict parameters relevant for quantum corrections to masses and 
mixings. From a larger point of view the experiments discussed in this
white paper might probe if a small value of $\theta_{13}$ is a numerical 
coincidence or the result of some underlying symmetry. 

%% file: baseline_lum.tex

\section{Optimal Baseline Distances, Luminosity Scaling, and the
Impact of Systematics}
\label{sec:location}

\subsection{Total Flux vs. Baseline}
\begin{figure}[htb]
\begin{center}
\includegraphics[width=12cm]{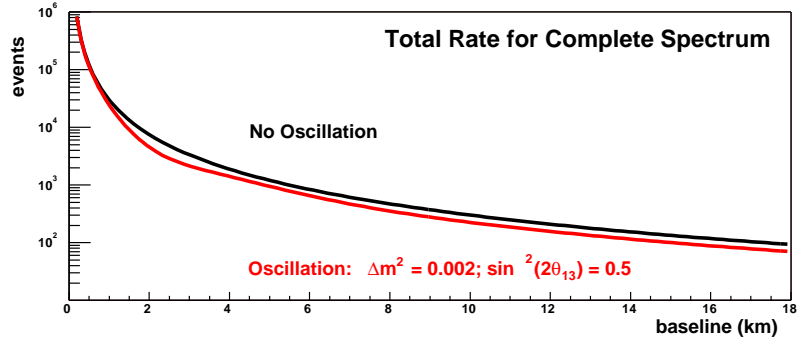}
\caption{Expected number of detected neutrino events as a function of 
baseline distance from the reactor core.  The two curves show the
expectations for the case of no oscillations and for an oscillation with 
$\Delta m^2 = 0.002$ and an amplitude of $\sin^2(2\theta_{13}) = 0.5$ 
which is 2.5 times the current limit.  The curves are 
calculated using the complete energy spectrum from a nuclear 
reactor and a luminosity of $\mathrm{600 \, t \, GW \, y}$.\label{fig:rateFull}}
\end{center}
\end{figure}

The equation for the survival probability of reactor neutrinos under
full three flavor mixing was previously described in Equation~(\ref{eq:p}).
It was pointed out by reference to Figure~\ref{fig:P} that a judicious 
choice of baseline distance 
could restrict one to oscillations dominated by one or the other
of the $\Delta m^2$ oscillations.  For this experiment, we are choosing
to focus on the shorter baseline, which corresponds to $\Delta m^2_{atm}$.
Thus, neglecting the other oscillation term, Equation~(\ref{eq:p}) reduces
to 
\begin{equation}
\label{eq:pAtm}
P(\bar\nu_e \rightarrow \bar\nu_e) =
1 - \sin^2(2\theta_{13})\sin^2(\Delta m^2_{atm}L/4E)
\end{equation}
where the current best estimate from Super-Kamiokande has
$\Delta m^2_{atm} = 0.002$ \cite{bib:reanal}.
To measure this oscillation effect, the optimal choice in baseline distance
will depend on the energy of the neutrinos.  As shown previously
in Figure~\ref{nuspec}, the detected spectrum for reactor 
neutrinos is between 1-10~MeV with a peak at about 3.8~MeV.
In addition, recall that the flux of neutrinos will fall as the square 
of the baseline distance.
\begin{figure}[htb]
\begin{center}
\includegraphics[width=12cm]{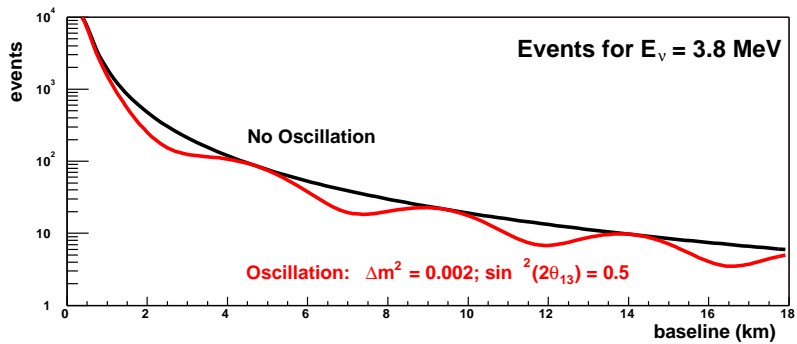}
\caption{Illustrative expected number of detected neutrino events as a function of 
baseline distance from the reactor core.  The two curves show the
expectations for the case of no oscillations and for an oscillation with 
$\Delta m^2 = 0.002$ and an amplitude of $\sin^2(2\theta_{13}) = 0.5$ 
which is 2.5 times the current limit.  The curves are 
calculated for a luminosity of $\mathrm{600 \, t \,GW \, y}$ and a mono-energetic
neutrino flux at 3.8~MeV. \label{fig:rateMono}}
\end{center}
\end{figure}
A comparison of the expected flux with and without oscillations is
shown in Figure~\ref{fig:rateMono} for a mono-energetic neutrino beam of
3.8~MeV.  Note that the amplitude of the oscillation shown is set
to $\sin^2(2\theta_{13}) = 0.5$ which is 2.5 times the current 
limit from CHOOZ in order to amplify the effect.  As would be expected, 
one sees the disappearance effect at regular intervals.
However, when the full energy spectrum, shown in Figure~\ref{nuspec}, is
folded in, the regular disappearance effect is washed out even with the
magnified amplitude (see Figure~\ref{fig:rateFull}).

\begin{figure}[b!]
\begin{center}
\includegraphics[width=12cm]{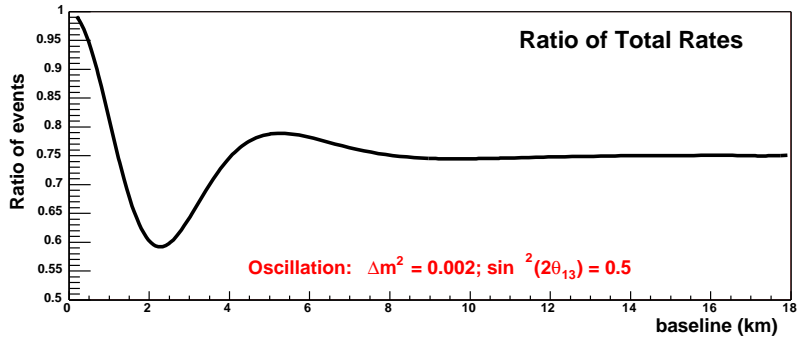}
\caption{The ratio of the expected number of neutrino events with and without 
oscillations as a function of distance from the reactor core.  This
calculation was made for a luminosity of $\mathrm{600 \, t \, GW \, y}$ 
and includes
the true neutrino energy spectrum.  The oscillation is assumed to 
have $\Delta m^2 = 0.002$ and an amplitude of $\sin^2(2\theta_{13}) = 0.5$ 
which is 2.5 times the current allowed limit.\label{fig:rateRatio}}
\end{center}
\end{figure}

In order to make the oscillation effect in Figure~\ref{fig:rateFull} visible,
the ratio of the two curves is shown in Figure~\ref{fig:rateRatio}.
Notice that the largest deviation occurs at a baseline distance of just over
2 kilometers.  This corresponds to the 
first oscillation for $E_{\nu} = 3.8$ MeV
as shown in Figure~\ref{fig:rateMono}.  This makes sense since this is the
peak of the neutrino energy spectrum and therefore has the most statistical
power.  However, it is clear that as baseline distance increases, the
effect of other parts of the energy spectrum being at their respective
maxima and minima of oscillation effectively neutralizes any ability
to detect a specific oscillation signature.

\subsection{Spectral Shape Information}
From Equation~(\ref{eq:pAtm}), it is clear that neutrinos of differing energies
oscillate with different frequencies.  
Figure~\ref{fig:rateRatio} shows that observable oscillation effects in
the total number of neutrinos detected wash out with increasing 
baseline distance.  But by looking at the specific energy 
distribution of the detected neutrinos, more information is available.
\begin{figure}[htb]
\begin{center}
\includegraphics[width=6cm]{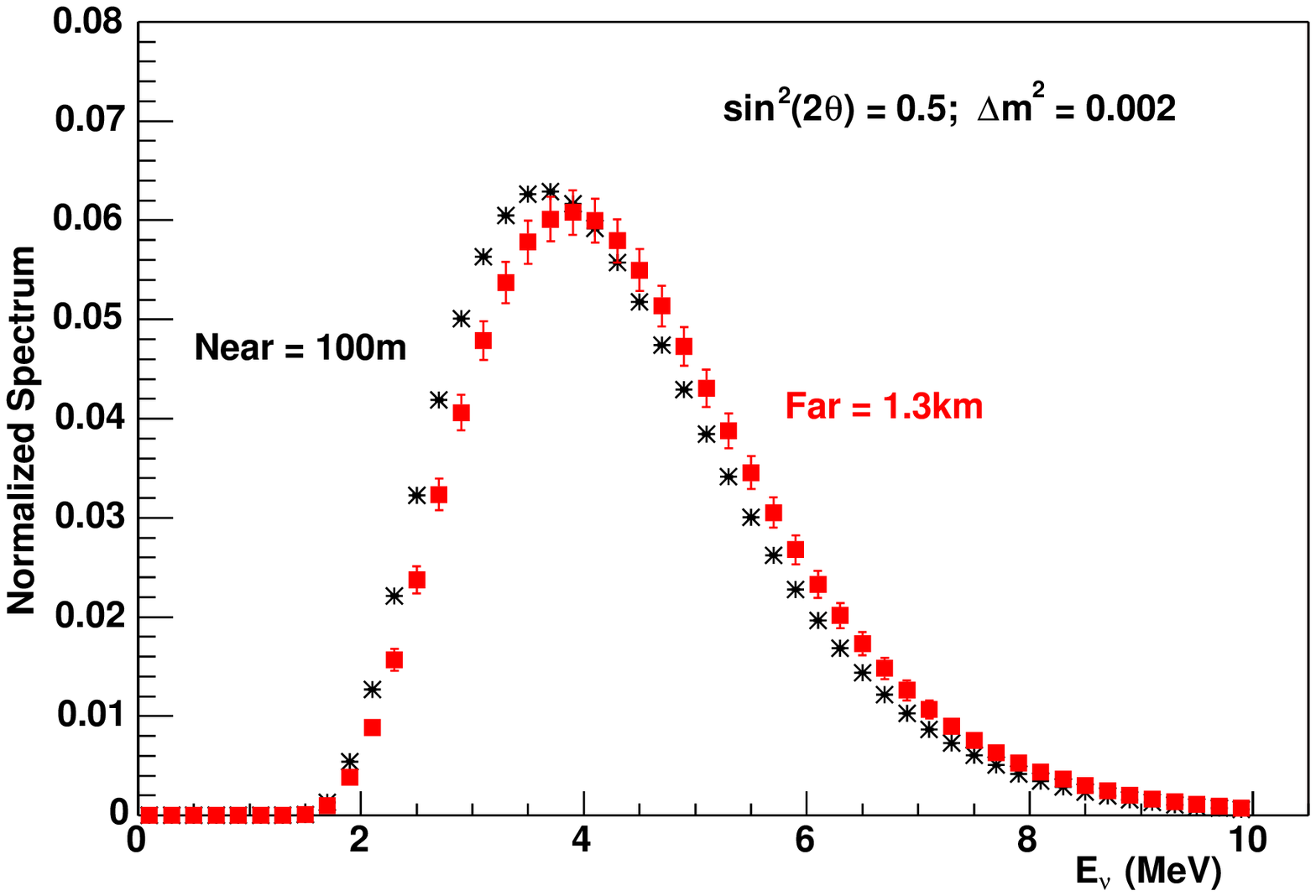}
\hfill
\includegraphics[width=6cm]{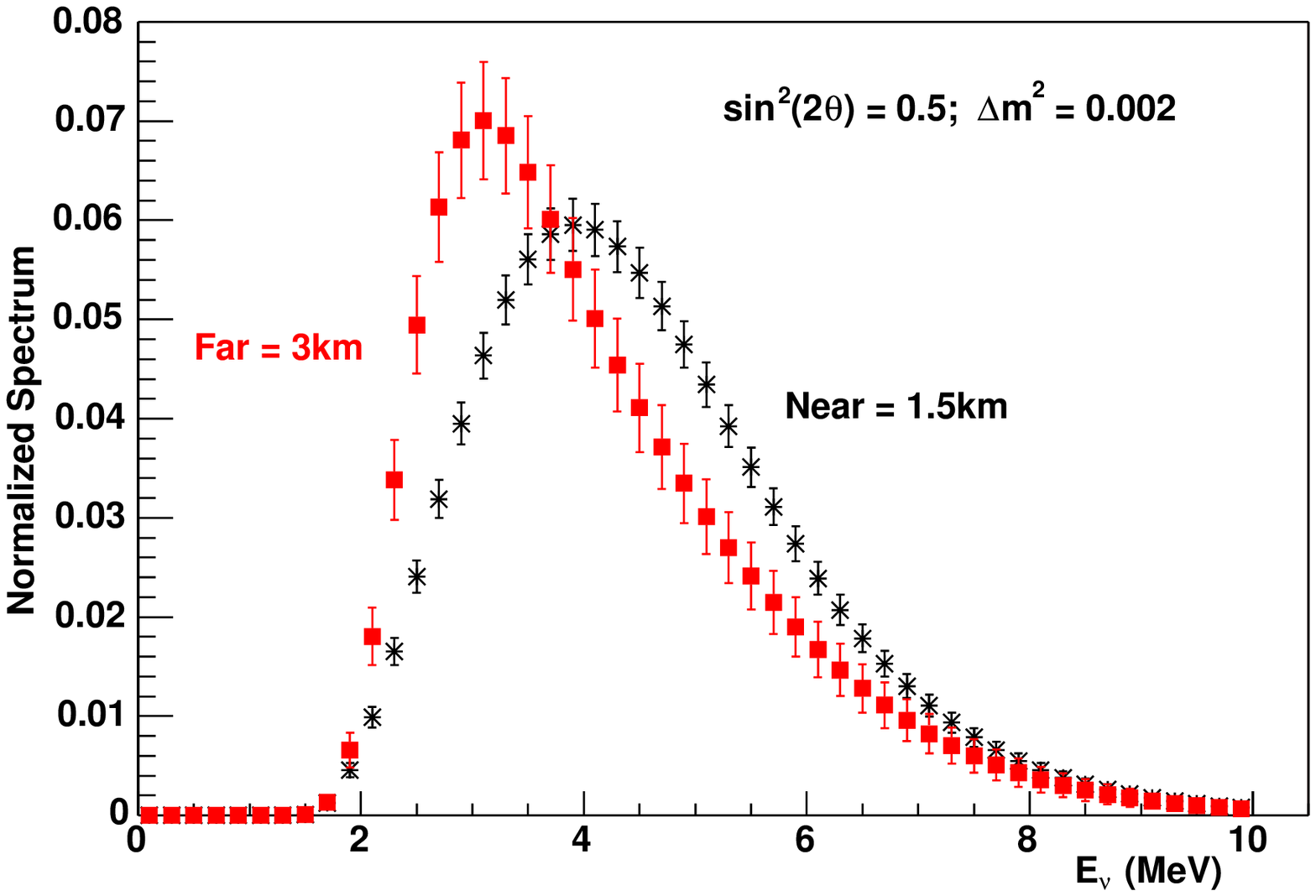}
\caption{Comparisons of the expected measured neutrino energy spectra
at various baseline distances in the case of oscillations.  The
oscillations are assumed to have  $\Delta m^2 = 0.002$ and an amplitude
of $\sin^2(2\theta_{13}) = 0.5$ which is 2.5 times the current
limit in order to magnify the effect.  These plots show the expected
statistics for a luminosity of $\mathrm{600 \, t \, GW \, y}$
 at the specified 
baseline distance.\label{fig:spect}}
\end{center}
\end{figure}
In Figure~\ref{fig:spect}, two comparisons of the normalized energy 
spectra are shown.
These plots show statistical errors only for 0.2~MeV bins and a luminosity 
of $\mathrm{600 \, t \, GW \, y}$ at each location. As with the previous plots, the
amplitude of the oscillation has been magnified by a factor of 2.5
($\sin^2(2\theta_{13}) = 0.5$) and a mass difference of $\Delta m^2 = 0.002$
has been used.  In addition, an energy resolution of 
10\%/$\sqrt{E}$ has been assumed.  The plot on the left compares the spectrum
at 100 meters, which is effectively unoscillated, with the spectrum at
1.3 kilometers.  One can notice that at this distance, the low energy
part of the spectrum is showing a deficit with respect to the near 
detector spectrum.  However, one could also confuse this with an
overall shift in the absolute energy scale between the two detector
locations.  

The right plot of Figure~\ref{fig:spect}, however, compares two oscillated
spectra from 1.5 and 3 kilometers.  Notice that the shapes of the spectra
are vastly different.  This arises from the fact that at 3 kilometers
it is the high energy part of the spectrum which is fully oscillated
away while the low energy part has returned to full presence.  It is
interesting to note that the statistical significance of the spectral
distortions in the two plots is nearly identical.  While the total number
of events is 
significantly higher in the plot with baselines of 100 m and 1.3 km, 
the spectral distortion is much more pronounced in the plot with
baselines of 1.5 and 3 km.  These two effects appear to compensate one
another.  

\subsection{Combining Shape and Rate Information}
It turns out that the most statistically significant spectral shape
distortion, given the assumed oscillation parameters above, is achieved
for a near detector at the closest possible location and a far detector
at about 1~kilometer.  Thus the spectral shape information has a
different optimal baseline than the depletion of the total
flux, which was previously observed to be maximal at 
just over 2~kilometers from the source.

Since the spectral shape measurement requires the use of a normalized
energy spectrum at each location, the statistical significance of
each measurement (each bin) is weighted by the total flux at that
location.  This gives a 1/L$^2$ reduction in the statistical sensitivity.
Therefore, from a strictly statistical perspective, the total flux 
measurement will have slightly more than a factor of two more power.
However, the total flux measurement is susceptible to systematic 
differences between the two detectors.  Since the normalized energy
spectra are normalized to the total number of events at that location,
all systematic effects, except those which will be uncorrelated
bin-to-bin within a detector, will be removed.  This additional
freedom from systematic effects implies that in the limit of infinite
statistics, a more precise measurement can be made with the information
from the energy spectrum.  

\begin{figure}[t!]
\begin{center}
\includegraphics[width=0.7\textwidth]{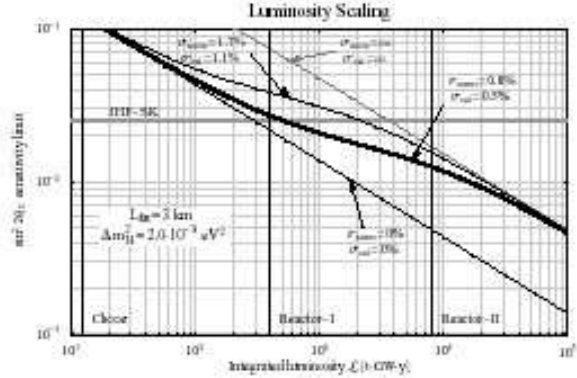}
\end{center}
\caption{\label{fig:luminosity} The sensitivity to $\sin^22\theta_{13}$ at 90\%
CL as a function of the integrated luminosity for different values of
the normalization error $\sigma_\mathrm{norm}$ and the energy
calibration error $\sigma_\mathrm{cal}$. The vertical lines mark the
luminosities of the CHOOZ experiment and our standard setups Reactor-I
and Reactor-II. We take $\Delta m^2_{31} = 2\times 10^{-3}$~eV$^2$ and
$L_{FD} = 3$~km.The horizontal line shows a typical sensitivity limit
obtainable by the JHF-SK superbeam experiment (also known as
JPARC-$\nu$) for the same parameter values.}
\end{figure}

This interplay between the two methods can be seen graphically by
referring to the plot in Figure~\ref{fig:luminosity} (see also the
discussion in Sec.~\ref{sec:lumi}). There, the sensitivity to
$\sin^2(2\theta_{13})$ is shown as a function of luminosity.  At low
luminosity ($\cal L < \mathrm{400 t \, GW \, y }$) the sensitivity is directly
proportional to statistics and is dominated by the total flux
measurement.  Then, for luminosities between 
$\mathrm{400}$ and $\mathrm{6000 \, t \, GW \, y}$,
additional statistics do not make significant gains in
the sensitivity.  In this region, the systematic uncertainties between
the two detectors (called $\sigma_{norm}$ in this plot) become
dominant.  However, beyond $\mathrm{6000 \, t \, GW \, y}$, notice that the
sensitivity again becomes proportional to the statistics.  This is
caused by the fact that enough statistics have been gained to allow
the spectral measurements to dominate over the systematically limited
flux measurement.

\begin{figure}
\begin{center}
\includegraphics[width=9cm]{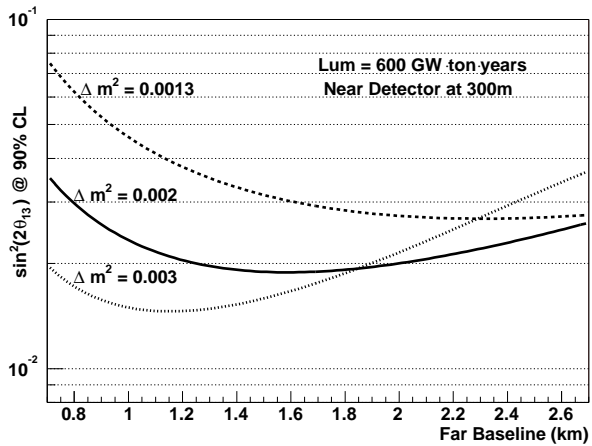}
\caption{Statistical sensitivity to $\sin^2(2\theta_{13})$ as a function
of baseline distance to the far detector.  The statistical power is
calculated using Equation~(\ref{eq:chi2}) with a luminosity of 
$\mathrm{600 \, t \, GW \, y}$
and a 1\% systematic limit bin-to-bin.  Curves are shown for three values
of $\Delta m^2$ representing the best fit and the upper and lower limits
of the 90\% allowed region from the Super-Kamiokande experiment.
\label{fig:baseline}}
\end{center}
\end{figure}

Realization of this interplay between the methods implies that the
optimal choice of baseline distance depends on the expected luminosity
of the experiment.  Since most of the discussion in this paper
does not expect a luminosity of greater than $\mathrm{6000 \, t \, GW \, y}$
(which 
would require kiloton sized detectors), we will focus on measurements
in which the total flux measurement is not systematically limited.
To make optimal use of the available statistics, we therefore
wish to combine the information from the total flux and energy
spectral measurements at both detectors.  One can rather simply
create a chi-squared comparison of the two statistical distributions
which takes into account both sets of information with the following
definition:
\begin{equation}
\chi^2 = \sum_i \frac{\left[N^{far}_i - 
                   \left(\frac{L_{near}}{L_{far}}\right)^2 N^{near}_i\right]^2}
                     {\left[N^{far}_i + 
                   \left(\frac{L_{near}}{L_{far}}\right)^4 N^{near}_i\right]}
\label{eq:chi2}
\end{equation}
where $L$ refers to the baseline distance to the near or far detectors and
$N_i$ refers to the number of events in the $i$-th bin of the measured
energy spectrum at that detector.
By using the definition of Equation~(\ref{eq:chi2}), one can determine the 90\%
confidence level limit on a measurement of $\sin^2(2\theta_{13})$ for a 
given luminosity.  This is shown as a function of baseline distance
in Figure~\ref{fig:baseline} assuming a luminosity of 
$\mathrm{600 \, t \, GW \, y}$.
For this estimation, the near detector is assumed to be fixed at
300 meters and a 1\% systematic limit has been used.  

One can see that for the current best fit value of $\Delta m^2 = 0.002$,
the optimal baseline distance for the far detector is at 1.6 kilometers.
However, given that the optimal distance depends quite strongly on 
the value of $\Delta m^2$, we also show the sensitivity plots for values 
of $\Delta m^2$ which match the upper and lower bounds of the 90\%
allowed region from Super-Kamiokande.  Notice that for each curve,
the sensitivity to $\sin^2(2\theta_{13})$ is relatively flat around the
optimum.  Therefore, a reasonable sensitivity can be reached, with 
flexibility to various values of $\Delta m^2$, for a far detector baseline
distance between 1.2 and 2.4 kilometers.


\subsection{Statistical Analysis and Luminosity Scaling}
\label{sec:lumi}

In this section some general analysis methods are proposed to investigate the
sensitivity to $\theta_{13}$ in a reactor experiment with two detectors and
one single reactor. The impact of the integrated luminosity, positions
of the near detector, various systematic errors, and a possible background
on the sensitivity limit are discussed. As a measure for the ``size'' of the
experiment the {\it integrated luminosity} $\mathcal{L}$ is useful, which is
defined as $\mathcal{L}$ = fiducial detector mass [tons] $\times$ thermal
reactor power [GW] $\times$ running time [years] (assuming 100\% detection
efficiency and no deadtimes). We define two benchmark setups Reactor-I
($\mathcal{L} = 400$~t~GW~y) and Reactor-II ($\mathcal{L} = 8000$~t~GW~y)
corresponding roughly to 31~500 and 630~000 reactor neutrino events for no
oscillations, respectively, assuming a PXE-based scintillator.

We take into account that spectral information is available in the near, as
well as in the far detector in form of $N_\mathrm{bins}$ bins in positron
energy. For the theoretical prediction for the number of reactor neutrino
events in the $i$th energy bin of the near ($A=N$) and far ($A=F$) detector,
respectively, we write
\begin{equation}\label{equ:theo}
  T^A_i = (1 + a + b^A + c_i) N_i^A  + g^A M_i^A \, ,
\end{equation}
and consider a $\chi^2$-function including the full spectral information
from both detectors:
\begin{eqnarray}
  \chi^2 &=& 
  \sum_{i=1}^{N_\mathrm{bins}} \left[ \sum_{A=N,F} 
      \frac{ (T_i^A - O_i^A)^2 }
           { O_i^A + \sigma_\mathrm{exp}^2 (O_i^A)^2 + 
             B_i^A + \sigma_\mathrm{BG}^2 (B_i^A)^2} +
      \left( \frac{c_i}{\sigma_\mathrm{shape}} \right )^2 
    \right] \nonumber\\
&+&
    \sum_{A=N,F} \left[
    \left( \frac{b^A}{\sigma_b} \right)^2 +
    \left( \frac{g^A}{\sigma_\mathrm{cal}} \right)^2
  \right] +
  \left( \frac{a}{\sigma_a} \right)^2 \,. \label{equ:chi2N+F}
\end{eqnarray}
Here, $N_i^A$ is the expected number of events in the $i$th energy bin of the
corresponding detector, which depends on the oscillation parameters.  $N_i^A$
is calculated by folding the reactor neutrino spectrum, the detection cross
section for inverse beta decay, the $\bar\nu_e$ survival probability, and the
energy resolution function. In the numerical calculations we assume an energy
resolution of $5\% / \sqrt{E_\mathrm{vis} [\mathrm{MeV}]}$, and we use
$62$ bins in the range between $E_{\bar\nu_e}=1.8\,\mathrm{MeV}$ and
$E_{\bar\nu_e}=8.0\,\mathrm{MeV}$, corresponding to a bin width of
$0.1\,\mathrm{MeV}$.\footnote{The results depend very weakly on our choices
for the energy resolution and the number of bins.} 
$O_i^A$ is the observed number of events.  In the absence of real data, we
take for $O_i^A$ the expected number of events for some fixed ``true values''
of the oscillation parameters, e.g., to calculate a sensitivity limit the
expected number of events for $\theta_{13} = 0$ is used. If the near detector
is so close to the reactor that no oscillations will occur, the $T_i^N$
will not depend on the oscillation parameters, and in that case one can set
$O_i^N = N_i^N$. 
The quantities $T_i^A$ and $O_i^A$ correspond only to reactor neutrino events.
If a certain background has to be subtracted from the actual number of events
it will contribute to the statistical and systematic errors. In
Equation~(\ref{equ:chi2N+F}) $B_i^A$ is the number of background events in the $i$th bin
of detector $A$, and we assume that it is known with an (uncorrelated) error
$\sigma_\mathrm{BG}$.

For each point in the space of oscillation parameters, the
$\chi^2$-function has to be minimized with respect to the parameters
$a,b^N$, $b^F$, $g^N$, $g^F$, and $c_i$ modeling various systematic
errors. 

\begin{enumerate}
\item\label{it:absnorm} The parameter $a$ refers to the {\bf error on the
 overall normalization} of the number of events common to near and
far detectors, and $\sigma_a$ is typically of the order of a few
percent. The main source for such an error is the uncertainty of the
neutrino flux normalization.
\item\label{it:relnorm} The parameters $b^N$ and $b^F$
 parameterize the {\bf uncorrelated normalization uncertainties} of
 the two detectors. Here contributes, for instance, the error on the
 fiducial mass of each detector. We assume that in this case an error
 below 1\% can be reached.
\item\label{it:cal} The {\bf energy scale uncertainty} in the two
detectors is taken into account by the parameters $g^N$ and $g^F$. To
this aim we replace in $N_i^A$ the visible energy $E_\mathrm{vis}$ by
$(1+g^A) E_\mathrm{vis}$, and to first order in $g^A$ we have $N_i^A
(g^A) \approx N_i^A(g^A = 0) + g^A \, M_i^A$. A typical value for
this error on the energy calibration is $\sigma_\mathrm{cal} \sim
0.5\%$.
\item\label{it:shape} In order to model an {\bf uncertainty 
on the shape of the expected energy spectrum}, we introduce a parameter $c_i$
for each energy bin, known with an error $\sigma_\mathrm{shape}$. This
corresponds to a completely uncorrelated error between different energy bins,
which is the most pessimistic assumption of {\it no} knowledge of possible
shape distortions.  However, we choose this error fully correlated between the
corresponding bins in the near and far detector (the same coefficient is used
for the corresponding bins in the two detectors), since shape distortions
should affect the signals in both detectors of equal technology in the same
way.
\item\label{it:exp} We include the possibility of an {\bf uncorrelated
experimental systematic error} $\sigma_\mathrm{exp}$.  In this way we assume
that the observed number of events in each bin and each detector $O_i^A$ has
in addition to the statistical error $\pm \sqrt{O_i^A}$ the (uncorrelated)
systematic error $\pm \sigma_\mathrm{exp} O_i^A$.  We call this uncertainty
``bin-to-bin error''. Taking it completely uncorrelated between energy bins as
well as between the near and far detectors corresponds again to the worst case
scenario. Values of $\sigma_\mathrm{exp}$ at the per mil level should be
realistic.
\end{enumerate}

Note that all the parameters describing the systematic errors are at
most at the percent level, which means that the linear approximation
in Equation~(\ref{equ:theo}) is justified. The 
following discussion of general features of such an analysis is based
on the results reported in Reference~\cite{Huber:2003pm}.

Let us first assume that the near detector is close enough to the reactor,
such that no oscillations develop ($L_{ND} \lesssim 200$~m). Furthermore, we
first assume that the background is negligible, and $\sigma_\mathrm{exp}
\lesssim 0.1\%$ and $\sigma_\mathrm{shape} \lesssim 2\%$, which means that
these errors can be neglected. Then the $\chi^2$-analysis can be significantly
simplified. In particular, it is not necessary to explicitly include the near
detector in the analysis and Equation~(\ref{equ:chi2N+F})
becomes~\cite{Huber:2003pm}
\begin{equation}\label{equ:chi2F}
  \chi^2_F = \sum_i 
    \frac{(T_i^F - O_i^F)^2}{O_i^F} +
    \left( \frac{g^F}{\sigma_\mathrm{cal}} \right)^2 +
    \left( \frac{\alpha}{\sigma_\mathrm{norm}} \right)^2 
\end{equation}
with
\begin{equation}\label{equ:sigma_alpha}
  \sigma_\mathrm{norm}^2 = \bar\sigma^2 + \sigma_b^2\,,\qquad
  \frac{1}{\bar\sigma^2} \approx 
  \frac{1}{\sigma_a^2} + \frac{1}{\sigma_b^2} \,. 
\end{equation}
For example, assuming typical values of $\sigma_a = 2\%$ for the flux
uncertainty and $\sigma_b = 0.6\%$ for the detector-specific
uncertainty, we obtain with Equation~(\ref{equ:sigma_alpha}) an effective
normalization error of $\sigma_\mathrm{norm} \simeq 0.8\%$. This is
the value which is used for the numerical calculations.

In Figure~\ref{fig:luminosity}, we show the sensitivity to $\sin^22\theta_{13}$
as a function of the integrated luminosity ${\cal L}$.  In this figure, the
lower diagonal curve corresponds to the idealized case of statistical errors
only, and shows just the expected $1/\sqrt{\cal L}$ scaling, whereas the
values $\sigma_\mathrm{norm} = 0.8\%$ and $\sigma_\mathrm{cal} = 0.5\%$ lead
to the thick curve. At a luminosity around $100 \, \mathrm{t \, GW \, y}$, we
detect a departure from the statistics dominated regime into a flatter
systematics dominated region. This effect is dominated by the error on the
normalization $\sigma_\mathrm{norm}$, whereas the energy calibration error
$\sigma_\mathrm{cal}$ only plays a minor role. In fact, we find that for all
considered cases (including Reactor-I and Reactor-II) the impact of the energy
scale uncertainty is very small, as long as the oscillation minimum is well
inside the observable energy spectrum. Hence, we will neglect this error in
the following.

At large luminosities $\gtrsim 10^4 \, \mathrm{ t \, GW \, y}$, the
slope of the curve changes, and we are entering again a statistics
dominated region with a $1/\sqrt{\cal L}$ scaling.  This interesting
behavior can be understood as follows.  For low luminosities the
sensitivity to $\sin^22\theta_{13}$ comes mainly for the total number
of events, hence the absolute normalization is important.  The
turnover of the sensitivity line into the second statistics dominated
region occurs at the point where the spectral shape distortion
becomes more important than the total rate, which implies that the
overall normalization, and consequently also the actual value of
$\sigma_\mathrm{norm}$, becomes irrelevant. We illustrate this by the
upper thin black line, which shows the luminosity scaling for the case
of larger systematic errors. As an example, we choose values of
$\sigma_\mathrm{norm} = 1.7\%$ for the normalization and
$\sigma_\mathrm{cal} = 1.1\%$ for the energy calibration. We find
that, in this case, the transition to the systematics dominated regime
occurs at much smaller luminosities. However, for large luminosities,
the same limit is approached as for the more optimistic case. The
diagonal gray curve shows the limit for no constraint at all on the
normalization and energy calibration.\footnote{Although we leave the
normalization free in the fit, we assume that the shape is known.}
Even in this extreme case, we obtain the same limit for high
luminosities.

\begin{figure}[t!]
\begin{center}
\begin{tabular}{cc}
\includegraphics[width=0.7\textwidth]{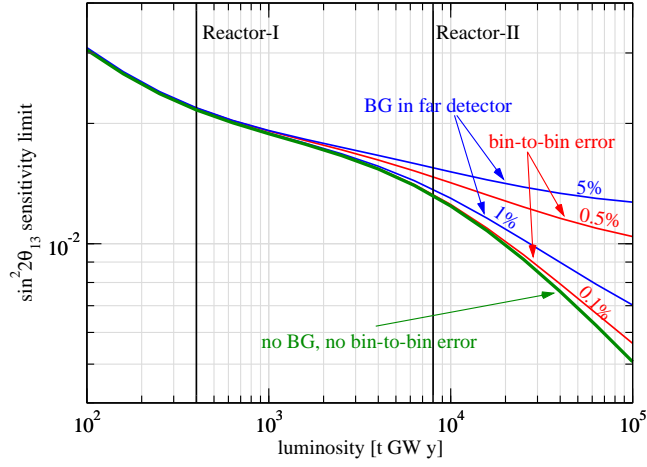}
\end{tabular}
\end{center}
\caption{\label{fig:background} Sensitivity to $\sin^22\theta_{13}$ at 90\% CL
as a function of the luminosity for an uncorrelated experimental systematic
error (``bin-to-bin error'') $\sigma_\mathrm{exp} = 0.1\%$ and 0.5\%, and
background levels in the far detector relative to the total number of events
for no oscillations of 1\% and 5\%.  Here $L_{ND} = 0.2$~km, $L_{FD} =
1.7$~km, $\Delta m^2_{31} = 2\times 10^{-3}$~eV$^2$, and
$\sigma_\mathrm{shape} = 2\%$. Identical detector masses are assumed for near
and far detectors.}
\end{figure}

\subsection{Systematics, Background, and the Position of the Near
Detector}
\label{sec:syst-BG-ND}

In the following we investigate the impact of various systematic
effects beyond the simple overall normalization. To this end we apply
the full $\chi^2$ function as given in Equation~(\ref{equ:chi2N+F}). In
Figure~\ref{fig:background} the luminosity scaling of the
$\sin^22\theta_{13}$-limit is shown for various choices for the
experimental bin-to-bin error $\sigma_\mathrm{exp}$ (see
item~\ref{it:exp} above) and background levels in the far
detector. For the sake of concreteness we assume a flat background in
each detector $B_i^A = B^A$ with an error of $\sigma_\mathrm{BG} =
10\%$. The size of the backround is measured by $f_\mathrm{BG}$, which
is defined as the fraction of the total number of background events
relative to the total number of reactor neutrino events for no
oscillations, i.e., $B^A = f_\mathrm{BG} \sum_i
N^A_{i,\mathrm{no\,osc}} / N_\mathrm{bins}$. From the figure we find
that Reactor-I is not affected by a bin-to-bin error up to 0.5\%, nor
by backgrounds in the far detector up to 5\%. In contrast, such
errors and backgrounds are
important to some extent for big experiments like Reactor-II. In that
case values of $\sigma_\mathrm{exp} \gtrsim 0.1\%$ start to
deteriorate the sensitivity limit, and the background in the far
detector should be smaller than 1\% of the reactor neutrino signal. We
note that backgrounds in the near detector up to a few percent do not
affect the result.  Regarding the huge number of reactor neutrino
events in the near detector it should be possible to obtain
backgrounds below 1\%.

\begin{figure}[t!]
\begin{center}
\includegraphics[width=0.7\textwidth]{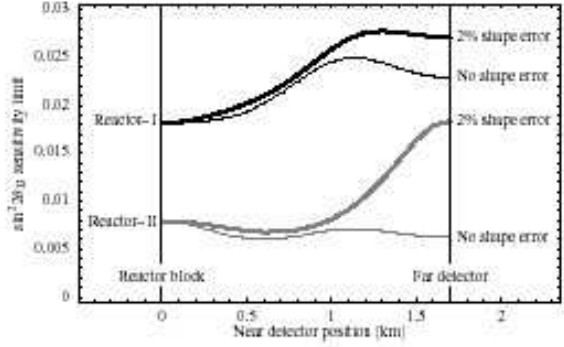}
\end{center}
\caption{\label{fig:neardet} The sensitivity to $\sin^22\theta_{13}$ at the
90\% CL for Reactor-I and Reactor-II as a function of the
near detector position. The far detector is situated at 1.7~km and we assume
identical detector masses and $\Delta m^2_{31} = 3\times 10^{-3}$~eV$^2$. 
Furthermore, the impact of an uncorrelated theoretical shape uncertainty 
$\sigma_\mathrm{shape} = 2\%$ is shown.}
\end{figure}

For practical reasons it might be hard to find a reactor station where
a near detector can be situated very close ($\lesssim 200 \,
\mathrm{m}$) to the core with sufficient rock overburden. Therefore,
it is interesting to investigate the impact of larger near detector
baselines on the $\sin^22\theta_{13}$ limit. In this case the
information provided by the near detector on the initial flux
normalization and energy shape is already mixed with some oscillation
signature. The results of such an analysis are presented in
Figure~\ref{fig:neardet}. We find that for the case of Reactor-I the
limit starts deteriorating around a near detector distance of 400~m,
whereas for Reactor-II the limit even improves slightly up to near
detector baselines of $\sim 1 \, \mathrm{km}$. Due to the high
statistics in the case of Reactor-II, flux normalization and shape are
very well determined by the near detector even in the presence of some
effect of $\theta_{13}$, and the additional information on
oscillations improves the limit a bit. Furthermore, we find from
Figure~\ref{fig:neardet} that the shape uncertainty becomes important
for near detector baselines $\gtrsim 1.1 \, \mathrm{km}$, especially
for Reactor-II. A reduction of this theoretical error would be helpful
in such a situation.  We note that assuming the shape error to be
completely uncorrelated corresponds to the worst case. A more
realistic implementation of the shape uncertainty including correct
correlations will lead to results somewhere in between the curves for
no and 2\% shape error in Figure~\ref{fig:neardet}. We have verified
that for Reactor-I the worsening of the limit comes mainly from the
fact that with increasing near detector baselines the number of events
decreases rapidly, i.e., it is statistics dominated, whereas for
Reactor-II the loss in sensitivity is driven by the systematic shape
uncertainty and cannot be compensated by larger near detectors.

\begin{table}[t!]
\begin{center}
\begin{tabular}{|ll|c@{\qquad}c|}
\hline
 & &Reactor-I &Reactor-II \\
\hline
Effective normalization  & $\sigma_\mathrm{norm} = 0.8 \%$ &
important & not important\\
Energy calibration & $\sigma_\mathrm{cal} \simeq 0.5\%$ &
not important & not important\\
Exp.~bin-to-bin uncorr.~error & $\sigma_\mathrm{exp} \lesssim 0.1\%$ &
not important & important\\
background in far detector & $f_\mathrm{BG} \lesssim 1\%$ &
not important & important\\
near detector baseline & &
$L_{ND} \lesssim 400$~m & $L_{ND} \lesssim 1$~km \\
theor.~shape uncertainty & $\sigma_\mathrm{shape} = 2\%$ &
\multicolumn{2}{c|}{important for $L_{ND} \gtrsim 1$~km} \\
\hline
\end{tabular}
\end{center}
\caption{\label{tab:error_summary} Relevance of various 
  systematic errors for the two reactor
  benchmark setups Reactor-I and Reactor-II.}
\end{table}

To summarize, for the case of the Reactor-I setup (400~t~GW~y) the
main information comes from the total number of events and the
systematic normalization error dominates. In order to obtain a
reliable limit, it should therefore be well under control. For large
luminosities, such as for the Reactor-II setup (8000~t~GW~y), the
sensitivity limit comes mainly from spectral shape information and is
independent of normalization errors. In that case a bin-to-bin
uncorrelated experimental systematic error should be below 0.1\% and
the background should be at the 1\% level. Furthermore, for the case
of Reactor-I-like experiments one should look for a site where the
near detector can be placed at a distance of at most 400~m from the
reactor. For large detectors, such as Reactor-II, near detector
baselines of up to 1~km will perform well. For near detector baselines
longer than about 1~km the correct treatment of the theoretical shape
uncertainty becomes important. These results are summarized in
Table~\ref{tab:error_summary}.


%% file: chooz.tex
\section{Previous Reactor Experiments}
\label{sec:previous}

Here we review the three most recent
previous neutrino experiments at reactors.  The best
current limit on $\quq$ comes from CHOOZ.  An experiment with similar
distance and running time, but smaller overburden, was conducted at Palo Verde.
Finally the KamLAND experiment was conducted with much larger overburden, and
larger size, but much longer baseline, which resulted in a reduced
sensitivity to $\quq$ but unprecedented sensitivity to $\theta_{12}$.
These previous experiments are reviewed in order to present several 
lessons which are needed to show that a future experiment can control
systematics to the level below one percent.

\subsection{CHOOZ}

 The CHOOZ experiment \cite{bib:chooz,Apollonio:1999ae,Apollonio:1998xe} was located close to the Chooz nuclear power plant, in
 the North of France, 10\,km from the Belgian border. The power plant consists of
 two twin pressurized water reactors (PWR), the first of a series of the newly developed
 PWR generation in France. The thermal power of each reactor is
 4.25\,GW (1.3\,GW
 electrical). These reactors started respectively in May and August 1997, just after
 the start of the data taking of the CHOOZ detector (April 1997). This opportunity
 allowed a measurement of the reactor-OFF background and a separation
 of individual reactor's contributions.

 The detector was located in an underground laboratory about 1\,km from the
 neutrino source. The 300\,mwe rock overburden reduced the external cosmic ray
 muon flux, by a factor about 300, to a value of $0.4 \, $m$^{-2} \,
 s^{-1}$. 
This was  the main criterion to choose this site. Indeed, the
 previous experiment at the Bugey reactor power plant  showed the
 necessity of reducing by 2 orders of magnitude the flux of fast neutrons
 produced by muon-induced nuclear spallations in the material surrounding the
 detector. 
 The neutron flux in CHOOZ was measured at energies larger 
than 8~MeV (endpoint of the
 neutrino flux from nuclear reactors) and found to be 
1/day, in good agreement with
expectation. 

 The detector envelope consisted of a cylindrical steel vessel, 
5.5\,m in diameter and
 5.5\,m in height.  The vessel was placed in a pit (7\,m diameter and
7\,m deep), and was surrounded  by 75 cm of low activity sand. It was
composed  of three concentric regions:
\begin{itemize}
\item{a central 5 ton target in a transparent plexiglass container filled with a
 0.09\,\% Gd-loaded scintillator}
\item{an intermediate 70 cm thick region, filled with non loaded scintillator and
 used to protect the target from PMT radioactivity and to contain the
gamma from  neutron capture on Gd. These 2 first regions were viewed by 192 PMTs}
\item{an outer veto, filled with the same scintillator.}
\end{itemize}
 
 The scintillator showed a degradation of the transparency over time, which
 resulted in a decrease of the light yield (live time around 250
 days). The event position was reconstructed by fitting the charge
 balance, with a  typical precision of 10 cm for the positron and 20
 cm for the neutron.  The time
 reconstruction was found to be less precise on source and laser tests, due to the
 small size of the detector. The reconstruction became more difficult
 when the event was located near the PMTs, due to the $1/r^2$ divergence
 of the  light collected (see Figure 31 of \cite{bib:chooz}).
   
 The final event selection used the following cuts\,:
 \begin{itemize}
\item{positron energy smaller than 8~MeV (only 0.05\,\% of the positrons have a
 bigger energy)}
\item{neutron energy between 6 and 12~MeV}
\item{distance from the PMT surface bigger than 30 cm for both positron and
 neutron}
\item{distance between positron and neutron smaller than 100 cm}
\item{only one neutron}
\item{time window between positron and neutron signals is from
2 to 100 $\mu$s.}
\end{itemize}

The 6~MeV cut on the gamma ray's total energy from a neutron capture on Gd
 cannot be computed from a simulation, because only  the global released energy
 is known. The number of gammas and their individual energies were very poorly
 known. The scintillating buffer around the target was important
to reduce  the gamma escape. This cut was calibrated with a
 neutron source (0.4\,\% systematic error).
 The 3 cuts on the distances were rather difficult to calibrate, due to the
 difficulty of the reconstruction described above. This created a tail
 of mis-reconstructed  events,  which was very difficult to simulate 
 (0.4\,\% systematic  error on the positron-neutron distance cut). 
 The positron threshold was carefully calibrated, as shown in 
Figure 39 of \cite{bib:chooz}.
 The value of the threshold depends upon the position of the event, due to the 
 variation of solid angles and to the shadow of some mechanical pieces such as
 the neck of the detector (0.8\,\% systematic error).
 The time cut relied on MC simulation. The time spectrum happened 
to be exponential
 to $ > 20 \, \mu$s, but there was no reason for
 this (the Gamow law, which allows to
  demonstrate an exponential behavior is wrong for Gd, whose capture
 cross section  is only epithermal). The corresponding systematic
 error was estimated  to be 0.4\%. 
     
The final result was given as the ratio of the number of measured events versus the
 number of expected events, averaged on the energy spectrum. It was:
\begin{center}
           R = 1.01 $\pm$ 2.8\,\% (stat) $\pm$ 2.7\,\% (sys)
\end{center}

 Two components were identified in the background\,:
  \begin{itemize}
 \item{a correlated one, which has a flat distribution for energies bigger than
 8~MeV, and is due to the recoil protons from fast spallation neutrons. It
 was extrapolated to 1 event/day.}
\item{an accidental one, which is obtained from the measure of the single rates.}
\end{itemize}
The background was measured while the reactor was off, and by extrapolating the
 signal versus power  straight line (see Figure 49 of \cite{bib:chooz}). It is in good
 agreement  with the sum of the correlated and accidental components
measured as $1.41 \pm 0.24$ events per day. 
These numbers have to be compared to a signal of 26~events/day at full power.

The systematic errors were due mostly to the reactor uncertainties (2\,\%), to
 the detector efficiency (1.5\,\%), and to the normalization of the
 detector, dominated by the error on the proton number from the H/C ratio in the liquid
 (0.8\,\%).
The resulting  exclusion plot is shown in Figure 58 of \cite{bib:chooz}. The corresponding
 limit on $\sin^22\theta$ is 0.14 for $\Delta m^2 = 2.6 \times 10^{-3} $eV$^2$, 
and 0.2 for
 $\Delta m^2= 2.0 \times 10^{-3} $eV$^2$. Due to specific source-detector
distance of about 1 km, no limit on $\mxang$ can be set for
$\dmsq = 0.8 \times 10^{-3} $eV$^2$, due to the limited distance from the between the
 cores and the CHOOZ detector.

%% file: paloverde.tex
\newcommand{\dm}       {\Delta m^2}
\newcommand{\sinq}      {\sin^2 2\theta}
\newcommand{\nuebar}      {\bar\nu_{\rm e}}
\subsection{Palo Verde}
The Palo Verde experiment was motivated by the discovery  of the atmospheric
neutrino anomaly~\cite{Fukuda:1994mc,Becker-Szendy:1992ym,Peterson:1999dc},
which could be explained by neutrino oscillations with large
mixing angle and a mass--squared difference in the range of 
$10^{-3} - 10 ^{-2}$~eV$^2$.
The Palo Verde experiment, together with the 
\textsc{CHOOZ} experiment~\cite{bib:chooz,Apollonio:1999ae,Apollonio:1998xe}
 with a similar baseline,
were able to exclude the  $\nu_\mu\rightarrow\nu_{\rm e}$
oscillations as the dominant mechanism for the atmospheric 
neutrino anomaly. 
While Palo Verde
pursued its goal of exploring the then unknown region of
small $\dm$, results from Super-Kamiokande~\cite{Fukuda:1998mi} were published
which
favored  the $\nu_\mu\rightarrow\nu_\tau$ oscillation channel
over $\nu_\mu\rightarrow\nu_{\rm e}$.  In this section, we provide a brief 
description of the Palo Verde experiment and its final results.  Details on the
experiment can be found in its physics publications~\cite{Boehm:2001prd,Boehm:1999gk,
Boehm:1999gl,Boehm:2000prd} and in technical publications cited below.

The Palo Verde experiment was carried out at the Palo Verde Nuclear Generating 
Station, located about 80~km west of Phoenix, Arizona.  
The largest nuclear
power plant in the Americas, Palo Verde 
consists of three identical
pressurized water reactors with a total thermal power of 11.63~GW.
The detector, containing 11.3~tons of liquid scintillator for the neutrino target, was located 
 at a shallow underground site, 890~m
from two of the reactors and
750~m from the third.
The 32~meter-water-equivalent overburden
entirely eliminated any hadronic component of cosmic radiation
and reduced the cosmic muon flux.   The collaborating institutions on the experiment
were the California Institute of Technology, Stanford University, University of Alabama,
and Arizona State University.   Data were collected from 1998 to 
2000.

\begin{figure}[tbh]
       \epsfxsize=5.0in
       \centerline{\epsffile{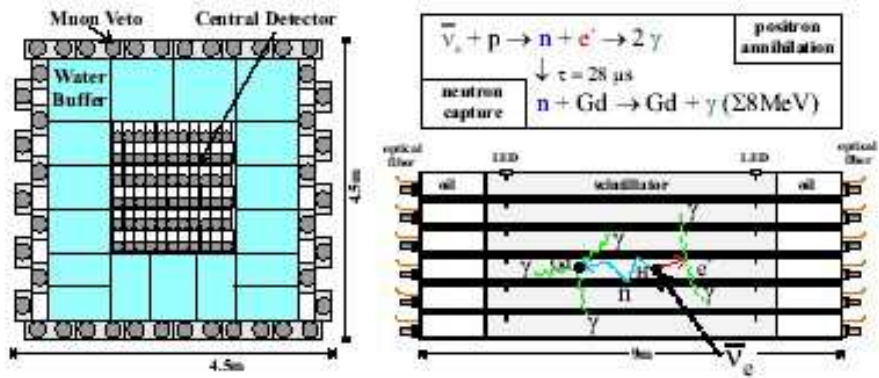}}
       \caption[View of the detector]
       {\parbox[t]{10cm}{Schematic view of the detector and the inverse $\beta$-decay
         reaction producing a {\em triple} coincidence pattern inside the detector.
        \label{pvdet}}}
\end{figure}

A schematic view of the detector is shown in Figure \ref{pvdet}.  
The central
detector was an 11 $\times$ 6 matrix of cells.  Each cell was 9 m
long, subdivided into a 740--cm central section filled with Gd--loaded liquid 
scintillator~\cite{Piepke:1999db}
and an 80--cm section of mineral oil at either end.  The cell was viewed at each end by a 5--inch PMT.   Surrounding the central detector along the long
sides were tanks providing a layer of water shielding 1 m thick.  The
water and mineral oil shielding sections attenuate gammas and neutrons
emitted from the laboratory walls and outer components of the detector,
e.g. the glass of the PMTs.
The detector
was fully enclosed by liquid scintillator detectors used to veto cosmic
muons. 

Electron antineutrinos were detected via inverse beta decay, manifested experimentally
as a prompt energy deposit due to the kinetic energy and annihilation energy of the
positron followed an average of $\sim$28~$\mu$s later by a gamma cascade of 8 MeV
total energy due
to capture of the neutron on Gd.
The central detector was segmented in order to improve the discrimination
between positrons from inverse beta decay and electrons, gammas and 
recoil protons.  The experimental signature required for a positron was
an energy deposit in one cell greater than $\sim$1 MeV (kinetic energy of
the positron) and energy deposits in adjacent cells consistent with 
those expected from back--to--back 511~keV annihilation gammas.

The event trigger was based on a so--called triple.
For triggering, the anode output of the PMT was split and sent
to two sets of discriminators, one set having a threshold corresponding
to an energy deposit of $\sim$50~keV (``LO'') in the cell and the other 
set having a threshold corresponding to $\sim$500~keV ('`HI''). 
The discriminator outputs were fed into a fast trigger processor~\cite{Gratta:1997cy}  which
generated a triple if there was a 
coincidence between at least 2 LOs and 
1 HI in any 5 $\times$ 3 cell submatrix in the detector.  The occurrence
of a triple initiated digitization of the associated event.  Readout
was carried out if two triples occur within 450 $\mu$s of each other.
Given the proximity in time of the ``prompt'' and ``delayed'' part
of a candidate event, two banks of Fastbus ADCs and TDCs had to be
used for digitization.  The trigger rates for triples and correlated triples were
approximately 50 Hz and 1 Hz, respectively. 

The muon veto hit rate was about 2 kHz.  A hit in the veto generated
5 $\mu$s of deadtime for the triple trigger processor.  Otherwise,
muon hits were only clocked and latched for readout, and the main $\mu$ veto
cuts were applied off--line.

Detector calibration for energy and position reconstruction 
was carried
out using $\gamma$ point sources, blue LED's, and a fiber optic flasher
system.  The detector simulation program used to estimate the triple
trigger efficiencies was tuned and 
checked against data taken with $^{22}\!\rm{Na}$ and $^{68}\!\rm{Ge}$ 
sources for the case of positrons and with a $^{252}\!\rm{Cf}$ source
and a tagged Am--Be source for the case of neutron capture.  Detector
stability between calibrations was monitored using the LED's and fiberoptic
flasher system.

\begin{figure}[htb]
\centerline{\epsfxsize=3.5in \epsfbox{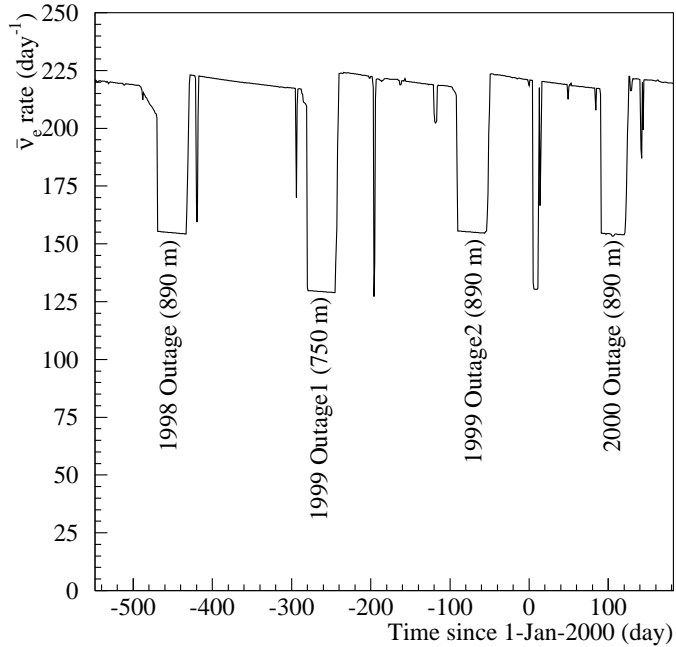}}
\caption{The calculated $\nuebar$ interaction rate in the detector
target for the case of no oscillations.  
The four long periods of reduced flux from reactor refuelings
were used for background subtraction.
The decreasing rate during the full power operation is a result
of the changing core composition as the reactor fuel is burned.}
\label{nurates}
\end{figure}

The expected $\nuebar$ flux was calculated from the reactor power and fuel 
composition.   The expected $\nuebar$
interaction rate in the whole target,
both scintillator and the acrylic cells, is plotted
in Figure~\ref{nurates} for the case of no oscillation from
July 1998 to July 2000.
Around 220 interactions per day are expected with all three units
at full power.
Four periods of sharply reduced rate occurred when one
of the three reactors was off for refueling,
the more distant reactors contributing each approximately
30\% of the rate and the closer reactor the
remaining 40\%. The short spikes of decreased rate are due to
accidental reactor outages, usually less than
a day. The gradual
decline in rate between refuelings is caused by fuel burn-up,
which changes the fuel composition in the core and the relative
fission rates of the isotopes, thereby affecting slightly the 
yield and spectral
shape of the emitted $\nuebar$ flux.

Inverse beta decay candidates were selected according to the following
criteria.
Each subevent (prompt and delayed) had at least one hit with energy
greater than 1 MeV and at least two additional hits with energy greater
than 30~keV.  The energy thresholds of this cut were chosen to select
events in the energy ranges where the triggers were efficient.  Any
event with hits greater than 8 MeV in either subevent was discarded.
The magnitude and pattern of energy deposits in the prompt subevent
 were required to resemble what was expected from the kinetic
energy of the positron and its annihilation.  The prompt and delayed 
subevents of the
event were required to be correlated in space and time.  To further 
suppress backgrounds, an event was accepted if it started at least 
150 $\mu$s after the
last veto hit and at least 3.5 MeV of energy was deposited in either 
the prompt or delayed subevent.  For the case of no oscillations,
the energy-dependent combined efficiency of the trigger and selection cuts on 
neutrino interactions is about 18\%.  The deadtime induced by the
veto--dependent  hardware and software cuts further reduced the efficiency to about
11\%.    The event rate of
$\sim$55 day$^{-1}$ after selection may be compared to an expected signal rate of
about
20 day$^{-1}$ for no oscillations. 
 Below, ``positron cuts'' refer to
cuts applied to the prompt subevent and ``neutron capture cuts''  to
the cuts applied to the delayed subevent.

Backgrounds surviving the event selection may be naturally classified as {\em{uncorrelated}}
and {\em{correlated}}.  {\em{Uncorrelated}} backgrounds are due to 
random coincidences between triple triggers within the delayed 
coincidence window.  The dominant source of {\em{uncorrelated}} events
is natural radioactivity.  {\em{Correlated}} background events are
events in which both subevents are due to the same process.  The
main source of this type of background are neutrons from muon 
spallation or capture.  These events are mainly comprised of 
{\em{proton--neutron}} events--in which a single neutron deposits
its kinetic energy by scattering from protons and is then 
captured--and {\em{double neutron}} events--in which two (typically
thermal) neutrons from the same spallation event are captured in
the detector.  The interevent time distribution for {\em{uncorrelated}} 
background 
events followed an exponential function with a time constant of 500
$\mu$s, as would be expected given the muon veto rate of $\sim$2 kHz
and the veto--dependent event selection requirements.  This time 
dependence is slow compared to that of signal and correlated 
backgrounds, hence the contribution of the 
uncorrelated background was  isolated and
studied by looking at long interevent times.   Based on these studies, the
contribution of uncorrelated backgrounds to the event rate after selection
was estimated to be about  7 day$^{-1}$.

To estimate the contribution of the correlated backgrounds, two different
approaches were used.  In the first approach, the so--called ``reactor power'' method, 
the correlation between reactor
power and observed event rate was analyzed: the signal rate would vary with
reactor power while the background rate is independent of reactor power.
In the second approach, the so--called ``swap'' method~\cite{Wang:2000prd},
the cuts on the prompt
and delayed subevent were interchanged and the resulting event rate was
subtracted from the event rate obtained with the standard cuts.    This approach
efficiently removed backgrounds such as uncorrelated events and double
neutrons, which are symmetric in the prompt and delayed subevents  while 
keeping most ($\sim$80\%) of the neutrino signal.    The remaining
important source of background, namely proton--neutron events, was estimated
from simulation.  Owing to the fact that only one reactor was refueled at a time
and the refueling time was short  (approximately 30 days every 6 months), the
swap analysis had more statistical power than the reactor power analysis, but
the  swap analysis had an additional contribution to the systematic error from
the uncertainty in the background.

Table~\ref{pvrates} summarizes the observed and corrected rates for the main
data taking periods of the experiment.    Data was accumulated at full reactor
power for four periods and one reactor out of three was off for four periods.

\begin{table*}
\caption{Data taking periods, efficiencies (including livetime), 
measured event rates $N_1$ and $N_2$ ($N_1$ - event rate
after applying the neutrino selection cuts, $N_2$ - event rate
obtained by applying the ``swapped" selection cuts),
$\Delta B_{\rm{pn}}$ - the residual contribution to ($N_1 - N_2$), 
mainly due to the proton-neutron component of the correlated
background, and estimates of the background.  $N_\nu$ and $N_{calc}$
are corrected measured neutrino event rate and calculated expected rate
for no oscillations.  Uncertainties are statistical only.}
\label{pvrates}
\begin{tabular}{|lcccc|} \hline
Period  & \multicolumn{2}{c}{1998} & \multicolumn{2}{c|}{1999-I}  
 \\
Reactor & on & 890m off & on & 750m off  \\
\hline
time (days)      & 30.4 & 29.4 & 68.2 & 21.8 \\
efficiency (\%)  & 8.0  & 8.0 & 11.5 & 11.6  \\
\hline 
$N _{1}$ (day$^{-1}$  & 39.6 $\pm$ 1.1 & 34.8 $\pm$ 1.1 & 54.9 $\pm$ 0.9 & 
45.1 $\pm$ 1.4 
\\
$N _{2}$ (day$^{-1}$ & 25.1  $\pm$ 0.9 & 21.8 $\pm$ 0.9 & 33.4 $\pm$ 0.7 & 32.0 $\pm$ 1.2 
\\
$\Delta B_{\rm{pn}}$ (day$^{-1}$) & 0.88 & 0.89 & 1.11 & 1.11 
\\ \hline
 & \multicolumn{4}{c|}{efficiency corrected results} \\
Background       & 292 $\pm$ 11 & 255 $\pm$ 10 & 265 $\pm$ 6 & 266 $\pm$ 10 
\\
$N_\nu$           & 202 $\pm$ 19 & 182 $\pm$ 18 & 212 $\pm$ 10 & 124 $\pm$ 17 
\\
$N_{calc}$      & 216 & 154 & 218 & 129  \\
\hline
\hline
Period  &  \multicolumn{2}{c}{1999-II} & \multicolumn{2}{c|}{2000} \\
Reactor  & on & 890m off & on & 890m off \\
\hline
time (days)       & 60.4 & 29.6 & 83.2 & 27.5 \\
efficiency (\%)   & 11.6 & 11.6 & 10.9 & 10.8 \\
\hline 
$N _{1}$ (day$^{-1}$  
 & 54.2 $\pm$ 0.9 & 49.4 $\pm$ 1.3 & 52.9 $\pm$ 0.8 & 43.1 $\pm$ 1.3   \\
$N _{2}$ (day$^{-1}$ 
 & 32.5 $\pm$ 0.7 & 32.6 $\pm$ 1.0 & 30.2 $\pm$ 0.6 & 30.4 $\pm$ 1.1  \\
$\Delta B_{\rm{pn}}$ (day$^{-1}$) &  1.11 & 1.11 & 1.07 & 1.07  \\ \hline
 & \multicolumn{4}{c|}{efficiency corrected results} \\
Background        & 256 $\pm$ 6 & 265 $\pm$ 9 & 249 $\pm$ 5 & 272 $\pm$ 9 \\
$N_\nu$           
 & 214 $\pm$ 11 & 161  $\pm$ 15 & 237 $\pm$ 10 & 129 $\pm$ 16 \\
$N_{calc}$      & 220 & 155 & 218 & 154 \\
\hline
\end{tabular}

\end{table*}

\begin{figure}[htb]
\centerline{\epsfxsize=3.7in \epsfbox{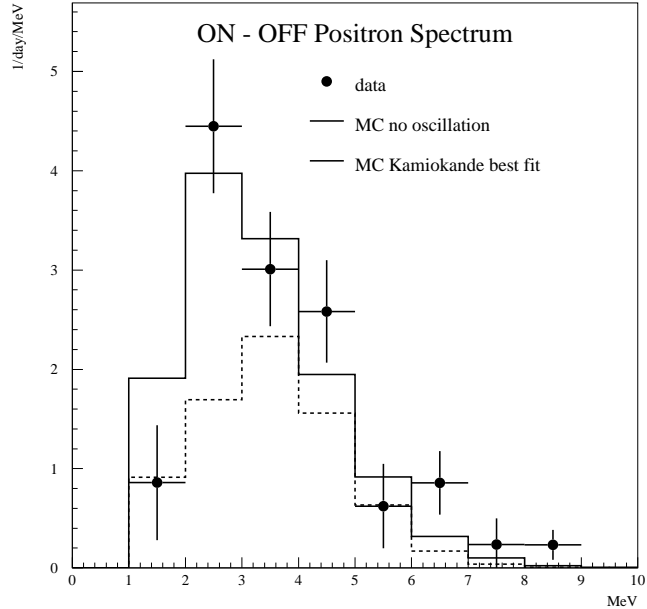}}
\caption{The prompt energy spectrum after {\em{on}}-{\em{off}}
subtraction averaged over the 4 pairs of 
{\em{on}}/{\em{off}} periods.
The histograms show the corresponding expectations for no oscillations
(solid line) and the Kamiokande best fit (dashed line).}
\label{pveonoff}
\end{figure}

Figure \ref{pveonoff} shows the energy spectrum of the neutrino candidates after 
background subtraction using the reactor power method.    Also shown are the expected
spectra for no oscillations and for oscillations based on the Kamiokande 
best fit (assuming it is due to $\nme$ transitions).  The
observed spectrum is consistent in shape and normalization with the hypothesis of
no oscillations.

Estimates of the systematic uncertainties in the reactor power and swap analyses
are presented in Table~\ref{pvsyst}.  The systematic uncertainty received contributions from the detection
efficiency and the flux calculation.  In addition, the 
reactor power method suffered a systematic error from background
variations, and the systematic uncertainty in the swap method 
had a contribution from the uncertainty in the estimate of 
$\Delta B _{\rm{pn}}$.   

\begin{table}
 \caption{Contributions to the systematic error of the {\em{reactor 
 power}} and {\em{swap analysis}}.} 
 \label{pvsyst}
  \begin{tabular}{lcc}
  Error source & on-off(\%) & swap(\%) \\ \hline 
  e$^+$ efficiency              & 2.0 & 2.0 \\ 
  n efficiency                  & 2.1 & 2.1 \\ 
  $\bar{\nu}_e$ flux prediction & 2.1 & 2.1 \\ 
  $\bar{\nu}_e$ selection cuts  & 4.5 & 2.1 \\ 
  B$_{pn}$ estimate             & N/A & 3.3 \\ 
  Background variation          & 2.1 & N/A \\ \hline
  {\bf Total}                   & {\bf 6.1} & {\bf 5.3} \\
  \end{tabular}
\end{table}

\begin{figure}[htb]
\centerline{\epsfxsize=3.7in \epsfbox{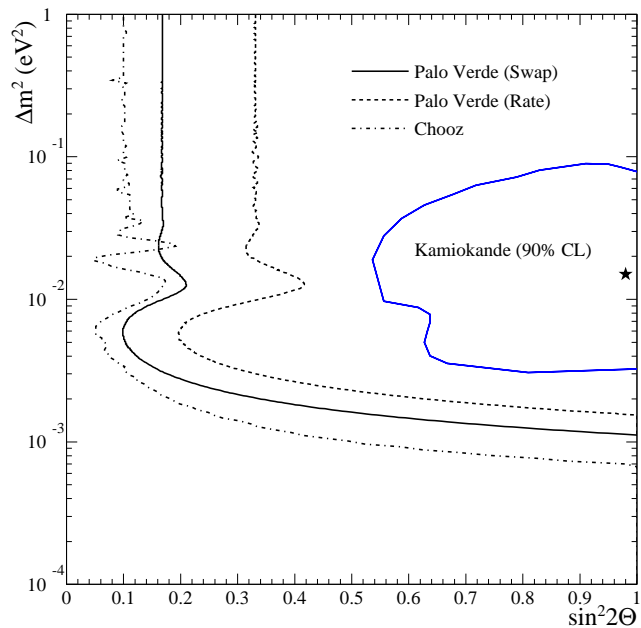}}
\caption{Regions of 
$\Delta m^2-\sin^22\theta$ plane (two flavor oscillations) excluded
at the 90\% CL by the {\em{reactor power}} analysis (dashed curve)
and {\em{swap}} analysis (solid curve).  Also shown is the Kamiokande
allowed region and best fit and the region excluded by the CHOOZ
experiment.}
\label{pvexcl}
\end{figure}

A $\chi ^2$ analysis, using the Feldman--Cousins prescription~\cite{Feldman:1998qc} for 
determining the 90\% CL acceptance region and taking into account 
statistical and systematic uncertainties, was performed on the data to determine
the regions in $\Delta m ^2 - \sin ^2 (2\theta)$ space excluded at the 90\% CL.
  The
analysis was performed for the reactor power method and for the swap method.
The results are shown in Figure \ref{pvexcl}.

%% file: kamland.tex
\subsection{KamLAND}
The KamLAND experiment, located in the underground Kamioka laboratory 
was conceived to test, in a 
wholly terrestrial experiment,
the MSW LMA neutrino oscillation solution to the solar neutrino problem by searching for 
disappearance of $\bar{\nu} _e$'s emitted by Japanese
nuclear power plants.
The power--weighted average distance between
Kamioka and Japanese nuclear power plants is approximately 175 km.
The total reactor power is such that, for
no disappearance, one would expect to see about 1.5 inverse beta
decay events per day at Kamioka in a 1 kton liquid scintillator detector.
Assuming two--flavor oscillations
with mass--squared difference $\Delta m ^2$ and mixing angle $\theta$,
the probability that a $\bar{\nu} _e$ of energy $E$ survives over a distance 
$L$ is given by 
$$ P = 1 - \sin ^2 2\theta \sin ^2 {\frac{1.27 \Delta m ^2 L}{E}}$$
for $\Delta m ^2$ in eV$^2$, $L$ in m, and $E$ in MeV.    
Given that the mean cross--section--weighted neutrino energy is about 5~MeV, KamLAND would thus
be sensitive to values of $\Delta m ^2$ below 10$^{-5}$ eV$^2$ for large
mixing angle, reasonable running time, and well--controlled backgrounds.

\par The KamLAND experiment is being carried out by a collaboration of
       universities and laboratories from the United States and 
       Japan.  The international character of the collaboration has 
       been essential for realizing an experiment which has the 
       the advantages of an excellent site, sophisticated
       detector technologies, and a team of physicists 
       with experience and expertise in mounting reactor neutrino 
       experiments.

In this section we briefly describe the KamLAND experiment and 
summarize its first published results~\cite{Eguchi:2003prl}.
While the design of the experiment allows it to measure other quantities, e.g.\ the $^7$Be component of
the solar neutrino flux, the focus of this section is exclusively on the reactor neutrino
measurement.  Further details on the detector and data analysis can 
be found in~\cite{Eguchi:2003prl}.

\begin{figure}[tbhp]
  \includegraphics[height=.3\textheight]{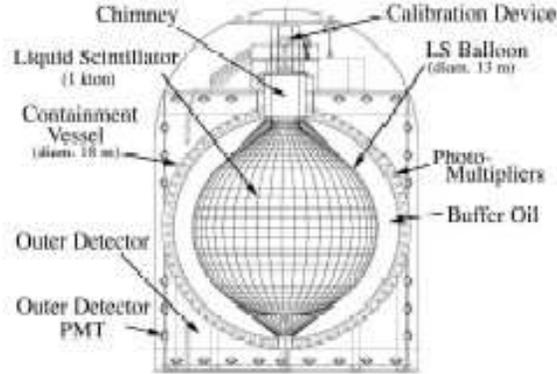}
\caption{Schematic view of the KamLAND detector \label{kldet}}
\end{figure}

A schematic view of the KamLAND detector is shown in Figure \ref{kldet}.
The experiment target consists of 1 kton of ultra pure liquid scintillator 
[20\%
pseudocumene, 80\% mineral oil (dodecane), and 1.52 g/l PPO] contained in a 13--m--diameter
transparent balloon.  The target is viewed by 1,879 PMTs, of which 1,325
are 17--inch and the balance 20--inch, providing a photocathode coverage
of about 34\%.    The PMTs are bolted to a 18--m--diameter  
stainless steel
sphere (containment vessel).  The volume between the balloon and stainless steel sphere is filled
with mineral oil, buffering the target against natural radioactivity in the PMTs,
stainless steel sphere, and surrounding rock and against fast neutrons 
generated by muon spallation outside the containment vessel.  A UV--transparent 
acrylic sheet 3~mm thick mounted just in front of the PMT faces 
acts as a barrier to radon emanating from the PMTs.  The stainless steel
sphere is enclosed by the outer detector, a water Cerenkov muon detector which consists
of a cylindrical tank of pure water  viewed by 225 PMTs.
The detector is 
located in the Kamioka Underground Observatory, in the cavern formerly
occupied by the Kamiokande Experiment.   The rock overburden exceeds
2,700~mwe, resulting in a muon rate through the experiment target of
about 0.34 Hz.

There are 16 commercial nuclear power plants in Japan, accounting for
97\% of the  neutrino flux at KamLAND.  The bulk of the 
flux--about 80\%--is due to
reactors 138--214 km away.   The mean $\bar{\nu} _e$ energy, weighted
by the inverse beta decay cross section, is about 5~MeV.  Records on 
thermal power, burn--up, and fuel--exchange are furnished to the
experiment on a continuing
basis by the plant operators.  The total reactor power 
varies by 20--30\% throughout the year (down to 50\% in 2003),
which provides a means of subtracting backgrounds by investigating the correlation of
the event rate with reactor power.

The primary trigger for the experiment 
presently requires 200 PMT hits, corresponding
to an energy threshold of about 0.7~MeV.  Following each primary trigger,
the threshold is lowered to 120 PMT hits for 1~ms to detect low--energy
delayed activity.  In addition to being discriminated for the trigger, the PMT
voltages are sampled and digitized by waveform analyzers (ATWDs).  There
are two ATWDs per PMT, allowing two--step sequential events to be fully recorded.
The ATWD sampling rate is about 630~MHz, and 128 samples are acquired per
waveform.   Signals in the outer detector are recorded
as part of the standard data  stream for offline analysis.  
The trigger rate
is about 20 Hz.  The amount of data recorded per day is about 150 GB.

Energy estimation, vertex reconstruction, and detection efficiency are calibrated
using gamma sources ranging in energy from 0.279~MeV to 7.7~MeV,
neutron sources (Am--Be), and light flashers (LEDs
and lasers).   Sources are deployed along the vertical axis of
the experiment by winch operated from a sealed glove box
at the top of the detector; in addition, there are 
blue LEDs permanently mounted
on the stainless steel sphere.  Besides detailed calibrations carried
out from time to time, the detector is monitored on a weekly basis with
gamma sources.  Natural sources, namely spallation neutrons,
cosmogenics, and natural radioactivity, are also used for detector
calibration.

The signal process is inverse beta decay
$$\bar{\nu}_e + p \rightarrow e ^+  + n.$$
The experimental signature is a prompt energy deposit of 1--8~MeV, due
to the positron kinetic energy and annihilation, followed an average of
200 $\mu$s later by emission of a 2.2~MeV gamma from neutron capture
on hydrogen.    Exploiting the delayed coincidence is key to controlling
backgrounds.

Following the first publication~\cite{Eguchi:2003prl}, we now describe event selection, 
background estimation, experimental uncertainties,
and interpretation of data in the context of $\bar{\nu} _e$ disappearance and 
neutrino oscillations.  

The cuts applied to select inverse beta decay candidates are the following:
\begin{enumerate}
\item Both the prompt and delayed subevent vertices lie within 5~m of the detector 
center.
\item The delayed subevent occurs within 0.5--660 $\mu$s after prompt subevent.
\item The prompt and delayed subevent vertices are separated by less than 1.6~m.
\item The delayed subevent energy lies between 1.8 and 2.6~MeV.
\item The delayed subevent vertex lies more than 1.2~m from the central vertical
axis.
\item  The  event occurs at least 2 s after a showering muon (energy deposit
greater than $\sim$3 GeV) and at least 2 s after, or more than 3~m away from,
any other muon track.
\item The prompt subevent energy is greater than 2.6~MeV.  
\end{enumerate}

Cut 1 is applied to control backgrounds due to natural radioactivity in the 
balloon system, the PMTs, and beyond and backgrounds due to muon 
spallation in the surrounding rock.  Cut 5
controls backgrounds from natural radioactivity in the thermometers suspended
in the central detector.  Cut 6 is made to suppress backgrounds
from cosmogenics, e.g. $^8$He and $^9$Li,  and spallation due to muons passing
through the central detector.  We have included Cut 7 to eliminate the low--energy
region in which terrestrial radioactivity is expected to contribute.

The efficiency of the Cuts 2--5 plus Cut 1 applied to the delayed subevent on
inverse beta decay events is 78.3$\pm$1.6\%.  The effect of Cut 6 is taken into
account in the livetime calculation.

\begin{figure}[tbhp]
  \includegraphics[height=.30\textheight]{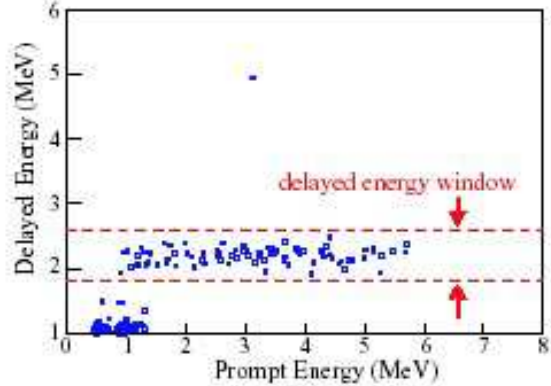}
\caption{The energies of the delayed subevents in KamLAND versus the energies of
the respective prompt subevents for events passing all selection cuts but
Cuts 4 and 7 (see text).
 \label{klcandidates}}
\end{figure}

Applied to the data set acquired March 4 -- October 6, 2002, which includes
in all 145.1 d of live time, 
54 events are selected.  Figure \ref{klcandidates} shows the energies of the delayed
versus prompt subevents after all cuts have been applied but Cuts 4 and 7.

The contribution to the event sample from accidental delayed
coincidences has been estimated by repeating the event selection with
an off--time delayed coincidence window.  The contribution is found to
be small, namely 0.0086 $\pm$ 0.0005 events.  

The residual cosmogenic background in the candidate sample has
been estimated by analyzing the time and spatial correlations between
muon tracks and event candidates.  The result of this analysis is that
the expected cosmogenic contribution to the sample is 0.94 $\pm$ 0.85 
events.

Fast neutrons from muon spallation can easily mimic inverse beta decay
events and thus are a potentially dangerous background.  Their contribution
is, however, suppressed by the fiducial volume cut (r $<$ 5~m).  The contribution
of this background has been estimated from a sample of muon events--where
the muon misses the inner detector--which contain an inner detector event
that passes the event selection cuts.    The vertex distribution for this sample
is extrapolated into the fiducial volume and then normalized to the muon track
reconstruction inefficiency.  This approach yields the result that less than 0.5
events are due to fast neutrons.

In summary, the total expected background in the event sample is 
0.95$\pm$0.99 events.

For comparing the observed events with the expected number of events
under different scenarios (no oscillations, two--flavor neutrino oscillations,
etc.), a number of quantities enter which carry systematic uncertainties.
These quantities and the estimated magnitudes of the systematic
uncertainties are: total liquid scintillator mass--2.1\%; fraction of mass within
fiducial volume--4.1\%; energy scale at 2.6~MeV--2.1\%; selection cut
efficiency--2.1\%; experiment live time--0.07\%; reactor power--2.0\%; fuel
composition--1.0\%; finite lifetime of fission products--0.3\%; neutrino spectra--2.5\%; and inverse beta
decay cross section--0.2\%.

Adding the individual contributions in quadrature, the total systematic 
uncertainty is estimated to be 6.4\%.

\begin{figure}[tbhp]
  \includegraphics[height=.35\textheight]{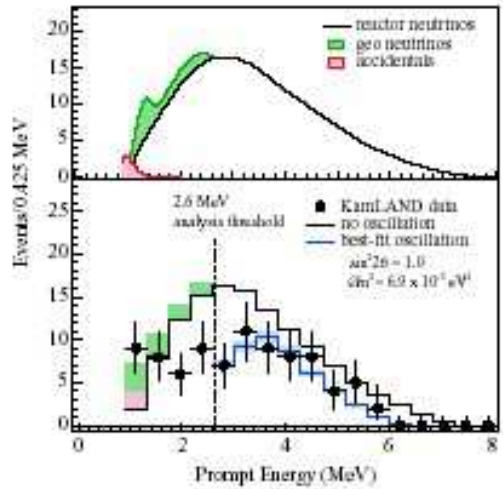}
\caption{Upper panel: Prompt KamLAND energy distributions for the expected
(no disappearance) case.  Lower panel: same as upper panel
except binned in energy and including energy spectrum of observed
events.  Also shown is the result of a fit to the observed events above 2.6
MeV in
terms of two--flavor neutrino oscillations; the shaded area superimposed
on this fit indicates the  systematic uncertainty in the best--fit result.
 \label{klecompare}}
\end{figure}

The integrated number of events expected for no disappearance is 86.8 $\pm$ 5.6
events.  Figure \ref{klecompare} shows, including 
accidental background and $\bar{\nu} _e$'s
from terrestrial radioactivity, 
both (a)~the expected prompt energy distribution for the case of no
disappearance and (b)~the corresponding distribution for the observed
events.

Accounting for the $\sim$1 background event, the 54 events observed above 
2.6~MeV 
are inconsistent
with the hypothesis of no disappearance at the 99.95\% confidence level.  

\begin{figure}[tbhp]
  \includegraphics[height=.40\textheight]{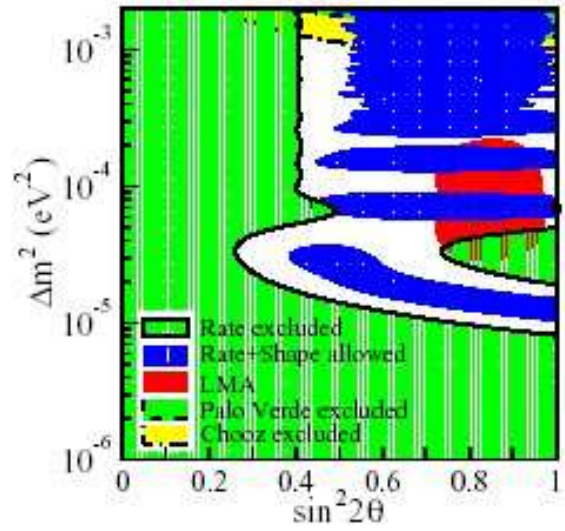}
\caption{Excluded regions of neutrino oscillation parameters for the rate analysis and allowed regions for the
combined rate and shape analysis from KamLAND at 95\% CL At the top 
are the 95\% CL excluded region from CHOOZ
and Palo Verde experiments, respectively. The 95\% CL allowed region of the LMA solution
of solar neutrino experiments is also shown. The solid
circle shows the best fit to the KamLAND data in the physical region
 \label{klfit}}
\end{figure}

One
may instead interpret  the observed events in terms of two--flavor neutrino
oscillations.  Figure \ref{klfit} shows the result for analysis of the total rate
and of the rate and shape of the energy spectrum.  The best fit to the rate
and shape gives $\Delta m ^2$ = 6.9$\times$10$^{-5}$ eV$^2$ and 
$\sin ^2 2 \theta$ = 1.0.  Assuming CPT invariance, it can be seen that
the KamLAND observation is consistent with the MSW LMA solution to the 
solar neutrino problem and further restricts the allowed LMA region.

In summary, since starting to take data in 
January 2002, KamLAND has observed reactor anti--neutrino disappearance with a 
very high level of confidence (99.95\%).  The MSW LMA solution is the 
only neutrino oscillation solution to the solar neutrino problem which is 
consistent with the KamLAND result and CPT invariance.  

KamLAND continues to take data and make improvements to the 
detector and data analysis.  These improvements include inclusion of the 20--inch tubes in the 
data analysis,
installation
of a 4$\pi$ deployment system for calibration sources, deployment of a $\beta$ source and 
additional gamma sources for better understanding of the detector energy response, and development
of a full detector simulation.  These 
improvements are expected to reduce significantly the uncertainties on 
the amount of mass
contained within the fiducial volume, the energy scale at 2.6~MeV, and the selection cut efficiency.
Improvement  in the accuracy to which the reactor--related quantities, e.g. 
reactor power, are known is also anticipated.  Achieving a total systematic error in the range of 3--4\% 
appears realistic, in which case the dominant source of error for the experiment after several years
of data taking will be statistical.

%% file: detector.tex
\section{Detector Design}
\label{sec:design}
From the discovery of the neutrinos by Reines and Cowan \cite{bib:reines}
at Savannah River to the evidence for $\bar{\nu}_e$ disappearance at
KamLAND \cite{Eguchi:2003prl}, reactor neutrino experiments have used the
same fundamental design.  A single detector, placed at a given baseline
distance from the reactor core, is used to measure the absolute flux and
energy spectrum of  $\bar{\nu}_e$ through the inverse beta decay process.
As detectors have moved further away from
the reactors over the years, it has become more important to shield them 
from background sources (primarily derived from cosmic muons) since the 
flux of reactor neutrinos falls with
the square of the baseline distance.  Experiments at baseline distances of one 
kilometer or 
more (Chooz, Palo Verde and KamLAND) have accomplished 
this by placing their detectors at significant depths underground.

\subsection{Detector Target and Buffer}
Any detector to measure reactor antineutrinos takes advantage of the
inverse beta decay reaction:
\begin{equation}
\bar{\nu}_e p \rightarrow e^+ n
\end{equation}
followed by neutron capture 
and measures a coincidence between the signal from the $e^+$ and
the neutron.  
The prompt positron will exhibit 1-8 MeV of visible energy when it
annihilates, with a minimum energy from the $e^+e^-$ masses.
The inverse beta decay reaction takes place on hydrogen,
an element which occurs in water and all forms of liquid
scintillator.   Liquid scintillator 
has been used in CHOOZ, KamLAND and earlier experiments.
Since scintillator consists of long organic molecules, 
one issue is to accurately determine the hydrogen fraction.

\par A neutron will capture on hydrogen and form deuterium giving
gamma rays with an energy 2.2~MeV.  This is the process measured
in the large KamLAND experiment.  Smaller experiments find it
advantageous to load the scintillator with about 0.1\% Gadolinium (Gd)
which has a very large capture cross section for neutrons
and also
leads to a higher energy gamma shower, 8~MeV.  In CHOOZ, 86.6 $\pm$ 1.0 \%
of the neutrons were captured in Gadolinium \cite{bib:chooz}.
However the Gadolinium lowers the optical attenuation length and, more
importantly, time dependent effects have been noticed regarding 
the optical
properties.  Time
dependent effects will need to be minimized or thoroughly understood, at
least in the way that they might manifest themselves differently in
multiple detectors.
Gd loaded targets are presently favored for a new reactor experiment,
but if the time dependent effects cannot be adequately controlled, then a
larger detector for a future 8000 t~GW~y experiment might consider not using
it.

\par There are some issues to be considered in the choice of scintillator.
In some forms, Gadolinium loaded scintillator is not compatible
with an acrylic vessel.  One example is pure PXE with 0.1-0.15\% Gd. 
The compound Gd-ACAC or
Gd-Carbolylate is being investigated at Max-Plank-Institut f\"ur Kernphysik
in Heidelberg.  The aromatic
component is $C_{16}H_{18}$  which has a low concentration of hydrogen.
The alternative is 
40\% PXE/PC with 60\% mineral oil and 0.1\% Gd, the cocktail already used by
Palo Verde.  It is compatible with acrylic, though there have
been problems of degradation of attenuation length versus time seen in
certain solutions of as much as 2~cm/day.  Aromatics and alkenes provide
more hydrogen per unit volume than pure PXE.  To compare, 10 cubic meters
of PXE weigh about 10~tons and contain $5.15~\times 10^{29}$ atoms of H,
compared with 10 cubic meters in CHOOZ, which weighed 8.54~tons and
contained $6.87~\times~10^{29}$, an increase of 33\%.

\par 
The glass in the phototubes themselves emits gamma rays from which the
fiducial volume must be shielded.  This will be accomplished with a
scintillating buffer, a non-scintillating buffer, or both.

\par The advantage of a scintillating buffer is that the positron
energy is fully contained.  No neutron will capture on the Gd in the
buffer (because there isn't any), but a positron which crosses the
target/buffer boundary will not lose any less visible energy if the
buffer is scintillating, so the minimum energy of 1~MeV is seen, i.e.
$E_{threshold} < E_{min}(e^+)$.  This reduces a systematic error which
was 0.8\% in CHOOZ to 0\% \cite{bib:kerretmc}.

\par
\begin{center}
\begin{figure}[thbp]
\vspace*{2.0mm} 
\includegraphics[width=8.3cm]{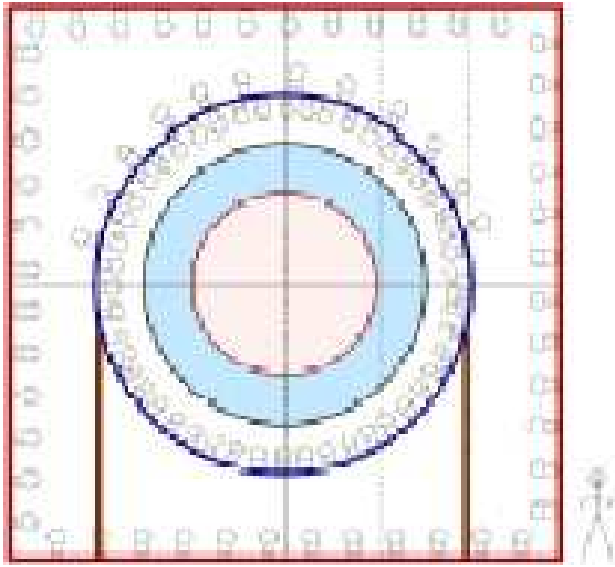} 
\caption{Conceptual detector design.  Inner volume is Gadolinium
loaded scintillator.  The second volume is non-doped scintillator.  
The third volume is
mineral oil.  Outside is veto region.}
\label{fig:det13}
\end{figure}
\end{center}

 The disadvantage of a scintillating buffer is that it contributes
to a greater accidental background from $^{40}$K radioactivity present
in the PMTs.  This would argue for the presence of both scintillating
and non-scintillating buffers.  One conceptual design of a detector
with two buffers is shown in Figure~\ref{fig:det13}.

\par
The inner volume of the conceptual 10 ton detector design shown in 
Figure \ref{fig:det13} is 1.4~m (radius) Gadolinium loaded scintillator
to serve as the neutrino target.  The size of this volume is determined
by the desired target mass, but may be constrained by the overall
size of the target hall.  This volume will serve to well define
the fiducial volume in which neutron capture takes place.  The 2nd volume
is 0.35~m scintillator without Gadolinium, the gamma catcher.  This 
distance is defined by the absorption length of gamma rays and assures
that all gammas from neutron capture in the fiducial volume are visible.
With a scintillating buffer, there is no need for a fiducial volume cut.
The 3rd
volume is 1.0~m mineral oil without scintillator, the PMT buffer.  
This boundary, a few photon interaction lengths, 
buffers the central region
against photons coming from radioactivity associated with the
phototubes.
The outside volume from 0.6 to 1.0 m thick is filled with water
has an independent
set of phototubes is used to veto cosmic ray muons and other entering
tracks.  Alternatively, there may be layers of active and passive
material to serve as a cosmic veto.
There may also be an additional meter of water in some directions
to serve as a neutron catcher.

\par The disadvantage of the second buffer is that it makes a detector
larger for a fixed fiducial volume.  The size of the detector will
affect the civil construction costs of the detector hall and tunnel,
which are expected to be the dominant costs for the experiment.  This
is especially true if movable detectors are considered, because the
path on which the detector moves will need to be that much larger.  Therefore
the advantage of the buffer will have to be evaluated quantitatively together
with the construction costs for any site.  For example, in a design with two 
buffers, 2.7 m of diameter is needed for the buffers.  This means that there 
is a maximum of 15 ton target volume if the detector needs to be moved through
a tunnel with a diameter of 6 meters.

\subsection{Mechanical Structure}
The structure will be a significant part of the mechanical design of the
detector.  There will need to be an optical structure, to separate the
target from any buffer regions and another one to separate
the buffers from each other.  
The optical structures will perhaps be made from acrylic.  KamLAND and Borexino have had
success using nylon balloons, but a flexible
container is probably not a good way
to maintain fiducial volumes to better than the 1\% accuracy required for this 
experiment.  
There will also need to be an
opaque structure to separate the buffers from the veto region and to support
a frame which holds the PMTs.
This will perhaps be made from steel.  The acrylic vessels will need 
to have access ports through which the volumes can be filled and also in which
sources can be put in and moved around.  All structures will need to be
designed so that proper buoyancy can be maintained during fill.

\subsection{Muon veto}
A sufficient volume of water is needed to shield the detector from neutrons
created by muon cascades outside the veto region.   The veto region will
need to have enough phototubes to efficiently tag all muons which enter
the muon region.  If movable detectors are chosen, it is not foreseen that
the veto detector would be movable.  But it would then need to be designed
so that the fiducial target could be removed.

\par Another design for a muon veto involves layers of passive shielding,
such as concrete, and a layer or layers of high efficiency planar detectors
to measure the incoming muon.  

\par There are reasons to also consider passive shielding outside the veto region,
depending on the conditions of local radioactivity and the orientation of
any shafts or tunnels.

\subsection{Liquid Handling System}
There will need to be a safe and careful system for handling and testing
the different liquids.  There will need to be very accurate measurements of
the volume of each liquid during fill, particularly in the target region.
Spills of liquid scintillator will need to be avoided, particularly with
regards to any contamination of ground water.  

The conceptual design for the detector shown in Figure~\ref{fig:det13} is
accompanied with the need to handle four different liquids:  Gd loaded
scintillator, scintillator, mineral oil, and water.  All will need to be
maintained with well-known optical properties.  One possible solution for
the issue of optical changes in Gd loaded scintillator is mixing through
recirculation, with or without filtering.  A recirculation system
will have to be designed to evenly affect each liquid throughout its volume.
Filtering might be designed to deal with optical degradation due
to organic compound production, but it may also remove Gd, and
a system which removes and precisely reloads solutions in 
the liquid may be difficult
to accomplish.  
Another concern about recirculation is the introduction of microbubbles,
which could both change the optical properties as a function of time and also
lead to small time-dependent density changes.  The advantages and disadvantages
of recirculating each volume will need to be carefully reviewed, probably
by extensive testing.

\subsection{Detector Shape}

\par The ideal detector has a spherical design.  This design
 offers the lowest ratio of surface to volume, which
implies the least number of photomultipliers per ton, the lowest surface
radioactivity, the most buffer material to absorb external neutrons,
the lowest
cross section to cosmic rays, and hence the lowest total mass per ton of target.
With SNO, KamLAND, MiniBooNE and Borexino, there is a great deal of
experience with the design and construction of spherical liquid 
scintillator detectors.  They also have the simplest parameterization of
optical effects, which will be even more important due to the low attenuation
length of Gd loaded scintillator.
\par
\begin{center}
\begin{figure}[thbp]
\vspace*{2.0mm} 
\includegraphics[width=8.3cm]{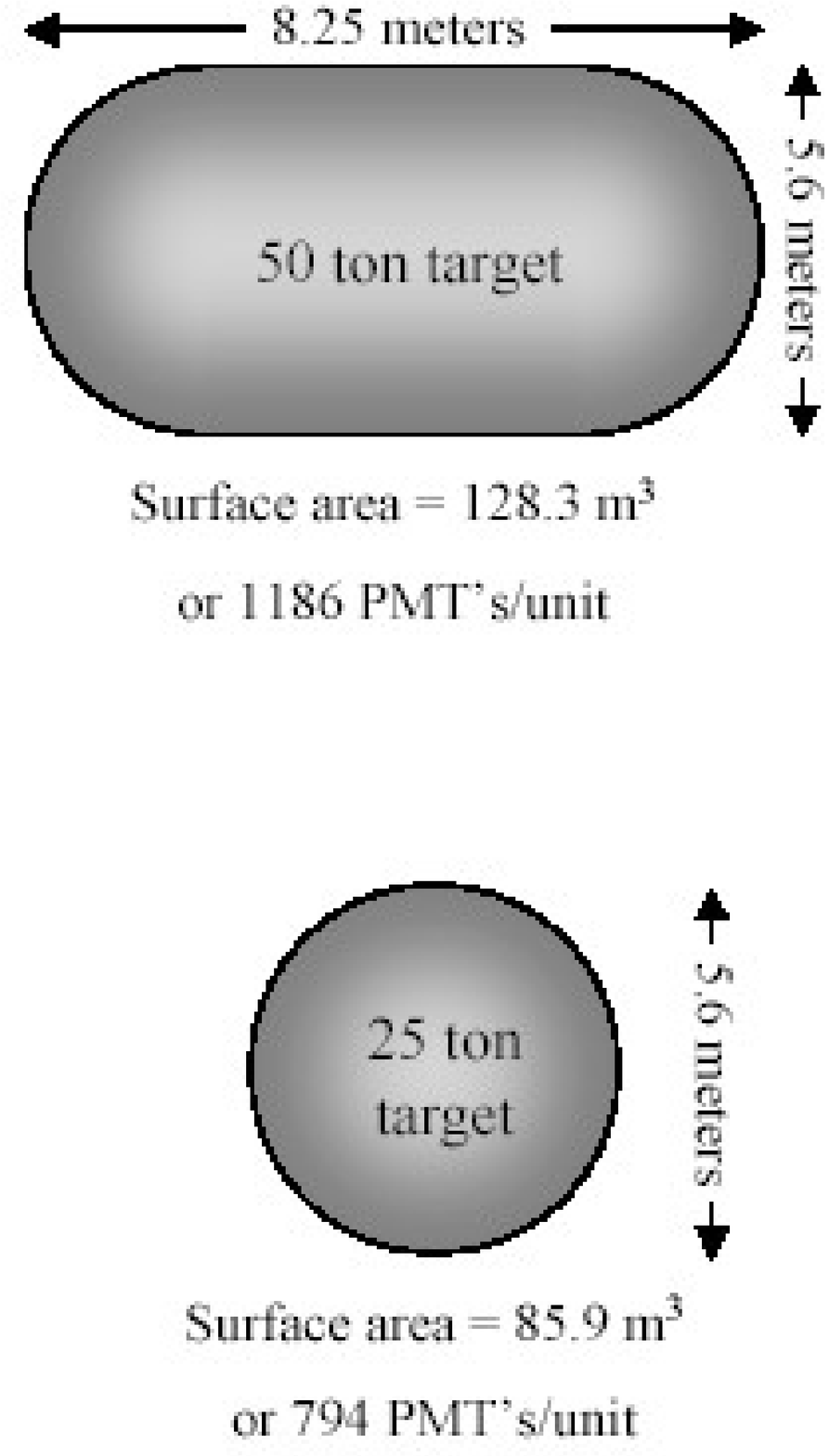} 
\caption{A comparison of spherical and cylindrical detectors}
\label{fig:detsphere}
\end{figure}
\end{center}

 Nevertheless, if movable detectors are chosen,
cylindrical detectors offer some advantages.
The
target mass must be moved in an area less than 6 meters squared (the 
maximum size of most economic tunnels).  To achieve a 50~ton or
greater mass, the detector must be cylindrical.  As a comparison,
in Figure \ref{fig:detsphere} there is a 50~ton cylindrical target and a
25~ton spherical one.  Both detectors have
a 70 cm pure oil buffer, and 20 cm region
to shield the PMTs, together with 20\% PMT coverage.  Assuming that
only one near detector is needed in each case, then for a 50~ton active
volume for the far detector, 2372 PMTs are needed for 2 cylindrical detectors
compared to 2382 for 3 spherical ones.   For 75~tons, the result is 3156
for 2 cylindrical detectors and 3176 for 4 spherical ones.  The costs of
the vessels cannot be ignored, but in the limit where the detectors costs
are dominated by PMT and channel counts, this would argue toward 
multiple spherical detectors over larger cylindrical ones.

\subsection{Movable Detectors}
\label{sec:move}
\par The purpose of designing movable detectors is to be able
to move the identical
far detector to the near detector location for part of the
running period and have
a head-to-head calibration of the relative number of events.
Assuming that there is no additional error introduced by
the act of moving the detectors, the
 uncertainty of this efficiency ($\varepsilon$) is
\begin{equation}
\sigma_\varepsilon = \varepsilon \sqrt{\frac{2}{N}}
\end{equation}
where N is the number of events measured during the calibration.

\par Designing the detectors to be movable will certainly have a great
effect on the detector design.  However, it will have an even larger
effect on the tunneling requirements.  Let's consider the options for
an experiment with two detectors and a flat overburden.  Three scenarios
are shown in Figures~\ref{fig:move1},~\ref{fig:move2},~\ref{fig:move3}. 
In Scenario One, there are two shafts and a tunnel connecting them in
which the detector can be moved.  The shaft for the near detector may
be placed away from existing infrastructure such as substations 
and parking lots, outside the security 
fence.  This is probably the easiest solution for detector design, but
the most expensive for civil construction.  The second solution takes
advantage of sloped access and the fact that the near detector does not
need to be as deep as the far detector.  The angle of approach for the
access ramp could be from any convenient direction.  In scenario three
there would be two shafts and a surface detector transport procedure.
The detector would be raised to the surface by a crane or elevator system
and transported on surface rails.  This may be the cheapest option, but
it may also be a more complex and risky procedure.
\begin{center}
\begin{figure}[thbp]
\vspace*{2.0mm} 
\includegraphics[width=12cm]{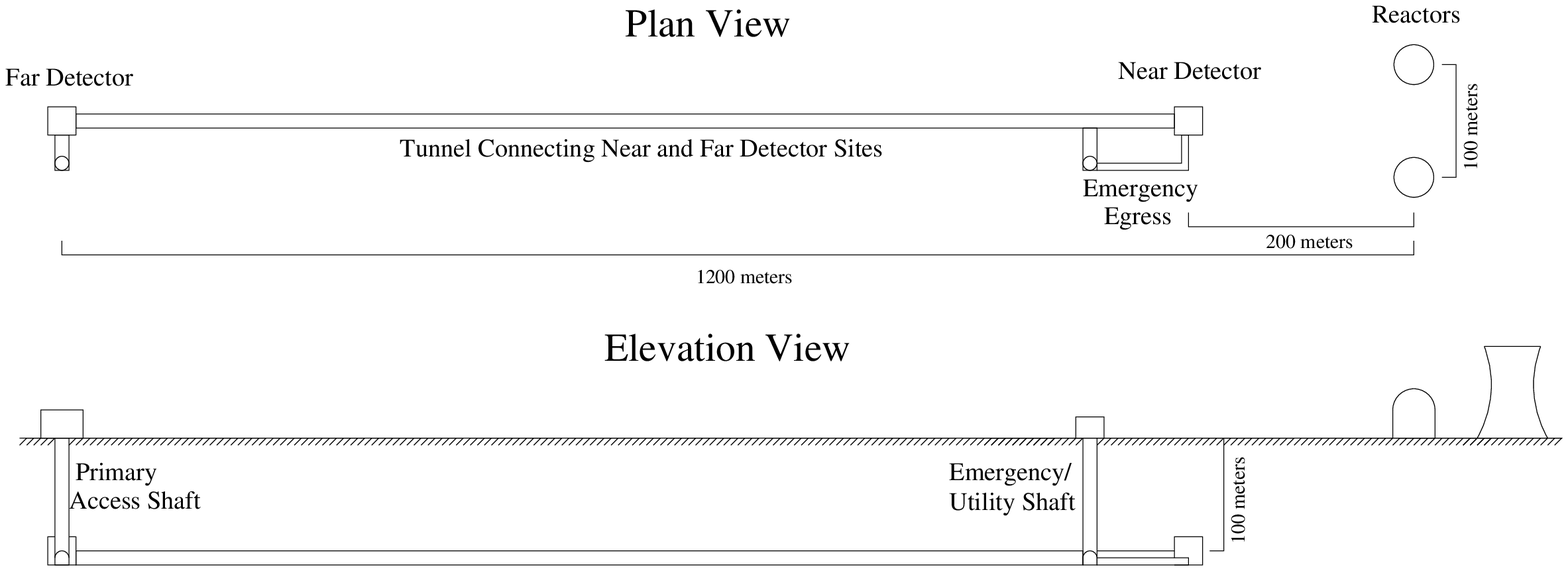} 
\caption{A scenario with two shafts and a connecting tunnel}
\label{fig:move1}
\end{figure}
\begin{figure}[thbp]
\vspace*{2.0mm} 
\includegraphics[width=12cm]{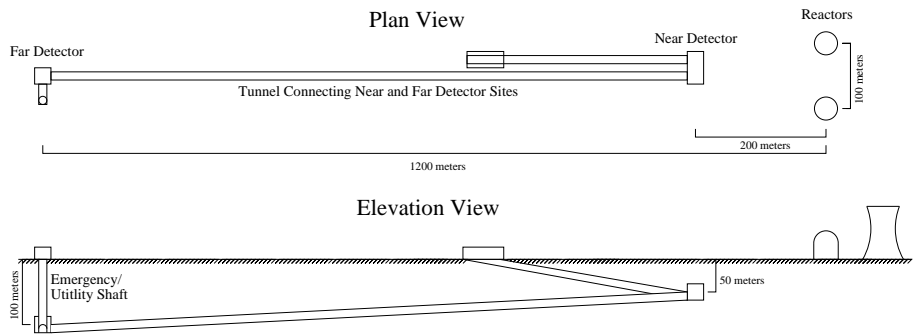} 
\caption{A scenario with sloped access}
\label{fig:move2}
\end{figure}
\begin{figure}[thbp]
\vspace*{2.0mm} 
\includegraphics[width=12cm]{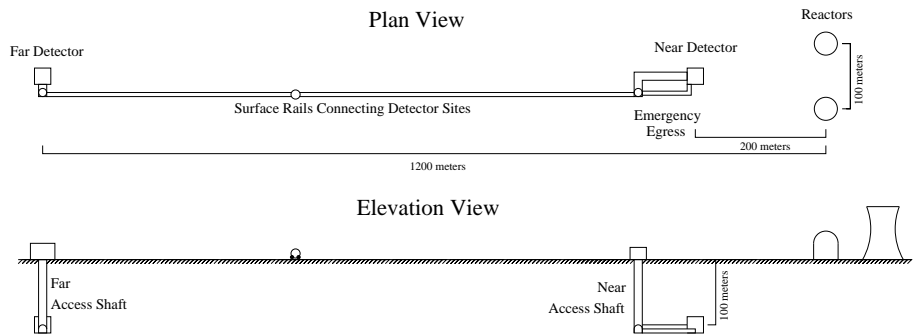} 
\caption{A scenario with two shafts}
\label{fig:move3}
\end{figure}
\end{center}

\subsection{Multiple Detectors}
\begin{center}
\begin{figure}[thbp]
\vspace*{2.0mm} 
\includegraphics[width=10cm]{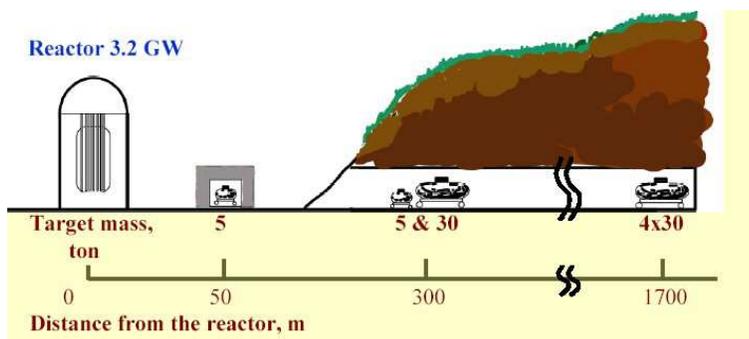} 
\caption{Possible layout for multiple detectors.}
\label{fig:detmult}
\end{figure}
\end{center}

Assuming that backgrounds and deadtime issues are under control,
the statistical power of the experiment is driven by the size of the 
far detector.  The near detector will have a larger event rate by
a factor of $(L_2/L_1)^2$, so for the same statistical power, it might
be reasonable to make the near detector smaller.  Any difference in
size between the two detectors, however, compromises the ability to
cancel systematic errors, and this experiment's goal is to measure the
neutrino flux with a much better systematic error than has previously
been achieved.  This dilemma has led to consideration of several other
detector configurations:
\\
{\bf Multiple Far Detectors}
It might be desirable, for example, to make a 20~ton near detector, and multiple identical
20~ton far detectors, perhaps five.  
This would be a way to get more fiducial mass at
the far detector, and also a possible way to stage some of the experiment.
The multiple far detectors would be a smaller fiducial mass than a single
larger detector with the same channel count.  However, they would provide
additional checks on some of the systematic errors of multiple detectors.
\\ {\bf Multiple Locations}
Another variation of this idea is shown in Figure \ref{fig:detmult} where
multiple identical detectors are used.  In this example, a small 5 kiloton
detector is placed near the reactor, and another one a few hundred meters away,
along with a larger 30~ton detector, which is repeated at 1.7 km.  Again,
there are multiple opportunities to study possible systematic errors and
staging possibilities.   This scheme provides additional opportunities for
cross calibration of each detector.
\par
One possible attraction of the multiple location idea is that the optimization of
distance for the 400~t~GW~y experiment could be at a different location
than for the more sensitive 8000~t~GW~y experiment.  This is because the
optimum location, discussed in Section~\ref{sec:location} is different
for the rate test and the energy shape test.  The rate test is more
powerful for the lower statistics and the shape test for the higher 
statistics experiment.  Unfortunately, the
optima discussed in Section~\ref{sec:location} does not seem to lend itself
to a solution that involves staging detectors in a natural way.

%% file: calibration.tex
\section{Calibration}

\label{sec:calibration}

\subsection{Introduction}

Calibration provides the information required to: (a)~reconstruct the energies
and positions of the prompt and delayed sub-events, (b)~tune the experiment's
simulation software to match as closely as possible the data without
introducing biases and so compute the detection probability as accurately as
possible, and (c)~reliably estimate the magnitude of uncertainties in detector
response. Although it will not be further discussed in what follows,
calibration should also enhance the capability to discriminate between signal
and background processes, e.g., through accurate characterization of detector
response to neutrons with energies typical of spallation products and through
development of a trigger to record muon-produced 
neutron capture in the experiment.

What calibration accuracy is required? Based on the record of previous
experiments similar in design to those being considered for a next generation
reactor neutrino experiment, the contribution from calibration to the overall
uncertainty for a single detector can be limited to a few percent. The
calibration program outlined should aim toward controlling the single
detector uncertainty at this level. To achieve an overall systematic
uncertainty on the relative far-to-near detector event rate at the percent
level, it is further required that the calibration of the near and far
detectors--as well as detector design, operation, and data analysis--be as
identical as possible.

\subsection{Calibration System Design Considerations}

A calibration program capable of achieving the accuracy specified above will
have the following characteristics:

\begin{itemize}
\item Calibration sources that are precisely deployable throughout the entire
active target region of each detector. At representative locations in each
detector, it should be possible for some calibration sources to pass through
and beyond the outer boundary of the active target. The requirement that
sources be deployable throughout the entire target region and somewhat beyond
its boundary at representative locations can be relaxed if events in the
boundary region are excluded from analysis. Whatever region is used for
analysis, it should be considered important that calibration sources can be
placed at any point in this region. The uncertainty in the position of the
calibration source should be about 5 cm or less, assuming the lineal
dimensions of the detectors are on the level of several meters; and the
deployment system and calibration sources should be designed to minimize
shadowing and absorption. \ These requirements require novel mechanical
designs for source insertion and removal that may be quite challenging.
\ Indeed, considerable R\&D may be required to establish that calibration
methods successfully demonstrated on the $\sim$12 ton CHOOZ\ detector
can be scaled up to 50+ ton devices being contemplated.

\item Multiple calibration sources that include the following:

\begin{enumerate}
\item Point gamma sources spanning the energy range from inverse beta decay
threshold to the highest achievable energy for calibrating energy estimation,
vertex finding, and detection efficiency. Readily available radio--isotopes
that would be suitable include Cs--137 (0.662 MeV), Ge--68 (2~$\times
$~0.511~=~1.022 MeV), Zn--65 (1.116 MeV), and Co--60
(1.333~+~1.173~=~2.506~=~MeV). Sources with energies in the range of 5--8 MeV
would also be desirable, but are typically difficult to fabricate and deploy
due to short half lives and low branching ratios for the gammas of interest. A
development effort to design and deploy gamma sources in this energy range may
well be worthwhile. At least one of the gamma sources should have its visible
activity known to $\sim$1\% for the purpose of measuring detection efficiency,
and a range of gamma energies is needed in order to help measure the effects
of quenching and Cerenkov radiation on light output.  The Ge--68 source is
particularly attractive because it probes the detector response at inverse
beta decay threshold.

\item $\beta$ source. Using gammas alone, it may prove difficult to determine
the absolute energy scale for positrons to better than 1\%. Some studies of
future reactor neutrino experiments have assumed that the absolute energy
scale is known to 0.5\%. To achieve this level of precision, use of a $\beta$
source, preferably a $\beta^{+}$ source, should be planned. Controlling
uncertainties in the $\beta$ energy spectrum from capsule shadowing and
absorption in the encapsulation materials will be a challenge, however. \ A
deployment of a Ge-68 source in solution with the liquid scintillator has been
suggested~\cite{bib:piepke}
to solve to this problem and it would also
provide a truly homogeneous and \textit{in situ} calibration for gamma rays.

\item Tagged neutron source for measurement of the neutron detection
efficiency. Am--Be is a good candidate for such a source. The 4.4~MeV gamma
can be used to tag single neutron emission. Corrections for the differences
between the Am-Be neutron energy spectrum and the inverse beta decay neutron
energy spectrum would be made using simulation. Important also is that such a
source provides additional gamma calibration points: (i)~2.225 MeV from
capture of neutrons on protons and (ii)~4.443~MeV provided that a significant
fraction of the neutrons can be moderated before they enter the scintillator.
Moderated Am--Be sources have been used successfully for energy calibration in
other experiments. \ The CHOOZ experiment also made good use of Cf neutron
sources. \ These produce a continuous neutron and gamma spectrum but offer the
advantages of higher rate for studies of homogeneity.

\item Variable intensity light flasher. Possible sources include UV lasers and
UV LEDs. As practically point-like sources of fast pulses of controllable
amplitude, such sources are valuable for calibrating vertex reconstruction
(which is usually based on PMT timing with charge-dependent corrections),
measuring PMT gain and interpolating the detector response between gamma
calibration points.
\end{enumerate}

\item Integration of the calibration system as an integral part of the data
acquisition, event reconstruction, and detector simulation software: \ A large
fraction of events read out over the course of the experiment will be various
calibration triggers. \ The acquisition of this data must not adversely affect
dead time for the experiment. \ Careful design of triggering and run control
is essential, and processing of calibration data must take place quickly to
help identify shifts in detector performance.

\item Simulations.  Because it is not possible to deploy a source that mimics inverse beta
decay in the reactor neutrino energy range, detector simulation ultimately
plays an important role in estimating the detector response to this process.
As such it requires an accurate and detailed description of the detector 
(regarding materials, geometry, and optics) and of the calibration sources and the
devices used to deploy them. Generation and transport of photons to the PMT
photocathodes must be modeled accurately, as should also be the digitization
of PMT pulses and effects of discriminator thresholds. Simulated data must be
reconstructed with the same software as real data. 
The event generators for inverse beta decay and the
calibration processes must be accurate and detailed.

\item Construction of identical calibration programs for the near and far
detectors in both hardware and software.   So far as is possible, the
calibration sources should be the same, or at least fabricated by identical
means; and the deployment systems should likewise be identical in design. If
the near and far detectors have different sizes, designing the deployment
systems to be the same to within overall scale should be a consideration.
\ The scope and magnitude of the calibration programs should be planned so
that any calibration carried out for one detector can also be carried out for
the other at nearly the same time.
\end{itemize}

\subsection{Concluding Remarks}
What is ultimately required for adequate calibration will not
be fully known until experience has been gained on the detectors
running under normal conditions.
However, given past experience and the experimental 
design and goals, it is clear that the baseline calibration program
should have the capability of precisely deploying well--understood 
calibration
sources at any desired position throughout--and preferably beyond,
in the case of gamma and neutron sources--the active volume of 
the detector.  Control of systematics will be aided by the ability to 
deploy the same sources in both the near and far detectors.   The
calibration program effort should be matched by an effort to develop
a full detector simulation tuned and checked against the calibration data;
this simulation program can then be used to extend understanding of the 
detector into areas for which calibration data is not available.

%% file: background.tex
\section{Detector Overburden and Backgrounds}
\label{sec:bac}
Backgrounds will be an important consideration in site selection and
detector design. They can be classified into two distinct sources:
(1)~internal backgrounds associated with radioactivity in the materials
of the detector and lab site, and (2)~external activity associated
with cosmic rays and their interaction daughters. The former is
a well-understood technology that has been solved many times over
several decades by a combination of materials selection, self-shielding,
and signal enhancement via addition of neutron absorbers that boost
the energy of the neutron capture reaction. It is not expected that this
will be a major problem.\\

Cosmic ray activity is typically more serious and in most cases requires
an underground location of at least at least a few tens of meters of
water equivalent (m.w.e.) for even rather small detectors of less than
1 ton. The limiting factor is typically the flux of neutrons associated
with muon nuclear interactions in the detector and in the surrounding
rock. Table~\ref{RS:T:RxSummary} is a brief summary of some previous
reactor experiments done over the last 25 years. As can be clearly
seen, longer distances have necessitated larger detectors, which have in
turn required underground sites to reduce the cosmic
ray muon flux and hence the flux of external neutrons.\\

\begin{table}[htbp]
\caption{Summary of some previous reactor experiments}
\label{RS:T:RxSummary}
\begin{center}
\begin{tabular}{lccccl}
Name & power$_{th}$ & mass & distance & depth & comments \\
 & (GW) & (tons) & (m) & (mwe) & \\ \hline
ILL~\cite{RS:R:Kwon0} & 0.057 & 0.32 & 8.76 & 7 & $^{3}He$ + scint. PSD \\
Gosgen~\cite{RS:R:Zacek0} & 2.8 & 0.32 & 38/46/65 & 9 & same as ILL\\
Rovno~\cite{RS:R:Afonin0} & 1.4 & 0.43 & 18 & & Gd scint.\\
Krasnoyarsk~\cite{RS:R:Vidyakin0} & 1.6 & 0.46 & 57/231 & 600 & $^{3}He$ only\\
Bugey 3~\cite{RS:R:Abbes0,bib:bugey} & 2.8 & 1.67 & 15/40/95 & 23/15/23 & $^{6}Li$ + scint. PSD\\
Savannah R.~\cite{RS:R:Greenwood0} & 2.2 & 0.25 & 18.2/23.8 & $\sim 10$ & Gd + scint. PSD\\
CHOOZ~\cite{bib:chooz} & 4.4 & 5 & 1000 & 300 & Gd + scint. \\
Palo Verde~\cite{Boehm:1999gl} & 11.6 & 11.3 & 800 & 32 & Gd + scint.\\
KamLAND~\cite{Eguchi:2003prl} & $\sim 80$ & 408 & $\sim 180,000$ & 2,700 & scint. \\
\end{tabular}
\end{center}
\end{table} 

The final design of this experiment will be a trade-off between several
conflicting goals: (1)~the desire to have the near and far detectors
be as identical as possible to allow cancellation of systematic errors
due to detector geometry, (2)~the desire to have the near
detector be small to avoid muon-related backgrounds and expensive
overburden requirements, and (3)~the desire for a far detector
of size large enough to give an acceptable event rate. Backgrounds
will be a major consideration in the final design and so we have developed
some general constraints based on the measurements of previous experiments.\\

\subsection{Types of Backgrounds}

The double coincidence afforded by 
$\overline{\nu}_{e} + p \rightarrow n + e^{+}$ followed by
$n + p \rightarrow d + \gamma$ requires
that two events appear within a given time and space window, typically
25-200 $\mu$s and $\leq 1$ m. The space requirement is usually enforced
either by vertex fitting or detector segmentation. Events that are 
inherently ``singles''
are therefore strongly suppressed as they must appear by accidental
coincidence in the correct window. Examples are internal radioactive decay,
stopping cosmic ray muons, and external gamma events.\\

Correlated backgrounds are more serious. Such events can mimic
the reactor anti-neutrino signal and there can be significant
uncertainties associated with the rate. The majority of correlated
backgrounds come from events associated with cosmic ray activity.
Examples are fast neutrons from muon nuclear interactions, muon
capture reactions from stopped muons, and 
excited nuclei from muon spallation reactions that may de-excite 
by neutron emission.\\

In many cases experiments have been placed deep enough underground to
sufficiently suppress the cosmic ray muon flux (e.g. KamLAND). In 
other cases a segmented detector was used to allow more precise
cuts on vertex location and partial detector vetoing (e.g. Palo
Verde and Bugey). In cases where the detector is small and close to the
core, only minimal shielding was required (e.g. Savannah River).
Table~\ref{RS:T:RxSummary4} shows the effective detection rate for
some previous experiments. This is the daily rate of neutrinos detected
(corrected to a distance of 100 m)
while the detector is operating per ton of fiducial target per
GW of reactor power~\footnote{There are
differences expected on the order of 5\% due to fuel composition}.
Variations of a factor of two are seen, mostly due to the overall
efficiency constraints imposed by the necessity to reject backgrounds.
Detectors which are segmented and near the surface (ILL, Gosgen, Bugey,
Palo Verde) tend to have less efficiency than monolithic ones (Savannah
River, CHOOZ, Rovno, Krasnoyarsk, KamLAND) with sufficient shielding
relative to the detector size.\\

\begin{table}[htbp]
\begin{center}
\begin{tabular}{lcc}
Name &  effective rate at 100 m \\
 & ($d^{-1} t^{-1} GW^{-1}$)\\ \hline
ILL & 15.7 \\
Gosgen & 14.4/14.5/13.6 \\
Rovno & 46 \\
Krasnoyarsk & 35/41 \\
Bugey 3 & 22/19/19 \\
Savannah River & 50/54 \\
CHOOZ & 51 \\
Palo Verde & 12 \\
KamLAND* & 61 \\
\end{tabular}
\caption{Effective $\overline{\nu}_{e}$ detection rate. (*)
The KamLAND rate is based on the expected rate of 86.8 events in 141 days. }
\label{RS:T:RxSummary4}
\end{center}
\end{table}

In the discussions below we use the experience of previous
reactor experiments and estimate depth and shielding requirements for both
a ``near'' ($100$~m) and a ``far'' ($1.5$~km) detector. It is assumed
that the basic detector module is roughly 10 tons, with the far site
requiring several such modules and the near site one.\\

\subsubsection{Accidental Coincidences}

Single, uncorrelated events may result from radioactive decay from
internal materials such as steel and glass, external gammas from
the shielding or surrounding laboratory, or from cosmic ray muons 
if the depth is shallow. The coincidence rate $R$ for two uncorrelated 
events with rates $R_{1}$ and $R_{2}$ to occur in a time window $w$ is
given by~\footnote{This assumes that $Rw \ll 1$}:\\

\begin{equation}
\label{RS:E:1}
R = R_{1}R_{2}w
\end{equation}

This rate must be compared with expected neutrino interaction rates
for a typical 10-ton scintillator detector. For the near (far) module this
would typically be about 1000 (4.5) events/day/GW with 100\% efficiency.
A typical reactor is assumed to be 3 GW thermal, giving a signal rate of
3000 (13.5) events/day. 
Using equation~\ref{RS:E:1} and setting a criterion that the accidental
rate should be less than 1\% of the actual event rate, restrictions
on various sources of ``singles'' can be estimated.\\  

\noindent {\em Cosmic Ray Muons:} Cosmic ray muons can produce
accidental coincidences in several ways: (1)~straight-through muons
mimicking positrons or neutron capture (unlikely), (2)~decay of
stopped muons giving an $e^{\pm}$ followed by an {\em accidental}
coincidence of a muon capture that gives a neutron, and (3)~a stopped
muon decaying to an $e^{\pm}$ followed by the accidental coincidence
of a neutron from spallation. In (2) and (3), the rate must be
less than the rate of accidentals from (1) since they are initiated
by somewhat rare reactions from (1). Correlated background from these types of events
is treated in a later section.\\

\begin{figure}[t]
{\psfig{file=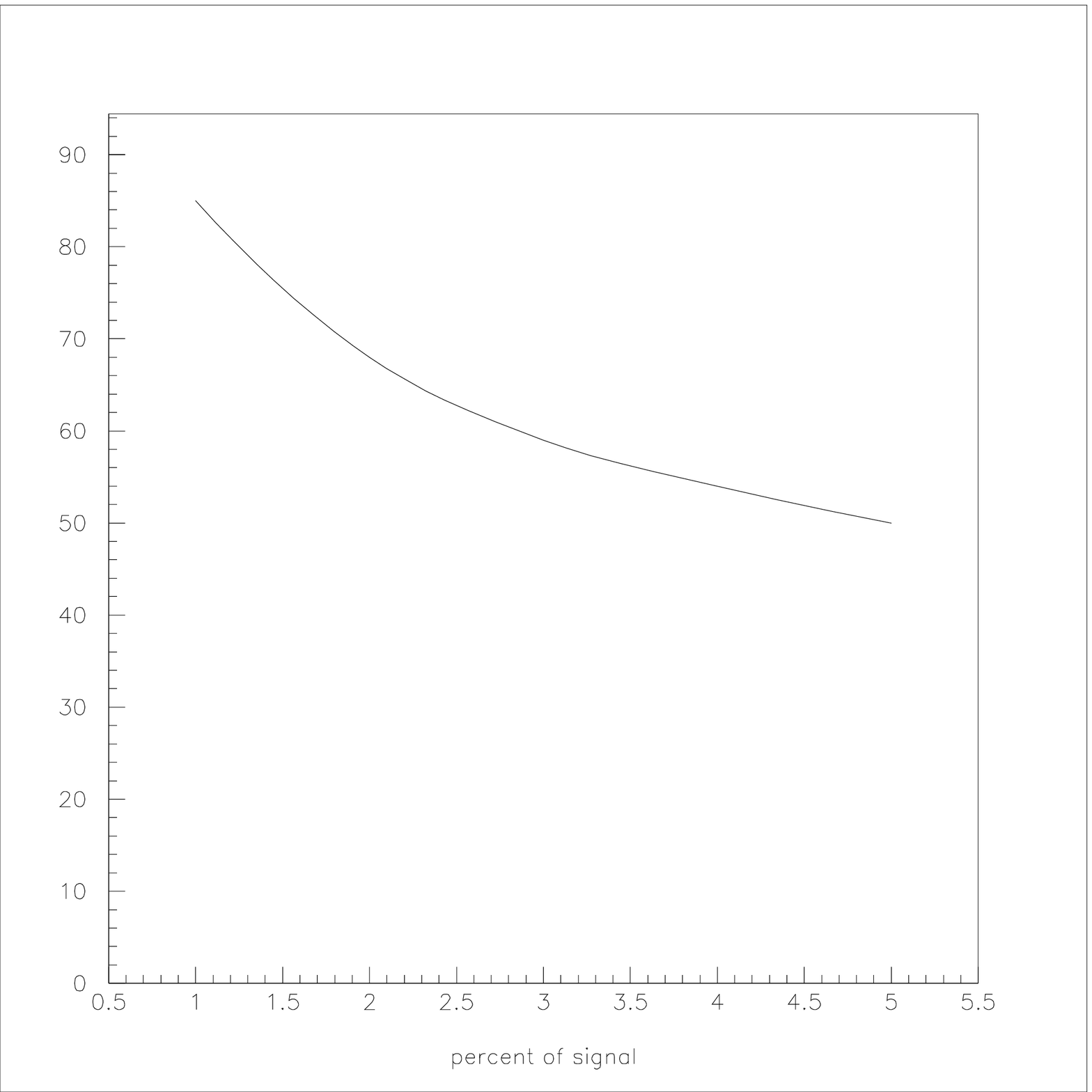,height=189pt,width=189pt}}
{\psfig{file=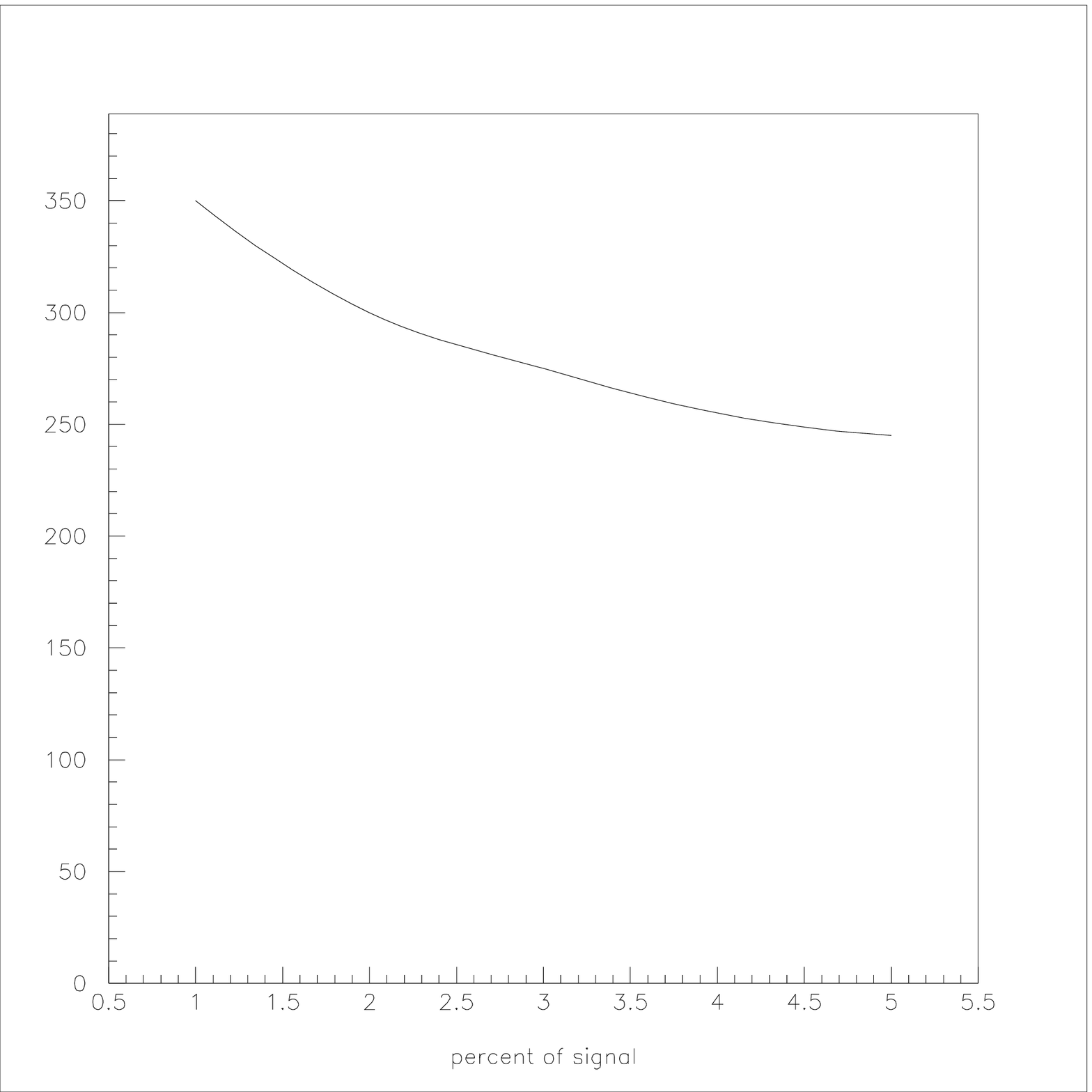,height=189pt,width=189pt}}
\caption{The rate of muon accidental coincidences as a fraction of the
expected signal as a function of depth. Left (right) is for the near (far) detector.}
\label{RS:F:mu}
\end{figure}

Taking $R_{1} = R_{2} = R_{\mu}$, where $R_{\mu}$ is the muon rate
through the detector, and a typical neutron capture time of $25 \mu$s which
would imply a window of roughly 100 $\mu$s for 98\% efficiency,
then allowed muon rate through the detector is 1.9 Hz (0.12 Hz) for the near
(far) detector. This is to be adjusted for muons that can be tagged.
Most all previous experiments had muon vetoes (see Table~\ref{RS:T:RxSummary3})
that were better than 95\% efficient in tagging muons. Therefore the acceptable
muon rates can be conservatively adjusted up by at least a factor of 20. This
would correspond to a depth of about 85 m.w.e. (350 m.w.e. for far) for a 
10 ton module. Figure~\ref{RS:F:mu} shows the percentage of the signal this
background would be as a function of depth.\\

This estimate is conservative in that it does not take into account the
low probability that a muon is misidentified as a positron and/or capture
gamma. In addition, vertex cuts may further reduce the possibility of a chance
coincidence. What can be said is that at these depths single muon accidentals will
not be a problem as long as the neutron capture time is kept short by
the addition of gadolinium or other absorber.\\

\begin{table}[htbp]
\caption{cosmic ray rejection}
\label{RS:T:RxSummary3}
\begin{center}
\begin{tabular}{lccc}
Name & muon rate & veto efficiency & dead time \\
 & ($s^{-1}$) & (\%) & (\%) \\ \hline
ILL & 250 & 99.8 & 8 \\
Gosgen & 260/300/340 & 99.8 & 8 \\
Rovno & 350 & 99 & 7 \\
Bugey 3 & & 99.5 & 2 \\
CHOOZ & 1 & & 2 \\
Palo Verde & 2000 & 96 & 2 \\
KamLAND & 0.3 & 92 & 11 \\
\end{tabular}
\end{center}
\end{table}

\noindent {\em Internal Radioactivity:}
Backgrounds from internal radioactivity come from 
many sources: (1)~$U/Th$ in detector materials 
(such as scintillator), (2)~$U/Th$ and $^{40}K$ activity from PMT glass, and
(3)~$U/Th$ in stainless steel, to mention a few. Radon incursions may also take
place from penetration through piping flanges, calibration device insertion,
or any place where detector scintillator contacts the radon-laden air common
in underground sites.\\

	The singles rate from these sorts of backgrounds is a strong
function of the detector threshold. This is because most long-lived beta activity
is below 2 MeV and energetic alphas are strongly quenched, typically with factors
of 0.1 or less such that they have visible energies of less than 1 MeV. The effect
of these events on the coincidence rate is therefore a strong function of threshold.
Table~\ref{RS:T:RxSummary2} shows the threshold for recent reactor experiments. 1 MeV
is clearly achievable if sufficient care is taken in materials selection, radon exclusion,
and the allowance of sufficient shielding from PMTs via a non-scintillating buffer region
and self-shielded fiducial volume. As an example, assuming a $^{238}U$ level of $10^{-14}$
 $g/g$ in the scintillator (about 100 times less stringent than the levels required by
Borexino) and only taking into account the 3 beta decays with $Q>1$ $MeV$ results in
a singles rate above 1 MeV of about 0.004 Hz in a 10 ton detector. This is conservative
in that it assumes all the betas are above threshold. This level is consistent with a
more detailed study done for the KamLAND detector in the U.S. proposal. In this study
similar rates are obtained for $^{232}Th$ and $^{40}K$ at levels that can be achieved by
reasonably careful control of detector materials and possibly scintillator washing
and/or filtration. Care is also taken to move the PMTs well back from the fiducial
volume.\\

Radon-laden air in enclosed spaces can sometimes reach as high as 1200 
$Bq/m^{3}$~\cite{RS:R:Radon}. Since the $^{222}Rn$ chain includes $^{214}Bi$(Q=3.272 MeV)
allowing such levels inside the scintillator would result in unacceptable background
levels. In addition, ambient radon levels are often seasonal and/or dependent on local
weather conditions. It is therefore necessary to control radon levels in the laboratory areas
and to take precautions to ensure radon cannot diffuse into the scintillator via use
of radon-free nitrogen blankets, radon impermeable o-rings and gaskets in all piping,
and the use of detector materials low in radon emissivity. For example, the radon level
in the Super-Kamiokande water is reduced to around $3 \times 10^{-3}$ $Bq/m^{3}$ by taking
such precautions. Similar low levels in a 10-ton detector would result in a coincidence rate
of around 0.003 $d^{-1}$, which is completely negligible.\\

\begin{table}[htbp]
\caption{$e^{+}$ energy measurement performance of past reactor experiments.}
\label{RS:T:RxSummary2}
\begin{center}
\begin{tabular}{lcccc}
Name & threshold & precision & resolution & variation \\
 & (MeV) & (\%) & (\% FWHM) & (\%) \\ \hline
ILL & 0.9 & 2 & 18 & 5 \\
Gosgen & 0.7 & 2 & 18 & 5 \\
Rovno & 0.7 & 1.5 & 30 & 1 \\
Bugey 3 & 1 & 1 & 8 & 1 \\
Savannah River & 2 & & & \\
CHOOZ & 1 & & 15 & \\
Palo Verde & 1 & & & \\
KamLAND & 0.9 & 2 & 8 & 1 \\
\end{tabular}
\end{center}
\end{table}

\noindent {\em External Gamma Activity:}
Gammas from $U/Th$ decay chains in the rock surrounding the detector can be a source
of background. Typical concentrations in rock can be as high as $10^{-4}g/g$. This
necessitates placing a buffer region around the detector which efficiently
shields these gammas. In addition, steel shielding may be necessary if the gamma
rates are very high. In the KamLAND detector the calculated external gamma rate
is about 0.2 Hz in a fiducial volume roughly 5 gamma attenuation lengths
from the rock wall. Assuming similar shielding for a 10-ton detector and scaling
by the surface area results in an external gamma rate of about 0.05 Hz, or a coincidence
rate of about 0.005 $d^{-1}$, which is negligible. Obviously, most of these events occur
on the edge of the detector and the rate changes dramatically with the amount of shielding.
Three attenuation lengths instead of five would increase the coincidence rate to about
14 $d^{-1}$. Care must be taken to measure the rate of external gammas at a potential
site and design the shielding accordingly.\\

\subsubsection{Correlated Backgrounds}

Correlated events that could be mistaken for a reactor neutrino event can come from
many different sources: (1)~direct cosmic ray hadrons, (2)~muon decay from stopped muons,
(3)~gammas from muon capture on carbon followed by neutron emission, 
(4)~spallation
products from muons, most notably those which may de-excite via neutron emission,
and (5)~fast neutrons which elastically
scatter off protons in the scintillator and are subsequently moderated and captured. All
these are dependent upon the cosmic ray rate and energy spectrum at depth.\\

\noindent {\em Direct Nucleons:} Cosmic ray protons and neutrons at sea level are not from the
primary flux but are all essentially from secondary interactions in the atmosphere. The flux
is highly peaked in the vertical direction ($\simeq \cos^{m}{\theta}$, where 
$m\simeq 8$)
and can be estimated by the expression~\cite{RS:R:neutron1}:\\
\begin{center}
$I_{n}(E) = (3 \times 10^{-5})(E/100)^{-\gamma}$
\end{center} 

where I is the differential vertical neutron intensity in $cm^{-2} s^{-1} sr^{-1} MeV^{-1}$,
$E$ is the kinetic energy in MeV, and $\gamma$ is a spectral index which is a weak function
of energy, about 2.4 in the range around 1 GeV and softening to 2.9 at higher energies.\\

The effective neutron attenuation length ($\Lambda$) has been determined for many types of
rock in association with geological dating studies that rely on production of rare elements
via neutron interactions~\cite{RS:R:neutron2}. Values range from about 
120 - 170 $g~cm^{-2}$.
Conservatively assuming all the flux to be vertical and taking the longest attenuation length
one can then estimate the flux of these neutrons underground above 10 MeV to be less than
$10^{-4} m^{-2}d^{-1}$ at 30 mwe.  Even taking a factor of ten for the uncertainty due to
secondary production, these nucleons are negligible for a 10 ton detector below this depth.\\

At depths less than 30 mwe it will be necessary to perform
a more detailed calculation, taking into
account secondary production, attenuation length of the actual shielding material, and the
flux expected at the particular geomagnetic cutoff.\\

\noindent {\em Stopped Muon Decay and Capture:} Stopped muons can contribute to correlated 
backgrounds in two ways: (1)~prompt muon ionization signal followed by muon decay, and
(2)~muon capture on $^{12}C$ to produce $^{12}B$.\\

\begin{figure}[t]
{\psfig{file=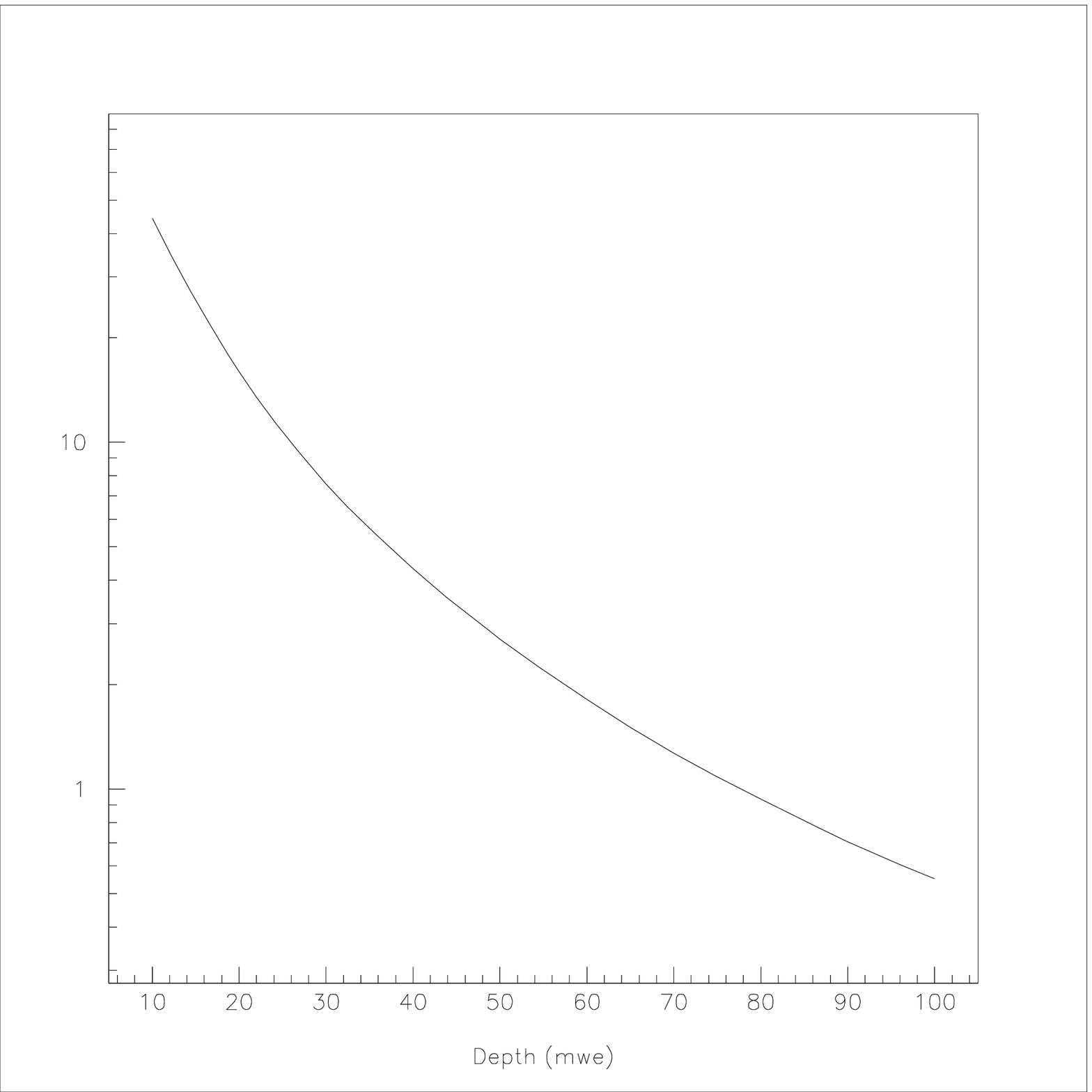,height=189pt,width=189pt}}
{\psfig{file=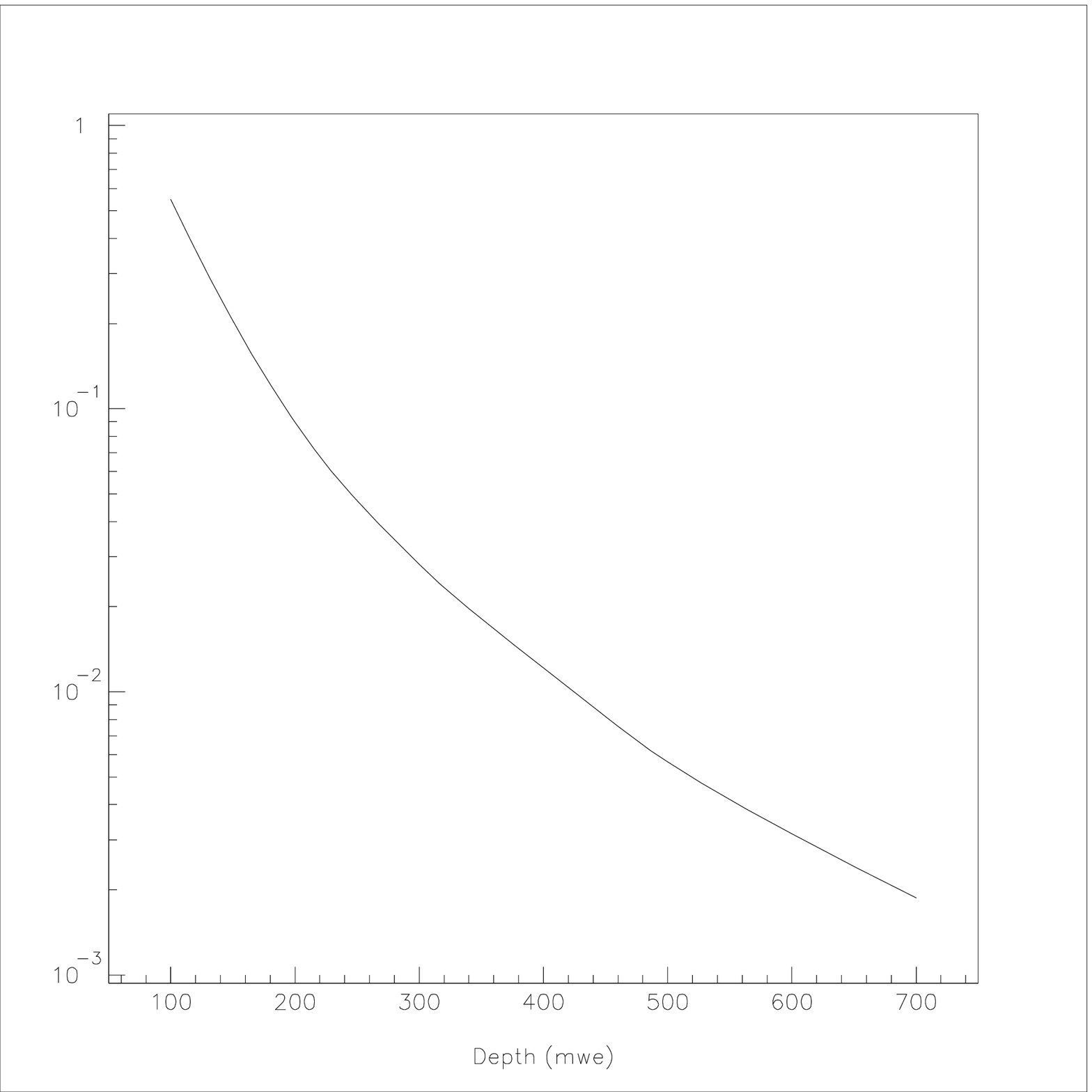,height=189pt,width=189pt}}
\caption{The rate of stopping muons in a 10-ton detector as a function of depth.
Left (right) is relevant for the near (far) detector.}
\label{RS:F:stop1}
\end{figure}

The stopping muon rate can be estimated as a function of depth by looking
at the change in the vertical muon rate. Figure~\ref{RS:F:stop1} shows the estimated
muon stopping rate in a 10-ton module as a function of depth. It is assumed that the
stopping power difference between shielding and oil
is a factor of two.\\

	The rate of untagged muons that decay in the detector is given by:\\
\begin{center}
$R = (1-\epsilon)R_{\mu} f_{dk}f_{E}e^{-t_{v}/\tau}$\\
\end{center}
where $R_{\mu}$ is the muon rate, $\epsilon$ is the 
efficiency of the muon veto, $f_{dk}$ is the fraction of those muons that stop and decay,
$f_{E}$ is the fraction of the decay electrons in the range of reactor neutrinos 
($\simeq 10\%$), $t_{v}$ is the veto time after any event, and $\tau$ is the effective
lifetime of muons in oil ($\simeq 2.126 \: \mu$s). For $t_{v} = 5 \: \mu$s and a 95\%
veto efficiency there would be about 330 untagged stopped muons/day. Clearly
one must improve the
muon recognition and also accept signal loss due to dead time. 30~mwe was roughly the 
depth of the
11-ton Palo Verde detector, which as Table~\ref{RS:T:RxSummary4} shows did
 have a poor efficiency
compared to other experiments. Going to 90 mwe would reduce this to
an acceptable 30~events/day, or 1\% of signal rate for a near
detector. To achieve a 1\% level
for the far detector would require
a depth of roughly 600 mwe.\\

In oil about 7.9\% of the $\mu^{-}$ will capture on $^{12}C$. Of these, only about 20\% will
go to the $^{12}B$ ground state, the rest going to particle unstable
 states that typically result
in neutron emission. Such neutron-producing reactions are a very dangerous background since
they are correlated with the initial muon and can look like a reactor 
neutrino event. We correct
the rate formula for untagged stopped muons (ignoring $f_{E}$) for the fraction of $\mu^{-}$ 
in cosmic rays ($=0.44$), and the capture fraction to particle-unstable states, 
$(0.44)(0.079)(0.8) =  0.028$. Thus at 30 mwe there will be roughly 190 untagged captures/day
that result in a correlated neutron. The rate at 50 mwe would be roughly 33/day, which is
acceptable for the near detector. For the far detector a depth of 400 mwe
 is needed to reach the
1\% level.\\

\noindent {\em Muon-Induced Spallation:} Most spallation products that can be
made by muons passing through scintillator are short-lived and/or do not make
double-coincidence events that can be misconstrued as reactor neutrino events.
In that case the coincidence rate must be less than the coincidence rate of
single, untagged muons, or they simply appear as a low level increase in the
overall singles rate. In either case they do not present a problem. There 
are two exceptions: (1)~$^{8}He$($t_{1/2}=119 ms$) 
$\rightarrow n + e^{-} + ^{7}Li$ with a branching ratio conservatively
estimated at 16\%; and (2)~$^{9}Li$($t_{1/2}=178 ms$) 
$\rightarrow e^-+^{8}Be + n$ followed by $^{8}Be \rightarrow
2\alpha$. In these cases the electron or two quenched alpha's
constitute the prompt signal and in both cases a $\simeq 1$ $MeV$ neutron
is produced to make a delayed signal (and also add to the prompt signal via
n-p elastic scattering).\\

The production rate of such events is very difficult to determine, as 
muon-induced spallation is not well-understood theoretically, though
there are experimental measurements from a muon beam at CERN \cite{bib:hagner}.
However,
most previous reactor experiments~\footnote{ The exception is the KamLAND
detector, which estimated roughly 2 untagged events/year in 400 tons at a 
depth of 2700 mwe.} have measured their correlated backgrounds
via reactor on/off data and such events would be lumped together with the
background from fast neutrons, described below.\\

\noindent {\em Fast Neutrons:} Fast neutrons come from muon-induced nuclear
reactions on either on the scintillator itself or on the rock. These neutrons
then elastically scatter off protons in the scintillator and produce an
energetic proton ``prompt'' signal. They subsequently capture in the usual
way, mimicking a reactor neutrino event. The spectrum of proton-recoil prompt
signals is fairly flat out to at least 50 MeV and has been measured by the
CHOOZ experiment at 300 mwe to be about $0.031$ $d^{-1} ton^{-1} MeV^{-1}$.
Extrapolation of this rate in the reactor prompt signal region from 1-8 MeV
would give a rate of 2.2 events/day for a 10 ton detector at 300 mwe of
similar geometry.\\

Comparison of different experimental results on neutron production per muon
per $g\cdot cm^{-2}$ of target ($R$) at depths ($x$) from 20-5000 mwe can be
roughly fitted by the empirical function~\cite{RS:R:KLneut}: 
$R=10.0 x^{0.417} \times 10^{-6}$ $n \cdot g^{-1}cm^{2}/\mu$. 
Thus one may extrapolate
the CHOOZ result to different depths. For the near detector a depth of 50 mwe
is sufficient to reduce the rate to roughly 30 events/day (assuming the
same resolution and efficiency of CHOOZ). For the far detector a depth of
1100 mwe would be required to reach 0.125 events/day. This is a very stringent
requirement which may be difficult to achieve in practice. Thus the background due
to fast neutrons will have to be well-understood at depths shallower than this.\\

\subsection{Summary of Overburden Requirements}

The final result of the experiment will depend on the measurement of the reactor flux
to high precision. To achieve a similar relative systematic error in the flux measurements
of the near and far detector they should have roughly the same background as a percentage
of the expected signal. Thus the requirement of similar S/N for the near and far detector
results in different overburden requirements. To keep the background from any single source
less than 1\% such that it is on the order of the required precision in the flux measurement
the near detector should be at a depth of 90 mwe or greater.\\

For the far detector, a depth
of 1100 mwe would be desirable, but may not be available in practice. 
At shallower depths
the understanding of the background from fast neutrons will likely
be a limiting factor. At depths less than 600 mwe accidental coincidences
from untagged stopped muons may become important. More restrictive cuts
than those described here or a more efficient veto will likely have to be
devised. At 400 mwe the correlated background due to $\mu^{-}$ capture
will begin to become important. This is potentially a serious background
which, as in the case of stopped muons, can be reduced by better muon
tagging.\\

%% file: systematics.tex
%
%
\section{Systematics}
\label{sec:sys}
\subsection{Overview}

A general discussion of systematic errors in a two-detector experiment has been
presented in Sections~\ref{sec:lumi} and \ref{sec:syst-BG-ND}.
For the sake of definiteness, we consider here a restricted class of
experiments, as follows: 
\begin{itemize}
\item There are at least two identical detectors
(one near, and one or more far). 
\item At least one detector might be movable
between near and far locations. 
\item There may be more than one reactor. 
\item The
total number of antineutrino events in the far detector(s), although
larger than in previous experiments, is not large enough that
spectral-shape information materially improves the result from total
rate information.
\end{itemize}
This is essentially a counting experiment, and we will
ignore errors in spectral shape that do not contribute to overall normalization.

As discussed in earlier sections, the great advantage of a near-far detector 
configuration is that systematic errors in knowledge of antineutrino flux
and antineutrino interaction cross-section are absent or greatly reduced,
and that detector-related systematics are reduced to those due to differences in
detectors. It may be possible to further reduce detector-related systematics
by moving one far detector to the near position so that rates can be compared
directly, as described in Section~\ref{sec:move}. However, the act of moving
may in itself be a source of systematic error due to changes in a detector 
during the move. These effects are best addressed after the sources of error
in a non-moving detector are fully understood. 

In this section, we will look at a particular detector configuration and
attempt to list and estimate the sources of systematic error. 
We consider the generic detector geometry shown in Figure
\ref{fig:det13}. The detector fiducial volume is monolithic (like CHOOZ
and KamLAND) rather than segmented (like Palo Verde), and liquid
scintillator is contained in hard-walled spherical or cylindrical
acrylic vessels rather than in soft-walled balloons. There is a muon
veto system surrounding each detector, which would remain in place if
detectors were moved, implying more than one veto setup for any movable
detector. The volume viewed by the inward-looking photomultipliers
(PMTs) that detect the positron from
antineutrino interactions and the subsequent neutron capture  has three 
concentric (or coaxial) regions: an
innermost volume (I)~containing Gadolinium-doped liquid scintillator,
whose light transmission will probably get worse with time; a
``gamma-catcher" region (II)~containing undoped liquid scintillator; and
a buffer region (III)~containing non-scintillating oil. Region III both isolates
the scintillating regions from radioactivity in the PMTs and also does
not permit scintillation light to be generated very close to any PMT.

For detector-related errors, we will be guided by the one-detector
experience of the CHOOZ experiment~\cite{bib:chooz} to identify the 
sources of systematic error that are potentially most serious, and will briefly 
discuss how much they might be reduced in a near-far experiment
with identical detectors.
We also postulate an event reconstruction method similar to that used by
CHOOZ~\cite{bib:chooz}: both the position and the energy of the prompt (positron)
and delayed (neutron-capture) subevents of each event are obtained
from a fit to the distribution of charge signals from the large number of
PMTs in the detector. This method relies on good knowledge of the calibration
of each PMT and of the time-dependent light attenuation in each detector region.

The detection efficiencies and their systematic uncertainties for 
the CHOOZ experiment
are listed in Table~\ref{tab:chooz_det_effs}; the combined systematic error
is 1.5\%, unacceptably large for  future near-far experiments. 

\begin{table}
\begin{center}
\begin{tabular}{||l||c||c||} \hline\hline
selection & $\epsilon(\%)$ & rel. error (\%) \\ \hline\hline
positron energy & 97.8 & 0.8\\ \hline
positron-geode distance & 99.9 & 0.1 \\ \hline
neutron capture & 84.6 & 1.0 \\ \hline
capture energy containment & 94.6 & 0.4 \\ \hline
neutron-geode distance & 99.5 & 0.1 \\ \hline
neutron delay & 93.7 & 0.4 \\ \hline
positron-neutron distance & 98.4 & 0.3 \\ \hline
neutron multiplicity & 97.4 & 0.5 \\ \hline \hline
combined & 69.8 & 1.5 \\ \hline \hline
\end{tabular}
\end{center}
\caption{Summary (from Reference~\cite{bib:chooz}) of the detection efficiencies in
the CHOOZ experiment and their systematic uncertainties.}
\label{tab:chooz_det_effs}
\end{table}

\subsection{Potential sources of systematic error}

For convenience, we list potential sources of systematic error in four
general categories: ``physical" errors, errors from triggering and data recording,
errors from event-selection cuts, and errors from background subtraction.

\subsubsection{Potential sources of ``physical" errors.}
This category includes errors in source and target parameters and in losses
of primary reaction products ($e^+$, $n$).

\paragraph{Potential errors related to reactor sources(s).} 
\begin{enumerate}

\item
{\em{Distance from each reactor core to each detector. }}
CHOOZ~\cite{Apollonio:1998xe}  quoted a precision of
$\pm 10$~cm.  If the near detector in a near-far experiment is at 250~m, a
similar precision would introduce 
a negligible uncertainty of 0.08\% in $1/r^2$.

\item
{\em{Relative flux from each reactor core.}}
This error cancels for the case of two cores and symmetric placement 
of detectors. The general case of multiple reactors and multiple detectors
is discussed in the following subsection. 

\end{enumerate}

\pagebreak
\paragraph{Potential errors in number of free protons.} 
\begin{enumerate}

\item
{\em{Mass of fiducial region. }}
It should be possible to weigh each kind of liquid 
added to each detector to $\sim 0.1$\%.
Each acrylic-walled region of each detector
would then have a thin standpipe to monitor the excess volume and its fluctuations
due to temperature change.

\item
{\em{Fraction by weight of free protons.}}
 CHOOZ found that determining this quantity from combustion measurements was not 
trivial, and quotes an uncertainty of $\pm 0.8$\%. With multiple detectors
the issue is whether the free proton content is the same in all detectors.
One could, for example, fill all detectors equally from each batch of mixed
scintillator. This relative uncertainty should be reducible to the $\sim 0.1$\%
level.

\item
{\em{Effective fiducial volume.}} 
The question here is whether the boundary of the innermost (Gd-doped)
detector region (I) actually delimits the fiducial volume. This would
happen, for example, if all detected $e^+$ subevents from region I were
accompanied by detected $n$ subevents, and no $e^+$ subevents from
region II were accompanied by detected $n$ subevents. This will
certainly not be true. However, to first order, ``spill-in equals
spill-out": the loss of $e^+ n$ events near the outside edge of region I
is compensated by a gain of $e^+ n$ events near the inside edge of 
region II. The difference in non-compensation among detectors should be
small. 

An alternative is to make an explicit cut on the fitted positions of the
$e^+$ and $n$ subevents, as was done for a single detector by CHOOZ (see
Section~\ref{selection_cuts} below). To bound this error, we will will
use their uncertainties resulting from this method. Whether using no
position cuts at all is a superior procedure is clearly a subject for 
detailed simulation studies. 

\end{enumerate}

\paragraph{Loss or absorption of reaction products.}
This category is coupled to the event-selection cuts
because of the multiple gammas
present in an event: two annihilation gammas from the positron, and typically
three from neutron capture on Gd. These deposit visible energy by successive
Compton collisions. The most likely symptom of gamma loss is thus a reduced
positron or neutron signal rather than none at all; this generates event
inefficiency via an energy cut. We consider here the rarer
situations in which positrons or neutrons are lost entirely.

\begin{enumerate}

\item
{\em{
Loss of positrons or neutrons by absorption in inert material (acrylic vessels)
or through openings such as chimneys or source pipes.}}
 One can  rely on
simulations to indicate that this effect is small, and if so, that it
essentially cancels when detectors are compared.

\item
{\em{
Loss of neutrons escaping beyond the active region before they are captured.}}
This effect should not be sensitive to small differences in detectors, and should
nearly cancel when detectors are compared.

\item
{\em{Capture of neutrons on the wrong material.}}
Neutrons can be captured on hydrogen instead of on Gadolinium. In this
case, the capture gammas have a total energy of only 2.2 MeV instead of
$\approx 8$~MeV, and would be rejected by typical energy cuts on the
delayed event (6 to 12 MeV). 

This is a serious source of systematic error in a single-detector
experiment. CHOOZ~\cite{bib:chooz} used a tagged $^{252}$Cf neutron source
in several positions in their detector to calibrate this effect. They
quote an efficiency for $n$-capture on Gd of $(84.6 \pm 0.85)$\%, a 1\%
systematic. 

In a multiple-detector experiment, in addition to measurements with a
tagged neutron source, one can compare  neutron counting rates
when the {\underline{same}} untagged neutron source is placed in
similar positions in each detector. The relative error among
detectors should be reducible to a few tenths of a percent. 

\end{enumerate}

\subsubsection{Potential errors from triggering and data recording}

\paragraph{Potential errors in the time domain.}
We assume that all detectors will be interlocked so that down-time due
to malfunctions such as high-voltage trips will be the same for all.

\begin{enumerate}

\item
{\em{Deadtime losses from the muon veto and from event recording.}}
These losses should be small and easily measurable. Note, however, that
if the muon-shielding overburdens of the detectors are very different, the
deadtime losses could be substantially different.

\item
{\em{Trigger time window for the delayed event.}}
Care should be taken that the hardware cut is looser
than any anticipated software cut.

\end{enumerate}

\paragraph{Potential errors in the pulse-height domain.}

CHOOZ's lowest-level trigger required a minimum analog sum of pulse
heights and a minimum number of PMTs firing at the $\approx 0.5$
photoelectron level. Care must be taken that such requirements do not
compromise clean cuts on fitted energy, especially since periodic
changes in trigger thresholds must be made due to increasing light
attenuation in the Gd-doped scintillator. Most of CHOOZ's quoted
systematic error due to positron energy (0.8\%) came from this effect.

\paragraph{Potential errors from differences in detector electronics.}
Just as multiple detectors should be physically identical, so must 
their electronics
be identical in performance. 
Identical pulser-driven ``pseudo-events" should be
delivered frequently to all detectors and analyzed as data.

\subsubsection{Potential errors  from event-selection cuts.}
\label{selection_cuts}
The CHOOZ analysis~\cite{bib:chooz} imposed six criteria for selecting events. Four
of these depend on the results of the position-energy fit to each event
described above. It is clear that in a near-far experiment, it is essential
to ensure identical behavior of this fit for all detectors.

\begin{enumerate}

\item 
{\em{Fitted positron energy $E_e<8$~MeV.}}
The presence of the region II gamma-catcher guarantees that even when
the positron has zero kinetic energy, the 1.1~MeV of energy from its
annihilation gammas will be seen. No lower cut on positron energy is
therefore needed, removing a potentially serious source of
systematic error. The
high-side energy cut $E_e<8$~MeV eliminates a negligible number of real
events, and generates an uncertainty of only $0.05\%$ even in a
single-detector experiment. 

\item
{\em{Fitted neutron energy $6<E_n<12$~MeV.}}
This cut is made to reduce background. Partial escape of capture gammas gave
it an efficiency of $(96.4\pm 0.4)\%$, as measured from the energy yield
of neutrons from radioactive sources. The relative error among detectors
due to the loss of the small low-energy tail should be much smaller than this.

\item
{\em{Fitted positions: minimum distance from photomultipliers of subevents.}}
(a) Fitted position of positron (prompt subevent) at least 30~cm from the ``geode"
of PMT faces ($d_{e^+} > 30$~cm; (b) fitted position of neutron (delayed subevent)
also at least 30~cm from the geode ($d_{n} > 30$~cm).
These requirements were motivated for CHOOZ in large part by the ungraceful 
performance of
the fitting procedure when light was deposited very near a PMT. It
should not be necessary at all in a detector geometry with a non-scintillating
oil buffer layer between the scintillating regions and the PMTs.
 CHOOZ assigned systematic
uncertainties of 0.1\% each to these cuts.  One might guess
that the the uncertainty in the difference in performance between
detectors should be at least a factor of two smaller than this. However, the
detector design should preserve a position resolution comparable to that
of CHOOZ ($\sigma \approx 4$~cm at 8 MeV). 

\item
{\em{Fitted positions: maximum distance between subevents.}}
Difference in fitted positions of positron and neutron less than 100~cm
($d_{{e^+}n} < 100$~cm).
This cut was  made to suppress  background, and it may well be
desirable in a future experiment. CHOOZ used Monte Carlo 
simulation to establish its efficiency at
$(93.7 \pm 0.4)\%$. The relative error among detectors in this quantity
should be several times smaller.

\item
{\em{Time window for neutron subevent.}}
Time delay of neutron subevent relative to prompt subevent
$2 < t_{{e^+}}n < 100$~$\mu$s.
CHOOZ imposed the  cut on the low end of this window because of overshoot 
problems in AC-coupled photomultipliers (which can be avoided in future
experiments); it caused a loss of $(1.6 \pm 0.2)\%$. 
The high-end cut covers about three
capture lifetimes in Gd-doped scintillator, and had an efficiency
of $(95.3 \pm 0.3)\%$. Since there is no reason for this cut to depend
strongly on details of detector performance, the relative error among
detectors should be considerably smaller.

\item
{\em{Neutron multiplicity: $N_n = 1$.}}
CHOOZ imposed this cut to reduce the correlated background from muon 
spallation, which typically generates more than one neutron. 
It rejects some real events when the accidental presence of a gamma ray
from natural background fakes the second neutron.
The  efficiency of this cut was  $(97.4\pm 0.5)\%$. 
Although the relative effects among identical detectors should be reduced,
care must be taken if the background environments of the detectors are 
substantially different.

\end{enumerate}

\subsubsection{Potential errors from background subtraction.}

The fraction of background in the near and far detectors will be different
because of the much higher real event rate in the near detector. Subtraction
of background in the far detector(s) must therefore be well understood.

A general discussion of backgrounds is given in Section~\ref{sec:bac}. 
Briefly, there are two types: accidental backgrounds, in which the prompt
and capture subevents are randomly associated; and correlated backgrounds,
in which the subevents are not random (but may not be in the proper order).
At 300 mwe, CHOOZ observed an accidental background rate of 
$(0.42 \pm 0.05)$~d$^{-1}$, and a correlated background rate of 
$(1.04 \pm 0.11)$~d$^{-1}$.
Their  real event rate at full reactor power was $24.7$~d$^{-1}$, so the
background fraction was 6\% and contributed a systematic uncertainty of 0.5\%.

It should be noted that due to reactor commissioning, CHOOZ had almost
as much reactor-off time to study backgrounds as they had reactor-on time, 
a circumstance unlikely for future experiments. It is also important
to note that their
background rates changed significantly whenever thresholds were
changed to accommodate increased light attenuation in Gd-doped
scintillator.

Let us assume for a moment that in a future experiment each far detector
will have about 5 times the fiducial tonnage of CHOOZ, but that they are
likely to be located about 1.7 times farther away from a source of
comparable power, and that background rate scales with detector mass.
Under these assumptions, the background-to-signal ratio will be $1.7^2 =
3$ times worse than for CHOOZ, $\sim 18\%$.
To keep the systematic error from background subtraction below 0.5\%
under these assumptions, it will be necessary to know the
background to better than 3\% of itself, more than three times as well as
CHOOZ did with ample reactor down-time. Serious thought must be given to
reducing background and to inventing ways to measure its spectrum and
magnitude.

\subsection{Summary - potential systematic errors}

For a near-far experiment, the systematic errors in rate comparison
between detectors associated with reactor sources and number of target
protons can be kept below a few tenths of a percent with reasonable attention.
Extrapolating from CHOOZ experience, the errors arising from
differences in detectors in performance and for
event-selection criteria can probably
be reduced about a factor of three from the CHOOZ figure of 1.5\%,
provided that sufficient care is taken in experimental design and execution.
The presence of gamma-catcher and inert buffer regions appear to be essential
elements of the design, and understanding the behavior of the position-energy
fit to each event under conditions of decreasing scintillator transparency is
crucial. Finally, the systematic error associated with background subtraction
could well be the hardest to keep under control because of the limited beam-off
time available to measure it.

%% file: systematics2.tex


\subsection{Multiple-reactor scenarios}
In the following 
parts of this section, the effect of multiple reactors  and multiple
detectors is considered.
The Kashiwazaki site, for example,  has seven reactors in two groups. 
It is shown that the uncertainty due to the nature of multiple 
reactors can be made small, if the near detector is placed in
such a way that the contribution from each reactor of the
cluster is approximately equal.

\subsubsection{Reactor induced systematics - multiple reactors, one detector}
 
\par To develop notation, we consider first the simplest case involving one
reactor and one detector.  The effective systematic error is given by
\begin{eqnarray}
\displaystyle
\sigma_{eff}^2=
\sigma_u^2+\sigma_c^2+(\sigma^{(r)}_c)^2+(\sigma^{(r)}_u)^2\nonumber
\end{eqnarray}
where
$\sigma_u$ is the uncorrelated error of the detector,
$\sigma_c$ is the correlated error of the detector,
$\sigma^{(r)}_c$ is the correlated error of the flux from the reactor
and $\sigma^{(r)}_u$ is the uncorrelated error of the flux
from the reactor.

\subsubsection{The case with one detector + multiple reactors}
When the number of reactors is larger than one,
$\sigma_{eff}^2$ becomes
\begin{eqnarray}
\displaystyle
\sigma_{eff}^2=
\sigma_u^2+\sigma_c^2+(\sigma^{(r)}_c)^2+(\sigma^{(r)}_u)^2
\sum_{a=1}^N\left({T_a \over T}\right)^2,
\label{sigmaeff1}
\end{eqnarray}
where $T_a~(a=1,\cdots,N)$ stands for the yield from the $a$-th reactor
and $T=\sum_{a=1}^N T_a$ is the total yield.
If the yield from each reactor is equal, i.e.,
$T_a=T/N$,
then the contribution of the uncorrelated systematic
error $(\sigma^{(r)}_u)^2$ from the reactor in Equation~(\ref{sigmaeff1})
becomes $(\sigma^{(r)}_u)^2
\sum_{a=1}^N\left(T_a / T\right)^2=(1/N)(\sigma^{(r)}_u)^2$,
so that we have
\begin{eqnarray}
\displaystyle
\sigma_{eff}^2=
\sigma_u^2+\sigma_c^2+(\sigma^{(r)}_c)^2+{1 \over N}(\sigma^{(r)}_u)^2.
\label{sigmaeff2}
\end{eqnarray}
Therefore, we see that the more reactors there are,
the smaller the contribution of the uncorrelated systematic
error $\sigma^{(r)}_u$ becomes.
This is because the average of independent $N$ fluctuations
is smaller than a single fluctuation.

\subsubsection{The case with multiple reactors and detectors \cite{paper1}}
\begin{figure}
\hglue 3.5cm
\epsfig{file=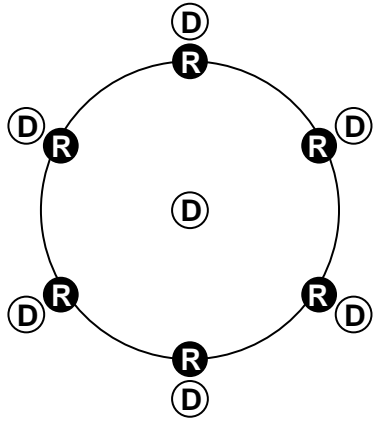,width=6.0cm}
\vglue -0.5cm 
\caption{
The ideal case with $N$ reactors (R) and $(N+1)$ detectors (D).
}
\label{fig:fig1}
\end{figure}

It has been known that the correlated error
is canceled in the case of a single reactor
experiment with near and far detectors.
Now a question arises: what happens to this
cancellation in the case of
an experiment with multiple reactors and detectors?
To answer this question, let us consider the ideal
case with $N$ reactors and $(N+1)$ detectors,
where each reactor has a near detector in its neighborhood
and it produces the same yield at a far detector (See Figure~\ref{fig:fig1}).

In this case the effective systematic error becomes
\begin{eqnarray}
\displaystyle
\sigma_{eff}^2=
{\left(1+1/N\right)\sigma_u^2 \over
1+(1/N)\left\{
1+(N+1)\left[
\sigma_c^2+(\sigma^{(r)}_c)^2+(\sigma^{(r)}_u)^2/N\right]
/\sigma_u^2\right\}^{-1}},
\label{sigmaeff3}
\end{eqnarray}
where the conditions $T^F_a=T^F/N$ and
$T^{N b}_a=\delta_a^b\,T^{N b}$ have been assumed;
$T^F_a$ and $T^{N b}_a$ are the yield at
the far and $b$-th near detector
which is close to the $b$-th reactor ($b=1,\cdots,N$)
from the $a$-th reactor, and $T^F$ and $T^{N b}$
are the total yield at the far and $b$-th near detector.
As in the case with one reactor,
the dominant contribution to $\sigma_{eff}$ comes
from the uncorrelated error $\sigma_u$, and
the contribution of the uncorrelated 
error $\sigma^{(r)}_u$ is reduced in (\ref{sigmaeff3})
by a factor of
$N$ due to the averaging over the independent $N$
fluctuations, but this reduction is irrelevant because such
an effect comes in the correlated error which is
almost canceled in the multi detector system.
To conclude, the answer to the question at the
beginning of this section is that
the cancellation of the correlated error occurs
also in the ideal case with $N$ reactors and $(N+1)$ detectors.
It should be noted that the number $N$ of the near
detectors in this case is sufficient but not necessary
to guarantee this cancellation, as we will see below in the
case of the Kashiwazaki plan.

\subsubsection{Kashiwazaki case study}
\begin{figure}
\hglue 2.5cm
\epsfig{file=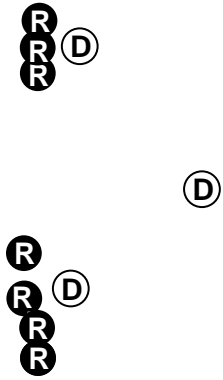,width=6.0cm}
\vglue -1.0cm 
\caption{
The Kashiwazaki plan with 7 reactors (R) and 3 detectors (D).
}
\label{fig:fig2}
\end{figure}

The next question which arises is:
what happens to the cancellation of the correlated error
in the Kashiwazaki plan.
The Kashiwazaki-Kariwa nuclear plant consists of two clusters
of reactors, and one cluster consists of four reactors while the
other consists of three (See Figure~\ref{fig:fig2}).
\begin{figure}
\vglue -1.cm 
\hglue 2.5cm
\epsfig{file=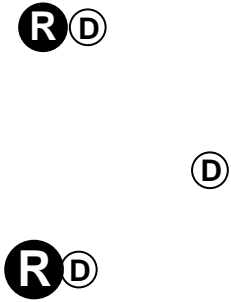,width=6.0cm}
\vglue -2.0cm 
\caption{
The ideal limit of the Kashiwazaki plan, where
one reactor has the power four times as much as one of the
Kashiwazaki-Kariwa reactors, while the other
has the power three times as much as that of the KK reactors.
}
\label{fig:fig3}
\end{figure}
Before we discuss the
effective systematic error for the actual Kashiwazaki plan,
let us consider the ideal limit, where the cluster
of the four reactors shrinks to one reactor and
the other of three reactors shrinks to another single
reactor (See Figure~\ref{fig:fig3}).
It can be shown analytically that
the effective systematic error for this ideal limit is
given by
\begin{eqnarray}
\displaystyle
\sigma_{eff}\simeq
{\sqrt{74} \over 7}
\,\sigma_u,
\nonumber
\end{eqnarray}
where the correlated error has been canceled due
to the near-far detector complex, and the reason
that we have the factor $\sqrt{74/49}$ instead of
$(1+1/N)^{1/2}|_{N=2}=\sqrt{3/2}$ is because
the yield  from the first reactor to that from the
second is 4:3 instead of 1:1.  Thus we see that
the cancellation of the correlated errors
occurs also in the ideal limit of the Kashiwazaki plan.
Furthermore, it can be shown 
from the numerical calculations that
the difference between $\sigma_{eff}$ for the actual
Kashiwazaki plan
and that for the ideal limit is very small:
\begin{eqnarray}
\displaystyle
{\left.\sigma_{eff}\right|_{actual~KK} \over
\left.\sigma_{eff}\right|_{ideal~KK}}
=1.04,
\nonumber
\end{eqnarray}
where $\sigma_u=0.6\%$, $\sigma_c=1.6\%$,
$\sigma^{(r)}_u=2.3\%$, $\sigma^{(r)}_c=2.5\%$ have been
used as reference values for the systematic errors.
This shows that the cancellation of the correlated errors
occurs also in the actual Kashiwazaki plan.

The systematic limit on $\sin^22\theta_{13}$ at 90\%CL
can be obtained from the effective systematic error by
\begin{eqnarray}
\left(\sin^22\theta\right)_{limit}^{sys~only}
={\sqrt{2.7}\,\sigma_{eff} \over
\mbox{"}\left\langle \sin^2\left({\Delta m^2L_{far} \over 4E}
\right) \right\rangle\mbox{"}
-\mbox{"}\left\langle \sin^2\left({\Delta m^2L_{near} \over 4E}
\right) \right\rangle\mbox{"} },
\label{limit}
\end{eqnarray}
where
\begin{eqnarray}
\hspace*{-20mm}
\displaystyle
\mbox{"}\left\langle \sin^2\left({\Delta m^2L_{far} \over 4E}
\right) \right\rangle\mbox{"}&\equiv&
\displaystyle\sum_{a=1}^7 \omega^{far}_a
\left\langle \sin^2\left({\Delta m^2L^{far}_a \over 4E}
\right) \right\rangle,\nonumber\\
\mbox{"}\left\langle \sin^2\left({\Delta m^2L_{near} \over 4E}
\right) \right\rangle\mbox{"}&\equiv&
\displaystyle{4 \over 7}\sum_{a=1}^7 \omega^{near-1}_a
\left\langle \sin^2\left({\Delta m^2L^{near-1}_a \over 4E}
\right) \right\rangle\nonumber\\
&+&\displaystyle{3 \over 7}\sum_{a=1}^7 \omega^{near-2}_a
\left\langle \sin^2\left({\Delta m^2L^{near-2}_a \over 4E}
\right) \right\rangle,\nonumber
\end{eqnarray}
stands for the averaged expectation value of the factor
$\sin^2\left({\Delta m^2L/ 4E}\right)$, $L^X_a$ ($X$=near-1,
near-2, far) is the distance between the detector $X$ and the $a$-th
reactor ($a=1,\cdots,7$), and $\omega^X_a$ ($X$=near-1, near-2, far) is
the fraction of the yield from the $a$-th reactor at the detector $X$
(=(power of $a$-th reactor)/$(L^X_a)^2$).  The expectation $\left\langle
\cdots\right\rangle$ is defined by
\begin{eqnarray}
\hspace*{-20mm}
\displaystyle
\left\langle \sin^2\left({\Delta m^2L \over 4E}
\right) \right\rangle&\equiv&
{\displaystyle
\int dE~\epsilon(E)f(E)\sigma(E)\sin^2\left({\Delta m^2L \over 4E}
\right) \over \displaystyle
\int dE~\epsilon(E)f(E)\sigma(E)},
\nonumber
\end{eqnarray}
where
$\epsilon(E)$, $f(E)$, $\sigma(E)$ stand for the detection
efficiency, the neutrino flux, and the cross section, respectively.
If $|\Delta m^2_{31}|=2.5\times10^{-3}$eV$^2$, then we have
\begin{eqnarray}
\mbox{"}\left\langle \sin^2\left({\Delta m^2L_{far} \over 4E}
\right) \right\rangle\mbox{"}=0.70
\nonumber\\
\mbox{"}\left\langle \sin^2\left({\Delta m^2L_{near} \over 4E}
\right) \right\rangle\mbox{"}=0.17
\nonumber
\end{eqnarray}
in the setup of the Kashiwazaki plan.
So, if $\sigma_u$=0.6\%, then Equation~(\ref{limit}) becomes
\begin{eqnarray}
\left(\sin^22\theta\right)_{limit}^{sys~only}
={\sqrt{2.7} \over 0.53}{\sqrt{74} \over 7}0.006=0.022.
\nonumber
\end{eqnarray}
Furthermore, if we include the statistical error
$\sigma_{stat}=1/\sqrt{60,000}$ in the case of 20 t$\cdot$yr, then we get
\begin{eqnarray}
\left(\sin^22\theta\right)_{limit}^{sys~only}
={\sqrt{2.7} \over 0.53}\sqrt{(74 / 49)0.006^2+1/60,000}
=0.025.
\nonumber
\end{eqnarray}

\begin{figure}
\hglue -1.cm
\epsfig{file=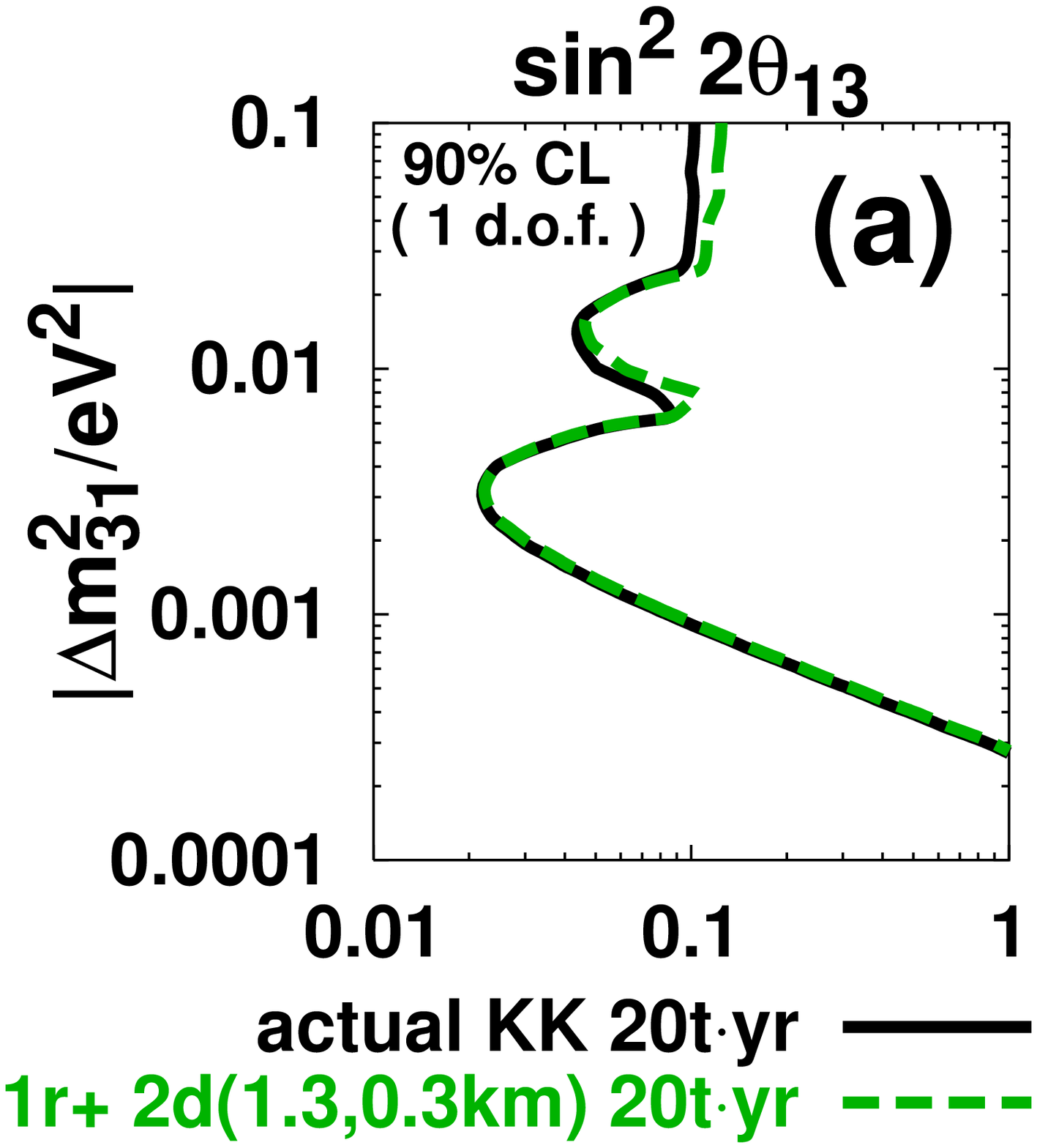,width=7.5cm}
\vglue -7.9cm
\hglue 6.5cm
\epsfig{file=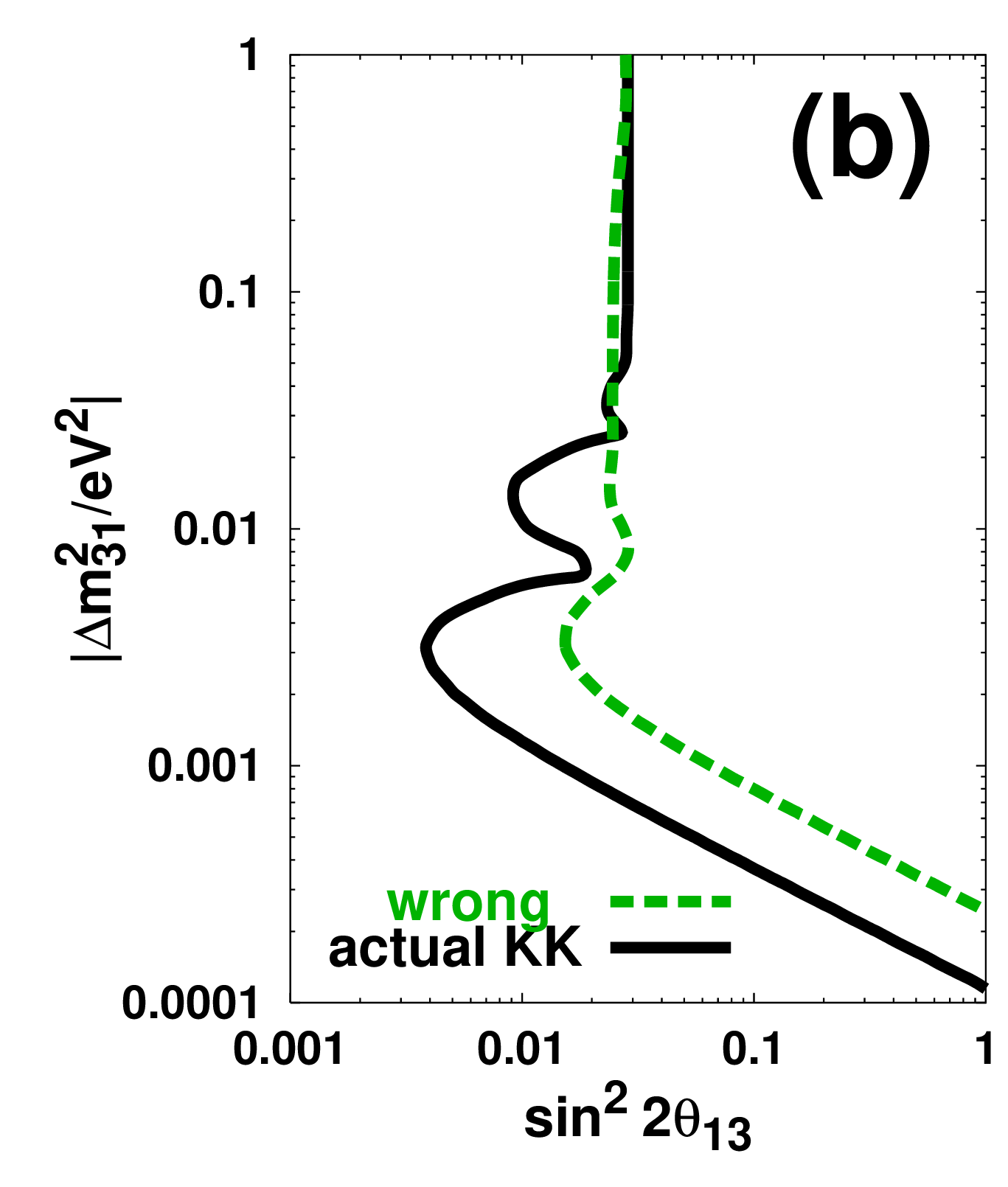,width=7cm}
\vglue -0.5cm
\caption{
(a) Comparison of the actual Kashiwazaki plan (the solid line)
and a hypothetical experiment (the dashed line),
which is depicted in Figure~\ref{fig:fig4},
with a single 24.3$GW_{th}$ reactor and
two detectors assuming the same uncorrelated systematic error
$\sigma_u$.
(b) Comparison of the actual Kashiwazaki plan (the solid line)
and a hypothetical experiment (the dashed line),
which is depicted in Figure~\ref{fig:fig5}, where the near
detectors are placed in wrong location.
The statistical errors as well as
all the systematic errors except $\sigma^{(r)}_u$,
which is assumed to be 2.3\%, are ignored for simplicity.
}
\label{fig:KK}
\end{figure}
\begin{figure}
\vglue -2.5cm
\hglue 2.5cm
\epsfig{file=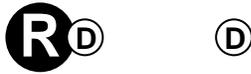,width=6.0cm}
\vglue -3.2cm
\caption{
The case where a single reactor has the power seven times as
much as one of the Kashiwazaki-Kariwa nuclear plant.
}
\label{fig:fig4}
\end{figure}

In Figure~\ref{fig:KK}(a) comparison is given between
the actual Kashiwazaki plan
and a hypothetical experiment, which is depicted in Figure~\ref{fig:fig4},
with a single reactor and
two detectors, where the same uncorrelated systematic error
$\sigma_u$ and the same data size (=20t$\cdot$yr) are assumed
(Only in this figure the statistical errors are taken into account).
\begin{figure}
\hglue 2.5cm
\epsfig{file=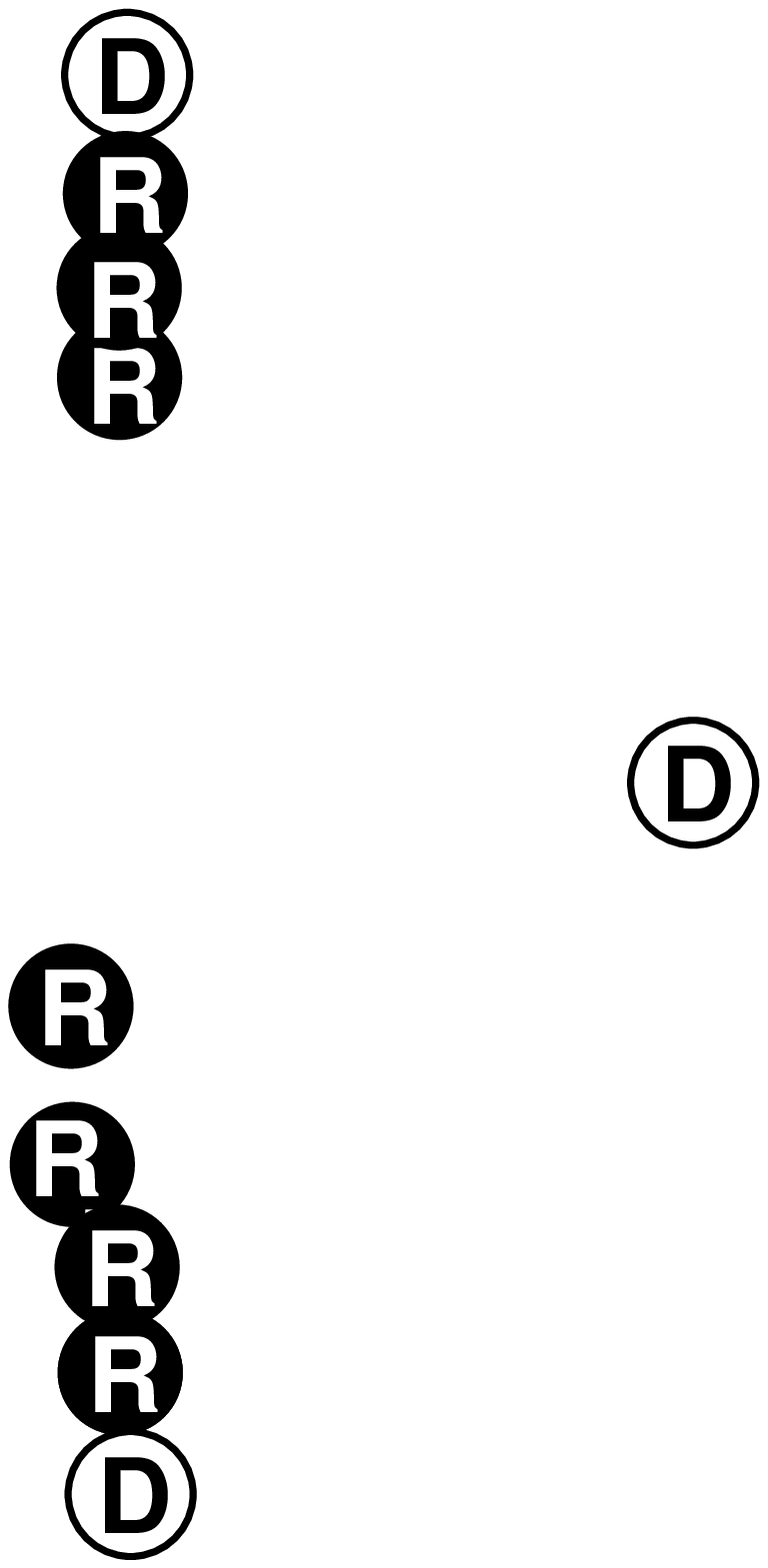,width=6.0cm}
\vglue -.5cm
\caption{
A hypothetical experiment with the same location of the reactors
as the Kashiwazaki plan but with stupid location
of the near detectors.
}
\label{fig:fig5}
\end{figure}
We observe that there is little difference between the sensitivities
of the actual Kashiwazaki plan and the hypothetical experiment
with a single reactor at $|\Delta m^2_{31}|=2.5\times10^{-3}$eV$^2$.
Also it is remarkable that the sensitivity of the actual Kashiwazaki plan
for higher value of $|\Delta m^2_{31}|$ is better than the
single reactor experiment.  This is exactly because of the
reduction of the uncorrelated error 
due to the nature of multiple reactors
(cf. Equation~(\ref{sigmaeff2})), where the near detectors
play a role as far detectors.
On the other hand,
to see how effectively the correlated errors
are canceled in the actual Kashiwazaki plan, comparison is given
in Figure~\ref{fig:KK}(b) between
the locations of the near detectors
in the actual Kashiwazaki plan and a hypothetical plan 
which is depicted in Figure~\ref{fig:fig5},
with one near detector very close to reactor \#1 in the
first cluster while the other detector is very close to reactor \#5
in the second.  In the hypothetical case, the sensitivity is
deteriorated because the correlated
errors do not cancel completely.  From these two figures
we see that the setup of the Kashiwazaki plan is almost optimized
and it does not suffer from an extra uncertainty because there are
more than one reactors.

\subsubsection{Energy spectrum analysis \cite{paper2}}
One can also derive a semi-analytic formula for the
effective systematic error in the analysis of
the energy spectrum, assuming that the uncorrelated
bin-to-bin error $\sigma_u$ is independent of bin.
The result is
\begin{eqnarray}
\displaystyle
\sigma_{eff}^2=
{{a_n \over n}\,\sigma_u^2 \over
\left[1+{\sigma_u^2 \over \sigma_u^2+2\sigma_{shape}^2}
+b_n\left({\sigma_u^2 \over \sigma_u^2+n\sigma_c^2}
+{\sigma_u^2 \over \sigma_u^2+2\sigma_{shape}^2
+n\sigma_c^2+2n(\sigma^{(r)})^2}\right)\right]},
\nonumber
\end{eqnarray}
where $n$ stands for the number of bins,
$\sigma_c$ is the total correlated error,
$\sigma_{shape}$ is an uncertainty of the
theoretical prediction for each energy bin which
is uncorrelated between different energy bins,
$\sigma^{(r)}$ is the uncertainty of the flux,
and the error of the energy calibration has been
ignored because it turns out to be small.
$a_n$ and $b_n$ are given by
\begin{eqnarray}
a_n&\equiv&{2 \over f_n}\nonumber\\
b_n&\equiv&{g_n \over f_n}\nonumber\\
f_n&\equiv&\displaystyle{1 \over n}\sum_{j=1}^{n-1}{1 \over j(j+1)}
\left(\sum_{k=1}^{j}c_{k}-(j+1)c_{j}\right)^2\nonumber\\
g_n&\equiv&{1 \over n^2}\left(
\displaystyle\sum_{j=1}^{n}c_{j}\right)^2\nonumber\\
c_j&\equiv&
\left\langle \sin^2\left({\Delta m^2L_{far} \over 4E_j}
\right) \right\rangle\quad(j=1,\cdots,n),\nonumber
\end{eqnarray}
and their numerical values are given in Table~\ref{table1}.
The integration over the energy in the
expectation value to define $c_j$ is from $E_j$ to $E_{j+1}$.
The systematic limit on $\sin^22\theta_{13}$ at 90\%CL
is obtained from $\sigma_{eff}$ by
\begin{eqnarray}
\left(\sin^22\theta\right)_{limit}^{sys~only}
=\sqrt{2.7}\,\sigma_{eff}.
\nonumber
\end{eqnarray}
For simplicity let us assume $\sigma_{shape}=0$ and
$\sqrt{n}\sigma^{(r)}\gg \sqrt{n}\sigma_c,~\sigma_u$.
Then we have
\begin{eqnarray}
\sigma_{eff}=\sqrt{{a_n \over 2n}}\,
\sigma_u\left(1+{b_n \over 2}{\sigma_u^2 \over \sigma_u^2+n\sigma_c^2}
\right)^{-1/2}.
\label{sigmaeff4}
\end{eqnarray}
If we further assume $\sigma_u=\sigma_c$ then (\ref{sigmaeff4})
is reduced to
\begin{eqnarray}
\sigma_{eff}=\sqrt{{a_n \over 2n}}\,
\sigma_u
\left(1+{b_n \over 2n+2}\right)^{-1/2},
\nonumber
\end{eqnarray}
and the numerical values of ${b_n /(2n+2)}$ are given in
Table~\ref{table2}, which shows that the contribution of
the correlated error is not negligible for lower value of $n$.
On the other hand, if we assume $\sqrt{n}\sigma_c\gg\sigma_u$, then
we have
\begin{eqnarray}
\sigma_{eff}=\sqrt{{a_n \over 2n}}\,\sigma_u,
\nonumber
\end{eqnarray}
and it indicates that the uncorrelated error gives
the dominant contribution to the effective systematic 
error.  The same conclusion is reached if
$\sqrt{n}\sigma_c\ll\sigma_u$ is assumed.
Therefore the realistic value of
the uncorrelated bin-to-bin error
$\sigma_u$ has to be estimated carefully in any case.

\begin{table}
\hglue 3.8cm
\begin{tabular}{|r|r|r|}
\hline
n & $a_n$ & $b_n$\\
\hline
2 & 151 & 44\\
4 & 76 & 20\\
6 & 60 & 15\\
6 & 51 & 12\\
16 & 41 & 10\\
$\ge 30$ &$\sim$ 40 & $\sim$ 9\\
\hline
\end{tabular}
\caption{The coefficients $a_n$ and $b_n$ which were
calculated numerically.}
\label{table1}
\end{table}
\begin{table}
\hglue 3.2cm
\begin{tabular}{|r|c|c|}
\hline
n & $(a_n/2n)^{1/2}$ & $b_n/(2n+2)$\\
\hline
2 & 6.2 & 7.3\\
4 & 3.1 & 2.0\\
6 & 2.2 & 1.1\\
6 & 1.8 & 0.7\\
16 & 1.1 & 0.3\\
62 & 0.6 & 0.1\\
\hline
\end{tabular}
\caption{The numerical values of the
coefficients $(a_n/2n)^{1/2}$ and $b_n/(2n+2)$
which appear in $\sigma_{eff}$ in the case of
$\sigma_c=\sigma_u$.}
\label{table2}
\end{table}

%% file: sites.tex
\section{Possible Sites}
\label{sec:site}
In this section we examine the power performance of commercial reactor 
sites all over the world and summarize the assets of sites that have been
identified as candidates to host a $\sin^22\theta_{13}$ experiment. 

Site selection will involve a comprehensive look at many criteria such as
potential for earth shielding, locations for near and far detectors,  
construction costs and reactor power.  Ultimately, choosing a site will require a
a comparison of cost and sensitivity for site specific proposals.
\subsection{Top Performing Reactors Worldwide}
One characteristic on which various reactor sites can be directly compared is
power. Tables~\ref{single}, \ref{double} and \ref{multi} list 
maximum and average power output for single, double and multi-reactor sites 
respectively.  The reactor sites are sorted by their average power output as 
measured over the 7 year period from 1996 to 2002.  The maximum rated power of 
each site is also listed.  Sites with maximum power ratings of less then 3 GW 
per reactor core (or 9 GW for multi-reactor sites) are considered only if they 
have been identified as potential host sites. 
\begin{table}[htbp]
\centerline{
\begin{tabular}{lccc} \hline
Reactor Site & Country & Avg $MW_{th}$ & Max $MW_{th}$\\ \hline
Brokdorf & Germany & 3900 & 4214\\
Emsland & Germany & 3892 & 4097\\
Grohnde & Germany & 3858 & 4184\\
Grand Gulf & US & 3505 & 3833\\
Grafenrheinfeld & Germany & 3357 & 3936\\
Wolf Creek & US & 3211 & 3565\\
Perry & US & 3199 & 3758\\
Callaway & US & 3176 & 3565\\
Leibstadt & Switzerland & 3130 & 3511\\
Waterford & US & 3152 & 3390\\
Watts Bar & US & 3049 & 3411\\
Unterweser & Germany & 3117 & 4126\\
Seabrook & US & 2924 & 3411\\
Vandellos & Spain & 2882 & 3181\\
Kruemmel & Germany & 2868 & 3851\\
Confrentes & Spain & 2858 & 3160\\
Hope Creek & US & 2794 & 3339\\
Fermi & US & 2750 & 3430\\
River Bend & US & 2676 & 3039\\
Trillo & Spain & 2672 & 3119\\
Columbia  & US & 2567 & 3486\\
Tokai & Japan & 2086 & 3219\\
Krasnoyarsk & Russia & 1600(?) & 2000(?) \\
\hline
\end{tabular}}
\caption{Power performance for single reactor sites around the
world~\cite{nureg-1350,nucleinics_week}.}
\label{single}
\end{table}

\begin{table}[htbp]
\centerline{
\begin{tabular}{lccc} \hline
Reactor Site & Country & Avg $MW_{th}$ & Max $MW_{th}$\\ \hline
South Texas Project & US & 6864 & 7600\\
Civaux & France & 6799 & 9135\\
Chooz & France & 6795 & 8872\\
Gundremmingen & Germany & 6734 & 7865\\
Braidwood & US & 6491 & 7172\\
Vogtle & US & 6456 & 7130\\
Byron & US & 6442 & 7172\\
Browns Ferry & US & 6377 & 6916\\
Limerick & US & 6365 & 6916\\
Isar & Germany & 6313 & 6985\\
Peach Bottom & US & 6290 & 6916\\
Sequoyah & US & 6209 & 6822\\
Penly & France & 6197 & 8088\\
Philippsburg & Germany & 6187 & 6976\\
Susquehanna & US & 6161 & 6978\\
Golfech & France & 6136 & 7977\\
Catawba & US & 6116 & 6822\\
Nogent & France & 6111 & 7977\\
San Onofre & US & 6061 & 6876\\
Diablo Canyon & US & 6043 & 6749\\
Comanche Peak & US & 5986 & 6916\\
St. Alban/St. Maurice & France & 5910 & 8082\\
Neckar & Germany & 5881 & 6452\\
McGuire & US & 5880 & 6822\\
Flamanville & France & 5879 & 8088\\
Biblis & Germany & 5528 & 7388\\
Asco & Spain & 5496 & 6013\\
Belleville & France & 5377 & 7977\\
Kuo-Sheng & Taiwan & 4749 & 5764\\
Angra & Brazil & 4547 & 5873\\
Indian Point & US & 4467 & 6096\\
La Salle & US & 4323 & 6978\\
Salem & US & 4281 & 6918\\
Ignalina & Lithuania & 3985 & 8778\\
D.C. Cook & US & 3281 & 6661\\
Millstone & US & 3271 & 6111 \\
\hline
\end{tabular}}
\caption{Power performance for double reactor sites around the 
world~\cite{nureg-1350,nucleinics_week}.}
\label{double}
\end{table}

\begin{table}[htpb]
\centerline{
\begin{tabular}{lcccc} \hline
Reactor Site & Country & Cores & Avg $MW_{th}$ & Max $MW_{th}$\\ \hline
Kashiwazaki-Kariwa & Japan & 7 & 20302 & 24029\\
Yonggwang & S. Korea & 6 & 16393 & 17264\\
Gravelines & France & 6 & 12458 & 16696\\
Zaporozhe & Ukraine & 6 & 12202 & 17557\\
Catternom & France & 4 & 12113 & 15942\\
Paluel & France & 4 & 11901 & 16176\\
Ohi & Japan & 4 & 11269 & 13782\\
Palo Verde & US & 3 & 10570 & 11552\\
Fukushima II & Japan & 4 & 10384 & 12875\\
Fukushima I & Japan & 6 & 10181 & 13741\\
Darlington & Canada & 4 & 9028 & 10932\\
Chinon & France & 4 & 8653 & 11166\\
Blayais & France & 4 & 8644 & 11131\\
Cruas & France & 4 & 8586 & 11190\\
Takahama & Japan & 4 & 8439 & 9925\\
Genkai & Japan & 4 & 8330 & 10177\\
Kori & S. Korea & 4 & 8314 & 9203\\
Ringhals & Sweden & 4 & 8307 & 10841\\
Tricastin & France & 4 & 8284 & 11178\\
Bruce & Canada & 4 & 8080 & 10710\\
Tihange & Belgium & 3 & 8075 & 9127\\
Hamaoka & Japan & 4 & 8031 & 10584\\
Forsmark & Sweden & 3 & 7773 & 9408\\
Dampierre & France & 4 & 7753 & 10967\\
Bugey & France & 4 & 7728 & 10897\\
Leningrad & Russia & 4 & 7642 & 11705\\
Balakovo & Russia & 4 & 7520 & 11705\\
Kozloduy & Bulgaria & 6 & 6618 & 11002\\
Kursk & Russia & 4 & 6577 & 11705 \\
\hline
\end{tabular}}
\caption{Power performance for multi-reactor sites around the 
world~\cite{nureg-1350,nucleinics_week}.}
\label{multi}
\end{table}

\subsection{Reactors Sites Under Consideration}
Many possible host reactor sites have been identified by groups around the
world.  In this section we list some of the relevant features from each site
mentioned for consideration.

\subsubsection{Angra, Brazil}
The Angra reactors are located on the coast of Brazil a few hours south of Rio de
Janeiro.  The site is ringed by sharply rising hill on three sides, with summits
of over 200 meters within 1.5~km.  

The president of the reactor company is a former particle physicist and groups at
several Brazilian institutions have shown an interest.

\subsubsection{Braidwood, Byron and La Salle, Illinois}
These reactor sites are all located within 100 km of both Fermilab and Argonne
National Laboratory, making them easily accessible to the
particle physics community. 
The topology of northern Illinois is generally flat, so obtaining cosmic 
shielding at these sites will require extensive excavation below the surface 
level.  Fortunately, the subterranean geology of this region is well
understood~\cite{Atkinson:2001ga}.  
At the Braidwood and La Salle sites there is a layer of dolomite, which is well 
suited for excavation, at a depth of 110 meters (approximately 300 mwe).  
At the Byron site, the western most of the three, the dolomite layer is much 
closer to surface.  Below the dolomite, starting at a depth of 50 to 100 
meters, is a layer of sandstone, which may be poorly suited for supporting 
excavated cavities. 

The Braidwood, Byron and La Salle reactors are owned and operated by the Exelon 
Corporation.  Currently, negotiations are ongoing with Exelon for the use of these
sites.

\subsubsection{Chooz, France}
This is the location of the previous 10 ton CHOOZ experiment, and a site for
a new near detector has been identified near the reactors.  The CHOOZ plant
has a total 8.4 thermal GW and the CHOOZ-far lab is located 1.05 km from
the two cores.

\subsubsection{Cruas, France}
The Cruas site is located in south eastern France.  The most likely near detector
baseline is about 1 km with a shielding potential of greater than 200 mwe.  A far
detector located at 1.8 km could get shielding greater than 400 mwe.

\subsubsection{Daya Bay, China}
There are two twin reactor cores located in Guangdong province near Hong Kong.
The total reactor power is 11.6 GW.  A third twin core is planned to be
online in 2010.  A near hall with an overburden of 400 mwe is potentially
located about 300~m from the core, and a far hall with 1200 mwe at a distance
of 1500-2000~m to the core.

\subsubsection{Diablo Canyon, California}
The Diablo Canyon site is located on the coast of California.  Within 1.5 ~km of
the reactors are hills of over 300 meters.  All the surrounding land is owned by
the utility, PG\&E, with which negotiations for the use of the site are ongoing.

\subsubsection{Kashiwazaki-Kariwa, Japan}
The Kashiwazaki-Kariwa reactor complex is the most powerful in the world.  The
site consists of 7 active reactors spanning approximately 700 meters.  The site is
relatively flat, so detectors would be located at the bottom of vertical shafts. 
Permission for the use of the site has been granted by the power company. A
full description of a proposal at this site is given in Appendix A.

\subsubsection{Krasnoyarsk, Russia}
The Krasnoyarsk reactor is part of a fully underground facility at a depth of 600
mwe.  Halls suitable for detectors exist at baselines of 115~meters and
1000~meters. Like the Wolf Creek site this single reactor site will allow for a
sensitive determination of the background rate.

Russian physicists have a long-standing relationship with the facility and have
mounted several neutrino experiments in the past~\cite{Vidyakin:1994ut}.

\subsubsection{Kuo-Sheng, Taiwan}
The Kuo-Sheng reactor complex is located in a hilly region of coastal Taiwan.  The
existence of a road tunnel within 2~km of the reactor indicates that tunneling is
viable in the hills surrounding the site.  Within 500 meters of the reactor there
are hills 50 meters or higher suitable for shielding a near detector.  Far 
detector shielding of 150 to 200 meters of rock is possible.

Local physicists have an existing relationship with the reactor operators and are
currently operating a neutrino magnetic moment experiment at a distance of
28~meters from one of the two reactor cores~\cite{Li:2002pn}. 

\subsubsection{Limerick and Peach Bottom, Pennsylvania}
These reactor sites are also owned and operated by Exelon.  The topology of
Pennsylvania is somewhat more hilly than in Illinois.  Therefore, the potential 
for significant earth shielding with horizontal access may exist
 at these sites. 
In particular, Peach  Bottom has elevation variations of greater that 60~meters 
within  1.5~km of the reactors.

\subsubsection{Penly, France}
The Penly reactor complex is embedded in 120~meter high chalk cliffs.

\subsubsection{Wolf Creek, Kansas}
The topology of this single reactor site is similar to the flat sites in Illinois. 
At a depth of 110 meters there is relatively wide layer of shale.  While the shale
is not ideal for tunneling it is likely sufficient to support the structures
required for this project. 

The main attraction of the Wolf Creek site is that as a single reactor facility it
will get full reactor off running  which can be used to measure the background
rate.

%% file: other.tex
\section{Other physics}
\label{sec:other}
\par The main goal of the experiment described in this white paper is
to search for or further limit the value of $\quq$.  It is reasonable
to explore whether there is other physics that can be done with this experiment,
or with modest enhancements.  In this section we briefly consider three
other physics topics: sterile neutrinos, $\theta_{12}$ and reactor physics.

\subsection{Sterile Neutrinos} 

The discovery of sterile neutrinos would have a revolutionary impact on
neutrino and particle physics. If there is non-negligible mixing of the
electron neutrino with sterile neutrinos, they would contribute to the reactor
antineutrino disappearance pattern.  An idea how to look for sterile neutrinos
at reactors along with $\sin{\theta}_{13}$ was proposed at the Kurchatov
Institute in 1998~\cite{bib:mikhaelyn}.

The notion of sterile neutrinos was originally introduced by
B.~Pontecorvo in 1967~\cite{bib:pont} and has been considered later by
many authors, e.g., D.~Caldwell and R.~Mohapatra~\cite{bib:caldwell},
J.T.~Peltoniemi, D.~Tommasini and J.W.F.~Valle~\cite{bib:valle},
S.~Bilenky, C.~Giunti and W.~Grimus~\cite{bib:bilenky}, K.~Benakli and
A.~Smirnov~\cite{bib:benakli}, B.~Kayser~\cite{bib:kayser}.
Information on the theory of sterile (and mirror) neutrinos and
references can be found in the recent paper by V.~Berezinsky,
M.~Narayan, F.~Vissani~\cite{bib:berezinsky}.

While solar, atmospheric, and laboratory (Super Kamiokande, SNO,
KamLAND) studies are understood in the framework of only 3-active
neutrino mixing (see, however, de~Holanda and 
Smirnov~\cite{deHolanda:2003tx}) they do not exclude some admixture of
sterile neutrinos~\cite{Maltoni:2003da}.
An experimental hint in favor of sterile neutrinos comes from the unconfirmed
observation of the LSND collaboration~\cite{bib:lsnd} on ${\nu}_{\mu}
\rightarrow {\nu}_e$ transitions. In particular the so-called (3+1) and (3+2)
neutrino schemes, which have been considered to explain the LSND signal,
predict reactor neutrino disappearance with a $\dmsq\sim$~eV$^2$ very close to
the current upper bound from the Bugey experiment (see References~\cite{rulingout}
and \cite{bib:3+2}).


\subsubsection{The effect of sterile neutrinos in $\theta_{13}$ reactor
experiments}

In the standard 3-active neutrino framework, antineutrino
disappearance at distances $L = 1000-2000$~m from a reactor source is
governed by $\dmsq_\mathrm{atm} \sim 2\times 10^{-3}$~eV$^2$ and by the
mixing parameter $\sin^2 2\theta_{13}$:
\begin{equation}
P_{\bar\nu_e \rightarrow \bar\nu_e} \approx 1 -
\sin^{2}2{\theta}_{13} \, 
\sin^{2}\left( \frac{\dmsq_\mathrm{atm} L}{4E}\right) \,.
\end{equation}

Let us denote the three standard neutrino mass states by $\nu_i$
($i=1,2,3$), such that the atmospheric and solar mass-squared
differences are given by $\dmsq_\mathrm{atm} = m^2_3 - m_1^2$ and
$\dmsq_\mathrm{sol} = m^2_2 - m_1^2 \sim 7 \times 10^{-5}$~eV$^2$, the
solar mixing angle is determined by $\tan^2 \theta_\mathrm{sol} =
|U_{e2} |^2 / |U_{e1}|^2 \sim 0.4$, and $\sin^2 2 \theta_{13} = 4
|U_{e3}|^2( 1 - |U_{e3}|^2 )$.
If we now assume that there are additional neutrino states $\nu_i$ ($i =
4,5,\ldots$) with masses such that\footnote{Note that any new mass-squared
difference with $\dmsq \ll 10^{-3}$~eV$^2$ will have no effect in reactor neutrino
experiments with $L \sim 2$~km.} $|\Delta m^2_{i1}| \gsim 10^{-3}$~eV$^2$ we
know from the Bugey, Palo Verde and CHOOZ experiments that the mixing of the
electron neutrino with these new mass states has to be small (see shaded
region in Figure~\ref{fig:sterile_limits}), i.e.,
\begin{equation}
|U_{ei}|^2 \ll 1 \quad\mbox{for}\quad i \ge 3 \,.
\end{equation}
To first order in these small quantities one obtains for the survival
probability at nuclear reactors
\begin{equation}\label{eq:Psterile}
P_{\bar\nu_e \rightarrow \bar\nu_e} \approx 1 -
4 \sum_{i\ge 3} |U_{ei}|^2 \, 
\sin^{2}\left( \frac{\dmsq_{i1} L}{4E}\right) \,.
\end{equation}
In the case of only one additional neutrino one can write
Equation~(\ref{eq:Psterile}) as
\begin{equation}
P_{\bar\nu_e \rightarrow \bar\nu_e} \approx 1 
- \sin^{2}2{\theta}_{13} \, 
\sin^{2}\left( \frac{\dmsq_\mathrm{atm} L}{4E}\right) 
- \sin^{2}2{\theta}_s \, 
\sin^{2}\left( \frac{\dmsq_\mathrm{new} L}{4E}\right) \,,
\label{eq:P1sterile}
\end{equation}
where $\sin^{2}2{\theta}_s \approx 4|U_{e4}|^2$ and $\dmsq_\mathrm{new} =
m_4^2 - m_1^2$ are the mixing parameters of the sterile neutrino, and
$\sin^{2}2{\theta}_{13}\approx 4|U_{e3}|^2$.

From Equation~(\ref{eq:P1sterile}) it is obvious that such a sterile neutrino would
have some impact on an experiment of the Kr2Det type~\cite{bib:mik2} or its
modifications (as discussed in 2002--2003 at the meetings in Paris, Alabama
and Munich), provided the associated mixing parameter $\sin^{2}2{\theta}_{s}$
is not too small and the mass-splitting is in the relevant range. 
\begin{itemize}
\item
If $\dmsq_\mathrm{new} \sim \dmsq_\mathrm{atm}$ it will be rather
difficult to disentangle $\bar\nu_e \to \bar\nu_{\mu/\tau}$
oscillations with $\sin^22\theta_{13}$ from $\bar\nu_e \to \bar\nu_s$
oscillations with $\sin^22\theta_s$ at a reactor experiment,
especially if the main information comes from the total rate
measurement. Only if enough spectral information is available and/or
$\dmsq_\mathrm{atm}$ and $\dmsq_\mathrm{new}$ differ sufficiently one
might be able to separate the two channels. If no effect is observed
both, $\sin^22\theta_{13}$ and $\sin^22\theta_s$ can be constrained.
\item
If $\dmsq_\mathrm{new} \gg \dmsq_\mathrm{atm}$ and the oscillations
with $\dmsq_\mathrm{new}$ are already averaged out at the near
detector position at $L\sim$~few~$\times 100$~m, no information about
sterile neutrino mixing can be obtained from the comparison of the far
and near detectors, and the transitions to the sterile state will
not affect the $\theta_{13}$ measurement. In that case information on
sterile neutrino mixing can be obtained from the near detector if
relatively precise information on the initial reactor neutrino flux is
available, or if a ``very-near'' detector at $L \sim 10$~m could be
installed (see below).
\end{itemize}

%
%



\begin{figure}[t]
\begin{center}
\includegraphics[width=0.8\textwidth]{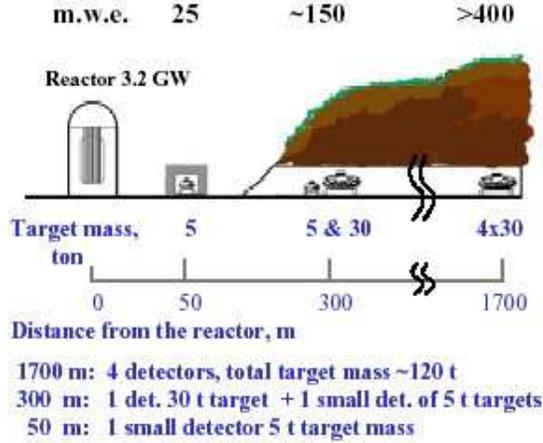}
\end{center}
\caption{\label{fig:sterile_layout} Example of a layout. Detector
  positions, target scintillator masses and overburden (m.w.e.) are
  shown.}
\end{figure}

\subsubsection{Example of layout}

Imagine that a tunnel is built near one 3.2~GW thermal power
reactor. We consider five identically designed 30 ton target
scintillator (movable) detectors, four of them stationed in the far
position at a distance of 1700~m from the reactor, and one  detector
stationed in the near position at 300~m from the reactor. To expand
the explored mass parameter region toward larger values two small
detectors are considered at 300 m and 50 m from the reactor. A
possible layout of such an experiment is illustrated in
Figure~\ref{fig:sterile_layout}. The expected neutrino detection rates
per 300 days are shown in Table~\ref{tab:sterile_rates}.

\begin{table}[t]
\begin{center}
\begin{tabular}{|c|c|c|}
\hline
Distance [m] & Target mass [t] & $\bar\nu_e$ rate/300day \\
\hline
50  & 5 & 1 100 000 \\
300 & 5 &   30 000 \\
300 & 30 &  190 000 \\
1700 & 4 x 30 & 24 000  \\
\hline
\end{tabular} 
\end{center}
\caption{\label{tab:sterile_rates}
Detector positions, scintillator target masses and
$\bar\nu_e$ detection rates per 300 days.}
\end{table}

We consider two types of data analysis: shape and rate. With only one reactor
as $\bar\nu_e$ source the shape analysis (as we already know) is independent
of the exact knowledge of reactor power, energy spectrum of the $\bar\nu_e$,
flux and spectral time variations, target volumes and proton concentrations,
and absolute efficiencies of $\bar\nu_e$ detection. Backgrounds can be
measured periodically during reactor-off periods. On the other hand, the
analysis based on the comparison of the far/near $\bar\nu_e$ total rates
requires good knowledge of the ratios of target volumes and detection
efficiencies. In both cases no exact information from the reactor services on
reactor power and fissile fuel composition is needed for the data analysis.

\subsubsection{Sensitivity}

\begin{figure}[t]
\begin{center}
\includegraphics[width=0.8\textwidth]{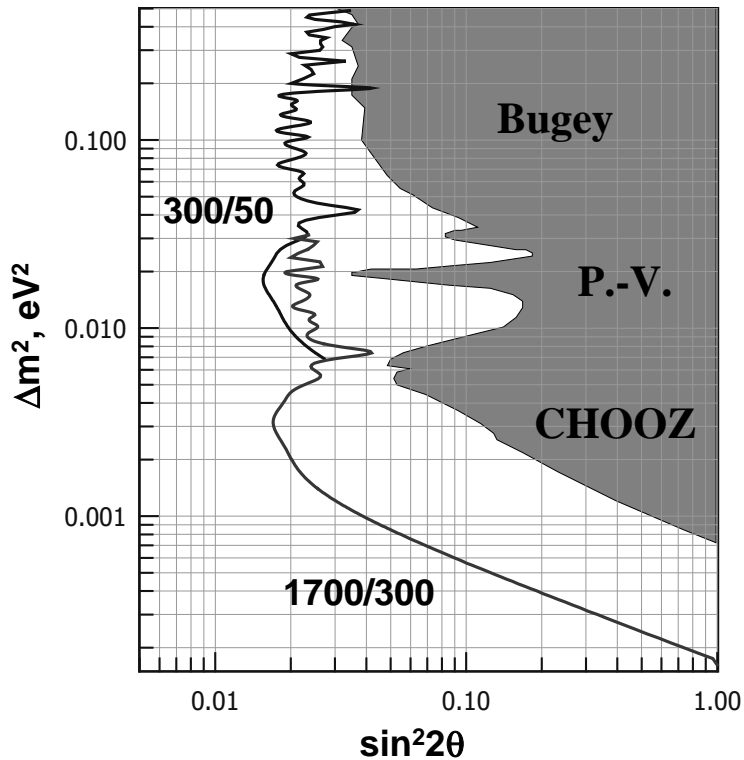}
\end{center}
\caption{\label{fig:sterile_limits} Expected 90\% CL limits from the
comparison of detectors at distances of 300/1700~m and 50/300~m. The shaded
region is excluded by the CHOOZ, Palo Verde and Bugey experiments.}
\end{figure}

In Figure~\ref{fig:sterile_limits} we show the sensitivity for
$\sin^22\theta_{13 / s}$ for the experimental configuration described
in the previous section. Within 3 years of data taking (300 days/year)
in a large part of the $\Delta m^2$ range from 0.001 to 0.5~eV$^2$ a
sensitivity of $\sin^{2}2{\theta}_{13 / s} \sim 0.01 - 0.02$ can be
obtained~\cite{Kopeikin:2003uu}, which is in general agreement with
the analysis performed by P.~Huber, M.~Lindner, T.~Schwetz and
W.~Winter~\cite{Huber:2003pm}. The limits shown in
Figure~\ref{fig:sterile_limits} were obtained by assuming an energy
resolution ${\sigma}_{E} = 0.08 \sqrt{E}$ and the systematics
${\sigma}_\mathrm{shape} = 0.5\%, {\sigma}_\mathrm{rate} = 1\%$. As
can be seen in the figure the CHOOZ limit on $\sin^{2}2{\theta}_{13}$
at ${\Delta}m^{2}_\mathrm{atm} = 2\times 10^{-3}$~eV$^2$ can be
improved by a factor 10.

%
%
%

We conclude that a search for sterile neutrinos at reactors does not require
much additional effort and can be done along with ${\theta}_{13}$. With one
reactor and a number of detectors a high sensitivity to ${\theta}_{13}$ and
sterile neutrino mixing can be reached. To increase the sensitivity in the
range $\dmsq \gg \dmsq_\mathrm{atm}$ [in particular to reach the
$\mathcal{O}$(eV$^2$) region indicated by the LSND evidence] a very-near
detector at $L\sim 10$~m is necessary.

\subsection{Solar Neutrinos and $\theta_{12}$}

\subsubsection{The present and near-future for $\theta_{12}$}

\newcommand{\be}{\begin{eqnarray}}
\newcommand{\ee}{\end{eqnarray}}
\def\nue{{\nu_e}}
\def\anue{{\bar\nu_e}}
\def\numu{{\nu_{\mu}}}
\def\anumu{{\bar\nu_{\mu}}}
\def\nutau{{\nu_{\tau}}}
\def\anutau{{\bar\nu_{\tau}}}
\newcommand{\st}{\mbox{$\sin^{2}2\theta$~}}
\newcommand{\br}{\mbox{$^{8}{B}~$}}
\newcommand{\ber}{\mbox{$^{7}{Be}$~}}
\newcommand{\cl}{\mbox{$^{37}{Cl}$~}}
\newcommand{\ga}{\mbox{$^{71}{Ga}$~}}
\newcommand{\chr}{\mbox{$\breve{\rm C}$erenkov~}}
\newcommand{\kl}{\mbox{KamLAND~}}
\newcommand{\bx}{\mbox{Borexino~}}
\newcommand{\thsol}{\mbox{$\theta_{12}$~}}
\def\ltap{\ \raisebox{-.4ex}{\rlap{$\sim$}} \raisebox{.4ex}{$<$}\ }
\def\gtap{\ \raisebox{-.4ex}{\rlap{$\sim$}} \raisebox{.4ex}{$>$}\ }
\newcommand{\sss}{\sin^2 \theta_{12}}
\def\ltap{\ \raisebox{-.4ex}{\rlap{$\sim$}} \raisebox{.4ex}{$<$}\ }
\def\gtap{\ \raisebox{-.4ex}{\rlap{$\sim$}} \raisebox{.4ex}{$>$}\ }
\newcommand{\ms}{\Delta m^2_{21}}
\newcommand{\ma}{\Delta m^2_{31}}
\newcommand{\ts}{\sin^2 2\theta_{12}}
\newcommand{\sch}{\sin^2 \theta_{13}}
\newcommand{\dma}{\mbox{$\Delta m^2_{\rm A}$ }}

The first data from the \kl experiment~\cite{Eguchi:2003prl} showed that the
reactor $\bar{\nu}_e$ undergo oscillations on a distance scale of ~$\sim
160$~km. Under the plausible assumption of CPT-invariance, this result
confirmed the Large Mixing Angle (LMA) solution of the solar neutrino
($\nu_{\odot}$) problem.  When combined with the solar neutrino data,
the \kl data split the allowed LMA solution region of $\nu_{\odot}$
oscillation parameters $\ms$ and $\sss$ into two sub-regions -- low-LMA
and high-LMA. The best fit values of $\ms$ and $\sss$ in the two
sub-regions read: $\ms = 7.2\times 10^{-5}$ eV$^2$, $\sss=0.3$
(low-LMA), and $\ms = 1.5\times 10^{-4}$~eV$^2$, $\sss=0.3$ (high-LMA)
\cite{kldata}. Adding the salt phase data from the SNO
experiment~\cite{bib:SNOsalt} the global analysis allows the high-LMA
solution only at the 99.13\% CL~\cite{salt}. Thus, the high-LMA
solution is disfavored by the current data compared to the low-LMA
one, although it is still not ruled out comprehensively.

The $\bar{\nu}_e$ survival probability in the reactor experiments of
interest, $P_{ee}$, depends on $\ms$, $\sss$, $\ma$ (driving the
atmospheric neutrino oscillations), the angle $\theta_{13}$ limited by
the CHOOZ and Palo Verde experiments, and on the type of neutrino mass
hierarchy \cite{SPMP02,th12hlma}.  The potential sensitivity of a
reactor experiment to each of these parameters depends crucially on
the baseline of the experiment.  Experiments with a baseline $L \sim
(1 - 2)$~km can be used to get information on $\sch$: on the indicated
distances only oscillations induced by $\ma$ are operative and
$\sin^22\theta_{13}$ determines their amplitude. For baselines $L
\gtap 50$~km, the $\bar{\nu}_e$ oscillations due to $\ma$ average out
and we have: $P_{ee} \approx [1 - \sin^2 2\theta_{12} \sin^2(\dm
L/4E)] \cos^4\theta_{13}$.  The oscillations generated by $\ms$ have
been seemingly detected in the KamLAND experiment.

In the absence of oscillations, the maximal contribution to the
signal in an experiment with reactor $\bar{\nu}_e$ detected via the
inverse $\beta-$decay reaction comes from $\bar{\nu}_e$ with energy $E \sim
3.6$ MeV.  For a fixed $\ms$, maximal sensitivity to $\sss$ can be
achieved if for $E \sim 3.6$ MeV, $L$ is ``tuned'' to a $\bar{\nu}_e$
survival probability minimum (SPMIN), i.e., if $\sin^2(\dm L/4E)
\approx 1$.  The latter is reflected in the $e^+-$spectrum, measured
in the experiment.  The corresponding minimum in the spectrum is
determined by $P_{ee} \approx 1 - \sin^2 2\theta_{12}$, and thus is
very sensitive to the value of $\sin^2 2\theta_{12}$.  If, in
contrast, $L$ is such that $\sin^2(\dm L/4E) = \epsilon \ll 1$,
$P_{ee}$ would have a maximum (SPMAX): $P_{ee} \approx 1 - \epsilon
\sin^2 2\theta_{12} \approx 1$.  In this case the sensitivity to
$\sin^2 2\theta_{12}$ is worse than in the preceding one. The
positions of the extrema in both cases are highly sensitive to the
value of $\ms$.
%
 
For values of $\ms$ from the low-LMA region, $\ms \sim 7.2\times
10^{-5}$ eV$^2$, and $E \sim 3.6$ MeV, the SPMIN and SPMAX take place
at $L \sim 70$ km and $L~\sim~160$~km.  For the \kl experiment, the
most powerful reactors (the Kashiwazaki complex) are located at a
distance of $\sim 160$ km, unfortunately close to the SPMAX for
low-LMA.  The values of $\ms$ and $\sss$, allowed at 99\% CL\ by
existing and prospective \kl data have been determined in
References~\cite{th12,nfproc}.  The uncertainty in $\ms$, determined using
only the $\nu_{\odot}$ data, reduces from 68\% to 30\% after the
inclusion of the first \kl data in the analysis, while that in $\sss$
does not change, remaining rather large (29\%).  The uncertainty in
$\ms$ would further diminish to 9\% (6\%) after 1 kTy (3 kTy) data
from KamLAND.  However, there will be little improvement in the
precision on the value of $\sss$ with the increase of \kl
statistics.
%

The $L$ best suited for measuring $\theta_{12}$ if $\ms$ lies in the
low-LMA region is $\sim 70$~km~\cite{th12}.  For a reactor complex
with a power of 24.6 GW (e.g., Kashiwazaki) and data of 3 kTy from a
KamLAND-like detector at $L \sim 70$ km, $\sss$ can be determined at 99\%
CL\ with a $\sim$10\% uncertainty~\cite{th12}.

A new reactor power plant, Shika-2, is expected to start operations in
Japan in March 2006. It will be located at $L \sim 88$ km from \kl and
will have a power of $\sim$4 GW. This baseline is close to the
``ideal'' one of $L \sim 70$ km.  The implications of the new source
of $\bar{\nu}_e$ on the \kl sensitivity to $\ms$ and $\sss$ were
studied in \cite{shika}.  It was concluded that due to
averaging effects of the $\anue$ fluxes from the Kashiwazaki and
Shika-2 reactors, the sensitivity of \kl to $\sss$ would not improve,
while its sensitivity to $\dm$ would diminish.

If contrary to the trend emerging in the solar neutrino experiments
the next \kl results conform to a point in the high-LMA region, one
would need an intermediate baseline reactor experiment with $L\sim
(20-30)$ km to get a SPMIN in the resultant $e^+-$spectrum
\cite{SPMP02,th12hlma}.  It was shown in \cite{th12hlma} that with an
experimental set-up at intermediate $L \sim (20 - 30)$~km from a
reactor with power of 5~GW and 3~kTy of statistics, one could measure
both $\ms$ and $\sss$ with a $\sim 3-7\%$ error at the 99\% CL. If in
addition the detector has a sufficiently high energy resolution and
$\sin^2\theta_{13} \gtap 0.03$, one could observe the $\ma-$driven
subdominant oscillations.  This could be used to measure also $\ma$
with a high precision, and even to get information on the neutrino
mass hierarchy~\cite{SPMP02,th12hlma}.

\subsubsection{Role of a new Experiment for $\theta_{12}$}

\par The goals of a reactor experiment at 70 km and 2 km from
a reactor are substantially different.  However, it is worthwhile
to consider the location of reactors when one is considering the
site of an experiment.  It might be useful to consider whether a
particular reactor has reasonable sites both 2 km and 70 km from
the core or multiple cores.  It is also reasonable to consider whether
multiple and movable neutrino detectors could have any role in a
future $\theta_{12}$ experiment.

\subsection{Reactor Physics}

Any near detector for an experiment to measure $\quq$ will measure the
flux and energy distribution of the reactor neutrinos with a greater
accuracy than has been done before.\footnote{This might be also of
some relevance for other reactor neutrino experiments without near
detectors like KamLAND. A precise determination of the reactor
neutrino spectrum could help to reduce the error implied by the flux
uncertainty for such experiments.} This will allow comparison with
both thermal power and reactor fuel loading measurements and
calculations. We are not currently aware that there are any important
checks that can be made on reactor design. However, we will continue
to work on this as a possibility and a possible benefit to the reactor
companies. Another application could be the direct check of nuclear
non-proliferation treaties.

\subsection{Supernovae Neutrinos}

A large vat of scintillator will be sensitive to antineutrinos from
a galactic supernova in the 10 to 50 MeV range.  This is higher energy 
than most neutrinos from a reactor.  A 50~ton detector is not large,
but could be sensitive to a portion of our galaxy.  Sixteen events would be
expected from a supernova at 10 kpc.
A search for these events will 
probably require
an accurate clock and a separate trigger.  If simulations show that a
supernova can be uniquely identified online with no background, a reactor
detector experiment could join the SuperNova Early Warning System, of
potentially invaluable use to astronomers. \footnote{See 
http://cyclo.mit.edu/snnet/.}

%% file: tunnel.tex
\section{Tunneling}
\label{sec:tun}
\subsection{Introduction}
This section provides guidance on the underground siting of the {$\quq$}
facilities. Requirements, design and construction issues are discussed and the basic phases in the planning and construction of a tunnel are outlined.
The near-term need for site-specific geotechnical data to support project 
development is emphasized. Site-specific geologic and geotechnical data is 
needed to support the identification and study of candidate alignment(s) 
at the different sites. This same data will also serve to support the early 
selection of construction methods and means and the development-associated 
work schedule(s) and cost estimate(s).  At present only a limited amount of 
geotechnical information has been gathered on individual sites so comments are 
necessarily general in nature. 
This section is limited to a discussion of the more general aspects 
of tunnel construction in rock.  Shaft excavation is 
discussed in Section~\ref{sec:shaft} and life safety issues are addressed 
in Section~\ref{sec:safety} of this White Paper.
\subsection{Factors Impacting Rock Tunnel Behavior}
A basic understanding of the factors that influence tunnel 
behavior is a prerequisite to the achievement of a 
practical and cost-effective underground design. A tunnel design developed 
with due regard to the constraints of the construction process will 
ensure realistic requirements setting, result in a cost effective design,
and ultimately help provide for a more affordable and lower risk 
construction product. Conversely, if the design requirements do not 
pay due attention to the particular constraints of the 
underground construction process, they are likely to place unreasonable 
demands on the contractor, mining methods and means and/or 
the rock mass. These unreasonable demands will translate to 
an increase in costs and risk that the project and/or funding agent 
may find prohibitive.
A discussion of some construction constraints is provided below. 
\subsubsection{Tunnel Size, Shape and Alignment Considerations}
The tunneled excavation, as designed, will satisfy the space demands of the end user while providing support to maintain an adequate safety margin against tunnel instability.  The potential for instability around the tunneled excavation is impacted by both the size and shape of the excavation.
Where tunnel instability is driven by the density and shear strength of 
natural fractures, the larger the opening excavated the greater 
the likelihood of more frequent, larger rock fall-outs occurring and 
the greater the density and size of the supporting structures 
needed to counter such fall-outs.
Where in situ stresses are relatively high compared to the 
strength of the rock mass, the tunnel profile may be modified to 
an elliptical or circular shape. Such cross-sections are selected to 
avoid corners, at which high stress concentrations occur. If in 
situ stresses are relatively low compared to the rock strength, the selection 
of the cross-section shape may be driven by the economic factors 
of the mining process itself (excavation, ground support and treatment).  
Water within the rock mass can have a deleterious impact 
on the stability of the excavation. Water pressure acting across 
planes of weakness or flushing through soil-like zones within 
the rock mass can instigate fall-out at the tunnel 
perimeter. Besides increasing the likelihood of fall-outs, the mere presence 
of water within the tunnel can have a significant detrimental impact 
on the efficiency of the tunneling operation. To reduce the degree 
of interference that water can have on the tunnel work, it 
is common practice to place tunnels on a slight gradient. This 
gradient provides for gravity drainage and collection of water 
at a sump located either at the base of a shaft or portal and 
away from the tunnel heading. Slopes are commonly kept 
below about 3\% where rail-mounted (steel wheels on steel track) operations are 
envisaged, but can be steeper if only rubber tired vehicle access is required.
\subsubsection{Tunneling Methods and Means}
In all but the weakest rock, three basic types of excavation methods 
are commonly considered to be feasible for tunnel work. Two 
of these mining methods rely on mechanical breakage of the 
rock, namely the tunnel boring machine-system (TBM-system), and the 
roadheader; the third method is ``drill and blast" (D\&B) which relies  
on the use of explosives to effect rock breakage.  
The TBM-system, which includes not only 
the cutterhead machine but also the ground support and muck 
evacuation systems, is often used to good effect in the 
excavation of longer, smaller diameter, relatively straight tunnels with 
uniform cross-sectional requirements, mined under more uniform rock 
mass conditions. In such applications, the TBM-system will have 
the ability to mine tunnel {\it faster and cheaper} than either of the 
other two mining methods cited. However, for many tunnel projects the 
TBM is not an automatic choice. It has a relatively high capital cost, mines 
a fixed, circular cross-section, cannot mine tight bends or corners 
and requires an extended period for fabrication/refurbishment, mobilization and 
demobilization. 
When mining shorter, larger and/or more complex tunnel layouts 
that demand greater flexibility from the mining equipment, including 
tight turns, steep gradients, multiple cross-sections and variations 
in ground support and treatment en route, roadheader or 
drill and blast excavation methods may be preferred. 
The roadheader and drill and blast methods have similar degrees of mining flexibility, but the roadheader has a limited range of economic application. Roadheader viability is severely compromised in harder, more abrasive and massive rock masses conditions where mining rate is reduced and the abrasive wear on the cutting tools increased significantly. 
Drill and blast tunnel excavation offers the user the 
most flexible excavation system that can be used economically 
in even the hardest, most abrasive, rocks. However, if 
drill and blast methods are used, the rock mass 
surrounding the tunnel will be subject to additional fracturing due to 
blast damage. Where explosives are used as the means of excavating 
the tunnel an increment in support should be anticipated over that 
required for the mechanically-mined tunnel.  
Even if the drill and blast method is ultimately not selected to mine the tunnel, the excavation of portals and shafts may require the use of explosives.
\subsubsection{Rock Support and Treatment}
In all but the most intrinsically stable rock mass, some ground 
support will be needed in the tunnel. Rock support will be installed to 
stabilize the tunnel periphery and ensure that the miners can work safely 
within the confines of the newly excavated tunnel. The support 
installed can serve either a temporary (during construction) or a 
permanent (for the life of the project) support role. 
Rock support installed at the heading will be adjusted locally 
in response to variations in the ``as-encountered" ground conditions. Rock supports 
may be supplemented by ground treatment work performed around and/or 
ahead of the face where necessary. Treatment of the rock mass 
may be needed to improve the tunneling conditions and reduce the 
impact of construction on the surrounding area (for example, water table 
draw-down or surface settlement). Treatment (freezing, grouting etc.) may 
be undertaken to achieve a temporary or permanent increase in rock 
mass strength or a reduction or increment in rock mass permeability.  Rock 
mass zones that may require significant amounts of such treatment should 
be identified early in the site investigation process in order that 
their presence, characteristics, extent and mining impacts can be 
fully evaluated during the siting process. Ground treatment, even if 
only required for a short stretch of tunnel along the alignment, can prove 
time consuming and costly.
\subsection{Underground Design Requirements}
At the conceptual stage of design, initial estimates of 
ground shielding, clearance envelopes (including tolerances), and general layout/environmental 
criteria required for the construction, installation, operation 
and maintenance phases of the Project should be defined 
in a drawing set that shows the tunnel(s) in plan and section 
(longitudinal and cross). The tunnel excavations should be laid-out to 
be compatible with selected construction methods and means.  
The tunnel(s) plan and section should show key geologic and 
hydrologic information. 
Some underground requirements that should be estimated 
during this conceptual period include:
\begin{itemize}
\item External loading of floors, wall and crown anchorages including detector supports, transportation and lifting system
\item Electrical, electronic, communications networks (cables and cable trays)
\item Heating, cooling, ventilation and air conditioning (duct work, fans, door and louvers, drip ceilings and underground chilling/heating units etc.)
\item Groundwater collection and evacuation systems (excavations, drains, pumps and pipes)
\item Survey controls (including stations and lines of sight)
\item Environmental requirements (spoil disposal, groundwater protection.)
\item Neighborhood issues (mitigation of construction/operation impacts on- and off-site). 
\end{itemize}
\subsection{The Phases of a Tunnel Project}
\begin{figure}
\vspace*{13pt}
\begin{center}
         \mbox{\epsfig{figure=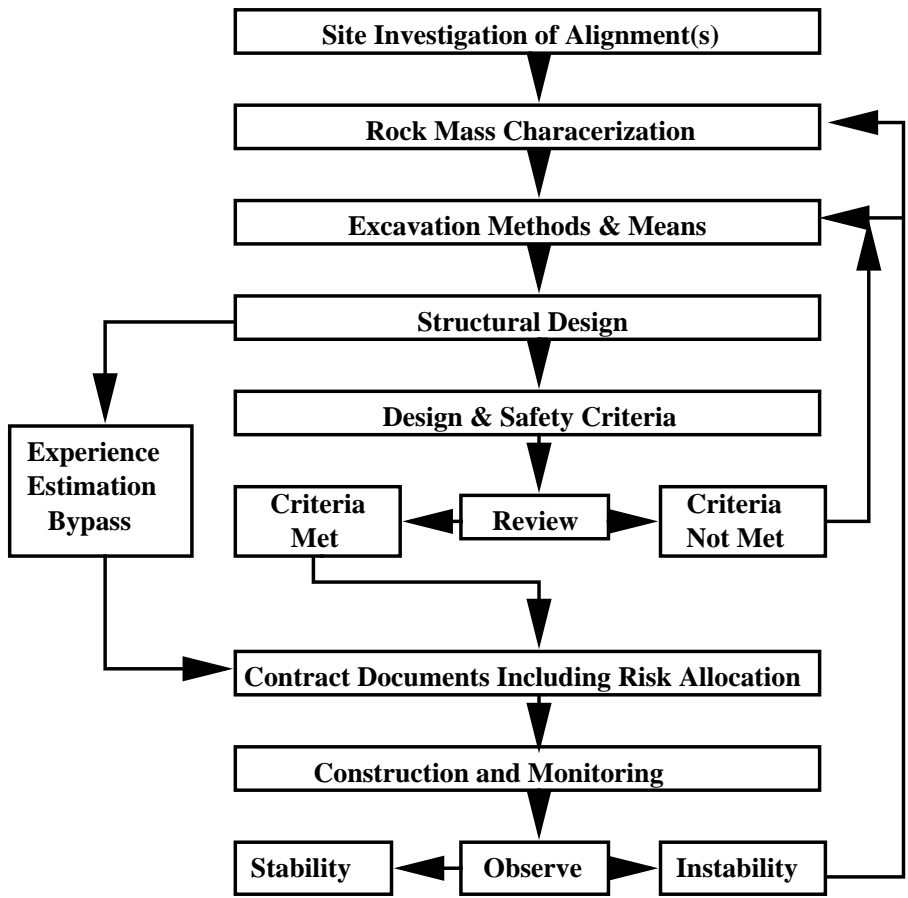,width=12.0cm}}
\caption{Flow Chart for Tunnel Design, after Reference~\cite{bib:lowe}
}
\label{fig:tunnel1}
\end{center}
\end{figure}

The main design and construction phases of a tunnel project are 
outlined in the flowchart in Figure \ref{fig:tunnel1}. The 
flowchart is modeled after the International Tunneling Association 
guidelines for tunnel design. The figure indicates a stepwise 
progression from site investigation through to construction and monitoring. In practice, the site investigation activity overlaps other planning activities to allow for the detailed investigation of design and the mapping of the excavated geology. 
In addition to the activities listed in the flowchart, there 
will also be a need for estimating and scheduling work. As 
the project progresses it is likely that periodic reviews will 
be held to evaluate progress, improve management confidence in 
budget and time goals and enhance the practicality and 
economy of the tunnel work itself.

The phases of the tunnel project are briefly discussed below.
\subsubsection{Site Investigation and Rock Mass Conditions}
To evaluate a site's suitability to be a ``host" for an underground facility 
both regional and location-specific geologic information will need 
to be gathered, including information on rock units, structural folds 
and faults, groundwater and stress regimes. This basic geological 
information will need to be interpreted to characterize the rock 
mass along the alignment(s) and provide input for the requisite constructibility 
and engineering analyses. The interpretation will be used to 
support critical decisions on alignment and selection of methods 
and means. Early acquisition of site investigation data can help quickly 
identify difficult or {\it showstopper} situations along an alignment 
and expedite the short-listing of the more serious alignment candidates. Much, 
if not all, of the site investigation data necessary to support initial 
decisions on alignment and methods and means choices can be obtained 
from desk studies using published reports and papers of the regional 
geology, relevant construction case history data sets and field observations 
at the sites.
As the design progresses from the conceptual to the alignment-specific stage, 
site-specific information will be needed to support the validation 
of methods and means for use in design and study rock mass issues 
that were noted as needing further investigation during the 
conceptual stage of the project.  At this stage a modicum 
of alignment specific data will need to be acquired typically based 
on the use of trench and borehole investigation and laboratory testing.
A general engineering description of the rock mass for tunneling purposes will typically include a geologic
 classification of the rock units (ideally with \% minerals), an estimate of the intact rock strength, and a description of the natural block structure (condition, roughness, orientation, size and shape).
The potential for the presence of atypical rock mass conditions 
also needs to be studied.  Atypical conditions that merit investigation 
include soil-like zones within the rock mass, zones of faulting, shears, 
open fractures, solution zones, hydrothermal alteration, weathering and 
buried valleys. Investigative efforts should also be made to evaluate 
the potential for encountering zones of high water inflow that may or 
may not be associated with soil-like zones of weakness within the rock mass. 
The potential for more pervasive rock-unit or regional adverse tunnel 
conditions, including the presence of relatively high in situ stresses, 
high ambient rock temperatures and more pervasive fluid/gas inflows should 
also be investigated. 
\subsubsection{Excavation Methods and Means and Structural Design}
Once a preferred alignment(s) has been identified and basic rock engineering characteristics determined, the selection of an initial set of baseline methods and means can be made for layout purposes. Throughout the planning period, and most notably in support of the selection of methods and means, contractor input is highly desirable. Practicing contractors are best positioned to provide state-of-the-industry input for selection of safe, practical and cost-effective methods and means for tunnel construction. 
The flowchart in Figure \ref{fig:tunnel1} identifies a discrete step for the structural design of a tunnel. This step may be eliminated if the rock mechanics or geotechnical engineer considers that a separate structural lining (reinforced cast-in-place concrete) is not required. The structural design phase may also be skipped or minimized if a similar design case history can be referenced.   
\subsubsection{Tunnel Contracting and Construction}
Even the most thorough site investigation of the most uniform geologic 
conditions will not be able to completely define the scope of an underground 
construction contract. Some surprises from the natural material should always 
be anticipated along the way. Risk analyses should be conducted in order to 
properly characterize the likelihood and severity of the impact of all such 
surprises on both the construction work and the project as a whole. 
Unacceptable risks should be mitigated by design, specification, contract 
provisions or insurance measures before the contract is let. 
The level of risk that tunnel construction brings to the overall project can be high, and is strongly influenced by factors, including:
\begin{itemize}
\item the complexity of the geology, 
\item the thoroughness of the site-specific investigation,
\item the amount and relevance of accumulated case history information,
\item the flexibility of the mining system and,
perhaps most importantly of all, 
\item by the skill-set of the owner's design and construction team that is assembled to plan and execute the work. 
\end{itemize}
Within this context, there is again a strong argument 
to be made for more actively involving the contractors 
in the design process.  Ironically, these key protagonists, who 
have the most relevant construction experience, are commonly excluded 
from the tunnel design.

\begin{figure}
\vspace*{13pt}
\begin{center}
         \mbox{\epsfig{figure=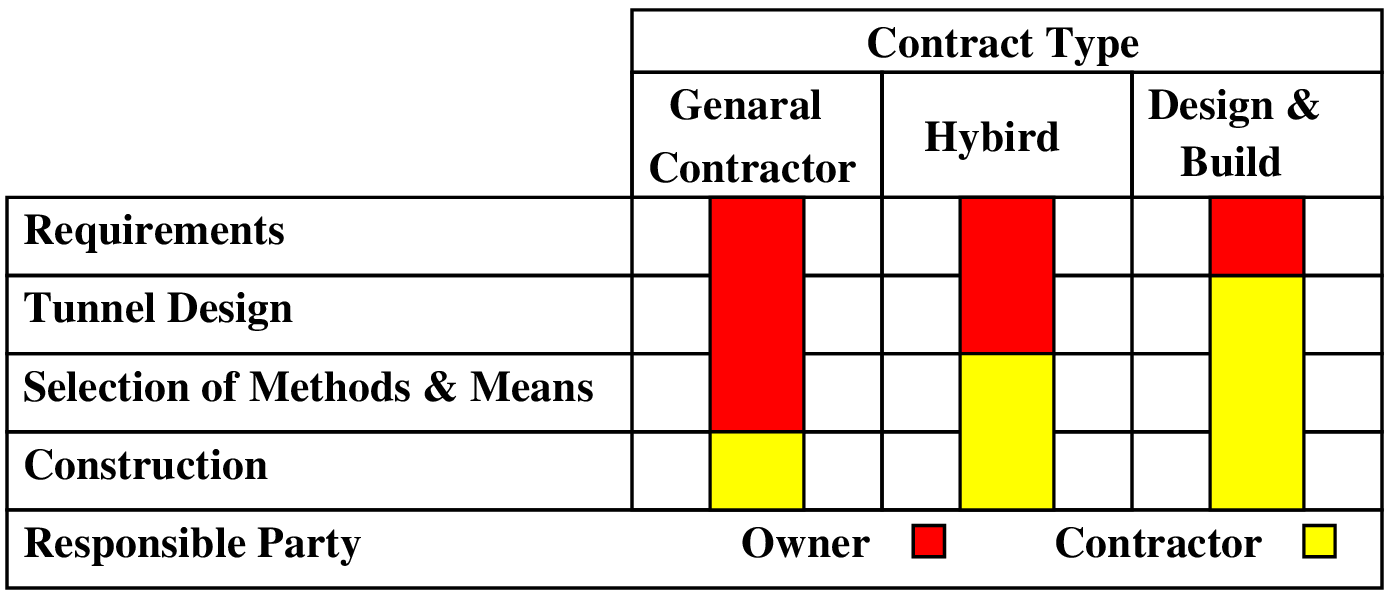,width=11.0cm}}
\caption{Responsibilities of Contractually-Bound Parties}
\label{fig:tunnel2}
\end{center}
\end{figure}

      Figure~\ref{fig:tunnel2} illustrates how, under 
conventional general contracting practices, the responsibility for 
design and selection of methods and means falls under 
the control of the owner or his representative(s).  The 
desire to involve the contractor more actively in the 
planning of the tunnel project favors the adoption of a 
more integrated design and build approach, where responsibility for 
design of the equipment (``hybrid") or the equipment and tunnel 
(design and build) passes to the contractor. Cording \cite{bib:cording}
notes ``The separation of design and specifications from the 
\, contractor's planning creates unnecessary impediments and adds 
unnecessary costs to the project."  Ultimately, a more 
integrated design strategy that involves the contractor can 
provide for a more integrated and 
innovative approach to tunneling \cite{bib:songer}.  
      The use of a design and build contract may result 
in cost and time savings to the Project, but would require the owner 
to freeze requirements at an earlier stage in the development of the project.
      However, the form of contract is a 
secondary issue compared to the need for the owner 
to assemble and manage a core project team that has a thorough 
understanding of both the end-user needs and tunneling constraints. In the 
case where the owner does not initially have all the 
requisite skill-set in-house, he/she may need to supplement such a 
team with outside contractors. Adequate care should be paid 
to pre-qualification and selection of such outside contractors 
(consultant-designers and/or builders). The pre-qualification and 
selection processes used by the project team should be 
project-specific. At a minimum, proposing outside contractors should 
be required to demonstrate a requisite level of individual 
and/or corporate technical expertise and provide work product for 
review that demonstrates the specific qualities that he/she/they can 
offer to the project.  Proposals should include a description 
of recent past experience working on similar jobs (similar 
requirements, geology, methods and means, etc.), and description of levels 
of project-specific responsibility on the relevant projects. References 
should be provided for follow-up \cite{bib:laughton}. During design and 
construction the owner's management team should take full 
responsibility for all aspects of project planning and be endowed 
with adequate responsibility and commensurate authority to be able 
to effectively administer all related design 
and construction activities \cite{bib:mccreath}.

\subsection{Shaft Hole}
\label{sec:shaft}
For a reactor $\theta_{13}$ experiment, the baseline is already known to
be around 1.5 km.  
The far detector should be placed at least several
hundred m.w.e. underground to suppress cosmic ray muons. 
If a tunnel is used, the reactor site is to  
be close to steep mountains with a few hundred meters height, depending
on the reactor power.
This condition puts restrictions when selecting a reactor site.  
The situation becomes more problematic for the near detector. 
The near detector is to be located within a few hundred meters from
the reactors, but still hundreds of mwe of overburden is necessary.  
If a shaft hole is used, it is possible to obtain necessary depth
underground at a reactor site which is located on a flat landscape. 
Thus a shaft hole gives added flexibility 
in the choice of a
reactor site. 

There are two techniques for digging shaft holes. 
One is blasting and excavating. 
In this case debris is winched out and dumped outside. 
Industrial needs to dig wide and deep shaft holes are growing.
For example, there has been intensive R\&D to construct 6.5 m
diameter and 300~m deep shaft holes as access tunnels to underground 
nuclear waste storage areas. 
If such a technique can be adopted, the shaft holes for our purposes are
expected to be constructed rather easily and cheaply.  

Another method is drilling. 
The drilling method is suited for soft soil as the case of
Kashiwazaki-Kariwa site which is located along a sea shore. 
The drilling is performed by filling the hole with water with higher head
pressure than outside water level  to prevent collapse of the shaft wall. 
The water helps to make the mud-rock   softer and to make the drilling 
easier. 
The debris generated by the drilling are forms of mud and are
pushed up to ground level by pressurized water, through the center pipe
of the drilling machine. 
When the drilling is finished, the water is pumped out and steel and
concrete rings are put in the hole. 
There exists a 6~m diameter drilling machine which can dig as deep as
300~m.  

The capability and cost of the hole construction very much depends on the
nature of the site and should be evaluated for each case.

\subsection{Summary}
Digging a hole is not as simple as it sounds. Cost and risk are potentially 
much higher than they are for equivalently sized conventional structures built 
on surface or as cut and cover structures using quality-controlled construction 
materials. 
Success in design will be largely determined by the owner's ability to 
properly integrate the end-user needs of the facility with the construction 
needs of the tunneler. During construction, the need for good active 
management by the owner will continue in order to ensure that the contract 
provisions are met and, in particular, that the ground conditions 
as-encountered are properly recognized, responded to and documented in an 
appropriate and timely fashion. 

%% file: safety.tex
\section{Safety}
\label{sec:safetysection}

\subsection{Safety Planning}
The purpose of this document is to give the present scientific case
for a new neutrino oscillation experiment at a nuclear reactor.  
If and when such an experiment is built and operated, safety must be
a paramount concern, and safety will be a paramount concern.  This will
be done by incorporating the following principles in the project
at every stage:

\begin{itemize}
\item Clear Management Responsibilities
\item Documentation
\item A Working Atmosphere of Safety
\item Integrated Safety Management
\end{itemize}

The collaboration and its management will perform and document a 
hazard/risk analysis for each phase of the project in order to 
systematically identify
the hazards that may be associated with it.  This review
will be intended to ensure that matters of environmental protection
and worker health and safety related to the project are identified and
that they will be thoroughly addressed in the design, construction and
operation of the project.  Also, at the appropriate time, a 
collaboration will be established in accordance with all applicable ES\&H
regulations, standards and good practices.
\subsection{Civil Construction}
\label{sec:safety}
The major cost driver for this project is likely to be
the construction of a tunnel or shaft in order to place the
neutrino detector underground.  This will 
require substantial civil construction.
An outside contractor will be responsible for building the underground
facility.  
\par
Civil construction of an underground site
offers special challenges.  Some of these involve use of heavy
construction equipment, while simultaneously dealing with
fire protection,
confined space issues and the potential for flooding.  When a specific 
site is chosen, a set of procedures for instituting safety rules
will have to be implemented and agreed to by a number of parties.  
These procedures will follow all applicable state laws, federal regulations, and
industry practices for the selected site.
\par In addition, there are a number of special considerations 
due to the fact that this facility will be built in 
conjunction with a nuclear reactor.  Nuclear reactor safety in the United
States is under the purview of the Nuclear Regulatory Commission.  It seems
clear that nothing involved in the neutrino experiment would have
any significant
consequence on the operation of the nuclear plant.   This lack of impact
may not be immediately obvious to everyone involved, so 
a careful analysis and documentation will be carried out.
\par Besides the normal safety requirements involved with the
handling of materials, there will be special security requirements 
involved in working near a nuclear reactor.  A major issue with regard
to siting of an experiment is the present boundary of the security fence
around each facility.   These issues are already being considered by the
security experts who work for nuclear reactor power companies as we 
approach them to discuss possible experiments.  In particular, there
will need to be careful regulation and procedures involving site access.
These will be worked out in cooperation with all appropriate
regulatory agencies.

\subsection{Safe Detector Construction and Operations}
As this project develops, there will be  a number of technical
reviews, cost reviews and safety reviews.   Safety will be an integral
part of documentation at all stages.  At an early stage of the project,
the collaboration will prepare a Preliminary Safety Assessment Document
which will deal with the the following issues:
\begin{description}
\item General Construction Safety
\item Fire Safety
\item Flooding Hazards
\item Mechanical Hazards During Installation
\item Electrical Hazards During Installation
\item Industrial Safety
\item Environmental Protection During Construction
\item Life Safety - Egress
\item Fire Protection
\item Electrical Hazards
\item Radiological Hazards
\item Mechanical Hazards
\item Hazardous/Flammable Materials
\item Cryogenics and Oxygen Deficiency Hazards
\item Emergency Preparedness
\item Emergency Communications
\item Conduct of Operations
\item Training
\item Qualification of Personnel
\item Waste Handing; storage and disposal
\end{description}

\par
Calibration of the detector often involves the handling of a number of
radioactive sources.  Physicists who participate in experiments
 at national laboratories
have a great deal of experience regarding the proper training and
documentation for handling, operating and storing radioactive sources with
a variety of compositions and strengths.  
\par Safety is an independent concern and priority for nuclear power
generation in the United States, under the auspices of the Nuclear
Regulatory Commission.  In the event of any kind of incident at the 
power plant, experimenters would be trained to cooperate fully with 
plant personnel.
\subsection{Quality Assurance}
It will be the policy of any reactor $\quq$ project that all activities
shall be performed at a level of quality appropriate to achieving the
technical, cost and schedule objectives of the project and at the same
time insuring that all related ES\&H considerations are properly 
addressed.  A Quality Assurance plan for the project will be 
developed in accordance with policies and recommendations from the 
Department of Energy and any other appropriate agencies.

%% file: outreach.tex
\section{Outreach and Education}
\label{sec:outreach}
\subsection{Goals of the Outreach \& Education Effort}
Outreach is a substantial part of any scientific enterprise.
The goals of the $\quq$ outreach efforts are:
\begin{enumerate} 
\item Providing meaningful education opportunities for the public.
\item Giving students and teachers of K-12 schools an experience with
modern science. 
\item Creating and maintaining a learning community between high-school and
college teaching professionals and the researchers and graduate
students involved in the experiments. 
\item Developing young people's interests in pursuing science or
engineering careers.
\item Increasing participation of members of underrepresented groups
in science activities.
\item Emphasizing the impact of pure science on the technological
capabilities of our society. 
\item Providing undergraduate and graduate students  and K-12 teachers 
with access to
valuable research experiences.
\item Involving local community colleges in the research effort.
\end{enumerate}

\subsection{Outreach and Education Opportunities at the $\quq$ Reactor Neutrino
Experiment}
In order to grasp the outreach opportunities we need
to look at our experiment from a more general perspective.

The experiment
measures neutrinos generated by a nuclear power plant to search for the
disappearance of neutrinos as a function of the distance from the
reactor. The neutrinos disappear as a consequence of 
oscillations. The flux is measured by detecting nuclear reactions
initiated by the neutrinos in large volumes of scintillator oil.
Due to the elusive nature of neutrinos the experiment has to be pushed
to the limits of detection sensitivity. This makes the experiment also
sensitive to background from cosmic rays and other  
astrophysical neutrino sources.

This makes it clear that the $\quq$ experiment 
is based on many different disciplines, all of
which will be a subject of our outreach activities. Our outreach efforts are
thus primed for interdisciplinary science
education which will happen within the
framework of an international collaboration of scientists.
The large and exciting list of topics that we can draw from reads as follows:
\begin{itemize}
\item {\bf History of the neutrino} The neutrino was postulated by
Pauli, its experimental discovery was worth a Nobel prize and even
John Updike wrote a poem about it \cite{updike_poem}. Particle and
nuclear physicists discovered that neutrinos have mass and that
they oscillate.
\item {\bf Particle Physics} While much of particle physics is well
established, but the neutrino sector is not. There are many open
questions regarding the nature of neutrinos. Are they
Dirac or Majorana particles? What are the parameters that define the
mixing in the neutrino sector? Can we observe CP-violation with neutrinos?
\item {\bf Nuclear Physics} Very rare neutrino-nucleon interactions
make the $\quq$ experiment possible. The chain of events that turns the
absorption of an elusive neutrino into a detectable signal is a
fascinating story.
\item {\bf Nuclear Engineering} Nuclear power generation and the
relation between neutrino flux and thermal power output broadens the
outreach efforts significantly. Why is it possible to draw so much
power from a relatively small amount of material, measured in
kilograms? This is Einstein's famous $E=mc^2$ equation put into
practice.
\item {\bf General Energy Management} We see the opportunity to directly
complement existing outreach efforts associated with nuclear energy
generation. Over one thousand nuclear reactors operate in the world,
many of which are subject to public scrutiny. Showing that
they contribute to the understanding of universal fundamental processes
could be helpful in the public discourse.	 
\item {\bf Astrophysics} The biggest producer of neutrinos in our
neighborhood is of course the sun. Fusion processes in the center of
the sun create the energy that ultimately fuels all activity on earth. In
the fusion process neutrinos are generated with a well-known flux of
$\overline{\nu}_e$ neutrinos. The fact that the amount of
$\overline{\nu}_e$ detected on earth is much smaller than expected
gave scientists the first evidence that neutrinos oscillate.
The oscillations were confirmed by studying neutrinos created 
in interaction of cosmic rays with the earth's higher atmosphere.
The list of topics is even richer because
there are also neutrinos expected to come from exotic astrophysical
objects. Supernovae for example are known to emit a huge burst of
neutrinos.
\end{itemize}

The outreach efforts we envision will consist of solid but palatable
explanations of each topic, descriptions of their common features, an
explanation of technology transfer, and generally an invitation to
participate, particularly to the young, in the science-based
activities of the collaborating nations.    

\subsection{Strategies for Outreach \& Education}
The strategic plan for outreach and education has to be paced with the
development of our experiment. We divide the $\quq$ experiment into
three phases:
\paragraph{Pre-construction phase}
Design of the experiment proceeds and the international $\quq$ 
collaboration forms. First outreach activities are started, such as
long-term cooperation between scientists and K-12 teachers  
to stimulate additional joint activities impacting K-12 education.
Universities and national laboratories 
are likely places to start outreach efforts by binding
$\quq$-specific activities into ongoing outreach efforts at those 
institutions. Quarknet \cite{quarknet} and 
similar activities might serve as examples. 
To coordinate and further outreach efforts an office of outreach will
be established within the $\quq$ collaboration 
(see also section \ref{outreachoffice}).
The first coordinated effort will be to develop a strong online
representation and to link it to several
education databases (e.g. \cite{edu_database}). 
The experiment will market itself, an effort that will
increase throughout the initial phase.
\paragraph{Construction phase}
With the K-12 teacher-researcher connections established in
the first phase and a web presence in place, 
the second phase of our experiment will increase our outreach
activities by targeting students and teachers 
to become actively involved. The construction of a
modern physics experiment is a very exciting time and this excitement
is likely to be contagious.
The engagement can be achieved by means of research experiences (REU,
Quarknet, Scientist in Residence) or workshops and seminars. This
phase will also be characterized by a large cooperation with the American
Physical Society (APS) and American 
Association of Physics Teachers (AAPT).
\paragraph{Data-taking phase}
As the detector installation is realized and data taking commences, 
the outreach core content will no doubt be polished. Outreach efforts
generally will increase.
We foresee the possibility to develop an on-site welcome center that provides 
inquiry-driven experiences in content areas in
which the materials developed for web-delivery are adapted for
live engagements. The welcome center will be tied into tours
of the power plant. We will actively work on attracting visitors
groups from schools and other institutions. The site visit program
will be complemented by a program designed to attract more
minority groups.   

We also intend to focus our attention on aggressively 
pursuing media opportunities in the towns and cities where the 
researchers/students/staff reside.	

\paragraph{International Activities}
The $\quq$ effort is an international one
with scientists from more
than a dozen nations being co-authors of this white paper.
The outreach efforts have the great opportunity to be 
internationally oriented, and a cooperation of outreach groups from
different regions (Europe \cite{outreach_europe}, Japan \cite{outreach_japan})
will enhance the efforts.
\subsection{\label{outreachoffice}Office of Outreach and Education}
The experiment has a large potential to further the education and the
interest of the general public in science. 
An office of outreach for the $\quq$
experiment would:
\begin{itemize} 
\item develop a rich palette of education
materials and establish an strong online representation,
\item provide support for the development of the media used in public
seminars and professional meetings.  For example, 3D rendered detector
images and photographs will be developed and cataloged. Delegating this
responsibility to the outreach office  
will minimize redundancy of effort and provide a central repository.      
\item work with
the media for the purpose of education and for marketing the
experiment,
\item coordinate outreach efforts within the international collaboration
\item coordinate $\quq$ activities with the various nuclear related
agencies (e.g. Office of Nuclear Energy, Science, and Technology, or
U.S. Nuclear Regulatory Committee, DOE)    
\item provide professional  
development opportunities for secondary and post-secondary science
teachers, and public seminars with the ultimate goal of establishing
an on-site welcome and outreach center.  
\end{itemize}
The office of outreach will also work to overcome any 
confusion that the $\quq$ experiment is specifically a nuclear physics 
project. A second challenge will be in the concern that any 
attention to an existing nuclear power plant is unwanted.  

It is important that the  $\quq$ experiment receive regular 
attention in the press, particularly in newspapers with a national
audience such as the Washington Post, 
and the New York Times, but also in the 
local newspapers near where research/students/staff reside.  The outreach office 
will operate as a communications office in order to ``advertise" the 
program and to provide relevant materials for the articles.	

We suggest the outreach office should also 
``market'' the experiment's education and scientific goals by encouraging and 
supporting introductory talks at professional meetings.  Here, we 
have an enhanced opportunity because of the multidisciplinary nature 
of the experiment.  Talks will be coordinated at particle, solar, nuclear, 
astrophysical, and engineering conferences as well as at teacher 
related conferences such as those sponsored by AAPT, the American
Astronomical Society (AAS) or the American Physical Society
\cite{aps_outreach_mtg}.
\subsection{Ideas for Web-Based Outreach Efforts}
Providing rich content is the most important aspect to consider 
for reaching the public with news and descriptions of the 
$\quq$ project.  The initial low-cost/low-risk approach to developing 
content, while maintaining the highest potential for impact, 
is to develop teaching materials for delivery on the 
Internet.  Thus, our initial strategy will
include a small team of content developers who have 
an ability to integrate technology in both a meaningful and artful way.

An obvious choice is to house the development office at a 
university where similar activities take place among students and 
where a director can have ready access to the experiment and 
its researchers who participate.  If selected carefully, software 
license costs, which can be significant, could be minimized 
if the host site were already participating in license agreements.

One interesting technology is Physlet \cite{physlet}. Physlets are Java-based
platform-independent applets with physics contents 
that provide a real interactive experience for a modest investment of 
resources and with the potential to be highly
visible and accessible to the public. This technology has been prominently
featured in AAPT workshops and is perfect for outreach activities.
\subsection{Ideas for Working with K-12 institutions}
Physics principles are often regarded as out of reach for 
many. The art of knowing will begin with self-study but, 
to that end, attention should be given to high school teachers who 
are willing to deliver discourse on this subject in their classrooms.  
Sample lesson plans will need to be constructed. Who better to do 
this but high school teachers themselves?

Teacher workshops could be organized at the summer AAPT meeting where
the activities of the $\quq$ experiment will be the principal
theme. However, because of its affiliation with the neutrino industry,
the four hour workshop will include an introduction to the neutrino,
an overview of particle/nuclear/astro physics, and sample lesson plans
that would be useful at both the high-school and college level.  We
expect this activity could be attended by 12-18 teachers on an annual
basis with only limited costs. 

Another suitable place for representation would be at area
planetariums where hands on teaching regularly takes
place. Constructing a display which features the $\quq$ experiment,
the adjoining nuclear reactor, and a description of the background 
including astronomy-related backgrounds would draw attention to the
experiment.

Direct interaction of K-12 education institutions in the community
surrounding the site of the experiment is crucial for the success of
the outreach effort. Programs like ``Scientist in Residence''
can help to achieve the goal. Another key to a successful outreach
program would be to offer research experiences to interested K-12
teachers, supplemented by visits of the nuclear power plant and the
$\quq$ welcome center.
\subsection{Ideas for Outreach to Community Colleges}
Local community colleges usually lack the funding to purchase research
equipment and also lack travel funds to work in remote research
facilities. Involving community colleges in the vicinity of the
reactor site helps to overcome these problems and constitute a genuine
opportunity for the community colleges to involve their students in
research of international standard. By targeting minority colleges,
an important goal of science outreach efforts can be met.

%% file: brazil.tex

\section{Appendix - The Angra reactor in Brazil}

This appendix will focus on the features of the site at Angra dos Reis
in Brazil.  This site has many desirable attributes including good 
surrounding topography, a single powerful reactor, and good relations
with the electrical and nuclear power companies.  In addition, the
local physics community has become very enthusiastic about this idea
and is beginning the process of developing a realistic proposal.

\subsection{The Reactor Site}
Angra dos Reis is located about 150~km south of Rio de Janeiro.  The
nuclear facility contains two operational reactors.  The Angra-I reactor
is an older low power (about 1.5 GW$_{th}$) reactor that is not frequently 
operated.  The Angra-II reactor, on the other hand, was brought on-line
in 2000 and is consistently operated at about 4.1 GW$_{th}$.  The 
reactors are located on the coast and the reactor company controls
a strip of land that stretches inland about 1-1.5~km and is approximately
4 or 5 ~km along the coast.  All experimental constructions which will
be considered here would be sited within the reactor companies site 
boundaries.  

Much of this terrain is mountainous granite with multiple peaks in the 
200-600~m region.  This allows good background reduction to be achieved
for an experimental hall with relatively cheap civil construction by
tunneling sideways into such a mountain.  Also within the site boundaries
there exists a town, Praia Brava, which houses most of the 2 or 3 thousand
people which work at the reactor facility and also contains a hotel and
stores.  Such already existing infrastructure could make using this 
facility more attractive.

\subsection{Communication with the Power Companies}
The company which runs Angra is state owned and operated. 
One of the unique features of attempting this experiment in Brazil
is that the presidents of both the electric power company and its
daughter nuclear power company are former particle physicists who
used to do experiments at CERN.  As a result, they are very receptive
to communications from members of the Brazilian physics community
and have been very helpful in providing resources and access to the
facility.  A one day site visit has already been performed to evaluate
the viability of performing the experiment there.  Significant assistance
was provided by the director of operations from the reactor facility
and significant time was spent with the director of civil construction
on the site.  With their help, possible
experimental site locations were explored.
The reactor
company has agreed to supply full detailed cost estimates of any civil
construction plan by using their knowledge of
the site geology and known contractors.

\subsection{The Experimental Design}
\begin{figure}[htb]
\includegraphics[width=\textwidth]{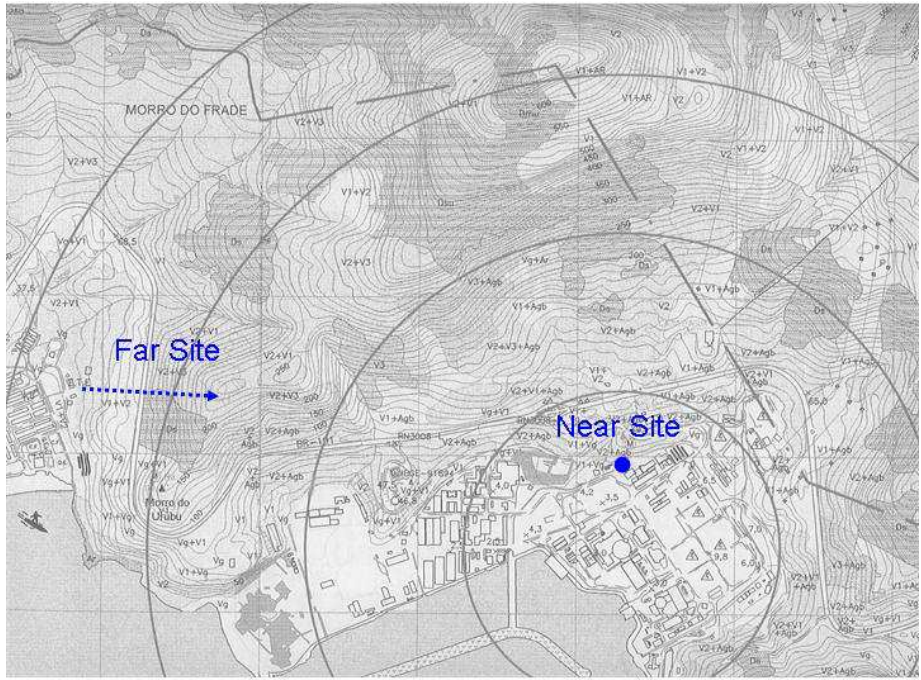}
\caption{A topographic map of the nuclear reactor site at Angra dos Reis.
The concentric circles are at 500 meter radial intervals from the 
core of Angra-II.  Proposed locations for the near and far detector 
experimental halls as well as the far detector access tunnel are shown.
\label{angraTopo}}
\end{figure}

A topographic map of the site is shown
in Figure~\ref{angraTopo}.  The concentric circles are at 500~m radial 
intervals from the
primary Angra-II reactor core.  The near site location is 300-350~m from 
the core.  It has the possibility to gain about 15-20~m of rock overburden
(30-50 meters of water equivalent).  The far site location would exist
under a 240~m hill at about 1350~m from the reactor core.  Access to the
far site would come from a 420~m tunnel which starts from the western
edge of the hillside.  This location is easily accessible from the 
town of Praia Brava and would be very near to the current location of 
their sewage treatment plant.

It is envisaged to place identical 50ton fiducial detectors at each
location.  Exact detector designs have not yet been developed, but
it is currently assumed that such detectors would build off of the 
developments from other groups.  Most likely a 3 volume detector
would be optimal: a central liquid scintillator volume that would
be doped with gadolinium (target); a surrounding volume of liquid
scintillator without gadolinium (gamma catcher); a non-scintillating
buffer to shield the radioactivity of the photo-tubes which would be
installed at the outer edge of this volume.  An active muon shield
would then be required to surround this system.  A spherical detector 
with an active target of 50 tons would have a total diameter of 
approximately 7.3 meters.  The access tunnels and experimental halls 
are being designed to accommodate these dimensions.

\subsection{Experimental Reach}
\begin{figure}[hbt]
\includegraphics[height=7cm]{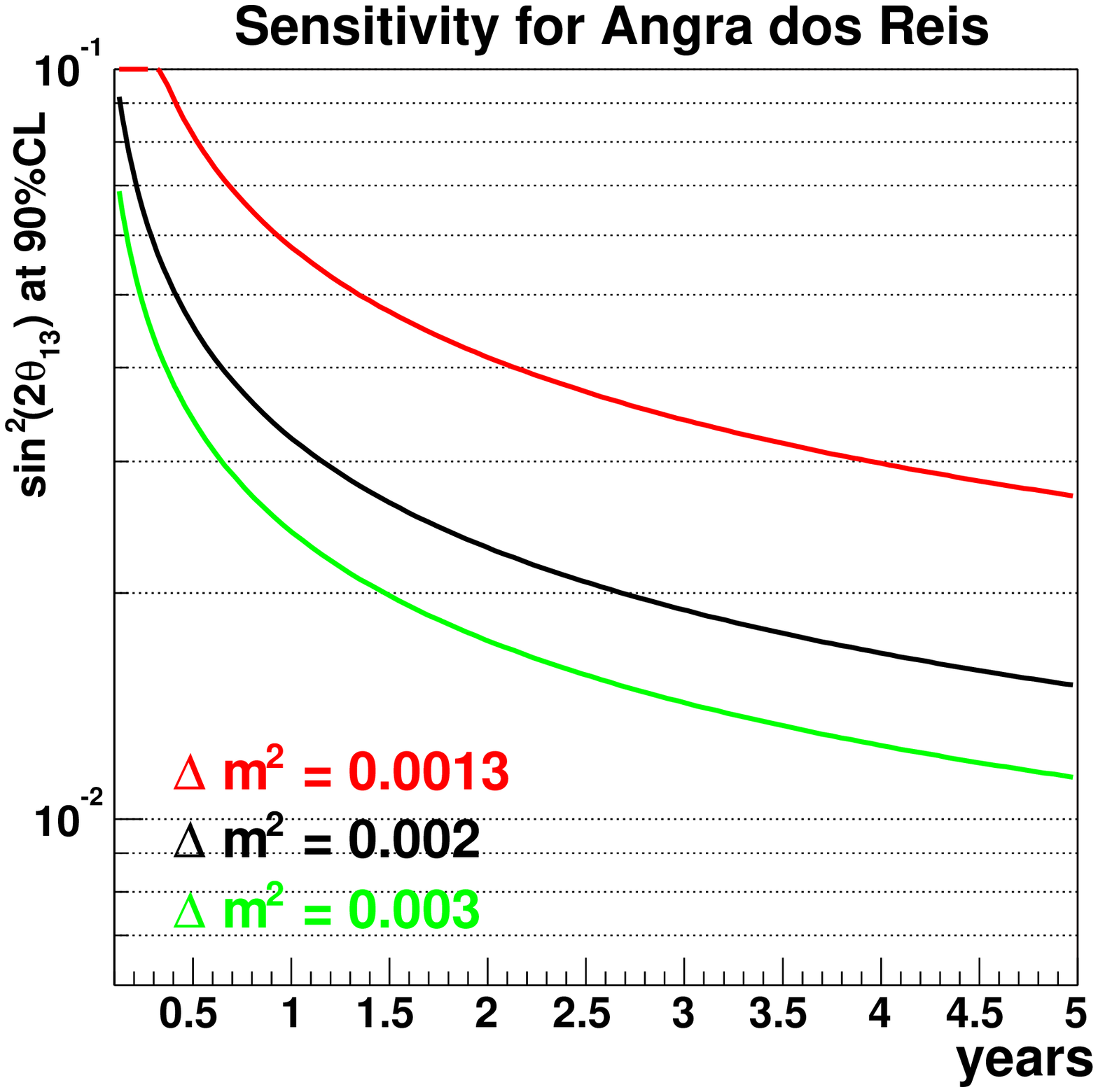}
\hfill
\includegraphics[angle=90, height=7cm]{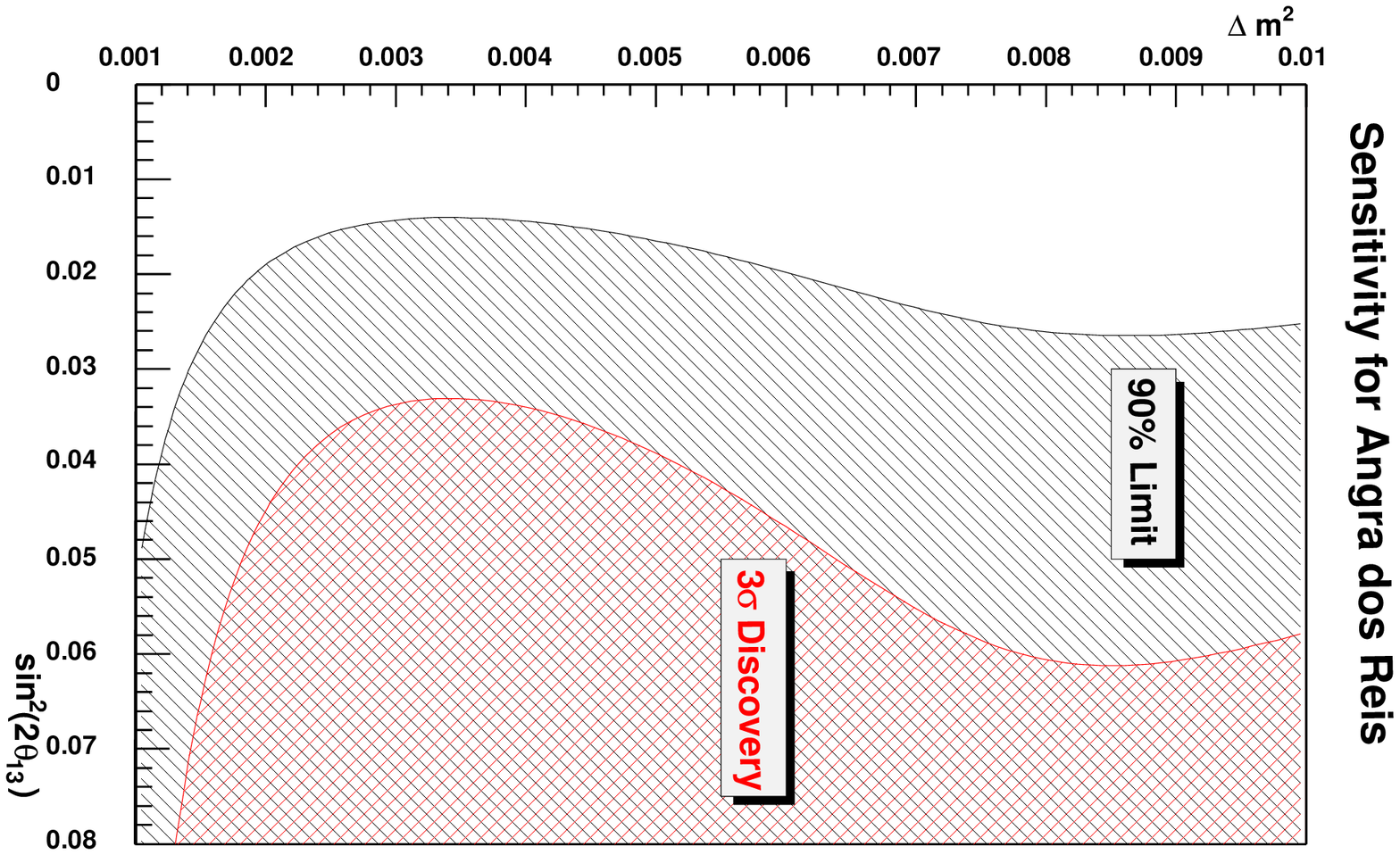}
\caption{Expected sensitivity to $\sin^2(2\theta_{13})$ which could be achieved
by an experiment at Angra dos Reis assuming a 1\% systematic error.  
The plot on the left shows 
the sensitivity as a function of years of running for three different
values of $\Delta m^2$.  On the right, the full coverage of $\Delta m^2$
vs. $\sin^2(2\theta_{13})$ is shown assuming three years of data taking.  
Curves for
both the limit at 90\% confidence level and the discovery at 3 $\sigma$ are
shown.  The current limit at 90\% confidence is $\sin^2(2\theta_{13}) < 0.2$.
\label{angraLimits}}
\end{figure}

Preliminary estimations have been performed of the signal and background
rates for the given detector configuration.  The detector at the far
location is expected to get about 120 signal events per day, while the
near site would be expected to receive about 3000. Some very preliminary
background estimates suggest that at the far detector, less
than 10~Hz uncorrelated backgrounds, which would easily be vetoed by
an active muon shield, would be expected and there would
be about 1-2 correlated background
events per day from muon induced radioactive isotopes.  Similarly
at the near detector there would be an uncorrelated background rate of
about 830~Hz (yielding an active live time of 63\% after muon vetoing)
and a correlated background of approximately 150 events per day.  Having
a signal to noise rate of about 100 in the far detector and 20 in the 
near detector should allow reasonable background rejection while maintaining
statistical sensitivity.  Figure~\ref{angraLimits} shows the expected 
statistical sensitivity as a function of time, for the best fit value and
90\% allowed limits of $\Delta m^2$ from Super-Kamiokande.  As can be seen, 
a limit of $sin^2(2\theta_{13})<0.02$ at 90\% confidence level can be
achieved within 3 years.  Also in Figure~\ref{angraLimits} is shown the 
complete limit and 3$\sigma$ discovery potential for a 3 year run over
all  $sin^2(2\theta_{13})$ and $\Delta m^2$.

\subsection{Brazilian Community and Support}

The Brazilian community has recently been having in-depth discussions
about this possibility.  There exists significant theoretical and
phenomenological support for neutrino oscillation work in Brazil already,
primarily located in the Universidade Estadual de Campinas (UNICAMP)
and the Universidade do S\~ao Paulo.
The director of the Centro Brasileiro de Pesquisas Físicas (CBPF) in
Rio de Janeiro has shown great enthusiasm for this project and has 
offered to make his institution the host of such an experiment.  There
exist strong experimental particle physics groups based mainly in Rio de
Janeiro and S\~ao Paulo.  They are already
strongly involved with the Pierre Auger project as well as experiments
based at Fermilab and CERN.  They have expressed great interest in
pursuing a project that is local to Brazil and several of the experimental
members have already started working on producing a more realistic
set of cost estimates and site plans.  

While it's not possible to put any real faith behind the numbers, 
some preliminary estimates have concluded that with the cheaper labor
costs in Brazil, the civil construction could probably be achieved
for 5-7 million dollars (US).  The local Brazilian community is currently
in consultation with the government to get a commitment to this
project and for covering the complete costs of civil construction.
For the detectors, a very conservative cost estimate concluded that they 
would require about 7 million dollars each.  Thus the total project
could probably be completed with an external contribution of less than 
15 million dollars (US).

%% file: chooz2.tex
\section{The Double Chooz Project}
Since the end of the year of 2002 a group of European (French, German,
 and Russian) physicists has been working on the possibility to measure
the last undetermined neutrino mixing angle $\theta_{13}$ with a two
detector
reactor neutrino experiment. During the year 2003, a site
investigation has been performed, and three sites have been selected to be
potentially suitable to performed such a project: the Penly site
(Normandie, France), the Cruas site (Ard\`eche, France), and the Chooz
site (Ardennes, France). Thanks to its existing infrastructure and to
the support of the EDF (Electricit\'e De France) power plant staff,
Chooz is the most reliable location to perform an experiment and we'll
only focus on it in the following.
\subsection{The Double-CHOOZ concept}
The new Chooz project will run two identical detectors of medium size,
containing 10 to 20 tons of liquid scintillator target.  The (still
existing) neutrino laboratory of the first Chooz
experiment, located at 1.05 km from the two cores of the Chooz powerful
nuclear plant (PWR, 8.4 thermal GW) will be used again. We label this
detector the
\emph{far} detector or \emph{CHOOZ-far}. The CHOOZ-far site is
shielded by about 300 mwe of $2.8 \, g/cm^{3}$ rocks. We plan to start to
take data
at CHOOZ-far within the first month of the year 2007.

In order to cancel the systematic errors originating from the nuclear
reactor (electron anti-neutrino flux and spectrum), as well as to
reduce the set of  systematic errors related to the detector and to the
event selection procedure, a second detector
will be installed  close to the nuclear cores, within a range from 100
to 200 meters. We label this detector the \emph{near} detector or
\emph{CHOOZ-near}. Since no natural hills or underground cavity exists
at this location, an artificial overburden of a few tens of meter
height has to be built. From the first discussions, this
construction has been allowed by the power plant company
authorities. A  pre-study of the civil engineering work to
be done for CHOOZ-near has been order and will be performed during the
end of 2003.
This pre-study which aims to determine the best
combination location-overburden as well as to optimize the cost is financed
by the French electricity power company EDF.

From the simulation, a new experiment at CHOOZ would reach a
sensitivity of $\sin ^2 (2 \theta_{13}) \, < \, 0.03 $, at $90\%$
CL for $\Delta m^2_{atm} \, = \, 2.0 \times 10^{-3} \, \, eV^{2}$ (latest
best fit value of the Super-Kamiokande experiment), after three years of
operation. A sensitivity of $0.05$ would be reached within the first
year of running with two detectors, improving the CHOOZ
bound by roughly a factor of five. These estimates are based on the
assumptions that the
relative normalization error between the near and far detectors could
be kept at $0.8 \, \%$, and that the backgrounds at both sites
amount to less than $1 \, \%$ of the anti-neutrino signals.

It is worth mentioning that the 1.05 km average baseline at CHOOZ is not
optimal (a
little too short) compared to the first oscillation maximum (for
anti-neutrino energy of a few MeV) if the
atmospheric mass splitting would remain at  $\Delta m^2_{atm} \, = \, 2.0 \times
10^{-3} \, \, eV^{2}$. A shorter baseline is
compensated by higher statistics for a fixed size detector,
however. A  value of $\Delta m^2_{atm} \, < \, 1.5 \times 10^{-3} \, \, eV^{2}$,
close to the $90 \, \%$ CL current lower bound,
would restrict the absolute potential of the double-CHOOZ experiment.
\subsection{Detector design}
The detector design foreseen is an evolution of the detector of the
first experiment, (\emph{CHOOZ-I}).
The first improvement  with respect to the CHOOZ-I sensitivity comes from an
increase of the exposure to get more than 40,000 anti-neutrino events at
CHOOZ-far. This condition translates to a statistical error lower than
$0.5 \, \%$. This increase of signal is obtained by using a
target cylinder of 120 cm radius and 280 cm height, providing a volume
of $14 \, m^3$, 2.5 higher than in CHOOZ-I. In addition, the
operation period will be extended to at least three years, and the
efficiency will be improved (the global load factor of the reactor is
about $80 \, \%$ while it was much lower for the CHOOZ-I
experiment due to the power plant commissioning).
In addition, the background level will be decreased in order to get a
signal to noise ratio greater than 100 (about 30 in CHOOZ-I).

The near and far detectors will be identical up to the PMTs surface.
This will allow to have a relative normalization error less
 than $0.8 \, \%$. However the outer shielding will not be identical since
the
 cosmic background varies between CHOOZ-near and CHOOZ-far, due to the
different overburdens. The overburden of the near detector has been
chosen in order to keep the signal to background ratio above  100;
with this condition,  even a poor measurement of the background keeps
the associated systematics  error well below $1 \, \%$.

The detector design was intensively studied and tested by Monte-Carlo
simulation,
using a code derived from the simulation of the CHOOZ-I experiment.
The one meter low radioactive sand shielding of CHOOZ-I, used to reduce the
external
radioactivity contamination, will be replaced by a 15 cm stainless steel
shielding.
The size of the liquid active buffer can then be increased. From
the center there will be:
\begin{itemize}
\item{a 120 cm radius 280 cm height acrylic target cylinder,
filled with $0.1 \, \%$ Gd loaded liquid scintillator}
\item{a 40 cm buffer of non-loaded liquid scintillator with the same
optical properties (light yield, attenuation length), to get the full
positron energy as well as most of the neutron energy released after
the neutron capture (this buffer is called the $\gamma$ catcher)}
\item{a 95 cm buffer of non scintillating liquid, to decrease the
level of the accidental background}
\item{a 60 cm veto filled with liquid scintillator.}
\end{itemize}

The spatial reconstruction is not affected by the cylindrical shape.
Each parameter of the detector is being studied by Monte-Carlo
simulation, to define the tolerance on the differences between the two
detectors. The inner volume dimensions as
well as the shape of the target vessels  are still preliminary and
could change prior to the completion of the proposal.
The aim is to reduce the number of analysis cuts with respect to the
CHOOZ-I experiment to push down the systematic errors related to the
anti-neutrino event selection in both detectors independently.
The non-scintillating buffer will reduce the singles rates in each
detector by two orders of magnitude with respect to CHOOZ-I. This
lowers the positron threshold down to 500~keV, well below the 1
MeV physical threshold of the inverse beta decay reaction.
 A very low threshold has three advantages: a)
the systematic error due to this threshold is then suppressed, b)
the background below the threshold can be measured, c) the beginning of
the positron spectrum provides an additional calibration point between
the near and far detectors.
This reduction allows the relaxation  or even suppression
of the localization cuts, such as the distance of an event to the PMT
surface and the distance between the positron and the neutron.  These
cuts,  used in CHOOZ-I, are difficult to calibrate and have to be
avoided  in a precision experiment.
The remaining event selection cuts will have to be calibrated between
the two detectors with a very high precision.
The main cut to calibrate is the energy selection of the delayed neutron
after its capture on a Gd nucleus (with a mean energy release of 8 MeV
gammas). The requirement is 50-100~keV on the precision of this cut
between both detectors, which is feasible with standard techniques using
radioactive sources at different positions and lasers. The main sensitivity
of  a reactor experiment of the Double-CHOOZ scale (400 GW t y)  
is the total number of events; the requirements on the
positron energy scale are then less stringent since the weight of the
spectrum distortion is low in the analysis. This is being studied by
simulation.
\subsection{Backgrounds}

The correlated background was measured in CHOOZ-I, and found to be
about one event per day. At CHOOZ-far the active buffer will
be increased by a factor of two, with a solid angle for the out-coming
background being unchanged. This together with a signal increased by a
factor of 2.5  will fulfill the requirement of a signal to noise ratio
greater than 100.

At CHOOZ-near, due to the shallow depth, the cosmic
background will be important. For a detector located at 150 meters
from the cores, with an overburden of 55-65 mwe, the signal will be
5,000 events per day, while the muon rate is expected to be a factor of
ten less. A dead time of 500 $\mu s$ will be applied after each muon,
leading to a global dead time of about $25 \, \%$. About 20 recoil
proton events per day, mimicking the anti-neutrino signal, are
expected  while the estimate of the muon induced cosmogenic events
($ ^{9}Li$ and $ ^{8}He$) is 15 per day with a large uncertainty (this
last point is being studied). This fulfills the requirement of a signal
to noise ratio greater than 100 at CHOOZ-near.
\subsection{Conclusion and outlook}
A new reactor neutrino experiment at the Chooz
nuclear power plant has the potential to reach a sensitivity of
$\sin ^2 (2 \theta_{13}) \, < \, 0.03$, at $90\% \,$ CL
 for $\Delta m^2_{atm} \, = \, 2.0 \times 10^{-3} \, \, eV^{2}$,
after three years of operation.
This potential is similar to what could be achieved by future long
baseline neutrino experiments (JPARC and NuMI-Off axis) at the
horizon 2013. Furthermore, both results are complementary.

The CHOOZ-far detector will be installed in the fall of 2006, in order
to start data taking in the beginning of 2007. Due to civil engineering
constraints, the CHOOZ-near detector will be installed one year after
CHOOZ-far, and will start data taking beginning 2008.
The Double-CHOOZ proposal will be written in the forthcoming months by
a team  of European physicists from three countries and five
laboratories (about 20 physicists); the international collaboration is
expected to grow during this period.
\par
We would like to thank the Electricit\'e de France
 power company for its technical and financial support to the
Double-CHOOZ project.

%% file: daya.tex
\section{Daya Bay}
\par
\begin{figure}[thbp]
\vspace*{0.5in}
\centerline{\epsfig{file=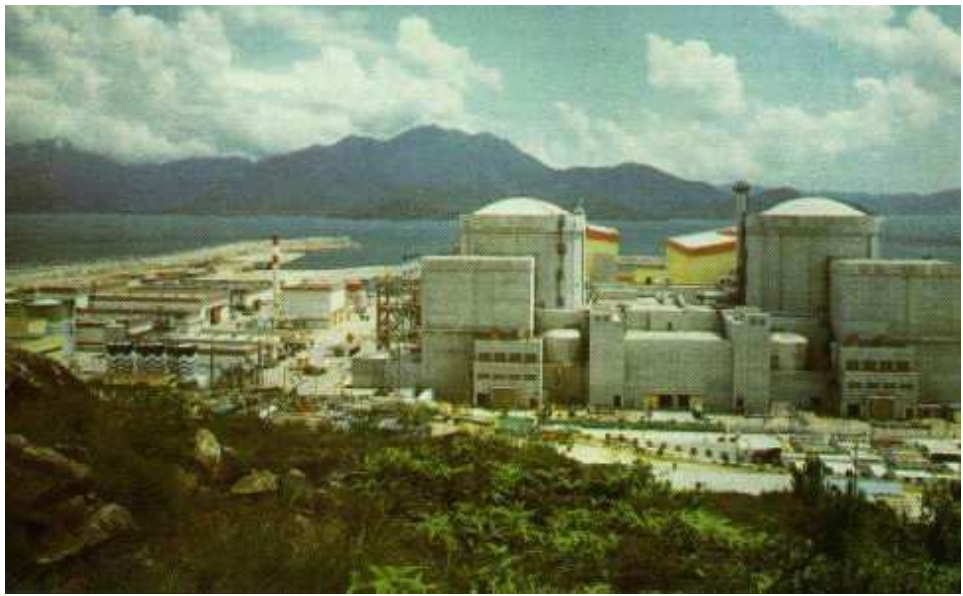,width=5.0in}}
\caption{The overview of the Daya Bay nuclear power reactors}
\label{fig:daya1}
\end{figure}

\begin{figure}[hbtp]
\vspace*{0.5in}
\centerline{\epsfig{file=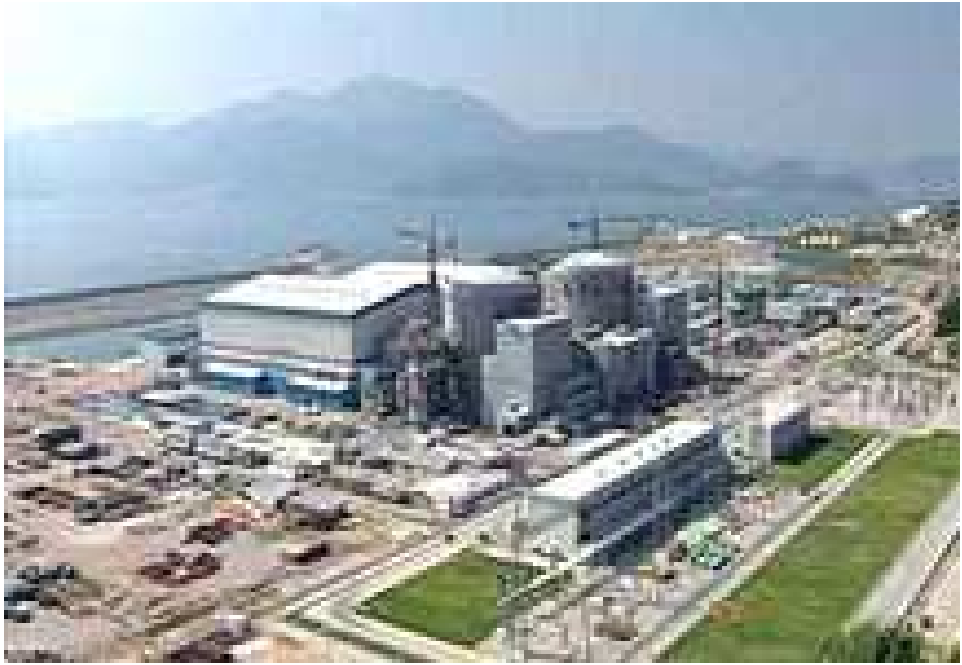,width=5in}}
\caption{The overview of the Lingao nuclear power reactors}
\label{fig:daya2}
\end{figure}

 
The Daya Bay nuclear power plant is the largest nuclear power generating 
station in China, and the tenth largest in the world. It consists of two 
twin reactor cores, one is called Daya as shown in Figure~\ref{fig:daya1}, and the other 
is called Lingao as shown in Figure~\ref{fig:daya2}, each core can generate a thermal power 
of 2.9 GW, making a total of 11.6 GW. All reactors are the type of the 
pressurized water. A third twin-core with a total thermal power of 5.8 GW, 
is planed to be online at about 2010. It will then be the second largest 
nuclear power plant in the world, only next to the Kashiwazaki in Japan. 

Daya Bay is located about 50 km from ShenZhen, Guangdong province, and it 
is only two hours drive from Hong Kong. The reactors are build on coast next 
to a mountain whose highest point is 700 m above the sea level. A group of 
Chinese physicist has inspected the site and it seems no problem to have a 
near experimental hall with an overburden of 400 mwe at a distance of about 
300 m to the core, and a far experimental hall with an overburden of 1200 mwe 
at a distance of about 1500-2000 m to the core. Initial discussions about the 
possible experiment with the authority of the power plant are positive and a 
conceptual design of the tunneling and experimental hall in collaboration with 
the power plant is underway.

A group of physicists has been working on the this possibility, including 
negotiation with the power plant, conceptual design of the experimental hall 
and tunneling, detector design and R\&D, Monte Carlo simulation etc.,.
Collaboration with all interested parts are welcome.

%% file: diablo.tex
%



\section{The Diablo Canyon Power Plant}


A number of commercial nuclear power plants worldwide are 
currently under evaluation as sites for a next-generation reactor neutrino 
oscillation experiment. An overview of these activities is given 
in this paper and references herein.  One of the candidate sites 
in the United States  is the Diablo Canyon Power Plant (DCPP) 
on the Central Coast in California. The Diablo Canyon Power Plant is 
located in San Luis Obispo County north of Santa Barbara and provides 
electricity to northern and central California. DCPP is owned 
by Pacific Gas and Electric Company (PG\&E) \cite{diablocanyon_pge}.  

A powerful reactor and overburden in excess of 300 mwe to shield the 
antineutrino detectors from cosmic rays and associated backgrounds are 
principal features of a suitable site for a reactor neutrino experiment. An underground detector facility or artificial overburden  are usually required to achieve sufficient shielding from cosmic rays. 

\begin{figure}

\begin{center}
\includegraphics*[width=0.8\textwidth]{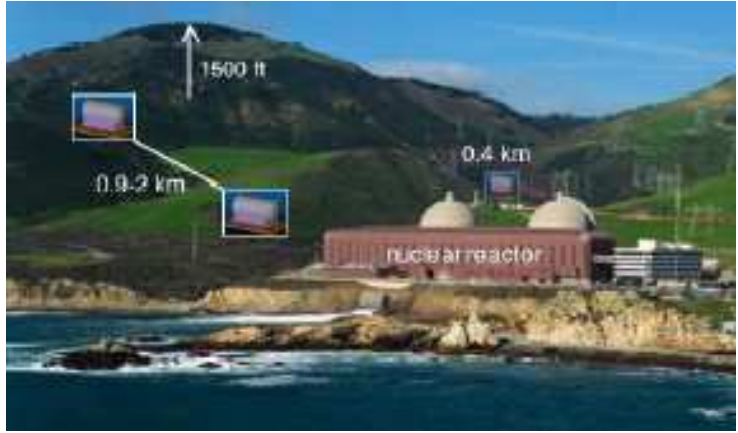}
	\caption{Picture of the Diablo Canyon nuclear power plant in San Luis Obispo County, California, USA. The local topography at Diablo Canyon allows the construction of an underground tunnel between  0.9-2~km for the placement of two or more neutrino detectors. A longer tunnel up to a distance of 3~km is possible. It may be possible to place an additional near detector at 0.4~km under artificial overburden.}
\label{fig:dc_picture}

\end{center}
\end{figure}

\begin{figure}

\begin{center}
\includegraphics*[width=0.8\textwidth]{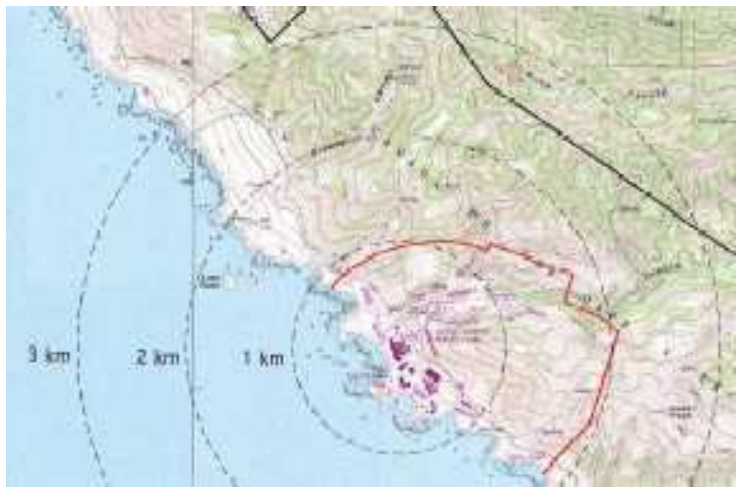}
	\caption{Topographic map of the site of the Diablo Canyon Power Plant. The land boundary (black) as well as the power plant site boundary (red) are indicated.}
\label{fig:dc_topo}

\end{center}
\end{figure}

The Diablo Canyon Power Plant consists of two 1.1~MW reactor cores separated by $\sim 100$ m with a total thermal energy of 6.2~GW$_{th}$. Nearby coastal mountains provide the opportunity for good overburden and make the plant an almost ideal site for a reactor neutrino 
experiment \cite{theta13lbnl}. The excavation of  a horizontal tunnel in the coastal mountains can provide overburden of up to 800~mwe with distances up to $\sim 3$~km. The general layout and topography of the site allows the construction of  a 1-2 km-long tunnel for two or more movable detectors. A picture of the Diablo Canyon Power Plant is shown in  Figure ~\ref{fig:dc_picture} and Figure ~\ref{fig:dc_topo} gives a topographic map of the area surrounding the DCPP site. 

Construction of a horizontal tunnel is required to place the neutrino detectors underground and shield them from cosmic rays and associated backgrounds. The local topography allows the construction of a tunnel at  distances between 0.9-2~km from the reactor. Longer tunnels up to a distance of 3~km are possible. A near detector within 500~m of the reactor cores requires the construction of a detector room and artificial overburden in excess of at least 50 mwe.

\begin{figure}

\begin{center}
\includegraphics*[width=0.9\textwidth]{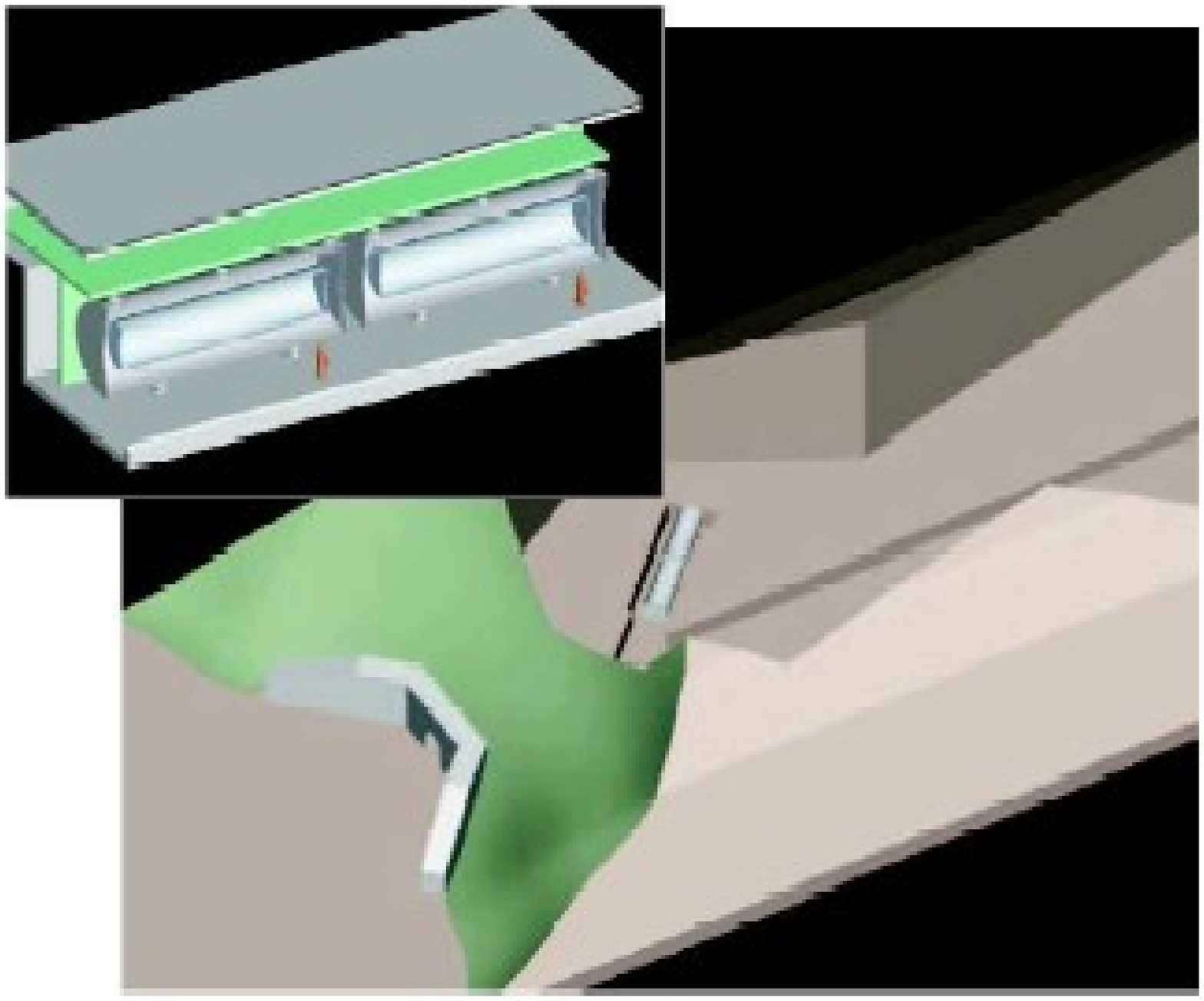}
	\caption{Horizontal access tunnel and side drifts for the the detector rooms. The side drifts will accommodate two neutrino detectors. (Drawings provided by D.~Oshatz, LBNL.)}
\label{fig:tunnels}

\end{center}
\end{figure}

The neutrino detectors  can be placed in detector rooms located in side drifts off the main tunnel (see Figure~\ref{fig:tunnels}). This allows the detectors to be surrounded by a hermetic muon veto and passive shielding  reducing the cosmic-ray induced neutron and external $\gamma$-background from the rock. The side drifts can provide a clean-room environment for the sensitive neutrino detectors and air control for the reduction of radon. Each side drift will have enough space to accommodate two neutrino detectors in order to allow for the relative calibration of the detectors' efficiencies at one particular distance from the reactor and under the same background conditions. Due to the natural topography of the site the overburden and hence the cosmic ray background will vary depending on the detector location and the distance from the reactor. 

A horizontal access tunnel will allow for the possibility to optimize the baseline of the detectors with respect to the reactor. The exact location of the side drifts can be chosen when the parameters that determine the neutrino oscillation length (such as $\Delta m^2_{atm}$) and hence the baseline of the experiment are better known. With multiple side drifts for the detectors one can imagine demonstrating conclusively the subdominant oscillation effect with measurements at various locations along the tunnel. 

\begin{figure}

\begin{center}
\includegraphics*[width=0.9\textwidth]{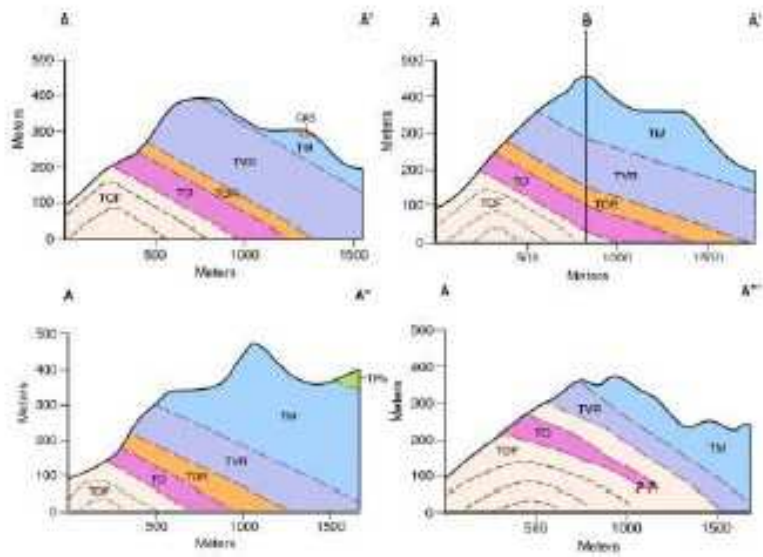}
	\caption{Geologic cross-sections of four possible tunnels at the DCPP site. The distance is given in meters from the tunnel portal. The tunnel portal is located at a distance of about 0.9~km from the reactor. A maximum overburden of up to 300~m of rock or $\sim 800$~mwe can be obtained. (Figure provided by C.~Onishi and P.~Dobson, LBNL.)}
\label{fig:xsections}

\end{center}
\end{figure}

Extensive geological studies of the site of the Diablo Canyon 
Power Plant have been done by PG\&E. A preliminary evaluation of 
the geology for this project is in progress at Berkeley Lab. The geology 
of the site appears to be suitable for tunneling and the excavation of 
the underground detector rooms. A tunnel may even offer the opportunity 
for interesting geo-science studies. The stratigraphy of the Diablo 
Canyon region is dominated by three Pliocene-Miocene marine sedimentary 
units: the Pismo Formation, the Monterey Formation and the 
Obispo Formation.  The rocks present in the area around the 
prospective tunnel sites at Diablo Canyon are predominantly the older 
sedimentary and volcanic rocks of the Obispo Formation, consisting of 
(1)~tuffaceous siltstones and claystones, and (2)~basaltic flows, dikes, and 
sills.  There are a variety of geologic hazards and 
issues associated with the construction and maintenance of 
a neutrino detector facility at Diablo Canyon: landslide hazards, rock quality issues, seismic hazards, and water quality. These issues are currently under investigation. The average bulk density of these rock formations varies between 2.2-2.8 g/cm$^3$. This leads to a maximum effective overburden of $\sim 800$~mwe at various locations along the tunnel. Geologic cross-sections of four possible tunnels at the DCPP site are shown in Figure~\ref{fig:xsections}. These tunnel cross-sections start at an elevation of $\sim 85$~m at a location approximately 0.9~km East-North-East from the Diablo Canyon Power Plant. 

The physics potential of a next-generation reactor neutrino measurement is discussed in great detail in the previous sections of this paper. The ultimate sensitivity of  such an experiment will strongly depend on the layout of the experiment including the distances from the reactor cores, the overburden, the size of the detectors, and our understanding of the relative detector response. The Diablo Canyon Power Plant is a site that offers the opportunity to design an experiment that meets these criteria and achieves an ultimate sensitivity of $\sin^2(2\theta_{13}) =$ 0.01-0.02. 

%% file: illinois.tex
\section{An Illinois Reactor Experiment}
The Illinois proposal is to locate the experiment at 
one of the Exelon nuclear reactors: Braidwood, Byron, or 
La Salle.  All three of these reactors are located 
in northern Illinois with 100~km of both Argonne National 
Lab and Fermilab.  The typical elevation of northern Illinois 
is about 250~meters and the topology is flat.  The baseline experiment 
design calls for three identical spherical detectors, each 
with a 25~ton gadolinium loaded scintillating target.  There is 
one near detector located at a distance of about 200 meters from the 
reactor cores.  The remaining two detectors are at a 
far location and share a baseline of 1400 to 1800~meters.  Both 
the near and the far detector halls are located below $\geq 100$~meters 
of rock ($\sim300$~mwe).  This shielding is achieved by digging 
shafts straight down to a depth of about 100 meters

In addition to the standard source and flasher calibration systems, the 
relative efficiency of the near and far detectors is measured head-to-head.  
To facilitate this measurement, the detectors are movable, and the 
near and far detector sites are connected by a tunnel 
(See Figure~\ref{fig:move1}).  The relative efficiency is 
measured with the two detectors side by side in the intense neutrino flux 
at the near detector site. Calibration running for 10\% 
of the experimental run results in an uncertainty on 
the relative efficiency that is smaller than the statistical error on the event rate in the far detector. 

\begin{figure}[t!]
\caption{Interaction of the Muon Veto Neutron Shield system with muon induced fast neutrons produced both inside and outside the shielding bunker.  }
\centerline{\epsfig{file=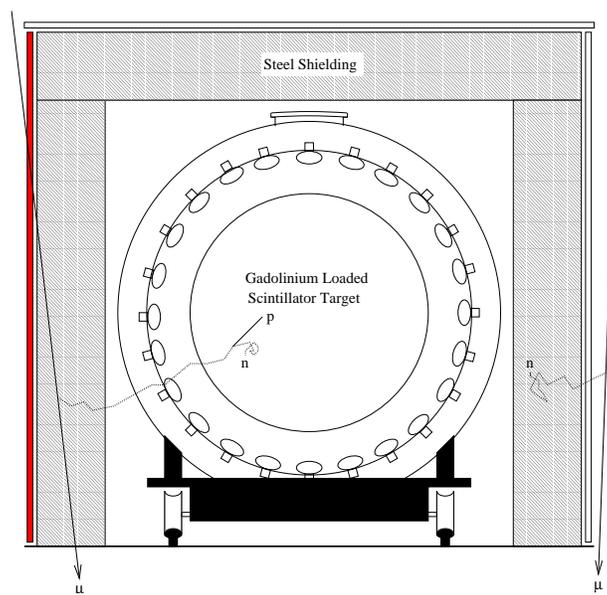,width=3.2in}}
\label{mvns}
\end{figure} 

Correlated coincident backgrounds, in which 
both parts of the event are generated by the same initial 
interaction, are the most difficult to account for.  Most 
correlated backgrounds are due to fast neutrons generated in 
cosmic ray $\mu$ spallation in the materials surrounding the 
detector.  Inside the detector a fast neutron can strike 
a hydrogen nucleus giving it enough energy to mimic the positron.  Alternatively, 
a neutron-carbon inelastic collision may result in the production 
of gamma rays.  In both cases the fast neutron will thermalize and 
capture with nearly the same spatial and time correlation as the 
neutrons from inverse $\beta$-decay.  

To deal with these correlated background events a Muon Veto Neutron Shield (MVNS) system is proposed.  In the MVNS system (shown in Figure~\ref{mvns}), the detector is housed in a bunker of dense material intended to range out neutrons. The outside of this bunker is covered in an array of plastic scintillator.  The scintillator array detects muons entering the bunker.  Muons in the bunker material may kick out fast neutrons and cause a correlated background event in the detector.  If a muon passes close to the bunker without passing through, neutrons generated by it in the surrounding rock are ranged out by the bunker. In addition, matching the energy distribution for muon events to the neutrino candidate events, outside the reactor energy range, the any surviving background events can be subtracted from the neutrino event sample. 

The MVNS bunker is fixed and does not move with the detector.  

The Illinois proposal is designed to control the dominant 
systematic errors, which are due to the relative efficiency 
of the near and far detectors and the uncertainty in the background 
rate.  By keeping these systematic errors low, a sensitivity of 
$\sin^22\theta_{13}<0.01$ at 90\% CL can be achieved in 
a three year run.

%% file: kash.tex
\section{The KASKA project}
\begin{center}
\begin{figure}[htbp]
 \includegraphics[width=7cm]{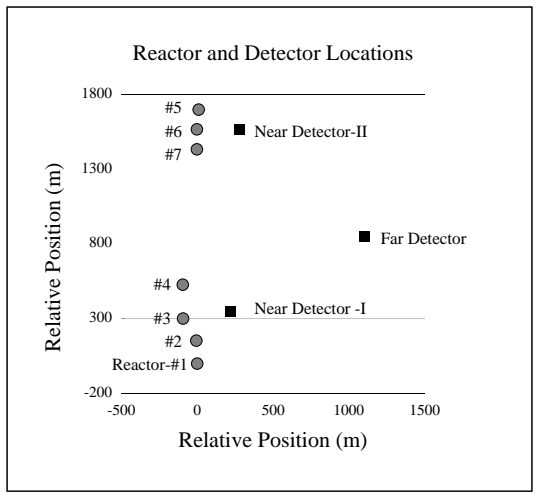} 
\caption{Reactor (circles) and detector (squares) relative locations. 
Reactor \#1 through \#4 form a cluster and \#5 through \#7 form another
 cluster.  
The two clusters separate about 1.3~km apart. 
Two near detectors will be placed at around 400~m from each cluster. 
The far detector will be placed at around 1.3~km from all the reactors. 
\label{fig:kash}}
\end{figure}
\end{center}

\begin{center}
\begin{figure}[htbp]
 \includegraphics[width=15cm]{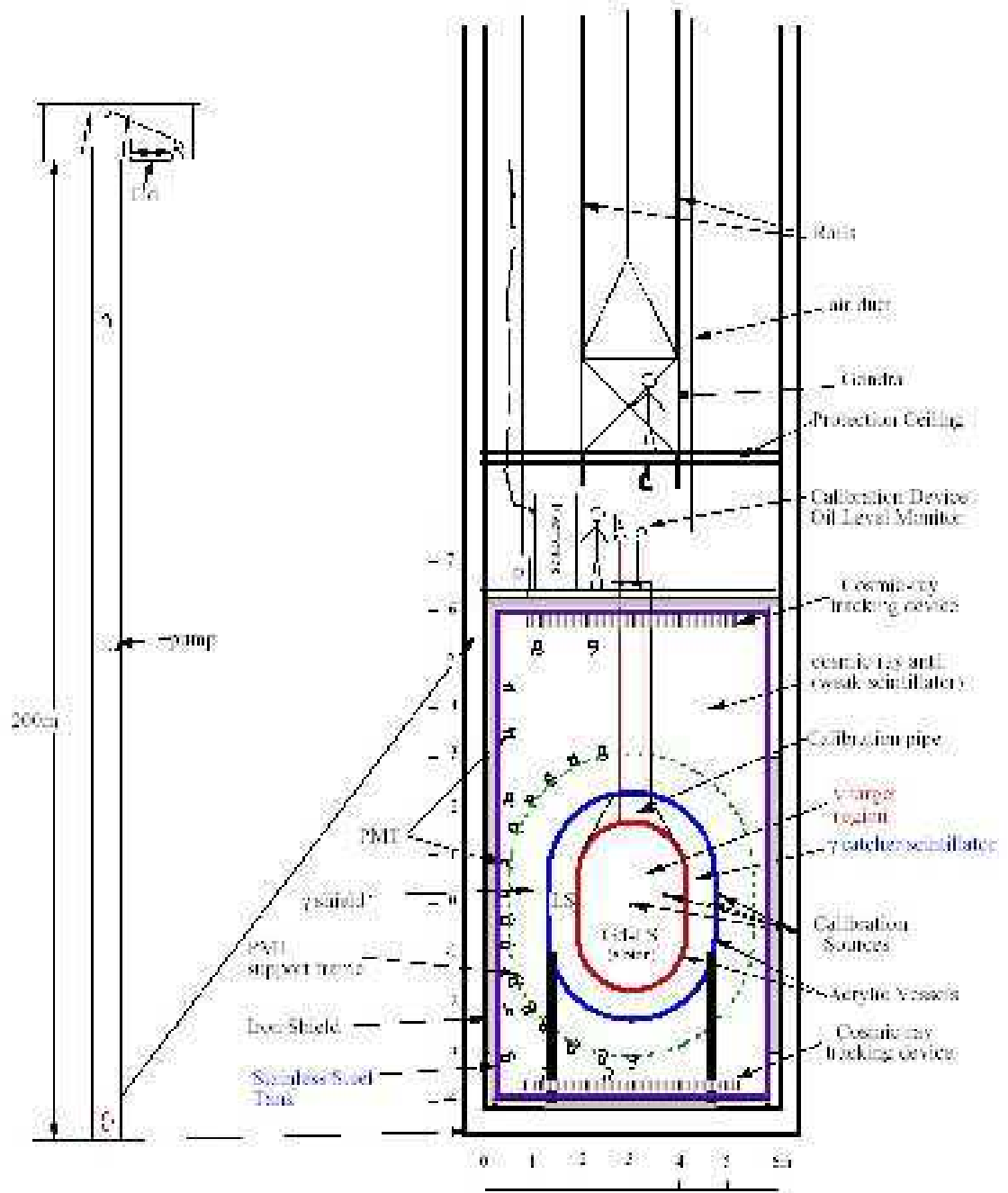} 
\caption{Schematic view of the detector.
\label{fig:detector_fig}}
\end{figure}
\end{center}

In the KASKA experiment \cite{Kashiwazaki}, three identical detectors are to be built in the site 
of Kashiwazaki-Kariwa NPP which has 7 reactors, producing total 
thermal energy of 24.3 GW.
This is the most powerful NPP in the world.
 Using a large-power nuclear power plant is profitable for obtaining not only
a high event rate but also a low background-to-signal ratio
at a given depth underground. 
The relative locations of reactors and detectors
are shown in the fig.-\ref{fig:kash}.
Although the far/near distance ratios between the reactors and
detectors are not unique, the effective uncertainty introduced from the variations
of the distances are estimated to be only 0.2\%.
The detector is a CHOOZ like detector as shown in the
Figure~\ref{fig:detector_fig}.  

The $\bar{\nu}_e$ target is 8 tons of Gadolinium loaded liquid
scintillator (GD-LS) contained in a acrylic vessel which is called
region-I. 
The component of the Gd-LS is the Palo Verde type, which
is proven to be stable in the acrylic container~\cite{Boehm:1999gl}.  

The Gd concentration is 0.1\% (0.15\% optional) and the neutron capture
efficiency on Gd is 88\%. 
The optional 0.15\% Gd concentration is intended to increase the neutron
capture efficiency and to reduce the systematic uncertainties
associating with the inefficiency. 
The reactor $\bar{\nu}_e$ is detected by the inverse $\beta$
decay reaction with proton. 
\begin{equation}
\bar{\nu}_e + p \rightarrow e^+ + n
\end{equation}
The kinetic energy of the positron is neutrino energy minus 1.8~MeV. 
The positron annihilates with electron after slowing down in the LS and then produces two 0.511~MeV $\gamma$'s.  
This process produces a prompt signal, whose energy is between 1~MeV
and 8~MeV. 

The energy threshold for prompt signal is set to be below the minimum
prompt signal energy of 1~MeV.  
In this way, no systematic ambiguities associated with threshold energy cut is
introduced.

The neutron produced in the inverse $\beta$ decay reaction thermalizes quickly and is absorbed by Gd, producing $\gamma$-rays whose total energy amounts to 8~MeV. 
The neutron absorption signal occurs typically 30~$\mu s$ after the prompt signal.

The neutrino signal is defined by the existence of correlated signals.
That is, when 8~MeV of energy deposit is observed after associating prompt signal whose energy is between 1~MeV and 8~MeV, this event is considered to be $\bar{\nu}_e$ event, regardless the positions of the signals. 
Because there is no position cut, the measured neutrino rate is free
from the position reconstruction error.
Requirements for the timing correlation between the positron signal and
the neutron signal suppress backgrounds severely.   

The main source of systematic error comes from the 
relative difference of the LS mass in the three detectors. 
The total volume of the LS will be controlled not by the size of the container
but by the liquid volume put in the container. 
The relative volume of the introduced LS will be calibrated before
putting in each detector. 
The total volume of the LS in the oval part of the vessel can be
measured precisely from the total volume of introduced LS and the 
liquid level in the thin calibration pipe even if there is a distortion of the vessel 
after the installation. 
The relative uncertainty of the LS mass between detectors 
will be less than 0.5\%. 
\par If neutrino events occur near the boundary of region-I, $\gamma$
rays from the positron annihilation and the 
neutron absorption may escape from region-I. 
In such cases, the visible energies of the signals may become less than
thresholds and this inefficiency introduces potential systematic error.
This is also true for the case of positron detection. 
\par In order to catch such $\gamma$ rays, a 
50~cm thick Gd unloaded LS (region-II) surrounds 
region-I acrylic container. 
The light output of region-II LS is adjusted to be same as region-I LS
to obtain accurate original energies accurately. 
It is estimated that most of the $\gamma$ rays can be contained 
within regions I and II and the inefficiencies of the positron and 
neutron signal is less than a few percent and the systematic error 
associated with this inefficiency is a fraction of a percent.
\par If the neutrino event occurs near the acyclic wall, the neutron 
may escape from region-I (spill-out effect).
This kind of events produces a loss of efficiency. 
However, the inefficiency due to the spill-out effect is estimated to be  
only $\sim$3\% and the systematic error due to the spill-out effect is
estimated to be on the order of 0.1\%.
On the other hand, there is the
opposite case that the inverse $\beta$ decay event occurs 
in region-II and then the resulting neutron spills into region-I 
(spill-in effect) and absorbed by Gd.
This event is also considered to be a neutrino event and it amounts to 20\% of the 
total events. 
However, the spill-in efficiency is insensitive to the 
hydrogen concentration in region-II LS because the 
spill-in event rate is proportional to the neutrino event 
rate and the neutron absorption length and both 
parameters have opposite dependence on the hydrogen concentration. 
Because most of the events are contained in regions I and II,
systematic error due to the spill-in effect is 
estimated to be less than 0.5\%.
\par The outer wall of region -II is also a acrylic container and the
region-II is surrounded by a 90~cm thick scintillator with 
very slight light output (region-III).
Region-III shields $\gamma$-rays from PMT glasses. 
Thanks to the region-III buffer, the singles rate is expected to
become less than 10 Hz. 

The outer most layer (Region-IV) is muon anti-counter filled with the same scintillator as region-III. 
These layers work as a shield of gamma rays and as cosmic ray anti counters.  
The slight light output is to detect low energy muons whose velocity is
below the Cerenkov threshold, and to obtain high muon tagging efficiency. 
There will be about 400 8-inch low background PMTs at the boarder of region-III and IV.
The wall between region-III and IV 
is just for light shield and no hermeticity is necessary. 
\par There are cosmic-ray tracking devices on both the top 
and bottom of region-IV. 
The trackers are used to measure cosmic-ray tracks
with accuracy of $\sim$10~cm.
The cosmic-ray spallation background rate, such as the $^9$Li rate, is
estimated from the excess at small distances in the distribution of 
distance between the neutrino event candidates and the cosmic-ray tracks.
In order to stop the hadronic and soft part of the direct cosmic rays, 
the thickness of the top part of region-IV will be made thick.  
Region-IV LS is contained in a stainless steel cylindrical hermetic
container. Outside the stainless steel container, there is a thick iron
shield to prevent $\gamma$-rays coming from the soil outside. 

The far detector will be placed at the bottom of a 200~m shaft hole with 
an inner diameter of 6~m. 
The cosmic-ray flux at the bottom is about 0.35~$m^{-2}s^{-1}$.
Although there is open space above the detector, 
the cosmic-ray rate directly 
down the open space is only a fraction of 
that which reaches the detector penetrating the soil. 
The near detectors will be placed at the bottom of a 70~m depth shaft hole.  
This depth is chosen to make muon/neutrino ratio approximately equal to or slightly larger than the far detector case.

The major component of the background comes from 
fast neutrons produced in nuclear interactions 
caused by cosmic rays going through the rock near by.
The visible energy distribution of the prompt signal in the fast neutron backgrounds was measured to be flat by CHOOZ group at the energy range
below 30~MeV~\cite{bib:chooz} and this kind of background rate can be
estimated by using the event rate within non-reactor-$\bar{\nu}_e$
energy range, such as below 1~MeV and above 10~MeV. 

The absolute background rate is expected to be less
than 2\% and the error associated with estimation of the background will be
less than 0.3\%. 

\par The systematic error in CHOOZ experiment was 1.7\% (detector associated)
$\oplus$ 2.1\%(neutrino flux associated).    
The systematic error associated with the neutrino flux is reduced to 0.2\% 
by comparing near and far detectors in the KASKA experiment.
The systematic error associated with the detector is 
estimated to be between 0.5\% and 1\%. 
Because the neutrino oscillation at the near detectors is small, 
the consistency check of the systematic error estimation can be
performed by comparing data of the two near detectors. 

\begin{center}
\begin{figure}[htbp]
 \includegraphics[width=10.cm]{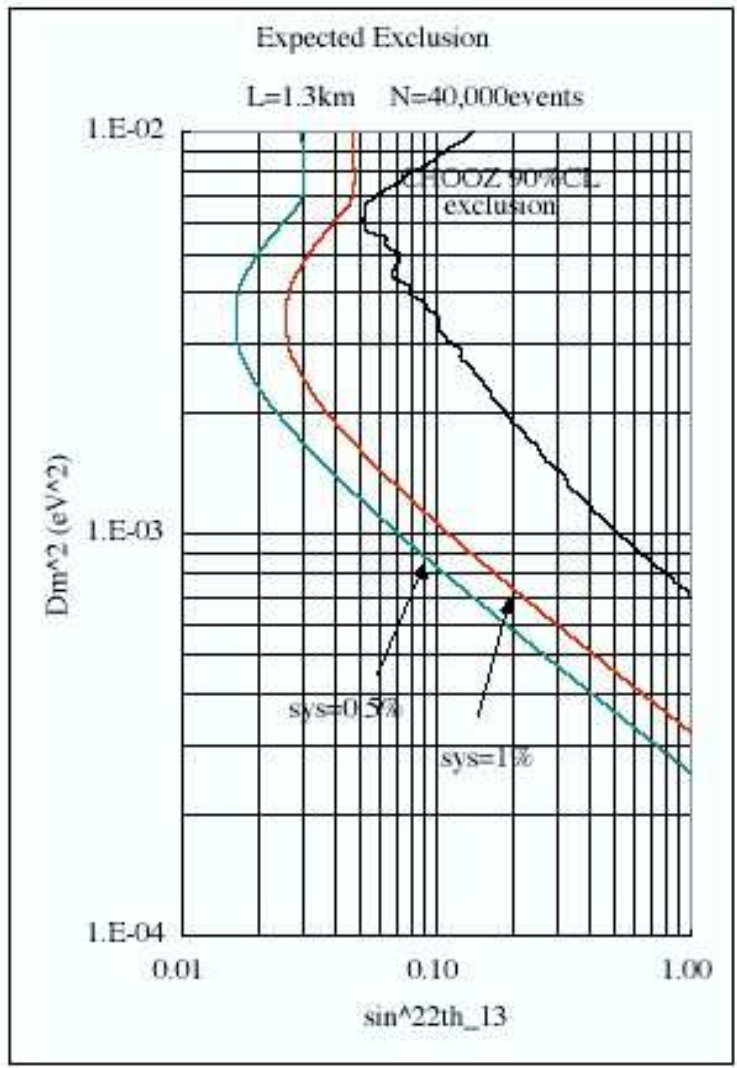} 
\caption{The expected 90\%CL exclusion region of this experiment for the
 case of $\sigma_{sys}$=1\% and 0.5\% obtained by rate only analysis.  
\label{fig:sensitivity}}
\end{figure}
\end{center}

In two years of operation, 40,000 neutrino events will be recorded in the far
detector and ten times more in the 
near detectors and the corresponding statistical error will be 0.5\%.
The 90\% CL sensitivity of this experiment is shown in the
fig.-\ref{fig:sensitivity}.
At $\Delta m^2 \sim 3\times10^{-3} eV^2$, the sensitivity of $\sin^22\theta_{13}\sim 0.02$ is expected. 
This is five times better limit than CHOOZ.

%% file: krproj.tex
\section{The Krasnoyarsk Reactor and KR2DET}
\begin{figure}[th]
 \vspace*{2.0mm} 
 \includegraphics[width=14cm]{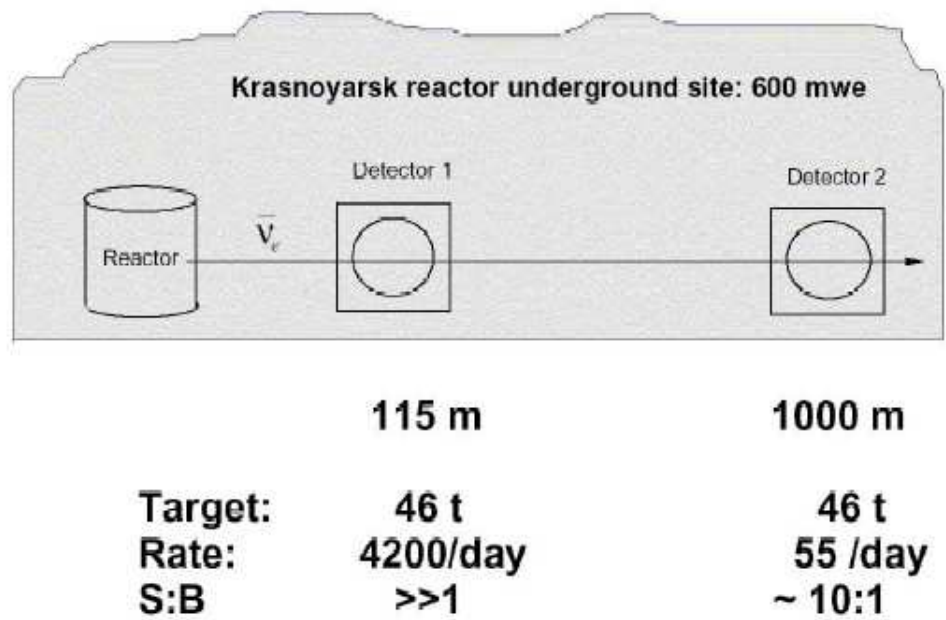} 
\caption{Scheme of the KR2DET experiment.}
\label{fig:kr3}
 \end{figure}

The KR2DET proposal is to place two
identical liquid scintillation spectrometers are stationed at distances $L_{far}\approx$ 1000 m 
(far position) and $L_{near}\approx$ 115 m from the underground 
Krasnoyarsk reactor. (Figure~\ref{fig:kr3}) 
The overburden at Krasnoyarsk is $\sim$ 600 m.w.e., which is twice as much as in the CHOOZ 
experiment. (At short distances form the reactor the one reactor - 2 detector approach was first 
probed at Rovno \cite{bib:ketov} and later successfully used 
at Bugey \cite{bib:declai}.
 
Two types of analysis can be used.
Analysis I is based on comparison of the shapes of positron spectra $S(E_e)_{far}$ and $S(E_e)_{near}$ 
measured simultaneously in two detectors. In no oscillation case the ratio 
$S(E_e)_{far}/S(E_e)_{near}$ is energy independent. Small deviations from the constant value of this 
ratio 
\begin{equation}
X_{shape}= C\cdot \frac{1 - \sin^{2}2{\theta}\cdot \sin^{2}\left(\frac{1.27{\Delta}m^{2}L_{far}}{E}\right)}{1 - \sin^{2}2{\theta}\cdot \sin^{2}\left(\frac{1.27{\Delta}m^{2}L_{near}}{E}\right)}
\label{eq:krshape}
\end{equation} 
are searched for oscillation parameters. 
 
\begin{figure}[th]
 \vspace*{2.0mm} 
 \includegraphics[width=14cm]{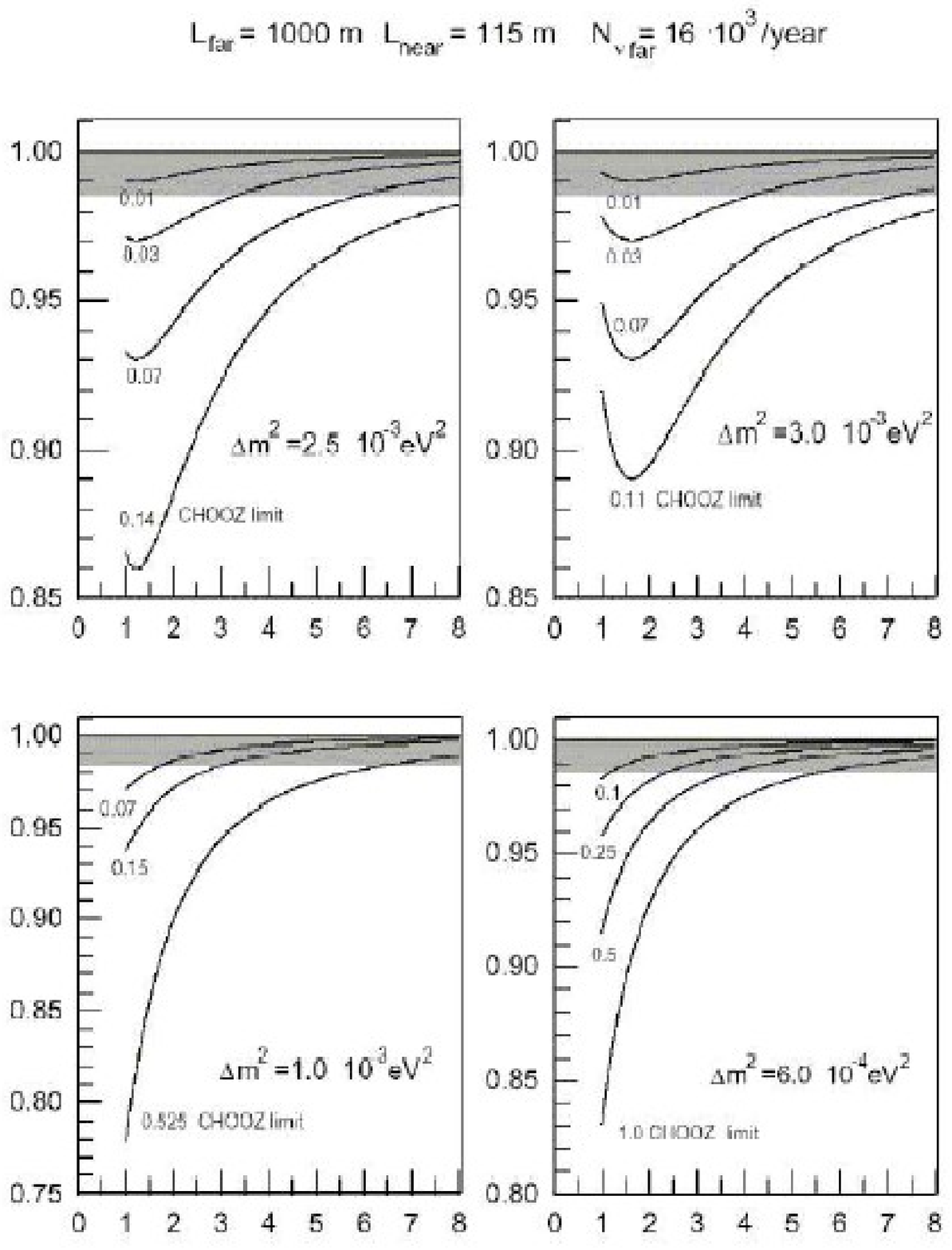} 
\caption{Calculated ratio of positron spectra 
$S(E_e)_{far}/S(E_e)_{near}$ for some oscillation parameters.  Values
of $\mxang$ are shown at the curves.}
\label{fig:kr4}
 \end{figure}

In the one reactor - two detector scheme
\begin{itemize}
\item Results of the Analysis I do not depend on the 
exact knowledge of the reactor power, absolute 
$\bar{{\nu}_e}$ flux and energy spectrum, burn up effects, absolute values of hydrogen atom concentrations, 
detection efficiencies, target volumes and reactor - detector distances. 
\item At Krasnoyarsk the detector backgrounds can be measured during reactor OFF periods, which 
periodically follow 50 day long reactor ON periods.  
\end{itemize}
Calculated ratios $S(E_e)_{far}/S(E_e)_{near}$ for a set of oscillation parameters are shown in 
Figure~\ref{fig:kr4}.

Analysis II is based on the ratio of the total number of neutrinos 
$N_{far}, N_{far}$ detected 
at two distances:
\begin{equation}
X_{rate}(\sin^{2}2{\theta},{\Delta}m^{2})= 
\left(\frac{L_{far}}{L_{near}}\right)^2\cdot 
\left(\frac{V_{near}}{V_{far}}\right)
\cdot\left(\frac{{\epsilon}_{near}}{{\epsilon}_{far}}\right)\cdot 
\left(\frac{N_{far}}{N_{near}}\right)   
\label{eq:krrate}
\end{equation}
$V_{far}, V_{near},{\epsilon}_{far},{\epsilon}_{near}$ are the target volumes and neutrino detection 
efficiencies.
In no oscillation case $X_{rate}$ = 1.
 
Analysis II is also independent of the exact knowledge of the reactor neutrino flux and energy 
spectrum. The absolute values of detection efficiencies are practically canceled, only their small 
difference is to be considered here while the ratios $(L_{far}/L_{near})^2$ and
$(V_{near}/V_{far})$ should be known accurately.

\begin{figure}[th]
 \vspace*{2.0mm} 
 \includegraphics[width=14cm]{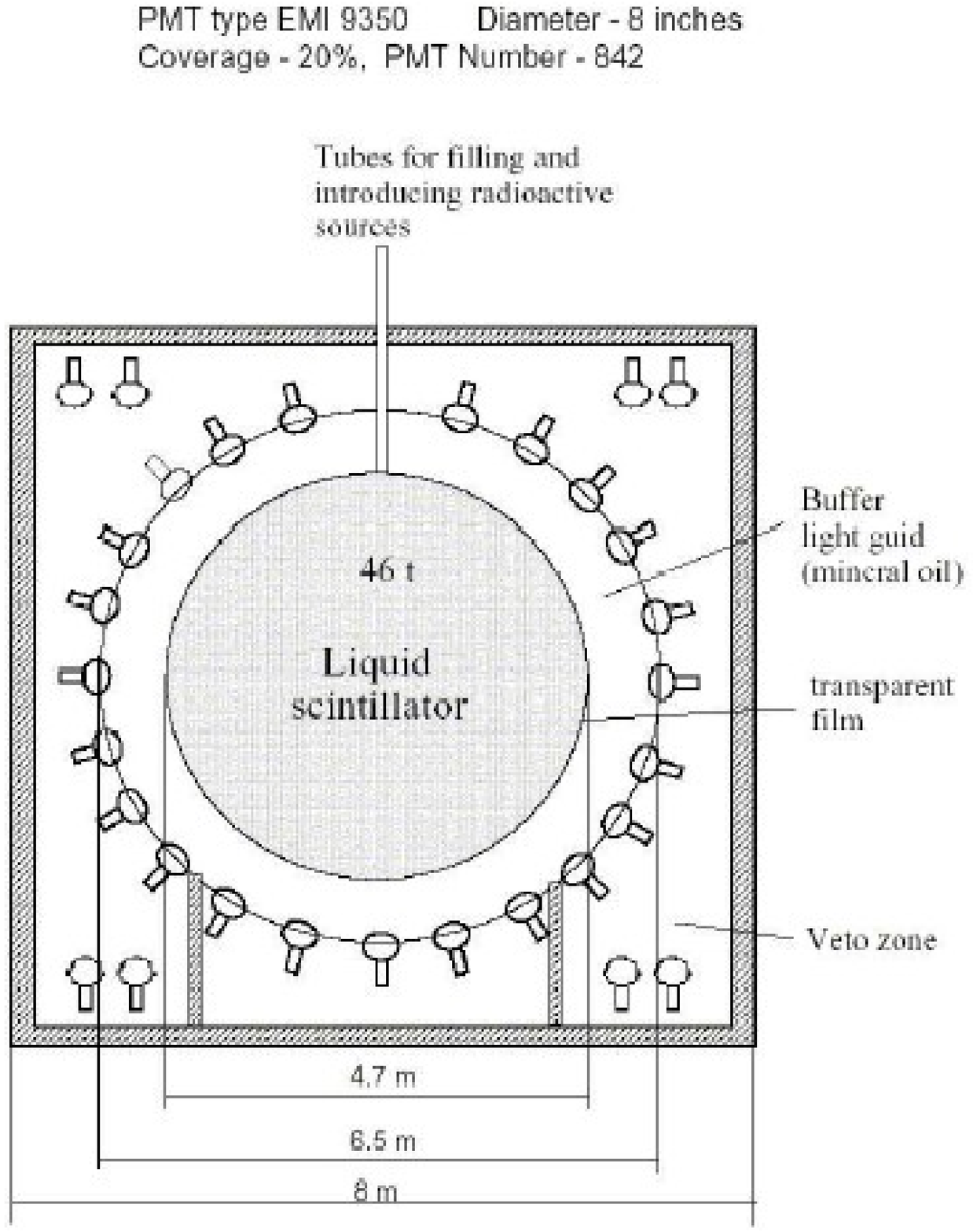} 
\caption{The KR2DET $\bar{\nu}_e$ spectrometer (schematic).}
\label{fig:kr5}
 \end{figure}

A miniature version of the KamLAND \cite{bib:piepk} and BOREXINO \cite{bib:alimo} and a scaled up version of the CHOOZ 
three - concentric zone detector design is chosen for the 
construction of the spectrometers (Figure~\ref{fig:kr5}). 
KR2DET plans a 4.7 m diameter liquid scintillator target, enclosed in transparent 
spherical balloon. The target is viewed by $\sim$800 8-inch EMI-9350 (9350 - 9356) photomultipliers 
trough $\sim$90 cm layer of mineral oil of the zone-2 of the detector. The PMTs of this type have 
successfully been used in the CHOOZ experiment and are used now in the BOREXINO and SNO 
detectors \cite{bib:baldi}. A 20\% light collection and 150 - 200 photoelectron signal is expected for 1 MeV 
positron energy deposition. The PMTs are mounted on the stainless steel screen, which separates 
external zone-3 from the central zones of the detector. The $\sim$75 cm thick zone-3 is filled with 
mineral oil (or liquid scintillator) and serves as active (muon) and passive shielding from the 
external radioactivity.

The ratio of measured positron spectra $S(E_e)_{far}/S(E_e)_{near}$ 
in Equation~(\ref{eq:krshape}) can be slightly distorted 
because of relative difference in response functions of the two ``identical" spectrometers. 

\begin{figure}[th]
 \vspace*{2.0mm} 
 \includegraphics[width=14cm]{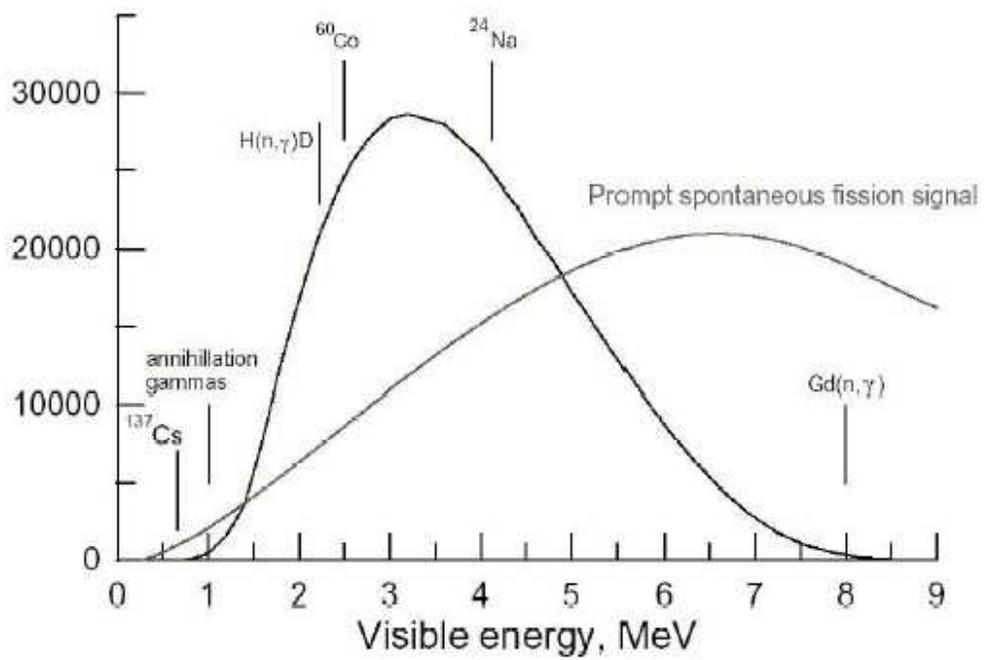} 
\caption{Positron visible energy spectrum.}
\label{fig:kr2}
 \end{figure}

The goal of calibration procedures is to measure this difference and introduce 
necessary corrections. This can be done by a 
combination of different methods. First there will be a 
periodic control of the energy scales in many points using ${\gamma}$-sources shown by arrows in 
Figure~\ref{fig:kr2}. A useful continuous monitoring of the scales at 2.23 MeV can provide neutrons produced 
by trough going muons and captured by the target protons during veto time. 

The second method uses small spontaneous fission $^{252}$Cf or $^{238}$U sources periodically 
placed in the detectors. These sources generates continuous energy spectrum due to prompt 
fission gammas and neutron recoils (the dashed line in Figure~\ref{fig:kr2}.). Deviation from unity of the 
measured spectra can be used to calculate relevant corrections.

The goal is that the
systematic uncertainty due to detector spectrometric difference essential for 
Analysis I can be controlled down to 0.5\%. 

In Analysis II the systematic uncertainty in the quantity 
$(L_{far}/L_{
near})^2\cdot (V_{near}/V_{far})\cdot ({\epsilon}_{near}/{\epsilon}_{far})$
in Equation~(\ref{eq:krrate}) can hopefully be kept within 0.8\%.

The choice of the scintillator has not been made so far. 
There should be progress in manufacturing Gd ($\sim$0.9 g/liter) loaded scintillators to improve 
the response to neutrons and suppress accidentals, which originate from U/Th gammas coming 
from surrounding rock. The Palo Verde Gd-scintillator showed better stability than the 
scintillator used in CHOOZ. The LENS project considers 
scintillators with rare earth contents 
as high as $\sim$50 g/liter.  One possibility is 
scintillator without Gd based 
on the mixture of isoparaffin or mineral oil 
and pseudocumene ($\sim$20\%) with $\sim$2g/liter 
PPO as primary flour. This scintillator 
has C/H ratio 1.85, density 0.85 kg/liter 
and $0.785\times 10^{29} $ H atoms per ton.

The neutrino events satisfy the following requirements: (i)~a time 
window on the delay between  
$e^{+}$ and neutron signals 2$-$600 ${\mu}$s, (ii)~energy window for the neutron candidate 
1.7$-$3.1 MeV and for $e^{+}$ 1.2$-$8.0 MeV, (iii)~distance between 
$e^{+}$ and neutron less 
than 100 cm. At this stage no pulse shape analysis to reject 
proton recoils is planned. 

Under these assumptions neutrino detection efficiency of 75\% was found and neutrino detection 
rate $N(e^{+},n)$ = 55/day calculated for the far detector. 

The time correlated background 0.1 per day per one target ton was found by extrapolation of 
the value 0.25/per day per target ton measured at CHOOZ:
\begin{equation} 
{\rm CHOOZ\ (300\ mwe),\ 0.25/day}\cdot{\rm ton} 
\rightarrow {\rm Kr2DET\ (600\ mwe),\ 0.1/day}\cdot{\rm ton}
\end{equation}

The accidental coincidences come from the internal radioactivity of detector materials and U 
and Th contained in the surrounding rock. The internal component of the background was estimated 
to be less 0.3/day, which is an order of magnitude smaller than the rate of the correlated 
background (see hep-ph/0109277). In contrast to the KamLAND and Borexino experiments three 
orders higher concentrations of U, Th, K and Rn can be tolerated in the liquids used in the 
Kr2DET case. 

First estimations of accidentals coming from the 
radioactivity of the rock showed however that 
external passive shielding of the detector should be increased in case scintillator without 
Gd is used as the neutrino target.
 
Calculated neutrino detection rates $N(e^{+},n)$ and backgrounds for scintillator with no Gd 
are summarized in Table~\ref{table:1}.
 
\begin{table}[htb]
\caption{}
\label{table:1}
\vspace{10pt}
\begin{tabular}{c|c|c|c|c|c|c}
\hline
Detector  & Distance, & Target, & $N(e^{+},n)$, & $N(e^{+},n)$, & \multicolumn{2}{c}{Backgr., day$^{-1}$} \\
\cline{6-7}
           &     m     & mass, ton & day$^{-1}$    & year$^{-1*}$     & correl. & accid.$^{**}$ \\
\hline
Far  & 1000 & 46 & 55 & $16.5\cdot 10^{3}$ & 5 & $\sim$0.3  \\
Near & 115 & 46 & 4200 & $12.5\cdot 10^{5}$ & 5 & $\sim$0.3  \\
\hline
\end{tabular}\\[2pt]
{\small $^{*}$ 300 days/year at full power.}\\
{\small $^{**}$ due to internal radioactivity of the detector materials only.}
\end{table}

\begin{figure}[th]
 \vspace*{2.0mm} 
 \includegraphics[width=14cm]{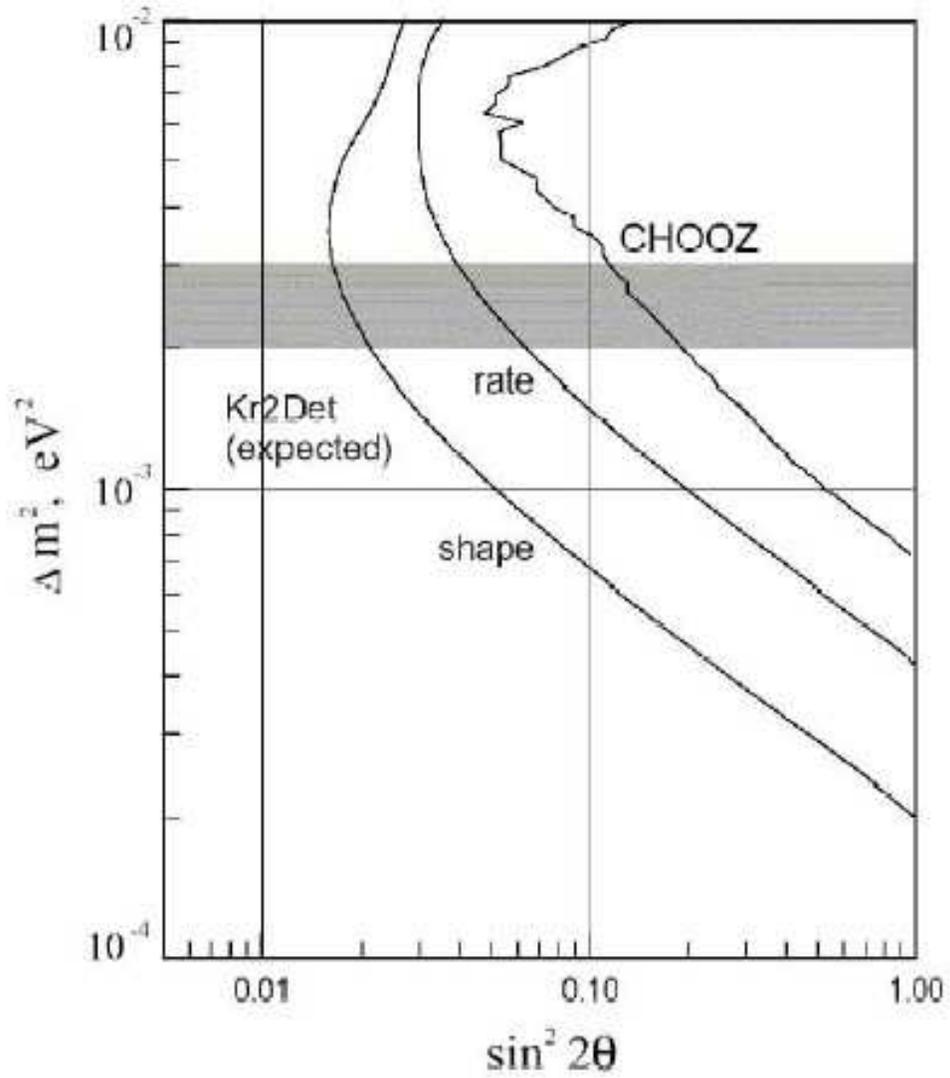} 
\caption{Reactor antineutrino oscillation plots.  Curves
``CHOOZ", ``KR2DET" (expected) ``shape" and ``rate" are 90\% CL $\bar{\nu}_e$
disappearance limits.  The KR2DET limits are obtained assuming 40,000
detected antineutrinos in the far detector, 10:1 effect to background
ratio and systematic uncertainties $\sigma_{shape}$ = 0.5\% and 
$\sigma_{rate}$ = 0.8\%.  The shaded area represents the most probable
atmospheric neutrino mass parameter region.}

\label{fig:kr1}
 \end{figure}

Expected 90\% CL constraints on the oscillation parameters (Figure~\ref{fig:kr1}, curves K2Det) are obtained for 
40000 detected $\bar{{\nu}_e}$ in the far detector (750 days of full power). The systematic 
uncertainties ${\sigma}_{shape}$= 0.5\% in the Analysis I (``shape") and ${\sigma}_{rate}$= 0.8\% 
in the Analysis II (``rate") have been assumed. The ``shape" analysis is somewhat more sensitive 
and can shift (at ${\Delta}m^2=2.5\times 10^{-3}$ eV$^2$ the $\sin^22{\theta}$ upper limit from 0.14 
(CHOOZ) to 0.017. 

The one reactor - two detector approach fully eliminates uncertainties associated with the reactor 
neutrino source inherent to the absolute method used at CHOOZ.

Small relative difference in conceptually identical detector properties can be minimized through 
calibration and monitoring procedures.

The detector backgrounds can be measured during reactor OFF periods, which periodically follow 
50 day long reactor ON periods.

Good signal to background ratio can be achieved due to sufficiently deep underground position of the 
detectors. 

High statistics can be accumulated in reasonably short time period using detectors with $\sim$45 ton 
targets, which are relatively small if compared to modern neutrino detectors.

Neutrino community has accumulated positive experience in building and running 3 concentric zone 
detectors similar to the Kr2DET detectors.

\subsection{Krasnoyarsk site details}
The reactor belongs to the Federal State-Owned Unitary Enterprise MINING \& CHEMICAL COMBINE (MCC) 
53, Lenin Str., Zhelezhnogorsk, Krasnoyarsk Territory, RUSSIA, 660972.

The Krasnoyarsk neutrino laboratory is built in the MCC underground territory.

There are two places to install the detectors. One of them at $\sim$115 m from the reactor is 10 m 
high $15\times15$ m square room. The other is a 125 m long, 11.5 high and 15 m wide corridor at 
$\sim$1000 m from the reactor. More information on neutrino 
at Krasnoyarsk can soon be found at \\
{\small http://www.lngs.infn.it/site/exppro/panagic/section$\_$indexes/frame$\_$particles.html 
}
\\
(click 
``Laboratories and experiments", then ``Underground and underwater laboratories" and go to 
``Krasnoyarsk neutrino laboratory")

Zhelezhnogorsk is located at about 70km from Krasnoyarsk on the bank of the Yenisei River. 
Zhelezhnogorsk is a very nice and clean town built in direct neighborhood to the Siberian taiga, 
rich of birds and animals. There is a beautiful large lake in the center of the town. Picturesque 
hills surround the town center. A musicale theater, hotel, rest home, restaurants, a lot of shops 
are in Zhelezhnogorsk.. The weather is comfortable; the number of sunny days is the same as in resort 
Sochi (at the Black Sea). Winter is cold but not so much compared with Moscow, air is dry. The summer 
and autumn are warmer and sunnier than in Moscow. 

Some information about tourism in Krasnoyarsk Territories is available at the site: 
$http://tlcom.krs.ru/kalinka/indexe.htm$, \\
tours $http://tlcom.krs.ru/kalinka/indexe.htm$

Every day there are flights from Moscow to Krasnoyarsk airport. Big comfortable airbus IL86 in 4.5 
hours time brings you from Moscow to Krasnoyarsk with good service of KrasAir company and a special 
minivan in 2 hours carries you from Krasnoyarsk airport Yemelianovo directly to the center of 
Zhelezhnogorsk.

MINING \& CHEMICAL COMBINE has two own rest homes; one of them is in the town territory near the 
forest and another outside of the town not far from it on the bank of Yenisei River. Both of them 
have conference halls, comfortable living rooms and dining rooms.

